%% file: ikthesis.tex
\documentstyle[12pt]{report}
\newcommand{\ed}{\end{eqnarray} \vspace{-6mm} \newline}

\newcommand{\bd}{\newline \vspace{-6mm} \begin{eqnarray}}

\newcommand{\veb}{\vspace{-17mm}}

\newcommand{\vs}{\vspace{24pt}}

\newcommand{\vsse}{\vspace{1mm}}

\newcommand{\mad}{\vspace{3mm}}

\newcommand{\lum}{\vspace{-4mm}}
  
\begin{document}
\include{ikpm}
\include{ikpt1}

\include{ikin}
\include{ikchap1a}

\include{ikchap2b}
\include{ikchap2c}
\include{ikchap3d}

\include{ikchap4e}

\include{ikcvtg}
\include{ikcur1h}

\include{ikfin1i}
\include{ikfmioncur}

\include{ikfrm1j}

\include{ikveck}

\include{ikrenl}

\include{ikBEin}

\include{ikltemfer}

\include{ikinterectn}

\include{iktopol}

\include{ikintrcs}

\include{ikveccs}

\include{ikGCS}

\include{ikapp1m}

\include{ikpic1}

\include{ikbibnew}
\end{document}

%% file: ikpm.tex
\begin{center}

{\Large \bf TEMPERATURE, TOPOLOGY} \\

{\Large \bf AND QUANTUM FIELDS}

{A Thesis Presented to The Faculty of the Division of Graduate Studies by} \\

{\Large  Igor Konstantinovich Kulikov}

{ In Partial Fulfillment of the Reqirements for the Degree Doctor of Philosophy in
Physics at the Georgia Institute of Technology, May, 1996} \\

{Copyright c 1996 by Igor K. Kulikov}

\vspace*{0.5cm}

{\Large \it  Dedicated to My Parents} \\
\end{center}

\vspace*{0.5cm}

This thesis uses Path Integrals and Green's Functions to study \\ 
Gravity, Quantum Field Theory and Statistical Mechanics, \\
particularly with respect to:  \\ 
finite temperature  quantum systems of different spin in  gravitational fields; \\ 
finite temperature interacting quantum systems in perturbative regime; \\   
self-interacting fermi models in non-trivial space-time of different dimensions; \\ 
non-linear quantum models at finite temperatures in a background curved space-time; \\ 
3-D topological field models in non-trivial space-time and at finite temperatures; \\ 
thermal quantum systems in a background curved space-time. \\
Results include:  NON-EQUIVALENCE of INERTIAL and GRAVITATIONAL Mass.  

%% file: ikpt1.tex
\vspace*{5mm}
\begin{center}
{\Large \bf CONTENTS} \\

\vspace*{5mm}

{page numbers may be approximate due to latex processing}
\end{center}

\hspace{13.5cm}Page
\begin{tabbing}
ACKNOWLEDGEMENTS     \` 8 \\

LIST OF FIGURES \`9 \\

CONVENTIONS AND ABBREVIATIONS \`  10  \\

SUMMARY \`   11  \\

PART I. LOCAL QUANTUM STATISTICS\\

AND THERMODYNAMICS  IN CURVED SPACE-TIME\\
 
Introduction \`   17 \\

CHAPTER I. QUANTUM FIELD METHODS\\

\hspace{25mm} IN STATISTICAL PHYSICS\\

 I.1 Equilibrium statistical mechanics \`   23  \\

 I.2. Statistical mechanics of  simple systems\\

 \hspace{7mm}  Formalism of second quantization \`   28  \\

CHAPTER II. PATH INTEGRALS\\

\hspace{27mm} IN STATISTICAL MECHANICS \`  34   \\

 II.1. Partition function  in path integral formalism \`   34  \\

 II.2. Partition function for bosons \`  38   \\

 II.3. Green's  function for boson  field \`   43  \\

 II.4. Notation \`   47  \\

 II.5. Partition function for fermions \`   49  \\

 II.6. Green's  function for fermi field  \`   53  \\

CHAPTER III. THERMODYNAMICS OF QUANTUM GASES\\

\hspace{28mm} AND   GREEN'S FUNCTIONS\\

 III.1. Thermal bosonic fields \`   58  \\ 

 III.2. Bosonic finite temperature\\ 

Green's function in the  Schwinger representation \`  63   \\

 III.3. Thermal fermionic fields \`   66  \\

 III.4. Fermionic finite temperature\\ 

Green's function in Schwinger representation \`   68  \\

CHAPTER IV. FINITE TEMPERATURE GAUGE FIELDS\\

 IV.1. Gauge theories: Pure Yang-Mills theory \`   73  \\

 IV.2. Ghost fields \`   77  \\

 IV.3. Effective action \`   80  \\

 IV.4. Propagator for vector field \`   82  \\

 IV.5 Partition function for gauge fields \`   84  \\

CHAPTER V. QUANTUM FIELDS IN CURVED SPACE-TIME\\

 V.1. Lorentz group and quantum fields \`   88  \\

 V.2. Fields in curved space-time \`  90   \\

 V.3. Spinors in general relativity \`   95  \\

CHAPTER VI. BOSE FIELD IN CURVED SPACE-TIME\\ 

 VI.1. Momentum-space representation of\\ 

the  bosonic Green's function\`   102  \\ 

 VI.2. The  Green's function and the  Schwinger-DeWitt method \`  111   \\

 VI.3. Connection between the two methods \`   115  \\

CHAPTER VII. FINITE TEMPERATURE\\ 

BOSONS IN CURVED SPACE-TIME \`  117   \\

CHAPTER VIII. FERMI FIELDS IN CURVED SPACE-TIME\\

 VIII.1. Momentum space representation\\ 

for the  Green's function of a fermion \`   124  \\ 

 VIII.2. The bi-spinor function in the  Schwinger-DeWitt 
representation \` 128  \\

CHAPTER IX. FINITE TEMPERATURE FERMIONS\\ 

IN CURVED SPACE-TIME\\

 IX.1.The  Helmholtz free energy of\\ 

 a fermi gas in curved space-time \`   130  \\

CHAPTER X. THERMODYNAMICS OF VECTOR BOSONS\\

 X.1 The Green's function of photons  \`   136  \\

 X.2. The thermodynamic  potential  of a  photon gas \`  141   \\

 X.3. Internal energy 

 and heat capacity of photon gas \`   143  \\

CHAPTER XI. RENORMALIZATIONS IN  LOCAL\\ 

  STATISTICAL MECHANICS\\

 XI.1. Divergencies of    finite temperature field models \`   145  \\

CHAPTER XII. LOCAL QUANTUM STATISTICS\\ 

  AND  THERMODYNAMICS  OF BOSE GAS\\

 XII.1. Density of Grand thermodynamical potential \`  151   \\

 XII.2. Statistics and  thermodynamics of bose gas \` 154    \\

 XII.3.  Bose-Einstein condensation \`   157  \\

CHAPTER XIII. LOCAL STATISTICS  AND THERMODYNAMICS\\ 

  OF FERMI GAS\\

 XIII.1. Grand thermodynamical potential and low\\ 

   temperature properties  of fermi gases \`  160   \\ 

PART II. INTERACTING FIELDS AT FINITE TEMPERATURE\\ 

 Introduction \`   167  \\

CHAPTER XIV. TWO LOOP RENORMALIZATIONS\\

 IN  $\lambda\phi^4$ MODEL \`  169   \\

CHAPTER XV. GREEN'S FUNCTION OF A  BOSON IN\\ 

    FINITE TEMPERATURE REGIME \`  184   \\

CHAPTER XVI. TEMPERATURE PROPERTIES OF A  BOSON\\

 XVI.1. Effective Hamiltonian of the  boson\\ 

 in non-relativistic approximation \`  197   \\

 XVI.2. Internal and gravitational masses of a  boson \`   198  \\

PART III. NON-LINEAR MODELS\\ 

IN TOPOLOGY NON-TRIVIAL SPACE-TIME\\

 Introduction \`   202  \\

CHAPTER XVII. NON-PERTURBATIVE EFFECTS\\ 

  IN GROSS-NEVEU MODEL\\

 XVII.1. Trivial case. Euclidean space-time \`   207  \\

 XVII.2. Non-trivial topology of space-time \`   209  \\

CHAPTER XVIII. $(\bar{\psi}\psi)^2$ NON-LINEAR  SPINOR MODEL\\

 XVIII.1. Dynamical mass and symmetry breaking  \`   212  \\

 XVIII.2. Model  with topologies $R_1 \times R_1\times S_1$\\ 

    and $R_1 \times Mobius~strip$ \`  215   \\

 XVIII.3. Torus topology $R_1\times R_1\times S_1$\\ 

    and topology  $R_1 \times Klein~bottle$ \`   219   \\

 XVIII.4. Non-linear spinor $(\bar{\psi}\psi)^2$ model\\

 in  Riemann space-time at finite temperature \`  220   \\

PART IV. TOPOLOGICALLY MASSIVE  GAUGE THEORIES\\

(non-trivial space-time and finite temperatures)\\

 Introduction \`   224  \\  
 
CHAPTER XIX. INTRODUCTION  TO\\ 

  TOPOLOGICAL FIELD MODELS \`   226  \\

CHAPTER XX. INDUCED CHERN-SIMONS  MASS TERM\\ 

  IN NON-TRIVIAL TOPOLOGICAL SPACE-TIME \\

 XX.1. Euclidean space-time. Trivial topology \`   240  \\

 XX.2. Non-trivial topology \`  245   \\

CHAPTER XXI. GRAVITATIONAL CHERN-SIMONS\\ 

  MASS TERM AT FINITE TEMPERATURE\\  

 XXI.1. Induced gravitational Chern-Simons mass term   \`  249   \\

 XXI.2. Induced gravitational Chern-Simons mass term\\ 

   at finite temperature \`   254  \\ 

APPENDIX\\

 I. Integral representations of modified Bessel functions I. \`  256  
\\

 II. Integral representations of modified Bessel functions II. \`  
260  \\

 III. Graphics \`  264   \\

BIBLIOGRAPHY	 \`  270   \\

\end{tabbing}

\newpage
\vspace*{5mm}
\begin{center}
{\Large \bf ACKNOWLEDGEMENTS}
\end{center}

I acknowledge and thank many people for their help in my work.
In particular I thank Professor David Finkelstein for  
encouraging me in my research, for his help and  great  contribution in my
professional growth. 
I thank my first graduate teachers Dr. Petr I. Pronin, Professor D.D. Ivanenko,
Professor V.N. Ponomariev and  Dr. G.A. Sardanashvilli,
who taught, supported and encouraged me during
my research work in Moscow State University.

Especially I am greatful to
Dr. Pronin whose physical intuition and deep knowledge of gravitation, field theory
 and quantum statistics helped to develop the subject of
local statistics and thermodynamics in curved space-time.
I thank him for his support during all the period of our collabaration.

I thank graduate students Frank (Tony) Smith, Jeffrey Hasty, Sarah Flynn, William 
Kallfelz, Zhong Tang and  Bereket Berhane for reading, correcting and discussing 
parts of the text. 
  
I thank  my family for patience and support, especially
my wife Yelena who encouraged me in my work  and helped to  correct the text.

I am thankful to  the  School of Physics of the Georgia Institute of Technology
for giving me the opportunity to complete this research.

\newpage
\vspace*{5mm}
\begin{center}
{\Large \bf LIST OF FIGURES}
\end{center}

Figure \hspace{12cm} page
\begin{tabbing}
I-1 Graphical expression of the function $g_{3/2}(z,R)$ \`  239 \\

I-2  Graphical solution for bosons  \` 240  \\

I-3 Chemical potential $\mu _{eff} (R)$ as a functional\\ 

of a  space-time curvature  \`  241 \\

I-4 Graphical solution for fermions \` 242  \\

II-1 One loop and counterterm contributions to the self\\

energy of the boson \`  150 \\

II-2 Feynman diagrams contributing to the vertex correction $\Gamma^{(4)}$ \`  153 \\

II-3 Counterterms and loop contributions\\ 

of the order $(\lambda ^2_R)$ to the self energy \`  156 \\

II-4 Green's function $D^{'}(p)$ of the boson\\ 

as a sum of proper self-energy insertions \`  172 \\

III-1 Topologies of cylinder and Mobius strip in $(x,y)$ space \` 243  \\

III-2 Topologies of torus  and Klein bottle  in $(x,y)$ space \`  244 \\
\end{tabbing}

\newpage   
\begin{center}
{\Large \bf CONVENTIONS AND ABBREVIATIONS}
\end{center}

The sign conventions of the metric of the Part  I. is $sign(+2)=diag(-,+,+,+)$ 
with Riemann tensor 
${R^\alpha}_{\beta \gamma \delta}=
\partial_\delta {\Gamma ^\alpha}_{\beta \gamma}-\partial_  \gamma{\Gamma
^\alpha}_{\beta\delta }+{\Gamma^\alpha }_{\xi \gamma }{\Gamma^\xi }_{\beta
\delta} -{\Gamma^\alpha }_{\xi\delta  }{\Gamma^\xi }_{\beta \gamma }$   
and Ricci  tensor  as contraction of the form 
$R_{\mu \nu} ={R^\alpha}_{\beta \gamma \alpha}$

In Part II.    metric with signature $sign(-2)=diag(+,-,-,-)$ is used   
in order to preserve standard formulation theory in real time formalism.

The units $\hbar=c=1$ and Boltzmann constant  $k=1$ are  used in the thesis. 

The following special symbols are used throughout:

$*$    complex conjugate

$+$    Hermitian  conjugate

$-$    Dirac  conjugate

$\partial_\alpha =\frac{\partial}{\partial x^\alpha}$ partial derivative

$\nabla_\alpha $ or $;$  covariant derivative 

${\Gamma ^\delta}_{\beta \gamma}$   Christoffel symbols

${\omega^\alpha}_{\beta \gamma}$ spin connection

$\mbox{tr}$ or $\mbox{Tr}$  are traces

$[a,b]=ab-ba$ commutator

$\{a,b\}=ab+ba$ anticommutator

$\sim$ order of magnitude estimate

$\simeq$ approximately equal

$\equiv$ defined to be equal to

\newpage
\begin{center}
{\Large \bf SUMMARY}
\end{center}

The   problems which are studied 
in this work  belong to  three different fields of theoretical 
physics:  gravity, quantum field theory and statistical mechanics.
Two well known methods of  modern quantum field theory, path integrals  and
Green's functions, allow one to connect these different branches of physics.
 It is possible to get a number of interesting physical
predictions by  combining these two  methods.  
This combination of path integral and Green's function methods is applied to:  
finite temperature  quantum systems of different spin in  gravitational fields;
finite temperature interacting quantum systems in perturbative regime;   self
interacting fermi models in non-trivial space-time of different dimensions; 
non-linear quantum models at finite temperatures in a background curved space-time;
 3-D topological  field models in non-trivial space-time and at finite
temperatures; and 
construction of the statistics and thermodynamics of thermal quantum systems in
a background curved space-time.\\
The thesis is divided into four parts.

The core of the thesis is Part I.
 It is concerned with the development of ideas of local quantum statistics 
and thermodynamics of ideal thermal quantum systems.
The goal is to apply the  methods of quantum field theory 
to thermal quantum systems, and so
to extend and to improve them as  convenient  mathematical tools for
describing the thermal behavior of quantum systems of different 
spins in external gravitational field. Particular attention is devoted to
the development of the  connection between Green's functions 
of the models  and their distribution  functions.
For the description of thermal quantum systems in an external curved space-time, 
Schwinger-Dewitt and momentum-space representations of Green's functions 
are introduced. The problem of introducing   temperature into field models in
curved space-time for these two methods is discussed.
The equivalence and difference between these two formalisms are studied.  
The aim of introducing   these different representations 
is to show how to introduce chemical potential in the 
momentum-space formulation of partition functions  of quantum systems
and to study their  thermal behavior in external gravitational 
fields. 
The concept of local thermodynamics is introduced, and densities of 
Helmholtz free energies  and  grand 
thermodynamical potentials for bose and fermi gases are computed.
It is shown how  average occupation numbers of  bosonic and fermionic
quantum system depend on   external gravitational fields.
Low temperature behavior of bose and fermi gases is explored, and 
the phenomenon of Bose condensation in thermal bose systems in a background
curved space-time is studied.     
Thermodynamical properties of a photon gas in a background 
gravitational field are considered.

In part II finite temperature  interacting quantum 
fields are explored from the  point of view of thermofield dynamics.
Two loop renormalizations of a  self-interacting $\lambda \varphi ^4$
scalar field  at zero and finite temperatures are discussed and analyzed.
On this basis, the  two loop renormalized finite temperature Green's function of
a boson and non-relativistic Hamiltonian of a boson in a heat bath 
in an external gravitational field are computed. 
The interesting problem of non-equality between inertial 
and gravitational masses of a boson is studied in this part.

In part III non-linear spinor models in 3-D and 4-D space-time are considered.
The Gross-Neveu model  and the 3-D Heisenberg model in   non-trivial space-times
with different topologies are studied. The problem of the dynamical mass 
of   composite fermions is discussed.
The 4-D Heisenberg model in a background curved space-time at finite temperature 
is explored. The influence of  curvature 
on the  dynamical mass generation of fermions is considered.

Part IV is devoted to the analysis of topological field models in 
 non-Euclidean 3-D space-time and finite temperature effects.
The properties of vector and tensor field models with Chern-Simons mass term 
are considered. 
The effective Chern-Simons terms of vector and tensor fields
appearing as the result of interaction of 3-D fermions  with external 
gauge vector  and tensor fields are discussed.
The ideas and methods developed  in part I are applied  to the analysis of  
topological models with non-trivial space-time structure.     
The  influences of the topology of 3-D space-time    
and of thermal behavior of interacting fermions on generation of
the effective Chern-Simons terms are studied and analyzed. 

\centerline{\Large New results contained in  thesis }

Finite temperature field theory in flat space-time has been studied 
in a  number of   journal publications (references in the text of part I)  
and  books 
[ Bonch-Bruevich et al 1962, Abrikosov et al. 1963, Popov 1987,  Kapusta 1989 et
al.]. Quantum field theory in curved space-time has been developed by many authors
(journal references in the text of part I)
and is  well described in the books [De-Witt 1965,
 Birrell \& Devies 1982,  Fulling 1989
et al.] Finite temperature field theory in curved space-time 
in the Schwinger-DeWitt  formalism [De-Witt 1965,1975] is considered
in many  journal publications  (c.f. references in the text of part I). 
Most important  are
the works of Dowker, Kennedy, Denardo et al.   
 [Dowker \& Kritchley 1977, Dowker \& Kennedy 1978, Denardo \& Spalucci 1983].
The momentum-space method in applications to a field theory in curved
space-time was discussed  in the works of Bunch, Parker, Panangaden  
[Bunch \& Parker 1979,  Panangaden  1981]. This method was used in  the study 
of the $\lambda \varphi^4$-model and the quantum  
electrodynamics in perturbative regimes in curved
space-time. 
In   Part I, on the basis of the  above mentioned  works, the author develops the 
theory of  local thermodynamics  and a statistics in curved space-time.
The author shows that the thermodynamical potentials 
(free energy, grand thermodynamical potential) of thermal quantum systems
may be computed  directly through the thermal Green's functions of corresponding 
quantum fields  in  chapter III (III.24) and (III.52); chapter VII (VII.22) and 
chapter IX (IX.15).
In chapters VII and IX the author shows that the densities of
free energy of thermal bose and  and fermi systems in curved space-time 
can be derived  with the Green's function in the  Schwinger-DeWitt formalism. 
The author shows the equivalence between the  Schwinger-DeWitt and momentum 
space methods
in applications to thermal quantum systems in  chapter VII (VII.25) and chapter IX
(IX.18). The author analyzes the difficulties of the computation of the 
grand thermodynamical potentials for bose and fermi systems
using the  Schwinger-DeWitt formalism (chapter IX) and computes 
the  grand thermodynamical potentials of ideal bose and fermi 
systems in gravitational fields  with momentum-space method in
chapter  XII (XII.1 and XII.2) and chapter XIII. 
The author  considers low temperature properties of these systems 
in chapters XII and XIII, and  studies the behavior of the chemical potentials 
of the bose and fermi systems  with respect to  space-time curvature
 (XII.19) and (XIII.20).   
The author analyzes the  phenomenon of Bose condensation  (section XII.3) and finds
the  variation of the  critical temperature of condensation with respect to
the  variation of the curvature (XII.25). 
The author studies the  thermodynamical properties of a photon gas in curved
space-time with momentum space-methods in chapter X. The author computes
Bose-Einstein  and Fermi-Dirac distributions  in curved space-time in
chapters XII (XII.14) and XIII (XIII.10)

The temperature properties of self-interacting  and gauge fields are studied in
imaginary and in real time formalisms in the  
works of Kislinger , Kapusta,  Donoghue et al., from the  point of view of their
renormalizability [Kislinger et al., 1976] and  finite temperature behavior of the
constant of the interaction   [Kapusta 1979, Fujimoto et al., 1987]. In the work of
Donoghue  the gravitational field  was  also introduced in the electrodynamics with
the Tolman's shift of temperature, and the  non-equality between inertial and
gravitational masses of fermion was also discussed [Donoghue et al.,1984].
In part II of this work
  the author computes  the  finite temperature effective mass and  renormalized  
Green's function of boson in a  heat bath  in two loop approximation
with a real time formalism in   chapter XV (XV.21), finds the  
non-relativistic finite temperature Hamiltonian of bosons in a  weak
gravitational field (section XVI.1),  and gets the ratio between the  inertial and 
the gravitational masses of a boson (section XVI.2) (XVI.9). 

Twisted fields were introduced   by Isham in 1978 and field
models in space-time with a non-Minkowskian topology were studied in the works
of Ford, Toms et al.,  [Ford 1980, Toms 1980]. Non-linear 3-D field models
with $\gamma_5$   symmetry   and  a large  number of flavors
and $(\bar{\psi}\psi)^2$ non-linear models in 3-D and 4-D space-time 
were studied in works  Gross \& Neveu 1974,
 Bender 1977,  Tamvakis,  Kawati \& Miyata  1981 et al.   
In part III the author applies the idea of twisted and untwisted spinor fields
to obtain the  dependence of the dynamical fermionic mass with respect to  
a  non-trivial topology of space-time in chapter  XVIII (section XVIII.2).
The author analizes the behavior of the dynamic fermionic mass with  respect to 
the  curvature of space-time and the  thermal behavior of interacting fields (section
XVIII.4).

The idea of computing the  induced Chern-Simons action for vector and tensor fields 
as the result of the  interaction of 3-D fermi fields with external vector and tensor
fields was studied  in the  works of Redlich,  Das, Ojima  et al.
Das  proposed the method of the  computation of finite temperature induced
Chern-Simons action of vector type  with the method of derivative
expansion [Babu \& Das 1987]. Ojima computed the Chern-Simons action of vector 
and tensor fields using the  path integral method [Ojima
1989].    In Part IV the author, using the  above mentioned works computes the 
induced Chern-Simons mass term of  a vector field in   space-times with  
topologies $R^2\times S^1$ and  $R^2\times mobius~strip$ in section (XX.2).     
In chapter XXI  the author computes the finite temperature  gravitational induced
Chern-Simons mass  term using  the  momentum space method developed in part I.

In the list of references the author includes only the publications most important
for undestanding the text.

%% file: ikin.tex
 \begin{center} 
\vspace*{8mm}{\LARGE \bf PART I} 
\end{center}
\vspace{2mm}
\begin{center} 
{\LARGE\bf{LOCAL QUANTUM STATISTICS}} 
\end{center}
\begin{center} 
{\LARGE\bf{ AND THERMODYNAMICS }} 
\end{center}
\begin{center} 
{\LARGE\bf{IN CURVED SPACE-TIME}} 
\end{center}
\vspace{1mm}
\begin{center} 
{\Large\bf  Introduction } 
\end{center}

\vspace{4mm}
The statistics and thermodynamics of 
quantum systems in gravitational fields have  the  attracted attention of physicists
for a long time. 
The  equilibrium distribution
 of a relativistic quantum 
gas in a gravitational field has been  studied using the kinetic Boltzmann equation,
 and it
was found that the solution of a functional Boltzmann equation is a relativistic Maxwell 
distribution for a certain type of gravitational field \cite{i1}.
Later,   different  variants of 
non-equilibrium statistical mechanics in classical general relativity
as a generalization of the standard statistical equations for 
classical systems  in curved space-time were proposed \cite{i2,i4}.

At the same time field-theoretical methods were developed and 
applied to statistical mechanics  and thermodynamics. 
Since the middle of the 50's,  significant progress was made in   quantum 
 theory  of many body systems, and 
  was connected with the  development of finite 
temperature quantum field methods in statistical physics.
 Feynman \cite{i5} studied the  $\lambda$-transition
in Helium using the partition function in the form of a path  integral
 in  quantum mechanics.
Martin and Schwinger \cite{i6} considered a many-particle system in the context of 
quantum field theory in order
to treat multiparticle system from the quantum field-theoretical point of view.
They described the microscopic behavior 
of a multiparticle system using 
 Green's functions, and  found that the finite temperature Green's  functions are 
 related to intensive macroscopic variables when the energy and number of 
particles are large.
 Matsubara  \cite{i7}  proposed  finite temperature perturbation 
techniques similar to perturbative quantum field theory
and calculated the grand partition function by introducing Green's 
functions in imaginary time formalism.
The works of Feynman, Matsubara, and  Martin \& Schwinger 
created the basis for  understanding the close connection between 
Euclidean field theory and statistical mechanics.

 In the  60's    convenient methods  using 
finite temperature calculations for interacting thermal systems 
 with a propagator formalism in the perturbative regime were proposed in   
\cite{i8}, \cite{key13}, \cite{i10},\\ 
and \cite{i11}. 
These works stressed the analogy between Euclidean Green functions and
distribution functions in statistical mechanics.
Analogies between statistical physics and  quantum field theory 
at finite temperatures were further strengthened in the study of infinite 
equilibrium systems of scalar and spinor fields \\
 \cite{i15}.  These works  provided an opportunity to construct the  
statistical mechanics and thermodynamics of quantum systems in the language of  
finite temperature field theory, and pointed the way towards a construction of  a 
statistical mechanics  and thermodynamics in curved space-time.   

In the  70's, interest in quantum statistical processes in general relativity 
was aroused 
by  the works of Hawking \cite{i21}, and Bekenstein \cite{i22} on the thermodynamics 
of black holes, where  an intimate connection between 
thermodynamics and  the structure of space-time was  pointed out. 

 Parallel to the research on the thermodynamics of black holes,   quantum effects 
present in the   early Universe  were thoroughly discussed by 
Guth et al.  \cite{i12,i13,i14}. 
From these works  one  may conclude that, for
  early time and high temperatures, 
the quantum statistical properties of matter and strong curvature of space-time 
become significant. 

With the development of a quantum field theory using functional integral methods
 at finite temperature  \cite{wein1,key1}, \cite{key9},\\
 \cite{key10} and of  
a formalism of  a quantum theory of field in curved space-time\\
\cite{dewitt1,key17},
 new possibilities for studing the behaviour 
of thermal systems with non-trivial 
geometry and topology were discussed. This approach was developed by
Denardo et al.  \cite{i16}, \cite{i18}.
 
 Hu  \cite{i19} introduced time dependent temperature  $T(t)=a(t)/T$
for an equilibrium gas of  scalar particles which is described by a conformal 
scalar field in a closed Robertson-Walker (RW) Universe in a different way, by   
 using the equilibrium temperature of a flat space-time $T$.

The problems  of  a  thermodynamical description of a  non-equilibrium 
thermal quantum gas  and conditions for thermal equilibrium 
in an expanding Universe were studied by Hu \cite{i20}.
However,  in general this problem has not been solved satisfactorily.  
Difficulties in the construction of a thermodynamics of thermal systems in dynamical
 space-time are connected with   such problems as  
the definitions of temperature, energy spectrum, and the vacuum state of a thermal
system.   

In Part I of this  work we extend the results of finite temperature 
field theory in order to construct a   statistical mechanics 
and thermodynamics of
bosonic and fermionic quantum systems  in an external curved space-time.

The methods developed here are based on using  the language of 
a finite temperature Green's 
function in  an arbitrary curved space time for the
definition of the thermodynamical potentials of thermal quantum systems. 

Green's functions of matter fields in curved space-time are nonlocal objects that are 
 well defined in a small range 
of the space-time manifold \cite{dewitt1}\\
\cite{i18}. 
Thermal properties
of quantum systems may be considered  the properties of the set of 
quasi-equilibrium sub-systems, which are  elements  of the whole quantum system 
\cite{i23}.
This suggestion will lead us  to the definition of  distribution functions of bose and
fermi systems, and  some interesting thermodynamical consequences.

The division  of the entire system into quasi-equilibrium sub-systems may be done 
in a simple way 
using  Riemann normal coordinates \cite{petrov1}. These coordinates permit us
 to rewrite Green's functions in a  momentum space representation, and
to introduce a local temperature in order to write thermodynamical potentials through
these Green's functions. As a result, the  thermodynamical potentials 
will be written as a series expansion in 
 powers of the curvature tensor at a selected point of the curved space-time 
manifold. Therefore,  at any point in space-time the coefficients of the series will 
 change because the curvature will change for all other points on the manifold.   

Therefore, in accordance with our suggestions, densities of thermodynamic 
potentials will be functionals of curvature, and they will be directly connected to the 
temperature Green's functions by a simple relation. The aim of this part will be 
to develop the  mathematical formalism of temperature Green's functions in 
external curved space-time and to use the formalism to construct a  quantum statistics 
and thermodynamics of thermal bosonic and fermionic  ideal quantum systems and a
thermodynamics of a photon gas in an arbitrary curved space-time. 

Part I is organized in the following way:
A short review of statistical mechanics is presented in chapter I.   
In chapter II functional integral methods are  applied to non-relativistic
and relativistic many-body systems in Euclidean space-time. 
The purpose will be 
to generalize  these methods for a  description of  statistical 
systems in curved space-time.
Finite temperature  Green's functions are introduced in chapter III
for computation of  the thermodynamical potentials of quantum gases. 
Finite temperature gauge fields are  studied in chapter IV.
An introduction to  quantum fields in  curved space-time is presented in chapter V.
The bosons and fermions in external gravitational fields are studied 
in chapters VI and VIII.
The thermodynamics of bose  and fermi gases in curved space-time 
is studied in chapters VII and IX.
The thermodynamics of vector bosons is considered in chapter X.
Renormalization problems are considered in chapter XI.
The phenomenon of Bose-Einstein condensation and the  low temperature properties 
of fermi gas are studied in chapters XII and XIII.

%% file: ikchap1a.tex
\chapter{QUANTUM FIELD METHODS}
\centerline{\Large \bf IN STATISTICAL PHYSICS}
\vs

The customary approach to many-body theory is 
  the method of second quantization \cite{key13,fw1}.
 This chapter is  a short introduction to 
statistical mechanics of simple quantum systems  using this method.  

\section{Equilibrium statistical mechanics}
\vsse

Statistical mechanics deals with three types of ensembles:

1) $The$ $microcanonical$ $ensemble$ is used to describe an isolated  system   
which has a fixed energy $E$, particle number $N$, and volume $V$.

2)  $The$ $canonical$ $ ensemble$ is used to describe a system in contact 
with a heat reservoir at temperature $T$.  The system can freely exchange 
energy with the reservoir and $T$, $N$, and  $V$ are fixed variables

 3)  $The$  $grand$ $canonical$ $ensemble$ is used to describe a system which can
 exchange particles  as well as energy with reservoir.  In this ensemble 
 $T$,  $V$ and the chemical potential $\mu$ are fixed variables.

The main aim of statistical mechanics is to derive the thermodynamic
 properties of macroscopic bodies starting from the  description of the 
motion of the microscopic components (atoms, electrons, etc.).

To solve the problem it is necessary to find the probability  
distribution of the microscopic components in thermal equilibrium
(after a sufficiently long time), and deduce the macroscopic properties 
of the system  from this  microscopic probability distribution.

Following this scheme let us consider a classical Hamilton system with 
$2N$ degrees of freedom in a box of  volume $V$.
The classical equations of motion  are:
\bd    
\dot{q}_i=\frac{\partial H}{\partial p_i},
\hspace{1cm}\dot{p}_i=-\frac{\partial H}{\partial q_i} \label{a1}
\ed  
where $H$ is the Hamiltonian, $q_i$ and $p_i$  $(i=1,2,...,N)$ 
are the set of coordinates  and  momenta of the system.

Let $A(q,p)$ be an arbitrary measurable quantity (an observable).
  The equilibrium average of this quantity  $\bar{A}$ is defined
by 
\bd
\bar{A}=\mathop{\lim}\limits_{t \to \infty}
\frac{1}{t}\int\limits_o^\infty
A(q(\tau),p(\tau))d\tau \nonumber
\ed
\veb
\bd    
=\int dqdp A(q,p)\varrho(q,p) \label{a2}
\ed  
where $\varrho(q,p)$ is the equilibrium probability.

This function is never negative and satisfies the normalization condition
\bd    
\int dqdp~\varrho(q,p) =1.\label{a3}
\ed  
The fundamental hypothesis of equilibrium statistical mechanics is that
$\varrho$ follows the canonical distribution
\bd    
\varrho(q,p) =Z^{-1} \exp[-\beta H(q,p)], \label{a4}
\ed  
where
\bd    
Z=\int dqdp \exp[-\beta H(q,p)] \label{a5}
\ed  
is the  partition function, and $\beta^{-1}=T$ is the absolute equilibrium
temperature.

It is  useful to find a characterization of the canonical 
distribution that 
distinguishes it from all other possible probability distributions.  It is 
convenient to introduce the entropy of a distribution $\varrho$
defined as follows:     
\bd    
S[\varrho]=-\int dqdp\varrho(q,p) \mbox{ln}\varrho(q,p)
=-<\mbox{ln}\varrho> \label{a6}
\ed  
Entropy has the following properties: 
the more ordered is the  system the smaller is the entropy (i.e., 
the more concentrated  the probability distribution in a restricted 
region of phase space);
the more disordered is  the  system (i.e., the more uniform the probability 
distribution), the larger is the entropy.

For relativistic quantum statistical systems the equation (\ref{a4}) 
will be turned into the equation for statistical operator
\bd    
\hat{\varrho}=Z^{-1} \exp[-\beta(\hat {H}-\mu\hat{N}] \label{a7}
\ed  
where $\hat {H}$ is the Hamiltonian and $\hat{N}$ is the number operator.
This operator is Hermitian and commutes with Hamiltonian $H$. 
The parameter $\mu$ 
is the  so called chemical potential.

The ensemble average of an operator $\hat{A}$  will be
\bd    
<A>=Z^{-1}\mbox{Tr}[\hat{A}\hat{\varrho}] \label{a8}
\ed  
The factor $Z$ will turn into a so called  grand canonical 
partition function of the form
\bd    
Z=\mbox{Tr} \exp[-\beta(\hat {H}-\mu\hat{N}] \label{a9}
\ed  
This function is the most important in thermodynamics.

The average value of the energy $U$ may be written from (\ref{a8}) as 
\bd    
U=<\hat{H}>=\mbox{Tr}[\hat{H}\hat{\varrho}] \label{a10}
\ed  
and the entropy as 
\bd    
S=-\mbox{Tr}[\hat{\varrho}\mbox{ln}\hat{\varrho}] \label{a11}
\ed  
From the equations (\ref{a9}) and (\ref{a11})  we get a useful relation
\bd
S=-\mbox{Tr}[\hat{\varrho}~\mbox{ln}\hat{\varrho}]  \nonumber
\ed
\veb
\bd
=-\mbox{Tr}[\hat{\varrho}(-\ln Z-\beta\hat{H}+\mu\hat{N})]  \nonumber
\ed
\veb
\bd
=\mbox{ln}Z+\beta U-\beta \mu N,  \nonumber
\ed
or
\bd    
(1/\beta)\mbox{ln}Z=U-S/\beta-\mu N. \label{a12}
\ed  
The quantity 
\bd    
\Omega=-(1/\beta)\mbox{ln} Z \label{a13}
\ed  
is the grand thermodynamical potential of the grand canonical ensemble.

For canonical ensemble ($\mu=0$) we can write the equivalent equation for 
thermodynamic potential $F$ (free energy). 

It is easy to find that  
\bd    
F-\Omega=\mu N. \label{a14}
\ed  
The grand partition function $Z=Z(V,T,\mu )$ is the most important 
function in thermodynamics.  All other standard termodynamic properties may be 
determined from this function.  For example, the pressure, particle number, 
entropy, and energy (in the infinite volume limit) are 
\bd
P=T\frac{\partial \mbox{ln} Z}{\partial V}, \hspace {2cm}
 N=T\frac{\partial \mbox{ln }Z}{\partial \mu}  \nonumber
\ed  
\veb
\bd    
S=T\frac{\partial ~T\mbox{ln}Z}{\partial T},\hspace {2cm}
E=-PV+TS+\mu N \label{a15}
\ed  

\section{Statistical mechanics of  simple systems}
\lum
\hspace{15mm}{\Large \bf  Formalism of second quantization} 
\vsse

Now we can apply  the formalism of statistical mechanics developed above  to  
non-interacting many body quantum systems in the  frame of  the method of second 
quantization.

\mad
\centerline{ \large \bf 1. Bosonic quantum system at finite temperature}
\mad

Let us study a bosonic quantum system. Each quantum state  $\epsilon$ of 
the system may be occupied by bosons.  Let $n$ $(n=0,1,2,3,...)$ be number of
 bosons in this state, and $|n>$ be  the wave function of this state.
 We will call a state $|0>$   a vacuum state.    
One  may introduce creation $a$ and annihilation $a^+$ operators with commutation 
relation
\bd
[a,a^+]=aa^+-a^+a=1 \label{a16}
\ed   
The action of these operators on  eigenstates is
\bd
a|n>=\sqrt n|n-1>, \hspace{2cm} a^+|n>=\sqrt {n+1}|n+1> \label{a17}
\ed   
and on vacuum state is
\bd
a|0>=0 \label{a18}
\ed   
From (\ref{a16}), (\ref{a17}) one can get the number operator $\hat{N}=a^+a$:
\bd
\hat{N}=a^+a|n>=n|n> \label{a19}
\ed   
We may build all states $|n>$ from vacuum $|0>$ by repeated applications
of the creation operator
\bd
|n>=(n!)^{-1/2}(a^+)^n|0> \label{a20}
\ed   
The Hamiltonian of the system in the state with energy $\epsilon$ may 
be constructed as a product $\epsilon$ with 
number operator (up to an additive constant) in the form
\bd
\hat{H}=\epsilon(\hat{N}+1/2)=\epsilon(a^+a+1/2) \label{a21}
=\frac{1}{2}\epsilon(a^+a+aa^+)
\ed   
The  additive term $\epsilon/2$ in (\ref{a21}) is the zero-point energy.
 Usually this term may be ignored and we get $\hat{H}=\epsilon \hat{N}$.

The  grand partition function  according to (\ref{a9})  will be
\bd
Z=\mbox{Tr} \exp[-\beta(\hat{H}-\mu\hat{N})]
=\mbox{Tr} \exp[-\beta(\epsilon-\mu)\hat{N}] \nonumber
\ed
\veb
\bd
\sum\limits_{n=0}^\infty <n|\exp[-\beta(\epsilon-\mu)\hat{N}]|n>
=\sum\limits_{n=0}^\infty \exp[-\beta(\epsilon-\mu)n] \nonumber
\ed
\veb
\bd
=(1-\exp[-\beta(\epsilon-\mu)])^{-1} \label{a22}
\ed   
Inserting (\ref{a22}) into (\ref{a15}) we get the average number of bosons 
in the system with
energy $\epsilon$
\bd
N=(\exp[\beta(\epsilon-\mu)]-1)^{-1} \label{a23}
\ed   
and the average energy in the form  $\bar{\epsilon}=\epsilon N$.

\mad
\centerline{ \large \bf  2. Fermionic quantum system at finite temperature}
\mad

For a fermionic quantum system  there are only 
two states $|0>$ and $|1>$ with energy $\epsilon$.

The action of the fermion creation and annihilation operators
on these states is
\bd
\alpha^+|0>=|1>,~~~\alpha|1>=|0>  \nonumber
\ed
and
\bd
\alpha^+|1>=|0>,~~~\alpha|0>=|1> \label{a24}
\ed   
Thus, these operators have the property that their square is zero
\bd
\alpha^+\alpha^+=\alpha \alpha=0 \label{a25}
\ed   
The number operator is $\hat{N}=\alpha^+\alpha$.  One has
\bd
\hat{N}|0>=\alpha^+\alpha|0>=0;~~~ \hat{N}|1>=\alpha^+\alpha|1>=|1> \label{a26}
\ed   
Creation and annihilation operators satisfy the anticommutation relation
\bd
\{\alpha,\alpha^+\}=\alpha\alpha^++\alpha \alpha^+=1 \label{a27}
\ed   
The Hamiltonian for this system may be taken in the form
\bd
\hat{H}=\frac{1}{2}(\alpha^+\alpha-\alpha \alpha^+)
=\epsilon(\hat{N}-1/2) \label{a28 }
\ed   
The partition function  for the  fermionic system will be  
\bd
Z=\mbox{Tr} \exp[-\beta(\hat{H}-\mu \hat{N})]
=\sum \limits_{n=0}^{1} <n|\exp [-\beta(\epsilon-\mu)\hat{N}]|n>  \nonumber
\ed
\veb
\bd
=1+\exp[-\beta(\epsilon-\mu)] \label{a29}
\ed   
The average  number of fermions is 
\bd
N=(\exp[\beta(\epsilon-\mu)]+1)^{-1} \label{a30}
\ed   
The average energy of the system in the state with energy $\epsilon$ is
$\bar{\epsilon}=\epsilon N$.

The next section describes an alternative approach, 
the method of functional integrals for studying  the behavior 
of statistical systems.

%% file: ikchap2b.tex
\chapter{ PATH INTEGRALS} 
\centerline{\Large \bf IN STATISTICAL PHYSICS }
\vs

Functional integration, introduced several decades ago \cite{fh1},
 is one of the most powerful methods of modern
theoretical physics. The functional integration approach to   systems with
an infinite number of degrees of freedom turns out to be very suitable for 
the introduction  and formulation of   perturbation theory
in quantum field theory and statistical physics.  This approach 
is simpler than  the  operator method.

Using functional integrals in statistical physics allows one 
to derive numerous interesting results more quickly than  
other methods. Theories of phase transitions of the second kind, 
superfluidity, superconductivity, plasma, and  the Ising model 
are examples of the problems for which   the functional 
integration method appears to be very useful.  If an exact solution exists,
 the functional integration method gives a simple 
way to obtain it.  In other cases, when exact knowledge is unobtainable,
 the application 
of functional integrals helps to build  a qualitative picture of the 
phenomenon, and to develop an  approximate solution scheme.   
  
The functional integration method is suitable for obtaining the diagram
techniques of the perturbation theory, and also for modifying the
perturbative scheme if such a modification is necessary.  The extension of 
functional integral techniques to   background curved space-time  
 allows one
 to take into account the gravitational field by considering the
 statistical and thermodynamical properties of the systems.

The  aim of this chapter is to introduce the finite temperature functional integral 
approach to statistical mechanics and local thermodynamics
in curved space-time.

\section{Partition function in path integral formalism}
\vsse

This paragraph discusses the  scalar field
which is described by the Schrodinger field operator $\hat{\varphi}(\vec{x})$ 
where  $\vec{x}$ is the spatial  coordinate.
We denote eigenstates of $\hat{\varphi}(\vec{x})$ by $|\varphi>$
\bd
\hat{\varphi}(\vec{x})|\varphi> = \varphi(\vec{x})|\varphi> \label{b1}
\ed   
where $\varphi(\vec{x})$ is a $c$-number function.  Let $\hat{\pi}(\vec{x})$
be its conjugate momentum operator.

The usual completeness and orthogonality conditions may be written as:
\bd
\int {d\varphi \left| \varphi \right\rangle }\left\langle \varphi \right|=1 \nonumber
\ed
\veb
\bd
\left\langle {{\varphi _a}} \mathrel
{\left | {\vphantom {{\varphi _a} {\varphi _b}}} \right.
 \kern-\nulldelimiterspace} {{\varphi _b}} \right\rangle
 =\delta \left[ {\varphi _a(\vec{x})-\varphi _b(\vec{x})} \right] \label{b2}
\ed   
where $\delta$-is the Dirac delta function.

Similarly, the eigenstates of the conjugate momentum field operator are labeled by 
$|\pi>$ and satisfy the equation
\bd
\hat{\pi}(\vec{x})|\pi> =\pi(\vec{x})|\pi> \label{b3}
\ed   
with eigenvalies $\pi(\vec{x})$. The completeness and orthogonality 
conditions are
\bd
\int {d\pi \left| \pi  \right\rangle }\left\langle \pi  \right|=1  \nonumber
\ed
\veb
\bd
\left\langle {{\pi _a}} \mathrel{\left | {\vphantom 
{{\pi _a} {\pi _b}}} \right. \kern-\nulldelimiterspace}
 {{\pi _b}} \right\rangle 
=\delta \left[ {\pi _a(\vec{x})-\pi _b(\vec{x})} \right] \label{b4} 
\ed
the overlap between an eigenstate of field operator and an  eigenstate of the momentum
operator is
\bd
\left\langle {\varphi } \mathrel{\left | {\vphantom {\varphi  \pi }} \right. 
\kern-\nulldelimiterspace} {\pi } \right\rangle 
=\exp \left[ {i\int {d^3x\pi (\vec{x})\varphi (\vec{x})}} \right] \label{b5}
\ed
For dynamics, one requires a Hamiltonian expressed as a functional of the field
and its conjugate momentum:
\bd
H=\int {d^3xH(\hat{\pi} ,\hat{\varphi} )} \label{b6}
\ed
Suppose that a system is in state $|\varphi_a>$ at a time $t=0$. After time $t_f$
it will evolve into $exp(-iHt_f)|\varphi_a>$. The transition amplitude for going 
from state $|\varphi_a>$ to some other state $|\varphi_b>$ after time $t_f$ is thus 
\bd
<\varphi_b| \exp (-iHt_f)|\varphi_a> \label{b7}
\ed
For statistical mechanics purposes, consider the case in which  
the system returns to its original state after time $t_f$.
Divide the time interval $[0,t_f]$ into $N$ equal steps of duration 
$\Delta t=t_f/N$. Then
\bd
\left\langle {\varphi _a} \right|\exp (-iHt_f)\left| {\varphi _a} \right\rangle
 =\mathop {\lim }\limits_{N\to \infty }
\int {\prod\limits_{i=1}^N {\left( {d\varphi _i{{d\pi _i} 
\over {2\pi }}} \right)}}  \nonumber
\ed
\veb
\bd
\left\langle {{\varphi _a}} \mathrel{\left | {\vphantom {{\varphi _a} 
{\pi _N}}} \right. \kern-\nulldelimiterspace}
 {{\pi _N}} \right\rangle \left\langle {\pi _N} \right|\exp (-iH \Delta t_f)
\left| {\varphi _N} \right\rangle \left\langle {{\varphi _N}}
 \mathrel{\left | {\vphantom {{\varphi _N} {\pi _{N-1}}}} \right. 
\kern-\nulldelimiterspace} {{\pi _{N-1}}} 
\right\rangle \left\langle {\pi _{N-1}} \right|\exp (-iH\Delta t_f)
\left| {\varphi _{N-1}} \right\rangle \times ... \nonumber
\ed
\veb
\bd
\times \left\langle {{\varphi _2}} \mathrel{\left | {\vphantom {{\varphi _2}
 {\pi _1}}} \right. \kern-\nulldelimiterspace}
 {{\pi _1}} \right\rangle \left\langle {\pi _1} 
\right|\exp (-iH\Delta t_f)\left| {\varphi _1} \right\rangle
 \left\langle {{\varphi _1}} \mathrel{\left | {\vphantom {{\varphi _1}
 {\varphi _a}}} \right. \kern-\nulldelimiterspace}
 {{\varphi _a}} \right\rangle \label{b8} 
\ed
It is known, that 
\bd
<\varphi_1|\varphi_a>=\delta(\varphi_1-\varphi_a) \label{b9}
\ed
and 
\bd
<\varphi_{i+1}|\pi_i>=exp\left[i\int{d^3x}
\pi_i(\vec{x})\varphi_{i+1}(\vec{x})\right] \label{b10}
\ed
For $\Delta t\rightarrow 0$ one can expand 
\bd
\left\langle {\pi _i} \right|\exp (-iH\Delta t)
\left| {\varphi _i} \right\rangle \approx \left\langle {\pi _i}
 \right|(1-iH\Delta t)\left| {\varphi _i} \right\rangle = \nonumber
\ed
\veb
\bd
\left\langle {{\pi _i}} \mathrel{\left | {\vphantom {{\pi _i} 
{\varphi _i}}} \right. \kern-\nulldelimiterspace}
 {{\varphi _i}} \right\rangle (1-iH_i\Delta t)
=(1-iH_i\Delta t)\exp \left[ {-i\int {d^3x\pi _i(x)\varphi _i(x)}} 
\right] \label{b11}
\ed
where $H_i=H(\pi_i, \varphi_i)$.
Inserting (\ref{b9}), (\ref{b10}) and (\ref{b11}) into (\ref{b8})  gives
\bd
\left\langle {\varphi _a} \right|\exp (-iHt_f)\left| {\varphi _a}
 \right\rangle =\mathop {\lim}\limits_{N\to \infty }
\int {\prod\limits_{i=1}^N {\left( {d\varphi _i{{d\pi _i} 
\over {2\pi }}} \right)}}\delta (\varphi _1-\varphi _a)\times  \nonumber 
\ed
\veb
\bd
\exp \left\{ {-i\Delta t\sum\limits_{j=1}^N 
{\int {d^3x\left[ {H(\pi _j,\varphi _j)-\pi _j
(\varphi _{j+1}-\varphi _j)\/ \Delta t} \right]}}} \right\}\label{b12}
\ed
where $\varphi_{N+1}=\varphi_1=\varphi_a$.

Taking the continuum limit of (\ref{b12}) , gives the result
\bd
\left\langle {\varphi _a} \right|\exp (-iHt_f)\left| {\varphi _a} \right\rangle
 =\int {D\pi \int\limits_{\varphi (\vec{x},0)
=\varphi _a(\vec{x})}^{\varphi (\vec{x},t_f)
=\varphi _a(\vec{x})} {D\varphi }}\times  \nonumber 
\ed
\veb
\bd
\exp \left\{ {i\int\limits_0^{t_f} {dt}
\int {d^3x\left[ {\pi (\vec{x},t){{\partial \varphi (\vec{x},t)} 
\over {\partial t}}-H(\pi (\vec{x},t),\varphi (\vec{x},t))} \right]}}
 \right\} \label{b13}
\ed
The integration over $\pi(\vec{x},t)$ is unrestricted, and integration over 
$\varphi(\vec{x},t)$ is such that the field starts at $\varphi_a(\vec{x})$
at $t=0$ and ends at $\varphi_a(\vec{x})$ at $t=t_f$.
Note that all operators are gone and r one can work only with 
classical variables \cite{key9}. 

\section {Partition function for bosons.}
\vsse

The partition function in statistical mechanics is expressed as
\bd
Z_\beta =\mbox{Tr}\exp \left\{ {-\beta \left( {H-\mu N } \right)} \right\}=  \nonumber
\ed
\veb
\bd
\int {d\varphi _a}\left\langle {\varphi _a} 
\right|\exp \left\{ {-\beta \left( {H-\mu N } \right)} \right\} \label{b14}
\left| {\varphi _a} \right\rangle 
\ed
where the sum runs over all states.

The  aim now is to express $Z_\beta$ in terms of a functional integral.

First   introduce the imaginary time $\tau=it$. The limits of 
integration on $\tau$ are $[0,\beta]$, then $-it_f=\beta$ and
\bd
Z_\beta =\int {D\pi \int\limits_{periodic} {D\varphi }}\times  \nonumber
\ed
\veb
\bd
\exp \left\{ {\int\limits_0^\beta  {d\tau }\int 
{d^3x\left[ {i\pi (x,t){{\partial \varphi (x,t)} 
\over {\partial \tau }}-H(\pi ,\varphi )
+\mu N(\pi ,\varphi )} \right]}} \right\} \label{b15}
\ed
Integration over the field is constrained so that
\bd
\varphi (\vec{x},0)=\varphi (\vec{x} ,\beta) \label{b16}
\ed
This is a consequence of the trace operation.

In the equation (\ref{b15}) make  the replacement  
\bd
H(\pi ,\varphi )\to H(\pi ,\varphi )-\mu N(\pi ,\varphi ) \label{b17}
\ed
where $ N(\pi ,\varphi )$ is conserved charge density, if the system admits 
some conserved charge.

Rewrite the equation (\ref{b15}) in a more convenient form.

For this purpose,  consider  a scalar field, which is described by 
the Lagrangian of the form
\bd
L=-{1 \over 2}(\partial \varphi )^2-{1 \over 2}m^2\varphi ^2-U(\varphi ) \label{b18}
\ed
where $U(\varphi)$ is a potential function.

The momentum conjugate to the field is
\bd
\pi ={{\partial L} \over {\partial (\partial _0\varphi )}}
={\partial_0 }\varphi \label{b19}
\ed
Here the Hamiltonian may be written as
\bd
H=\pi {\partial_0 }\varphi-L= \nonumber
\ed
\veb
\bd
{1 \over 2}\pi ^2+{1 \over 2}(\nabla \varphi )^2
+{1 \over 2}m^2\varphi ^2+U(\varphi ) \label{b20}
\ed
For evaluating the partition function,   return to the discretized 
version of (\ref{b15}):
\bd
Z_\beta =\mathop {\lim }\limits_{N\to \infty }
\left( {\prod\limits_{i=1}^N {{\int\limits_{-\infty}^{\infty}
{{d\pi _i} \over {2\pi}}}
\int\limits_{periodic} {d\varphi _i}}} \right)
\exp \left\{ {\sum\limits_{j=1}^N {\int {d^3x\left[ {i\pi _j} \right.}}}
 \right.(\varphi _{j+1}-\varphi _j)- \nonumber
\ed
\veb
\bd
\left. {\left. {-\Delta \tau \left( {{1 \over 2}\pi _j^2
+{1 \over 2}(\nabla \varphi _j)^2
+{1 \over 2}m^2\varphi _j^2+U(\varphi )} \right)} \right]} \right\} \label{b21}
\ed
Using the equation for a Gaussian integral
\bd
{1 \over {\sqrt {2\pi i}}}\int\limits_{-\infty }^\infty  
{dx\exp \left\{ {{i \over 2}ax^2+ibx} \right\}
={1 \over {\sqrt a}}\exp \left\{ {-{i \over 2}{{b^2} \over a}} \right\}} \label{b23}
\ed
one can write 
\bd
\int {{{d\pi _j} \over {2\pi }}}\exp 
\left\{ {i\pi _j(\varphi _{j+1}-\varphi _j)
-{{\Delta \tau } \over 2}\pi _j^2} \right\} \nonumber
\ed
\veb
\bd
={1 \over {\sqrt {2\pi \Delta \tau }}}\exp 
\left\{ {-{{(\varphi _{j+1}-\varphi _j)^2} 
\over {\Delta \tau }}} \right\} \label{b22}
\ed
and after momentum integrations in (\ref{b21}) one gets
\bd
Z_\beta =N^{'}\mathop {\lim }\limits_{N\to \infty }
\int {\left( {\prod\limits_{i=1}^N {d\varphi _i}} \right)}
\exp \left\{ {\Delta \tau \sum\limits_{j=1}^N 
{\int {d^3x\left[ {-{1 \over 2}\left( {{{(\varphi _{j+1}-\varphi _j)} 
\over {\Delta \tau }}} \right)^2} \right.}}} \right.- \nonumber
\ed
\veb
\bd
\left. {\left. {-{1 \over 2}(\nabla \varphi _j)^2
-{1 \over 2}m^2\varphi _j^2-U(\varphi )} \right]} \right\} \label{b24}
\ed
Returning to the continuum limit, one obtains:
\bd
Z_\beta =N^{'}\int\limits_{periodic} {D\varphi \exp 
\left( {\int\limits_0^\beta  {d\tau 
\int {d^3xL(\varphi ,\partial \varphi )}}} \right)} \label{b25}
\ed
Equation (\ref{b25}) expresses $Z_\beta$ as a functional integral in 
time interval $[0,\beta]$. The normalization constant $N^{'}$ is irrelevant, 
since it does not change the thermodynamics.

Now it is  seen  how  this method works in application to a non-interacting 
 scalar field.
Let the  Lagrangian of the scalar field be 
\bd
L(\varphi ,\partial \varphi )=-{1 \over 2}(\partial \varphi )^2\label{b26}
-{1 \over 2}m^2\varphi ^2
\ed 
Define the finite temperature action of the Bose system at temperatute $T$
in a large cubic volume $V=L^3$ as
\bd
S_\beta =-{1 \over 2}\int\limits_0^\beta  
{d\tau \int \limits_{V}{d^3x\varphi \left( {\partial _\tau ^2 
+\partial _i^2+m^2} \right)}}\varphi \label{b27} 
\ed
Due to the constraint of periodicity of the field  
$\varphi(\vec{x},\tau)$ $(\vec{x}\in V,\tau\in[0,\beta])$ (\ref{b16}), it  can be  
expanded  in a Fourier series as
\bd
\varphi (\vec{x},\tau )=(\beta V )^{-1/2}{\sum\limits_{n=-\infty }^\infty } 
{\sum\limits_{\vec{k}} \varphi _n(\vec{k})
\exp i(\vec{k}\vec{x}+\omega _n\tau )} \label{b28}
\ed
where $\varphi _n(\vec{k})$ is the Fourier coefficient,
$\omega_n=2\pi n/\beta$, and $\vec{k}=2\pi m/L$  $n$, $m$  are integer numbers.

Using the identity
\bd
\int\limits_0^\beta  {d\tau }\exp i(\omega _n-\omega _m)\tau 
=\beta \delta _{n,m} \label{b29}
\ed
one obtains the equation for the action in terms of the Fourier coefficients
\bd
S_\beta =-{1 \over {2}}{\sum\limits_{n=-\infty }^\infty}  
{\sum\limits_{\vec{k}}{\varphi _n^*(\vec{k})}
(\omega _n^2+{\vec{k}^2+m^2}})\varphi _n(\vec{k}) \label{b30}
\ed
where $\varphi _{-n}(-\vec{k})=\varphi _n^*(\vec{k})$ goes from 
the reality of the field $\varphi$. 

Then the partition function (\ref{b25})  may be written as 
\bd
Z_\beta =N^{'}\int {\prod\limits_n {\prod\limits_{\vec{k}} {d\varphi _n(\vec{k})}}}
\exp \left\{
-{1 \over 2}\varphi _n^*(k)(\omega _n^2+{\vec{k}}^2+m^2) d \varphi _n(k)
\right\} \label{b31}
\ed
where the explicit form of the measure of the functional integration (\ref{b31}) is  
\bd
D\varphi =\prod\limits_n {\prod\limits_{\vec{k}} d{\varphi _n(\vec{k})}} \label{b32}
\ed
So, integration over the coefficients $\varphi_n(\vec{k})$  in (\ref{b31}) 
gives the following equation for the logarithm of the partition function
[Bernard 1974]
\bd
\mbox{ln} Z_\beta =-{1 \over 2}
\mbox{ln} Det(\omega _n^2+\epsilon _{\vec{k}}^2)
=-{1 \over 2}\sum\limits_n {\sum\limits_{\vec{k}}} 
\mbox{ln}({\omega _n^2}+\epsilon _{\vec{k}}^2) \label{b33}
\ed
where $\epsilon _{\vec{k}}^2={\vec{k}}^2+m^2$ is the energy of a one particle 
state with a certain momentum.
Summation in respect with $\omega_n$ in (\ref{b33}) may be done with the help of 
the basic equation 
\bd
\sum\limits_{n=-\infty }^\infty  {{z \over {z^2+n^2}}}
={\pi  \over 2}\coth \pi z \label{b34}
\ed
The derivative of the sum gives 
\bd
{\partial  \over {\partial \epsilon }}
\sum\limits_n {\ln \left( {{{4\pi ^2n^2}
 \over {\beta ^2}}+\epsilon ^2} \right)}
=\sum\limits_n {{{2\epsilon }
 \over {4\pi ^2n^2/ \beta ^2+\epsilon ^2}}=}  \nonumber
\ed
\veb
\bd
=2\beta \left( {{1 \over 2}+{1 \over {\exp \beta \epsilon -1}}} \right) \label{b35}
\ed
After integration with  respect to  $\epsilon$ we will have
\bd
\sum\limits_n {\ln \left( {{{4\pi ^2n^2} \over {\beta ^2}}
+\epsilon ^2} \right)}=2\beta 
\left[ {{\epsilon  \over 2}
+{1 \over \beta }\ln (1-\exp[- \beta \epsilon]) } \right] \label{b36}
\ed
The expression for the  thermodynamic potential will have the standard form
\bd
F=-{1 \over \beta }\ln Z_\beta ={1 \over \beta }
\sum\limits_{\vec{k}} {\ln (1-\exp[- \beta \epsilon _{\vec{k}}])} \label{b37}
\ed
Here the infinite, temperature independent part of the function $Z_\beta$
is dropped

\section {Green's function of boson field}
\vsse

Introduce the two point Green function (or propagator) of boson field at 
finite temperature  as the  thermal average of a time-ordered product 
of two scalar field operators\footnote{In quantum field theory 
two-points Green's function is introduced as 
the vacuum expectation  value of 
a time-ordered product of two field operators.}.
\bd
G_\beta (x-y)=<\mbox{T}\varphi(x)\varphi(y)>_\beta \nonumber
\ed
\veb
\bd
=\mbox{Tr}\left[ {\exp (-\beta H)\mbox{T}\varphi (x)
\varphi (y)} \right]/ \mbox{Tr}\exp (-\beta H) \label{b38}
\ed
One  may express the Green's function as the sum 
\bd
G_\beta(x-x^{'})=\theta(\tau-\tau^{'})G_\beta^+(x-x^{'})
+\theta(\tau^{'}-\tau)G_\beta^-(x-x^{'}) \label{b39}
\ed
where $G^{\pm}$ are Wightman functions
\bd
G_\beta^+(x-x^{'})=<\mbox{T}\varphi(x)\varphi(x^{'})>_\beta \nonumber
\ed
\veb
\bd
G_\beta^-(x-x^{'})=<\mbox{T}\varphi(x^{'})\varphi(x)>_\beta \label{b40}
\ed
and
\bd
\theta (\tau )=\left\{ \matrix{1,\tau >0,\hfill\cr
  0,\tau <0.\hfill\cr} \right. \label{b41}
\ed
is the step function.
 
In  the interval $[0,\beta]$ 
\bd
G_\beta(x-y)_{|{x^{4}}=0}=G_\beta^+(x-y)_{|{x^{4}}=0} \nonumber
\ed
and
\bd
G_\beta(x-y)_{|{x^{4}}=\beta}=G_\beta^-(x-y)_{|{x^{4}}=\beta} \label{b42}
\ed
Using the fact that  $exp(-\beta H)$ and time ordering
operation  commute and taking into account the cyclic property of 
the trace, one  gets an important property of the thermal Green's function
\bd
G_\beta(x-y)_{|{x^{4}}=0}=G_\beta^+(x-y)_{|{x^{4}}=0} \nonumber
\ed
\veb
\bd
=\mbox{Tr}\left[ {\exp (-\beta H)\varphi (\vec{x},\tau)
\varphi (\vec{y},0)} \right]/ \mbox{Tr}\exp (-\beta H) \nonumber
\ed
\veb
\bd
=\mbox{Tr}\left[ {
\varphi (\vec{y},0)\exp (-\beta H)\varphi (\vec{x},\tau) }\right]
/ \mbox{Tr}\exp (-\beta H) \nonumber
\ed
\veb
\bd
=\mbox{Tr}\left[ {
\exp (-\beta H)(\exp (\beta H)\varphi (\vec{y},0)
\exp (-\beta H))\varphi (\vec{x},\tau) }\right]
/ \mbox{Tr}\exp (-\beta H) \nonumber
\ed
\veb
\bd
=\mbox{Tr}\left[ {\exp (-\beta H)\varphi (\vec{y},\beta)
\varphi (\vec{x},\tau)} \right]/ \mbox{Tr}\exp (-\beta H) \nonumber
\ed
\veb
\bd
=G_\beta^-(x-y)_{|{x^{4}}=\beta}=G_\beta(x-y)_{|{x^{4}}=\beta} \label{b43}
\ed
Thus one  has a periodicity condition
\bd
G_\beta(x-y)_{|{x^{4}}=0}=G_\beta(x-y)_{|{x^{4}}=\beta} \label{b44}
\ed
This relation leads  to the Green's function in Euclidean quantum
field theory.

Since the Green's function $G_\beta$ is periodic (\ref{b44}) and depends only 
on coordinate differences $G_\beta(x^0-y^0, \vec{x}-\vec{y})$, 
it may be represented by Fourier series  and integral as
\bd
G_\beta (x-y)= (1/ \beta )
\sum\limits_{n} {\exp [i\omega _n(x^4-y^4)]}
\int{{d^3k} \over {(2\pi )^3}}\exp [i\vec{k}(\vec{x}-\vec{y})] \nonumber
\ed
\veb
\bd
\times G_\beta (\omega _n,\vec{k}) \label{b45}
\ed
with $\omega_n=2\pi n/\beta$.

In compact form one  can write this equation as 
\bd
G_\beta (x-y)=\int\limits_k^- {\exp [ik(x-y)]  G_\beta (k)} \label{b46}
\ed
where
\bd
\int\limits_k ^-{f(k)}={1 \over \beta }\sum\limits_n \int {{{d^3k}
 \over {(2\pi )^3}}}f(\omega_n,\vec{k}) \label{b47}
\ed
The Green's function satisfies the equation
\bd
\left( {-\partial _x^2+m^2} \right)G_\beta (x-y)=\delta (x-y) \label{b48}
\ed
with periodic conditions (\ref{b44}) and $\delta$ -function in the form:
\bd
\delta (x)=\sum\limits_{n}\int{d^3k\over{(2\pi)^3}}
\exp[i(\omega_nx^4+\vec{k}\vec{x})] \label{b49}
\ed
From  (\ref{b46}) and  (\ref{b48}) it follows that $G_\beta(k)$ is
\bd
G_\beta (k)=\frac{1}{k^2+m^2} \label{b50}
\ed
and 
\bd
G_\beta (x)=\int\limits_k ^-{\frac{\exp (ikx)}{k^2+m^2}} \label{b51}
\ed

\section{ Notation} 
\vsse

To work with the standard quantum field theory of  signature  $(+2)$, 
 return to imaginary time formalism, on the time variable 
$t=-i\tau$ defined in the interval $[0,-i\beta]$. Therefore the equation 
(\ref{b44}) may be rewritten as
\bd
G_\beta(x-y)_{|x^0=0}=G_\beta(x-y)_{|x^0=-i\beta} \label{b52}
\ed
It leads to the replacement   $\omega_n\rightarrow i\omega_n$ and
\bd
\int\limits_k {f(k)}={i \over \beta }\sum\limits_n \int {{{d^3k}
 \over {(2\pi )^3}}}f(\omega_n,\vec{k}) \label{b53}
\ed
with $\omega_n=2\pi in/\beta$.

The equation for the Green's function becomes 
\bd
G_\beta (x)=-\int\limits_k {\frac{\exp (ikx)}{k^2+m^2}} \label{b54}
\ed
This expression looks like the  expression for the  Green's function 
in a common field theory.

%% file: ikchap2c.tex
\section{Partition function for fermions}
\vsse

The previous section  discussed a quantization scheme for Bose fields 
in the functional integral formulation. However, the  methods of path integral may 
be applied in the same way to finite temperature Fermi systems. This 
section  develops these methods.

Consider a free spinor field which is described by the action 
\bd
S=-\int {d^4x}{\bar{\psi}(x) (i\gamma \cdot \partial +m)}\psi(x) \label{c1} 
\ed 
where $\{\gamma_\mu\}$ is a set of Dirac matrices, which are defined by
\bd
\{\gamma_\mu,\gamma_\nu\}=2g_{\mu\nu} \label{c2}
\ed
metric $g_{\mu \nu}=diag(-,+,+,+) $and, by notation, 
$\bar{\psi}=\psi^{+}\gamma^0$ is hermition conjugate.

The action (\ref{c1}) has  a global $U(1)$ symmetry  and that is associated with 
conserved current
\bd
j^\mu(x)=\bar{\psi}(x)\gamma^\mu\psi(x) \label{c3}
\ed
The total conserved charge is
\bd
Q=\int{d^3x}j^0=\int{d^3x}\psi^+\psi \label{c4}
\ed
 To apply  the formalism developed in chapter II.2  treat the field 
$\psi$ as the basic field, and the momentum congugate to this field  
\bd
\pi=\frac{\partial L}{\partial(\partial\psi/\partial t)}=i\psi^+ \label{c5}
\ed
Thus  consider  $\psi$ and $\psi^+$ to be  independent entities in the  
Hamiltonian formulation.

The Hamiltonian density is found by the standard procedure as
\bd
H=\pi {{\partial \psi } \over {\partial t}}-L
=\psi^+ \left( {i{{\partial \psi } \over {\partial t}}} \right)-L
=\bar{\psi}(-i\gamma \partial +m)\psi \label{c6} 
\ed
Introduce the partition function  as
\bd
Z=\mbox{Tr} \exp\left[ -i\beta(H-\mu Q)\right] \label{c7}
\ed
Follow the steps leading up to the equation (\ref{b21}). For the partition 
function the intermediate equation will be:
\bd
Z_\beta \propto \mathop {\lim}\limits_
{N\to \infty }\int {\left( {id\psi ^+{{d\psi } \over {2\pi }}} \right)}\times \nonumber
\ed
\veb
\bd
\exp \left\{ {-i\Delta t\sum\limits_{j=1}^N 
{\int {d^3x\left[ {H(\psi ^+,\psi )-\psi _j^
+(\psi _{j+1}-\psi _j)\/ \Delta t} \right]}}} \right\} \label{c8}
\ed
Taking into account the limit of this equation,  find
\bd
Z_\beta =N^{'}\int {id\psi ^+}\int {d\psi } \nonumber
\ed
\veb
\bd
\exp \left\{ {i\int\limits_0^{t_f}
 {dt\int {d^3x\left( {i\psi ^+{{\partial \psi }
 \over {\partial t}}-H(\psi ^+(x,t),\psi (x,t))} \right)}}} \right\} \label{c9}
\ed
For the time variable $\tau$,  get
\bd
Z_\beta =N^{'}\int {id\psi ^+}\int {d\psi } \nonumber
\ed
\veb
\bd
\exp \left\{ 
\int\limits_0^\beta  {d\tau} 
\int {d^3x\psi^+} \left( \mu-\frac{\partial}
{\partial \tau}+i\gamma ^0\vec{\gamma} \cdot \vec{\nabla}
-m\gamma ^0 \right)\psi  
\right\} \label{c10}
\ed
The quantization of a Fermi system can be obtained as a result of integration 
over the space of anticommuting functions 
$\psi(\vec{x},\tau)$ $(\vec{x}\in V,\tau\in [0,\beta])$,
 which are the elements of an infinite Grassman algebra.  

To obtain the correct result it is neccessary to impose on $\psi$, $\bar\psi$
the antiperiodicity conditions in the variable $\tau$:
\bd
\psi(\vec{x},\beta)=-\psi(\vec{x},0),
~~\bar{\psi}(\vec{x},\beta)=-\bar{\psi}(\vec{x},0) \label{c11}
\ed
The Fourier series for $\psi$, $\bar\psi$ in Fermi case (\ref{c11}) are
\bd
\psi (x,\tau )=(\beta V)^{-1/ 2}\sum\limits_n {\sum\limits_{\vec{k}}} 
{\exp (i[\omega \tau +\vec{k}\vec{x}])\psi _n(\vec{k})} \nonumber
\ed
\veb
\bd
{\psi}^+ (x,\tau )=(\beta V)^{-1/ 2}\sum\limits_n {\sum\limits_{\vec{k}}} 
{\exp (-i[\omega \tau +\vec{k}\vec{x}]){\psi}^+_n(\vec{k})} \label{c12}
\ed
where $\psi _n(\vec{k})$ and $\psi^+ _n(\vec{k})$ are 
the generators of Grassmann algebra. 

Inserting (\ref{c12}) into (\ref{c10})  get
\bd
Z_\beta =N^{'}\left[ {\prod\limits_i {\prod\limits_n 
{\int id\psi _n^+d\psi _n}}} \right]\times  \nonumber
\ed
\veb
\bd
\exp \sum\limits_n {\sum\limits_{\vec{k}} {\psi _n^+
\left[ {(-i\omega _n+\mu )-\gamma ^0\vec{\gamma} \cdot{\vec{ k}}
-m\gamma ^0} \right]\psi _n}} \label{c13}
\ed
Calculation of the Gaussian functional integral over the fermi fields 
in (\ref{c13}) leads  to the following equation
\bd
 Z_\beta=Det[(-i\omega _n+\mu )-\gamma ^0\vec{\gamma} \cdot{\vec{ k}}
-m\gamma ^0] \label{c14}
\ed 
or using useful relation  
\bd
\mbox{ln}DetD=\mbox{Tr}\ln D \label{c15}
\ed
 find
\bd
 \ln Z_\beta=\mbox{Tr}\ln[(-i\omega _n+\mu )-\gamma ^0\vec{\gamma} \cdot{\vec{ k}}
-m\gamma ^0]= \nonumber
\ed
\veb
\bd
=2\sum\limits_n {\sum\limits_{\vec{k}} {\ln \left[ {(\omega _n
+i\mu )^2+\epsilon _{\vec{k}}^2} \right]}} \label{c16}
\ed 
Since both positive and negative frequencies are summed over, (\ref{c16}) 
can be written in the form
\bd
\mbox{ln} Z_\beta=\sum\limits_n \sum\limits_{\vec{k}} 
\left\{\ln \left[ {(\omega _n
-\mu )^2+\epsilon _{\vec{k}}^2} \right]
+\ln \left[ {(\omega _n
+\mu )^2+\epsilon _{\vec{k}}^2} \right]\right\} \label{c17}
\ed 
Summation (\ref{c17}) over $\omega_n$ leads to the folowing equation
\bd
\ln Z_\beta=  \nonumber
\ed
\veb
\bd
2V\sum\limits_{\vec{k}} {\left[ \beta {\epsilon_{\vec{k}}} 
+(1+\exp \{-\beta (\epsilon_{\vec{k}} -\mu )\})
+(1+\exp \{-\beta (\epsilon_{\vec{k}} +\mu )\}) \right]} \label{c18}
\ed
The contributions for particles $(\mu)$ and antiparticles $(-\mu)$,
 and, also, the zero-point energy of vacuum, have now been obtained. [Kapusta 1989] 

\section{Green's function of fermi field}
\vsse

The finite temperature fermionic Green's function  (or fermionic propagator)
may be introduced in the similar way to scalar field.

Consider that
\bd
G^F_\beta (x-y)=<T\psi(x),\bar{\psi}(y)>_\beta \nonumber
\ed
\veb
\bd
=\mbox{Tr} [\exp (-\beta H)\mbox{T}\psi(x),\bar{\psi}(y)]/\mbox{ Tr}~ \exp (-\beta H) \label{c19}
\ed
For fermions  the analog of (\ref{b39}) is  in the form 
\bd
G^F_\beta(x-y)=\theta(x-y)G^{+F}_\beta(x-y)+\theta(y-x)G^{F}_\beta(x-y) \label{c20}
\ed 
where
\bd
G^{+F}_\beta(x-y)=<\psi(x)\bar{\psi}(y)>_\beta,~~x^4>y^4 \nonumber
\ed
\veb
and
\bd
G^{-F}_\beta(x-y)=-<\bar{\psi}(y)\psi(x)>_\beta,~~~~x^4<y^4 \label{c21}
\ed
The desired boundary conditions now follow from 
\bd
G^{-F}_\beta(x-y)=-\mbox{Tr}[\exp (-\beta H)\bar{\psi}(y^4,\vec{y})
\psi(0,\vec{x})]/\mbox{Tr}~\exp (-\beta H) \nonumber
\ed
\veb
\bd
=-\mbox{Tr}[\exp (-\beta H) \exp (\beta H)
\psi(0,\vec{x}) \exp (-\beta H)\bar{\psi}(y^4,\vec{y})]/\mbox{Tr}~ \exp (-\beta H) \nonumber
\ed
\veb
\bd
=-\mbox{Tr}[\exp (-\beta H)
\psi(\beta,\vec{x})\bar{\psi}(y^4,\vec{y})]/\mbox{Tr} \exp (-\beta H)
=-G^{+F}_\beta(x-y) \label{c22}
\ed
It leads to antiperiodic boundary conditions   
\bd
G^F_\beta(x-y)_{|x^4=0}=-G^F_\beta(x-y)_{|x^4=\beta} \label{c23}
 \ed
and this means that the fields $\psi$, $\bar{\psi}$ are antiperiodic  
\bd
\psi(\vec{x},0)=-\psi(\vec{x},\beta) \label{c24}
\ed
The Fourier series of the fermionic propagator is written as
\bd
G^F_\beta(x-y)=(1/\beta)\sum\limits_n \exp~[i\omega_n(x^4-y^4)]
\int{\frac{d^3k}{2\pi^3}} \exp[i\vec{k}(\vec{x}-\vec{y})]
G^F_\beta(\omega_n,\vec{k}) \nonumber
\ed
\veb
\bd
=\int\limits_k ^- \exp [ik(x-y)]  G^F_\beta(k) \label{c25}
\ed
with $k^\mu=(\omega_n,\vec{k})$,~~$\omega_n=({2\pi}/{\beta})(n+1/2)$. 

The Green's function of the fermionic field satisfies the equation 
\bd
(i\bar{\gamma}\cdot\partial+m)G^F_\beta(x-y)=\delta(x-y) \label{c26}
\ed
and has the following form
\bd
G^F_\beta(x)=\int\limits_k^-\frac{\bar{\gamma}\cdot k+m}{k^2+m^2}
\exp (ikx) \label{c27}
\ed
where $\{\bar{\gamma}_\mu\}$ is the set of Euclidean gamma matrices.

\section{ Notation } 
\vsse

As in the case of the scalar field one can work with a  quantum model in Minkowski
space-time, restoring the time variable 
$t=-i\tau$ defined in the interval $[0,-i\beta]$. It gives the following
antiperiodic conditions for Green's function  
\bd
G^F_\beta(x-y)_{|x^0=0}=-G^F_\beta(x-y)_{|x^0=-i\beta} \label{c28}
\ed
and the rule of integration 
\bd
\int\limits_k {f(k)}={i \over \beta }\sum\limits_n \int {{{d^3k}
 \over {(2\pi )^3}}}f(\omega_n,\vec{k}) \label{c29}
\ed
with $\omega_n=(2\pi i/\beta)(n+1/2)$.
In the Minkowski metric (signature $(+2)$) the equation 
for Green's function is written as [Dolan, Jackiw 1974]
\bd
G^F_\beta(x)=\int\limits_k ^{-}\frac{{\gamma}\cdot k+m}{k^2+m^2}\exp (ikx) \label{c30}
\ed

%% file: ikchap3d.tex
\chapter{THERMODYNAMICS OF QUANTUM GASES}
\centerline{\Large \bf AND GREEN'S  FUNCTIONS}
\vs

This section  develops a method for calculation of thermodynamic potentials
 directly from finite temperature Green's functions. For this purpose 
 the   Schwinger proper time formalism \cite{schw1} has been applied,
to write down  generating functionals of quantum fields in terms of Green's 
functions of these fields. The  finite temperature generalization of this formalism
leads  to thermodynamical potentials which are written through
finite temperature Green's functions.   

\section{Thermal bosonic fields}
\vsse

In standard form the generating functional $Z$ of the scalar field $\varphi$
with the Lagrangian (\ref{b26}) is written as : 
\bd 
Z[J]=\int {D\varphi \exp \left\{ {-(i/ 2)\int 
{d^4x\varphi (x)K_{xy}\varphi (y)}} +i\int{d^4xJ(x)\varphi(x)}\right\}} \label{d1}
\ed 
The  symmetric operator
\bd 
K_{xy}=\left( {-\partial _x^2+m^2} \right)\delta (x-y) \label{d2}
\ed 
can  formally be treated 
as a symmetric matrix with continuous indices $(x,y)$. It has the following
 properties:
\bd
\int{d^4y}K^{1/2}_{xy}K^{-1/2}_{yz}=\delta(x-z) \nonumber
\ed
\veb
\bd 
\int{d^4y}K^{1/2}_{xy}K^{1/2}_{yz}=K_{xz} \label{d3}
\ed 
The functional $Z$ gives the transition amplitude from the initial $|0^->$
and final $|0^+>$ vacuum in the  presence of the source  $J(x)$, $(Z=<0^+|0^->_J)$
which is producing 
particles. 
The Green's function may be treated as the solution of the equation
\bd 
\int {d^y}K_{xy}G(y,x^{'})=\delta (x-x^{'}) \label{d4}
\ed 
with operator $K_{xy}$.
Now   rewrite (\ref{d1}) in  a convenient form changing the integration 
variable from $\varphi$ to
\bd 
\varphi^{'}(x)=\int {d^4 y}K_{xy}\varphi(y) \label{d5}
\ed 
Then the quadratic form of (\ref{d1}) may be recast as
\bd
-(1/2)\int{d^4x}\left[\varphi^{'}(x)-\int{d^4y}K^{-1/2}_{yx}J(y)\right] \nonumber
\ed
\veb
\bd 
-(1/2)\int{d^4x}{d^4y}\varphi(y)G(x,y)\varphi(y) \label{d6}
\ed 
Substituting (\ref{d6}) into (\ref{d1}) and performing an integration of the
Gaussian type integral,  get
\bd 
Z[J] \propto (detK^{1/2})^{-1}\exp
\left[-(i/2)\int{d^4x}{d^y}\varphi(y)G(x,y)\varphi(y)\right] \label{d7}
\ed 
where the Jacobian  arises from the change of variable (\ref{d5}).

The functional determinant in (\ref{d7}) may be written with Green's function $G(x,y)$:
\bd 
(DetK^{1/2})^{-1}=(DetG(x,y))^{1/2}=\exp \left[(1/2)\mbox{Tr} \ln G(x,y)\right] \label{d8}
\ed
The Green's function is found  from  functional differentiation of $Z$
 with respect to source $J$
\bd 
(i)^2\left(\frac{\delta \ln Z}{\delta J(x)\delta J(y)}\right)_{J=0} 
=<0|\mbox{T}\varphi(x)\varphi(y)|>=G(x,y) \label{d9}
\ed
Now  pay  attention to the functional determinant in (\ref{d7}) because 
of its important role in the applications to statistical mechanics. 

One can introduce the heat kernel for operator $K$ as the solution of the partial 
differential equation
\bd 
i\frac{\partial}{\partial s}\Im(x,x^{'};is)
=\int{d^4z}K_{xz}\Im(z,x^{'};is) \label{d10}
\ed
with the boundary conditions $\Im(x,x^{'};0)=\delta(x-x^{'})$

The Green's function $G$ may be written with $\Im$ in the form
\bd 
G(x,x^{'})
=\int\limits_0^\infty {ids}\Im(z,x^{'};is) \label{d11}
\ed
and the logarithm of functional determinant of the operator $K$ as
\bd 
\ln Det K=\int \limits_0 ^\infty {ids}(is)^{-1} \mbox{tr} \Im(z,x^{'};is) \label{d12}
\ed
To get the equation for the heat kernel $\Im$,
 find proper time representation of Green's function.

Using the useful equation  
\bd
(k^2+m^2)^{-1}=i\int\limits_0^\infty ds \exp\{-is(k^2+m^2)\} \label{d13}
\ed   
write  the equation for Green's function in the form
\bd
G(x,x^{'})=\int\frac{d^4k}{(2\pi)^4}\frac{1}{k^2+m^2}\exp(iky) \nonumber 
\ed
\veb
\bd
=i\int\limits_0^\infty ds \int\frac{d^4k}{(2\pi)^4}\exp\{iky-is(k^2+m^2)\} 
\ed
where $y=x-x^{'}$.

After the integration with respect to  momentum, we  get the Schwinger 
representation of the Green's function
\bd
G(x,x^{'})=\frac{i}{(4\pi)^2}\int\limits_0^\infty 
ids(is)^2\exp(-ism^2-\sigma/2is) \label{d14} 
\ed   
The  variable $\sigma$ equals half the square of the distance between $x$ and $x^{'}$
\bd
\sigma=(x-x^{'})^2/2 \nonumber
\ed
The equation for the heat kernel directly follows 
from the comparision (\ref{d11}) and (\ref{d14}):
\bd
\Im(x,x^{'};is)=\frac{i}{(4\pi is)^2}
\exp(-ism^2-\sigma/2is) \label{d15} 
\ed   
For future calculations it is better to work with the generating 
functional of connected  Green's functions  $W[J]$
which is  connected with  generating functional $Z$ by means of equation
\bd
Z[J]=\exp\{iW[J]\} \label{d16}
\ed   
From (\ref{d12}) and (\ref{d16}) we  find
\bd
W[0]=(i/2)\ln Det K \nonumber
\ed
\veb
\bd
=-(i/2)\int\limits_0^\infty ids(is)^{-1}\mbox{tr}\Im (x,x^{'};is) \label{d17}
\ed   
Inserting (\ref{d15}) into (\ref{d17}) and taking into account (\ref{d14}) 
we get an  important equation
\bd
W[0]=-(i/2)\int{d^4x}\int\limits_{m^2}^\infty dm^2\mbox{tr}G(x,x^{'}) \label{d18}
\ed   
Now  find finite temperature functional $W_\beta$ inserting
Green's function at finite temperature
\bd
W_\beta=-(i/2)\int\limits_\beta{d^4x}\int\limits_{m^2}^\infty 
dm^2\mbox{tr}G_\beta(x,x^{'}) \label{d19}
\ed   
Now
\bd
W_\beta=-(\beta/2)\int\limits_V{d^3x}\int\limits_{m^2}^\infty 
dm^2\int\limits_k(k^2+m^2)^{-1} \nonumber
\ed
\veb
\bd
=(\beta/2)V \int\limits_k \ln (k^2+m^2) \label{d20}
\ed   
The  Helmholtz free energy may be treated as effective potential of the 
finite part of $W_\beta$, therefore
\bd
F(\beta, V) =-iW_\beta. \label{d21}
\ed   
and
\bd
F (\beta, V) =-(i\beta/2)V \int\limits_k \ln (k^2+m^2) \nonumber
\ed
\veb
\bd
=(1/2)V\int\frac{d^3k}{(2\pi)^3}\sum\limits_n \ln(\omega^2_n+\epsilon^2_k) \label{d22}
\ed   
After making the summation over the frequencies $\omega$, 
this equation will coincide with (\ref{b37}) 
(if  the infinite, temperature independent term is dropped).

So one  can conclude that the density of 
Hemholtz free energy may be written as [Kulikov \& Pronin 1987] 
\bd
f(\beta)=(i/2)\int\limits_{m^2}^\infty 
dm^2\mbox{tr}G(\beta,x-x^{'}), \label{d23}
\ed   
where $G(\beta,x-x^{'})$ is temperature contribution 
in $G_\beta(x,x^{'})$.

\section{Bosonic finite temperature Green's} 
\lum
\hspace{20mm}{\Large \bf function in the  Schwinger representation}
\vsse

Start from the finite temperature Green's function (\ref{b51}) in the limit 
of coincidence  $x=x^{'}$: 
\bd
\mathop{\lim}\limits_{x\to x^{'}}
G_\beta (x,x^{'})=\int \limits _{k}^{-}\frac{d^4k}{(2\pi)^4}\frac{1}{k^2+m^2} \nonumber
\ed
\veb
\bd
=-\int\limits_0^\infty ds (1/\beta)\sum\limits_n\int\frac{d^3k}{(2\pi)^3}
\exp\{-is(-\omega^2_n+\vec{k}^2+m^2)\}
\ed
After the integration with respect to momentum   find
\bd
\mathop{\lim}\limits_{x\to x^{'}} 
G_\beta (x,x^{'})=\int\limits _0 ^\infty{(4\pi is)^{3/2}}\exp(-ism^2)
(1/\beta)\sum\limits_n \exp(-4\pi^2n^2/\beta^2), \label{d24} 
\ed   
Rewrite the sum in  (\ref{d24}) with the help of the  equation \cite{dit1} 
\bd
\sum\limits_n \exp[-\alpha (n-z)^2]=\sum\limits_n 
(\pi/\alpha)^{1/2}\exp(-\pi^2n^2/\alpha-2\pi izn) \label{d25} 
\ed   
In bosonic case  $(z=0)$  (\ref{d24}) has the form: 
\bd
G_\beta (x,x)=\frac{i}{(4\pi)^2}\sum\limits_n \int\limits_0^\infty
\frac{ids}{(is)^2} \exp(-ism^2-n^2\beta^2/4is) \label{d26}
\ed   
Selecting the  temperature independent $(n=0)$ part of the equation (\ref{d26})
we  find the finite temperature contribution in Green's function of scalar field:
\bd
G_\beta(x,x)=G(x,x)+G(\beta,x-x) \nonumber
\ed
\veb
\bd
=\frac{i}{(4\pi)^2}\int\limits_0^\infty
\frac{ids}{(is)^2} \exp(-ism^2) \nonumber
\ed
\veb
\bd
+\frac{i}{(4\pi)^2}2\sum\limits_{n=1}^\infty \int\limits_0^\infty
\frac{ids}{(is)^2} \exp(-ism^2-n^2\beta^2/4is) \label{d27}
\ed   
Inserting $G(\beta,x-x)$ of the equation (\ref{d27}) into (\ref{d23}) 
we  find the equation for free energy in the form
\bd
f (\beta)=-(1/(4\pi)^2)\int\limits_{m^2}^\infty dm^2
\sum\limits_{n=1}^\infty \int\limits_0^\infty
\frac{ids}{(is)^2} \exp(-ism^2-n^2\beta^2/4is) \label{d28}
\ed
The integral over $(s)$ occuring in (\ref{d28}) can be found  with
\bd
\int\limits_0^\infty dxx^{\nu-1}\exp\left(-\alpha\frac{1}{x}-\gamma x\right)=
2\left(\alpha/\gamma\right)^{\nu/2}K_\nu(2\sqrt{\alpha\gamma}),\label{d29}
\ed
where $K_\nu$ is a modified Bessel function.
Then 
\bd
\int\limits_0^\infty  ids(is)^{j-3}\exp (-ism^2-\beta ^2n^2/ 4is)
=2\left( {\beta n/ 2m} \right)^{j-2}K_{j-2}(\beta mn),\label{d30}
\ed
and the equation for $f$ will be 
\bd
f(\beta)=-(m^2/2\pi^2\beta^2)\sum\limits_{n=1}^\infty(1/n^2)K_2(\beta mn) \label{d31}
\ed
The integral representation of the sum of the modified Bessel function (\ref{bb14})
allows  the standard equation for the  Helmholtz  
free energy to be written as 
\bd
f(\beta)=(1/\beta)\int{\frac{d^3k}{(2\pi)^3}} \ln (1-\exp(-\beta\epsilon_k)) \label{d32}
\ed
directly  from (\ref{d31}).

\section {Thermal fermionic fields}
\vsse

Let us now develop the same formalism for fermionic fields. 

Add to the fermionic Lagrangian the terms with anticommutative 
sources $\eta(x)$  and $\bar{\eta}(x)$
\bd
L\rightarrow L=L_0+\eta(x)\bar{\psi}(x)+\psi(x)\bar{\eta}(x) \label{d33}
\ed
and write the generating functional
\bd
Z[\eta,\bar{\eta}]=\int {D\psi}{D\bar{\psi}} 
\exp \left\{  i \int{d^4x}
\left[L_0+\eta(x)\bar{\psi}(x)+\psi(x)\bar{\eta}(x)\right]\right\},\label{d34}
\ed
where the  functions $\psi$ and $\bar\psi$ are considered  as
Grassman variables.

After integration in respect with these Grassmann variables   obtain
\bd
Z[\eta,\bar{\eta}]\propto (Det K) \exp\left[ -i\int{d^4x}\int {d^4y}
\bar{\eta}(x)K_{x y}\eta(y)\right] \label{d35}
\ed
where 
\bd
K_{xy}=(i\gamma\cdot\partial_x+m)\delta(x-y) \label{d36}
\ed 
and bi-spinor $G_F$ satisfies the equation which is equivalent to (\ref{d4}). 

In operator form \cite{schw1}  it is written
\bd
KG_F=\hat1 \label{d37}
\ed 
At zero sources the generating functional $W[0]$  is 
\bd
W[0]=-i\ln Z[0]=-i\ln Det (K) \label{d38}
\ed 
On the other hand it is
\bd
W[0]=-i\ln Det(K) \nonumber
\ed
\veb
\bd
=(i/2)\int\limits_0^\infty ids(is)^{-1}\mbox{tr} \hat{\Im}(x,x^{'};is) \label{d39}
\ed 
where the kernel $\hat{\Im}(x,x^{'};is)$ is the solution of the equation 
\bd
i\frac{\partial}{\partial s}\hat{\Im}(x,x^{'};is)
=(K\hat{\Im})(x,x^{'};is), \label{d40}
\ed 
The equation for the bi-spinor Green's function is written as
\bd
G_F(x,x^{'})=\int\frac{d^4k}{(2\pi)^4}\frac{\hat1}{k^2+m^2} \exp(iky) \nonumber
\ed
\veb
\bd
=i\int\limits_0^\infty ds \int\frac{d^4k}{(2\pi)^4}
\exp\{iky-is(k^2+m^2)\}\label{aud1}
\ed
and after integration over the momentum, we  get
\bd
G_F(x,x^{'})=-\frac{\hat1}{(4\pi)^2}\int\limits_0^\infty 
ds(is)^2 \exp(-ism^2-\sigma/2is), \label{d41} 
\ed   
As follows from (\ref{d11}) the heart kernel has the  form
\bd
\hat{\Im}(x,x^{'};is)=-i\frac{\hat 1}{(4\pi)^2}
(is)^2 \exp(-ism^2-\sigma/2is), \label{d42} 
\ed
Inserting (\ref{d42}) into (\ref{d39}) and using (\ref{d41})  we  find
\bd
W[0]=(i/2)\int{d^4x}\int\limits_{m^2}^\infty dm^2 \mbox{tr} G_F(x,x^{'}) \label{d43} 
\ed   
For non-interacting fermionic field this equation is divergent  
and does not give   any useful information but, 
in the finite temperature case we can get some interesting results
connected with temperature properties of quantum Fermi gas. 

\section{Fermionic finite temperature Green's}
\lum
\hspace{20mm}{\Large \bf   function in Schwinger representation}
\vsse

In the limit of coincidence  $(x \to x^{'})$
the finite temperature fermionic  bi-spinor $G_F$ is written as: 
\bd
\mathop{\lim}\limits_{x\to x^{'}}
G_F(x,x^{'})=\int\frac{d^4k}{(2\pi)^4}\frac{\hat{1}}{k^2+m^2} \nonumber
\ed
\veb
\bd
=-\int\limits_0^\infty ds (\hat{1}/\beta)\sum\limits_n\int\frac{d^3k}{(2\pi)^3}
\exp\{-is(-\omega^2_n+\vec{k}^2+m^2)\}
\ed
where $\omega_n=(2i\pi n/\beta)(n+1/2)$.

After the integration in respect with momentum we get
\bd
\mathop{\lim}\limits_{x\to x^{'}} G_F(x,x^{'})= \nonumber
\ed
\veb
\bd
\int\limits_0^\infty {(4\pi is)^{3/2}}\exp(-ism^2)
(\hat{1}/\beta)\sum\limits_n \exp[-4\pi^2(n+1/2)^2/\beta^2] \label{d44} 
\ed   
Using the equation (\ref{d25}) with $z=-1/2$, we find  
\bd
\sum\limits_n \exp[-(2\pi /\beta)^2 (n+1/2)^2]=\sum\limits_n(-1)^n 
[\beta/(4\pi i s)^{1/2}] \exp(-\pi^2n^2/4is),\label{d45} 
\ed   
Then the equation (\ref{d44}) will be
\bd
\mathop{\lim}\limits_{x\to x^{'}}
G(x,x^{'})=i\frac{\hat 1}{(4\pi)^2}\sum\limits_n (-1)^n\int\limits_0^\infty
\frac{ids}{(is)^2} \exp(-ism^2-n^2\beta^2/4is).\label{d46}
\ed   
Now we can select the temperature independent  part of the equation (\ref{d46}):
\bd
G_F^\beta(x,x)=G_F(x,x)+G_F(\beta,x-x) \nonumber
\ed
\veb
\bd
=i\frac{\hat 1}{(4\pi)^2}\int\limits_0^\infty
\frac{ids}{(is)^2} \exp(-ism^2) \nonumber
\ed
\veb
\bd
+i\frac{\hat 1}{(4\pi)^2}2\sum\limits_{n=1}^\infty(-1)^n \int\limits_0^\infty
\frac{ids}{(is)^2} \exp(-ism^2-n^2\beta^2/4is) \label{d47}
\ed   
Inserting  (\ref{d47}) into the equation [Kulikov \& Pronin 1987]:  
\bd
f=(-i/2)
\int \limits_{m^2}^\infty dm^2 \mbox{tr} G_F(\beta,x-x^{'})) \label{add1}
\ed
we find an  expression for the  free energy of Fermi gas
in the form
\bd
f(\beta)=-(1/(4\pi)^2)\mbox{tr} (\hat{1})\int\limits_{m^2}^\infty dm^2
\sum\limits_{n=1}^\infty(-1)^n \int\limits_0^\infty
\frac{ids}{(is)^2} \exp(-ism^2-n^2\beta^2/4is) \label{d48}
\ed
The integral over $(s)$ occuring in (\ref{d48}) can be written  as 
\bd
\int\limits_0^\infty  ids(is)^{j-3}\exp (-ism^2-\beta ^2n^2/ 4is)
=2\left( {\beta n/ 2m} \right)^{j-2}K_{j-2}(\beta mn), \label{d49}
\ed
and the equation for density of free energy of fermionic field $f$ will be 
\bd
f(\beta)=\mbox{tr} (m^2/2\pi^2\beta^2)(\hat{1}) \sum \limits _{n=1}^\infty
\frac{(-1)^n}{n^2}K_2(\beta mn) \label{d50}
\ed
The standard equation for Helmholtz  free energy follows from (\ref{aa14})
\bd
f(\beta)=-(4/\beta)\int{\frac{d^3k}{(2\pi)^3}}\ln (1+\exp(-\beta\epsilon_k)) \label{d51}
\ed
Now we can describe the thermal behavior of massless vector fields 
in the Schwinger proper time formalism  to prepare the 
mathematical foundation for further computations in curved space-time.

%% file: ikchap4e.tex
\chapter{FINITE TEMPERATURE GAUGE FIELDS}
\vs

In the present day  progress in quantum field theory is 
to a great extent due to the development of   gauge fields 
\cite{key18,key19,hooft1}. These fields open 
up new possibilities 
for the description of interactions of the elementary particles.
Gauge fields are involved in most modern models of strong, weak and 
electromagnetic interactions.  They are extremely attractive for the  unification 
 of all interactions into a single universal interaction. On the other hand the 
functional formulation of the gauge fields helps us to get statistical and 
thermodynamical results connected with finite temperature properties of the
particles which are described by these fields. In this connection we will 
develop the  finite temperature formalism for gauge fields 

\section{Gauge theories: Pure Yang-Mills theory}
\vsse

The basic idea of gauge field relies on the  local gauge invariance principle 
of the quantum field theory \cite{key18,key19}. 
Let us consider a gauge tansformation  parametrized by 
the functions $\omega^a(x)$  
\bd
U(x)=\exp[-i\omega^a(x)\tau^a] \label{e1}
\ed    
where the generators of the  Lee algebra obey the equation
\bd
[\tau^a,\tau^b]=if^{abc}\tau^c   \label{e2}
\ed    
Numbers $f^{abc}$ are the structure constants of the group.

Let the vector (gauge) fields $A_\mu(x)=A_\mu(x)^a\tau^a$
transform according to 
\bd
A_\mu(x) \to A^\omega_\mu(x)=UA_\mu(x)U^{-1}+(i/g)U\partial_\mu U^{-1}  \label{e3}
\ed   
For infinitesimal transformations we have
\bd
(A^{\omega})^a _\mu
=A^a_\mu (x)+f^{abc}A^b_\mu(x) \omega^c(x)+(1/g)\partial_\mu \omega^a(x) \nonumber
\ed
\veb
\bd
=A^a_\mu (x)+\delta A^a_ \mu (x)  \label{e4}
\ed   
The transformations (\ref{e1}) preserve the Maxwell type Lagrangian
\bd
L=-(1/2)\mbox{Tr} (F_{\mu\nu}(x))^2=-(1/4) F^a_{\mu \nu} F^{a \mu \nu}  \label{e5}
\ed   
where
\bd
F_{\mu \nu}=\partial_{\mu} A_{\nu}
 -\partial_{\nu} A_{\mu}-ig[A_{\mu} , A_{\nu}] \nonumber
\ed
\veb
\bd
=(\partial_{\mu} A^a_{\nu} -\partial_{\nu} A^a_{\mu}
+gf^{abc}A^b_{\mu} A^c_{\nu})\tau^a  \label{e6}
\ed   
is the strength of the vector field $A_\mu$.

The path integral over the field $A_\mu$ with the Lagrangian (\ref{e5}) in the form   
\bd
\int DA_\mu \exp (iS[A_\mu,\partial_\nu A_\mu]) \label{e7}
\ed   
is undefined because of the gauge degrees of freedom.
 Namely, the gauge transformations  create 
the "orbit" of the  field $A^\omega_\mu$ in the functional field space
and the functional integration  (\ref{e7}) over this space
 overcounts the degrees of  freedom of the theory. 
To improve the situation we have to eliminate  all unphysical degrees of 
freedom which origin from local gauge invariance.
 So, we have to "slice" the orbit 
once so that we do not have this infinite overcounting.  
This is the origin of the gauge fixing problem. 
To solve this problem one can 
select  the surface in this functional space. 
The surface is good if it intersects the orbit of any given field 
under gauge transformation once and only once.
The equation of this surface may be 
written as  
\bd
F(A_\mu)=0  \label{e8}
\ed   
To  eliminate  the unphysical degrees of freedom 
we may  insert the factor $\delta[F(A _\mu)] $ in the functional integral.

We can do it inserting   
\bd
1=\Delta_{FP}(A_\mu)\int D \omega \delta [F(A^\omega _\mu)]  \label{e9}
\ed   
into (\ref{e7}).
This expression can not change the measure of integration in (\ref{e7})
so we will have
\bd
\int DA_\mu  \exp\left\{i\int d^4x L\right\} \nonumber
\ed
\veb
\bd
=\int DA_\mu \left(\Delta_{FP}(A _\mu)
\int D\omega\delta[F(A^\omega _\mu)]\right)
\exp\left\{i\int d^4x L\right\}, \label{e10}
\ed   
where $\delta_{FP}$ is  Faddeev-Popov determinant,
and $D\omega=\prod\limits_{a}\prod\limits_{x} d\omega^a(x)$ 
is the invariant group measure \cite{sf1}.

One can show that it is gauge invariant \cite{key4}
\bd
\Delta_{FP}(A_\mu)=\Delta_{FP}(A^\omega_\mu) \label{e11}
\ed   
therefore the equation (\ref{e10}) is 
\bd
\int DA_\mu 
\int D\omega \Delta_{FP}(A^\omega _\mu) \delta[F(A^\omega _\mu)]
\exp\left\{i\int d^4x L\right\}. \label{e12}
\ed   
Making the gauge transformation from $A^\omega _\mu$ to $A_\mu$
and taking into account that action is gauge invariant too,
we get, dropping multiplicative divergent factor $\int d\omega$,
the expression for generating functional of the vector field $A_\mu$
\bd
Z[J]=\int DA_\mu 
\Delta_{FP}(A_\mu) \delta[F(A _\mu)]
\exp\left\{i\int d^4x L+i(J,A)\right\} \label{e13}
\ed   
Our task now is  to calculate an explicit result of the 
Faddeev-Popov determinant $\Delta_{FP}(A_ \mu)$.

To do this we denote the numerical value of the function $F(A_\mu)$  at $x$ as
\bd
F(A_\mu)=F^a(x) \label{e14}
\ed   
Then one can write the equation (\ref{e9}) in the form
\bd
\Delta ^{-1}_{FP}(A)=\prod\limits_x \prod\limits_a \int d\omega^a(x)~
\delta (F^a(x)) \nonumber
\ed
\veb
\bd
=\prod\limits_x \prod\limits_a \int dF^a(x)~
\delta (F^a(x))\frac{\partial(\omega_1(x)...\omega_N(x))}
{\partial(F_1(x)...F_N(x))} \nonumber
\ed
\veb
\bd
=\prod\limits_x  
\left\| {{{\partial \omega ^a(x)} \over {\partial F^b(x)}}} \right\|_{F=0}
=Det \left(\frac{\delta \omega}{\delta F}\right)_{|F=0} \label{e15}
\ed   
The Faddeev-Popov determinant 
\bd
\Delta _{FP}(A)=Det \left(\frac{\delta F}{\delta \omega}\right)_{|F=0}\label{e16}
\ed   
is the  functional determinant of the continuous matrix  
$\parallel \delta F^a(x)/\delta \omega^b(y) \parallel$ with rows labelled by $(a,x)$ 
and columns by $(b,y)$.

The generating functional (\ref{e13})  will have the form
\bd
Z[J]=\int DA_\mu(x)Det\left(\delta F^a/\delta \omega^b\right)
\delta[F^a(A_\mu)] \exp \left\{ iS+i(J,A)\right\} \label{e17}
\ed   
Now we can make the last step to rewrite our expression in a form convenient for
practical calculations introducing Faddeev-Popov (ghost) fields. 

\section{Ghost fields}
\vsse

For the calculation of generating functional (\ref{e17}) we must select a 
gauge constraint
and compute the  FP-determinant (\ref{e16}) in this gauge. 

Let us select the  Lorentz gauge constrain 
\bd
F^a(A_\mu)=\partial^\mu A^a_\mu(x)=0 \label{e18}
\ed   
We can find $Det(\delta F^a/\delta \omega^b)$ from the  Teylor expansion 
of the $F^a$ with gauge transformation $\omega ^b$: 
\bd
F^a\to (F^ \omega)^a=F^a+Det(\delta F^a/\delta \omega ^b)\omega^b+... \label{e19}
\ed   
For our gauge conditions (\ref{e18}) it is 
\bd
\partial^\mu A^a_\mu(x) 
\mathop \to \limits^\omega 
\partial^\mu A^a_\mu(x)
+f^{abc}\partial(A^b_\mu (x)\omega^c(x))+(1/g)\partial^2 \omega^a(x) \nonumber
\ed
\veb
\bd
=\partial^\mu A^a_\mu (x)
+((1/g)\partial ^2 \delta ^{ac}+A^{ac}_\mu (x)\partial ^\mu )\omega ^c(x) \nonumber
\ed
\veb
\bd
=\partial^\mu A^a_\mu(x)
+\int dy\sum\limits_c 
\left[(1/g)\partial ^2 \delta ^{ac}\delta (x-y)
+A^{ac}_\mu(x)\partial ^\mu \delta (x-y)\right]\omega ^c(y) \label{e20}
\ed   
where $A^{ac}_\mu=f^{abc}A^b_\mu(x)$.

Then we can find, that
\bd
Det\left[\delta F/ \delta \omega\right]=Det
\parallel<x,a|\delta F/ \delta \omega |y,c>\parallel \nonumber
\ed
\veb
\bd
=Det\left[((1/g)\partial ^2 \delta^{ac}
+ A^{ac}_\mu (x)\partial^\mu)\delta (x-y)\right] \label{e21}
\ed 
One can  rewrite Faddeev-Popov determinant (\ref{e21}) in terms of a path integral
over anti-commuting  fields $(\bar{c}, c)$\footnote{The functional determinant 
of anti-commuting fields $\bar{c}$, $c$ is
$$\int D\bar{c}Dc \exp\{i\bar{c}^aM^{ab}c^b\}=Det(M)$$}.
They are  called Faddeev-Popov ghosts. the fields  $\bar{c}$, $c$ are independent 
anti-commuting scalar fields which obey Fermi statistics.
  
Then the  determinant (\ref{e21}) may be written in the form 
\bd
Det\left[\delta F/ \delta \omega\right]=
\int Dc D\bar{c}\exp\left[i\int d^4x L_{ghost}(x)\right]  \label{e22}
\ed
Lagrangian $L_{ghost}(x)$ is
\bd
L_{ghost}(x)=\bar{c}^a(x)(\partial ^2 \delta^{ab}
+ gA^{ab}_\mu (x)\partial^\mu)c^b(x)  \label{e23}
\ed
The first term in the Lagrangian (\ref{e23}) is the pure  ghost part 
\bd
L^{(0)}_{ghost}(x)=\bar{c}(x)(\partial ^2 \delta^{ac})c(x)  \label{e24}
\ed
and the ghost  propagator  $G_0$ directly follows from (\ref{e24}). 
The equation for this propagator is 
\bd
\partial^2_xG_0(x-y)=\delta (x-y), \label{e25}
\ed   
In the momentum space  $G_0(k)$ is written
\bd
G_0(k)=1/(k^2+i\epsilon). \nonumber
\ed
The second term  in (\ref{e23})  describes interaction ghosts and
the  vector field
\bd
igf^{abc}\partial_\mu  \label{e26}
\ed   
For better understanding of the ghost contributions 
let us study an effective action of the ghost fields.

\section{ Effective action}
\vsse

The functional determinant (\ref{e21}) may be rewritten as
 \bd
Det (\delta F/\delta \omega)
=Det (\partial ^2 \delta^{ac}+gA^{ac}_\mu\partial_\mu) \nonumber
\ed
\veb
\bd
=Det((1/g)\partial^2)Det\left(1+g(1/\partial^2)
A^{ac}_\mu\partial_\mu\right)  \label{e27}
\ed
Ignoring the first determinant\footnote{This term is  important 
in the  finite temperature limit.} 
\bd
Det (\partial^2)=Det(G^{-1}_0), \label{e28}
\ed
we can write  an effective action in the form
\bd
S_{eff}=S-i Tr\ln (1+L), \label{e29}
\ed   
where the element $L$ is written as
\bd
<x,a|L|y,c>=g<x,b|\left[
(1/\partial^2)A^{ac}_\mu(x)\partial^\mu_x \delta (x-y)\right]|y,c> \nonumber
\ed
\veb
\bd
=g\int dzG_0(x-z)A^{ac}_\mu(x)\partial^\mu_z \delta (z-y). \label{e30}
\ed   
Now we can expand the determinant with 
\bd
Det(1+L)=\exp\left\{\mbox{Tr}\ln (1+L)\right\}= \nonumber
\ed
\veb
\bd
=\exp\left(\sum\limits_{n=1}^\infty \frac{(-1)^{n-1}}{n}\mbox{Tr} L^n\right)  \label{e31}
\ed   
Write the contributions  $\mbox{Tr}(L^n)$ in (\ref{e31}).

The first expression of the order $g$ is
\bd
\ln L = \mbox{tr}\int dx \int dz 
\left\{G_0(x-z)gA^{ac}_\mu (z)\partial_{z}\delta (z-x)\right\} \label{e32}
\ed   
because   $\mbox{tr}(A^{ac}_\mu(x))=\mbox{tr}(f^{abc}A^b_\mu(x))=0$.

The second expression of the order $g^2$ will be 
\bd
\ln (L^2) =g^2 \mbox{tr}\int dx \int dy\int dz\int dt
 G_0(x-z)\times \nonumber
\ed
\veb
\bd
\times A^{ac}_\mu(z) \partial_{z}\delta (z-x)G_0(y-t)
A^{ca}_\nu (t) \partial_t \delta (t-x) \nonumber
\ed
\veb 
\bd
=\int dz\int dt \left\{gA^{ca}_\nu(t)\partial_t G_0(t-z)\right\}\times \nonumber
\ed
\veb
\bd
\times \left\{gA^{ac}_\mu(z)\partial _z G_0(t-z)\right\} \label{e33}
\ed   
As we can see from (\ref{e33}) and (\ref{e31}), for 
$n>2$ the trace  of (\ref{e29}) equals 
the integral  with integrand in the form of the cycle multiplication of the terms
\bd  
\left\{gA^{ac}_\mu(z)\partial _z G_0(t-z)\right\} \label{e34}
\ed
namely,  we can write 
\bd
\mbox{Tr} (L^n)=\int dz_1...\int dz_n
\left\{gA^{ac_1}_\mu(z_1)\partial _{z_1} G_0(z_1-z_2)\right\}\times... \nonumber
\ed
\veb
\bd  
\times...\left\{gA^{c_na}_\mu(z_n )\partial _{z_n}
 G_0(z_{n-1}-z_n)\right\} \label{e35}
\ed
These terms give non-trivial corrections due to ghosts to the effective action
$S_{eff}[A]$ 

Therefore the determinant (\ref{e31}) may be treated as contributions 
of the closed loops with internal  ghost lines, and external lines of 
which correspond to vector fields. 

\section {Propagator of  vector field}
\vsse

In the previous section we found out how to compute the Faddeev-Popov determinant in
the generating functional (\ref{e17}). Now we must treat the second 
important term the  delta function arising from  the  gauge constraint. For this 
we will rewrite the delta function as \cite{vl1}  
\bd
\delta(\partial^\mu A^a_\mu)
=\mathop {\prod}\limits_{x}\delta(\partial^\mu A^a_\mu(x)) \nonumber
\ed
\veb
\bd
=\mathop {\lim}\limits_{\alpha \to 0}\mathop {\prod}\limits_{x}
(-2i\pi \alpha)^{-1/2}
\exp\left[ -\frac{i}{2\alpha}(\partial^\mu A^a_\mu (x))^2 \right] \nonumber
\ed
\veb
\bd  
\sim\mathop {\lim}\limits_{\alpha \to 0}
\exp\left\{ -\frac{i}
{2\alpha}\int d^4x \left[\partial^\mu A^a_\mu (x)\right]^2 \right\} \label{e36}
\ed
The argument of the exponent (\ref{e36}) may be  combined with 
the  quadratic part of the vector Lagrangian (\ref{e5}):
\bd  
i\int d^4x\left[ -\frac{1}{4}(\partial_\mu A^a_\nu-\partial_\nu A^a_\mu)^2
 -\frac{1}{2\alpha} \left(\partial^\mu A^a_\mu (x)\right)^2\right] \label{e37} 
\ed
The second term of the expression (\ref{e37})fixes the  gauge.

As a result we have the quadratic part over the vector fields in the form
\bd  
-\frac{i}{2}\int d^4xA^a_\mu(x)
\left[ (\eta^{\mu\nu}\partial^2
+\left(1-\frac{1}{\alpha}\right)\partial_\mu \partial_\nu)
\right]A^a_\nu(x) \label{e38} 
\ed
The propagator of the vector field may be found from the equation
\bd  
\left[ (\eta^{\mu\nu}\partial^2
+\left(1-\frac{1}{\alpha}\right)\partial_\mu \partial_\nu)\right]
G_{\nu \lambda}(x-y)=-\delta_{\nu \lambda}\delta(x-y).\label{e39}
\ed
The solution of this equation is
\bd  
G_{\mu\nu}(x)=\int\frac{d^4k}{(2\pi^4)}\left[\eta_{\mu\nu}
+(1-\alpha)\frac{k_\mu k_\nu}{k^2}\right]\frac{\exp(ikx)}{k^2+i\epsilon} \label{e40}
\ed
The finite temperature propagator of the vector field in the  Feynman gauge $(\alpha =1)$ 
has the simple form
\bd  
G_{\mu\nu}(x)=\int\limits_{k} \frac{\eta_{\mu \nu}}
{k^2+i\epsilon} \exp(ikx) \label{e41}
\ed
where the definition of the  integral $\int \limits_{k}$ is the same as in (\ref{b54})

\section {Partition function for Gauge fields}
\vsse

Let us study thermal gauge fields in the axial gauge $(A_3=0)$.

The standard expression for the  partition function is 
\bd
Z=\mbox{Tr} \exp[-\beta H] \nonumber
\ed
where $H=H(\pi^a_i,A^a_i)$ is the Hamiltonian of the Yang-Mills field.  

According to section II (\ref{b15})  we can write
\bd
Z=(\mbox{Tr}\exp[-\beta H])_{axial}
\ed
\veb
\bd
=N\int \prod\limits_a d{\pi^a_1}d{\pi^a_2}
\int \limits_{periodic}d{A^a_1}d{A^a_2}\times \nonumber
\ed
\veb
\bd  
\times \exp\left\{\int \limits_0^\beta d\tau
\int d^3x[i\pi_i\dot{A}^a_i-H(\pi^a_i,A^a_i)]\right\} \label{e42}
\ed
The Hamiltonian  in this gauge has only two degrees of freedom 
for each vector field: $A^a_{1,2}$. The conjugate momenta 
for these fields are $\pi^a_{1,2}$.

After Gaussian integration over the momenta, we get
\bd 
Z=N\int \limits_{periodic}D{A^a}\prod\limits_a\delta[A^a_3]
\exp\left\{\int \limits_0^\beta d\tau
\int d^3xL(x)\right\} \label{e43}
\ed
The result is similar to the expression for the  Faddeev-Popov ansatz in the 
axial gauge.
Calculations of   (\ref{e43}) give the correct  result for the  partition 
function of a photon gas. The same   result
 will also be obtained for Coulomb gauge $(\vec{\nabla} \vec{A}=0)$.

For the  Feynman gauge the simple calculation of the type (\ref{e42}) leads 
to an incorrect result. But we may correct the result  taking into
consideration the ghost contribution 
and rewrite the  partition function (\ref{e42}) in the form of (\ref{e13}). 
 \bd 
Z=N\int \limits_{periodic}D{A^a}Det(\delta F^a/\delta \omega^b)
\delta [F^a] \exp\left\{\int \limits_0^\beta d\tau
\int d^3xL(x)\right\} \label{e44}
\ed
Since under the gauge transformations 
$\delta A^a_\mu=-\partial_ \mu\omega^a$, we have
\bd 
Det(\delta (\partial_\mu A^{a \mu})/\delta \omega ^b)=Det (\partial^2) \label{e45}
\ed
that coincides  with (\ref{b28}) at order $\sim 0(g)$.
For the  Feynman gauge $(\alpha=1)$ 
\bd 
\int \limits_{periodic}D{A^a} 
\exp\left\{\int \limits_0^\beta d\tau
(-1/2)\int d^3x\left[A^a_\mu(x)\partial^2A^{a\mu(x)}\right]\right\}
=Det(G_{\mu\nu})^{1/2} \label{e46}
\ed
At finite temperature  $Det(G_{\mu\nu})$ may be written in the form 
\bd
Det(G_{\mu\nu})=\exp(-1/2)\mbox{Tr} \ln (G_{\mu\nu}) \nonumber
\ed
\veb
\bd 
=\exp(-2)\sum\limits_n\int\frac{d^3k}{(2\pi)^3}
\ln (\omega_n^2+\vec{k}^2) \label{e47}
\ed
and the temperature contribution of  $Det (\partial^2)$ in the form
\bd
Det(\partial^2)=\exp \mbox{Tr}\ln (\partial^2)=\exp \mbox{Tr} \ln (G) \nonumber
\ed
\veb
\bd 
=\exp\sum\limits_n\int\frac{d^3k}{(2\pi)^3}\ln (\omega_n^2+\vec{k}^2) \label{e48}
\ed
where $G$ is just the propagator of the  scalar field.

The logarithm of the finite temperature generating functional  $Z[0]$,
according to previous results, is 
\bd 
\ln Z[0]=\ln Det(G)-(1/2)\ln Det(G_{\mu\nu}), \label{e49}
\ed
or
\bd 
\ln Z[0]=-\sum\limits_n\int\frac{d^3k}{(2\pi)^3}
\ln (\omega_n^2+\vec{k}^2). \label{e50}
\ed
After summation the   free energy is found  in the form
\bd
 f(\beta) =-\frac{1}{\beta}\ln Z_\beta
=\left(\frac{2}{\beta} \right) \int\frac{d^3k}{(2\pi)^3}
\ln (1-\exp[-\beta \epsilon]), \label{e51}
\ed
where $\epsilon^2=(\vec{k})^2$.

So, we got the correct answer \cite{key21} by introducing  the  finite temperature 
ghost contribution.
The index $(2)$ in the equation (\ref{e51}) reflects the two degrees of freedom of 
the massless vector field. 
The generalization to Schwinger proper time formalism is the same as 
in the previous section.

%% file: ikcvtg.tex
\chapter{QUANTUM FIELDS} 
\centerline{\Large \bf IN CURVED SPACE-TIME}
\vs

In this chapter we introduce the basic formalism of quantum fields 
 in curved space-time which was  studied in
 works of DeWitt, Fulling, Bunch et  al. \cite{dt2}, \\\cite{key17}, 
\cite{pr1}, \cite{key16}. 

\section{Lorentz group and quantum fields}
\vsse

Field theory in Minkowski space-time has been well studied from the group point 
of view
by many authors \cite{key3},\cite{{ramond1}}.
 Here we will consider the  connection of fields of different spin   with 
the Lorentz group. 

In flat space the spin of the field is 
classified  according to the field's properties under infinitesimal
 Lorentz transformations 
\bd
\bar{x}^\alpha=\Lambda^\alpha_\beta x^\beta
=(\delta ^\alpha_\beta+\omega ^\alpha_\beta)x^\beta  \label{g1}
\ed   
with 
\bd
\omega _{\alpha\beta}=\omega_{\beta \alpha} \nonumber
\ed
which preserve the length of the coordinate vector $x^2$.

Under   Lorentz transformations the general multicomponent 
field 
\bd
F^{\alpha \beta...\lambda} \nonumber
\ed
transforms according to
\bd
F^{\alpha\beta...\lambda} 
~\mathop \to \limits^\Lambda~ 
[D(\Lambda)]^{\alpha\beta...\lambda}_{\alpha^{'}\beta^{'}...\lambda^{'}}
F^{\alpha^{'}\beta^{'}...\lambda^{'}} \label{g2}
\ed   
where
\bd
[D(\Lambda)]=1+(1/2)\omega^{\alpha\beta}\Sigma_{\alpha\beta}.\label{g3}
\ed   
In order  for the  Lorentz transformations to form a group, the antisymmetric 
$\Sigma_{\alpha\beta}$ group generators are constrained to satisfy 
\bd
\left[ {\Sigma _{\alpha \beta },\Sigma _{\gamma \delta }} \right]
=\eta _{\gamma \beta }\Sigma _{\alpha \delta }
-\eta _{\alpha \gamma }\Sigma _{\beta \delta }
+\eta _{\delta \beta }\Sigma _{\gamma \alpha }
-\eta _{\delta \alpha }\Sigma _{\gamma \beta } \label{g4}
\ed   
We may write down $\Sigma _{\alpha \beta }$ for different types of fields: 

1) It is  easy to see that the  scalar field  has the following form of $\Sigma$
\bd
\Sigma _{\alpha \beta }=0 \label{g5}
\ed
2) A vector field $F^\alpha$ transforms as 
\bd
F^{\alpha} 
~\mathop \to \limits^\Lambda~ 
\Lambda^{\alpha}_{\alpha^{'}}F^{\alpha^{'}} \label{g6}
\ed   
so  from (\ref{g3}) and (\ref{g1}) we get  an  expression for $\Sigma$ 
in vector case  in the form
\bd
{[\Sigma_{\alpha\beta}]^\gamma}_\delta=
\delta _\alpha ^\gamma \eta _{\beta \delta }
-\delta _\beta ^\gamma \eta _{\alpha \delta }.\label{g7}
\ed   
3) For a  spinor field it may be written as  
\bd
{[\Sigma_{\alpha\beta}]^\gamma}_\delta
=(1/ 4){[\gamma _\alpha ,\gamma _\beta]^\gamma}_\delta, \label{g8}
\ed
where $\gamma$ are Dirac matrices.

Now we can  generalize this concept   to curved  space-time. 

\section{Fields in curved space-time}
\vsse

Let us consider   general coordinate transformations.
We will define a general coordinate transformation as an arbitrary
reparametrization of the coordinate system:
\bd
{x^{'}}^\mu={x^{'}}^\mu (x^\nu) \label{g9}
\ed   
Unlike Lorentz transformations, which are global space-time 
transformations, general coordinate transformations are 
local. 
Under reparametrizations, a scalar field transforms simply as
follows:
\bd
{\varphi^{'}}({x^{'}})=\varphi(x), \label{g10}
\ed  
and vectors $\partial_\mu$ and $dx^\mu$ as:
\bd
{\partial^{'}}_\mu=\frac{\partial x^\nu}{\partial{x^{'}}^\mu}\partial_\nu \nonumber
\ed   
\veb
\bd
d{x^{'}}^\mu=\frac{\partial {x^{'}}^\mu}{\partial x^\nu}dx^\nu \label{g11}
\ed   
As we can see, the transformation properties of (\ref{g10}) and (\ref{g11})
are  based on 
the set of arbitrary real $(4 \times 4)$ matrices of the group $GL(4)$.  
Now we can give the abstract definition of covariant and contravariant tensors 
and write the general form of transformation
\bd
{A^{'}}^{\mu_1...}_{\nu_1...}(x^{'})
=\frac{\partial {x^{'}}^{\mu_1}}{\partial x^{\xi_1}}...
\frac{\partial x^{\lambda_1}}{\partial {x^{'}}^{\nu_1}}...
A_{\lambda_1...}^{\xi_1...}(x)  \label{g12}
\ed   
Let us  introduce a metric tensor $g_{\mu\nu}$ which allows us to 
write the four interval 
\bd
ds^2=g_{\mu\nu}dx^\mu dx^\nu \label{g13}
\ed   
The tensor $g_{\mu\nu}$ transforms under the general coordinate transformation 
as genuine tensor. The derivative of 
the scalar field ${\partial}_\mu\varphi(x)$ is a genuine tensor,
 but the derivative of  a vector is not. To create a vector from 
${\partial}_\mu A_\nu (x)$ and $({\partial}_\mu A^\nu (x))$ we introduce new fields,
connections, that absorb unwanted terms.
The  covariant derivatives are written then as 
\bd
\nabla_\mu A_\nu=\partial_\mu A_\nu +\Gamma^\lambda _{\mu\nu} A_\lambda \nonumber
\ed   
\veb
\bd
\nabla_\mu A^\nu=\partial_\mu A_\nu -\Gamma^\nu _{\mu\lambda} A^\lambda \label{g14}
\ed   
From the restriction 
\bd
\nabla g_{\mu\nu}=0 \label{g15}
\ed   
we get the expression for connection $\Gamma ^\lambda _{\mu \nu }$:
\bd
\Gamma ^\lambda _{\mu \nu }=(1/ 2)g^{\lambda \sigma }
(\partial _\mu g_{\nu \sigma }
+\partial _\nu g_{\mu \sigma }
-\partial _\sigma g_{\mu \nu }) \label{g16}
\ed
which are  named  the Christoffel symbols.

Another very important object in curved space-time is the
Riemann curvature tensor $R^\xi _{\mu \nu \lambda}$, 
which arises from the commutator 
of the covariant derivatives
\bd
[\nabla_\mu,\nabla_\nu] A^\lambda = R^\xi _{\mu \nu \lambda} A_\xi
\label{icor1}
\ed
The Riemann  tensor is written in the form
\bd
R^\xi _{\mu \nu \lambda }=\partial _\lambda \Gamma ^\xi _{\nu \mu }
-\partial _\nu \Gamma ^\xi _{\mu \lambda }
+\Gamma ^\xi _{\sigma \nu }\Gamma ^\sigma _{\mu \lambda }
-\Gamma ^\xi _{\lambda \sigma }\Gamma ^\sigma _{\mu \nu } \label{g17}
\ed
Contracting the indicies of the Riemann tensor one can 
get the Ricci curvature tensor 
\bd
R_{\mu \nu}=R^\lambda _{\mu \nu \lambda } \label{g18}
\ed
and scalar curvature 
\bd
R=g^{\mu\nu}R_{\mu \nu} \label{g19}
\ed
The transformation properties of the volume element and the square root of the 
metric tensor are
\bd
d^4 x^{'}=Det \left(\frac{\partial {x^{'} }^\mu}{\partial x^\nu}\right)d^4x \nonumber
\ed
\veb
\bd
\sqrt{g^{'}(x^{'})}=Det \left(\frac{\partial x^\mu}{\partial {x^{'}}^\nu}\right)
\sqrt{g(x)} \label{g20}
\ed
From (\ref{g20}) we get that the product of these two is invariant 
\bd
\sqrt{g(x)}d^4 x=inv. \label{g21}
\ed
Now we can construct the  Einstein-Hilbert action  as an  invariant  of the form
\bd
S_g=-(1/2k^2)\int d^4 x\sqrt{g(x)}R, \label{g22}
\ed
and the  action for  a scalar field 
\bd
S_\varphi=-(1/2)\int \sqrt{g(x)}
\left(g^{\mu\nu}\partial_\mu \varphi\partial_\nu\varphi
+m^2\varphi^2 \right), \label{g23}
\ed
where the scalar matter couples to gravity via the interaction 
\bd
\sqrt{g(x)}g^{\mu\nu}\partial_\mu \varphi\partial_\nu\varphi
\sim g^{\mu\nu}T_{\mu\nu}. \nonumber
\ed
The coupling of the gravitational field to the vector field 
is also straightforward
\bd
S_v=-(1/4)\int \sqrt{g(x)}g^{\mu\nu}g^{\lambda \xi}
F^a_{\mu\lambda}F^a_{\nu\xi} \label{g24}
\ed
For the  gauge-fixing term the generalization to curved space-time is
expressed by
\bd
S_{gf}=-(1/2)\alpha^{-1} \int d^4x\sqrt{g(x)}
\left(\nabla_\mu A^\mu \right)^2 \label{g25}
\ed
For  the  ghost term it is of  the form:
\bd
S_{gh}= \int d^4x\sqrt{g(x)}
\left(-\partial_\mu c^\alpha \partial^\mu \bar{c}^\alpha\right) \label{g26}
\ed
However, the coupling of gravity to spinor fields leads to a difficulty,
 because there are no finite dimensional spinorial representations of $GL(4)$.
This prevents a naive incorporation of spinors into general relativity.
The method we may use for these constructions involves vierbein formalism.

\section{Spinors in General relativity}
\vsse

The  vierbein (tetrad) formalism utilizes  the fact that we can construct  a flat 
tangent space to the curved manifold and introduce the spinorial
representation of the Lorentz group in each point of the manifold in this 
tangent space. Spinors can then be defined at any point on the curved 
manifold only if they transform within the flat tangent space. 

Let us erect normal coordinates ${y^\alpha_{(X)}}$ at each  
point $X$ of the space-time manifold $M$.
To preserve the connection with the previous sections we will label the flat 
tangent indices with letters $\alpha$, $\beta$, $\gamma$, $\delta$,...  from the
beginning of the Greek alphabet, as we did it before, and general coordinate 
transformation indices with letters $\lambda$, $\mu$, $\nu$, $\xi$,...    
from the end of the Greek alphabet. 

Introduce the vierbein as the mixed tensor (matrix)
\bd
h^\alpha_\mu(X)=\left(\frac{\partial y^\alpha_{(X)}}{\partial x^\mu}\right)_{x=X}
~~~\alpha=0,1,2,3 \label{g27}
\ed
Note that the label $\alpha$ refers to the local 
inertial frame associated with 
  normal coordinates $y^\alpha_{(X)}$ at the point $X$, while $\mu$
is associated with the general coordinate system $\{x^\mu \}$.

The inverse of this matrix is given by $h_\alpha^\mu(X)$.
\bd
h^\alpha_\mu h_\beta^\mu=\delta^{ab} \label{g28}
\ed
The vierbein can be viewed as the "square root" of 
the metric tensor $g_{\mu\nu}$ :
  \bd
g_{\mu\nu}=h^\alpha_\mu h^\beta_\nu \eta_{\alpha\beta} \label{g29}
\ed
 For general coordinate transformations $x^\mu=x^\mu (x^\mu{'})$ we can consider
the effect of changing the $x^\mu$ while leaving the $y^\alpha_{(X)}$
fixed. Then the verbein transforms as 
\bd
h^\alpha_\mu
\to{h^{'}}^\alpha_\mu= 
\frac{\partial {x^{'}}^\mu}{\partial x^\mu} h^\alpha_\mu \label{g30}
\ed
We also can transform the $y^\alpha_{(X)}$ arbitrarily at each point $X$
\bd
y^\alpha_{(X)}
\to {y^{'}}^\alpha_{(X)}= 
{\Lambda(X)}^\alpha_\beta y^\beta_{(X)} \label{g31}
\ed
In this case $h^\alpha_\mu (X)$ transforms as a Lorentz covariant vector
\bd
h^\alpha_\mu (X)
\to {h^{'}}^\alpha_\mu (X)= 
{\Lambda(X)}^\alpha_\beta h^\beta_\mu (X) \label{g32}
\ed
which leaves the metric  (\ref{g29}) invariant.

If a  covariant vector $A_\mu$ is contracted into $h_\alpha^\mu$,
the resulting object  
\bd
A_\alpha=h_\alpha^\mu A_\mu \label{g33} 
\ed
transforms as a collection of four scalars under a  general coordinate 
transformations, while under  local Lorentz transformations (\ref{g1}) it
behaves as a vector.

Thus, by use of tetrads, one can convert general tensors into local,
 Lorentz-transforming tensors, shifting the additional space-time 
dependence into the tetrads. 

Now we may construct the generally covariant Dirac equation.

We introduce a spinor $\psi(x)$ that is defined  as a scalar under general 
coordinate transformations and an ordinary spinor under flat tangent 
space Lorentz transformation:

Coordinate transformations:  $\psi \to \psi$

Lorentz transformations: $\psi \to  D[\Lambda(x)]\psi$

It is important to note that we have introduced local Lorentz transformations
in flat tangent space, so $\omega_{\alpha\beta}$ is a function of the space-time.
This means that the derivative of the spinor is no longer a genuine tensor.
Therefore we must introduce a connection field $\omega_\mu^{\alpha\beta}$
that allows us to gauge the Lorentz group. The covariant derivative for 
gauging the Lorentz group may be written as 
\bd
\nabla_\mu\psi=(\partial_\mu+(1/2)\Sigma_{\alpha\beta}
\omega_\mu^{\alpha\beta})\psi \label{g34}
\ed
Let $\{\gamma^\alpha\}$ be a set of Dirac matrices with
\bd
\{\gamma_\alpha,\gamma_\beta\}=2\eta_{\alpha\beta} \label{g35}
\ed
in the tangential space-time.

The Dirac matrices $\gamma^\alpha$ can be contracted with vierbeins:     
\bd
h_\alpha^\mu(x)\gamma_\alpha=\gamma_\mu(x) \label{g36}
\ed
Then 
\bd
\{\gamma_\mu(x),\gamma_\nu(x)\}=2g_{\mu\nu}(x) \label{g37}
\ed
In the result (\ref{g32}) and (\ref{g34}) the generally covariant Dirac 
equation is given by
\bd
(i\gamma^\mu(x)\nabla_{\mu,x}+m)\psi(x)=0 \label{g38}
\ed
and hence the action for Dirac particle interacting with gravity is given by: 
\bd
L=\sqrt{g}\bar{\psi}(x)(i\gamma^\mu(x)\nabla_{\mu,x}+m)\psi(x) \label{g39}
\ed
where $\sqrt{g}=Det(h^\alpha _\mu)$.

We can construct a new, alternative, 
version of the curvature tensor by taking the commutator of two covariant 
derivatives:
\bd
\left[ {\nabla _\mu ,\nabla _\nu } \right]\psi =
{1 \over 2}R^{\alpha \beta }_{\mu \nu }\Sigma _{\alpha \beta }\psi \label{g40} 
\ed
Written out, this curvature tensor is generally covariant in $\mu$, $\nu$,
but flat in $\alpha$, $\beta$:
\bd
R_{\mu \nu }^{\alpha \beta }=\partial _\mu \omega _\nu ^{\alpha \beta }
-\partial _\nu \omega _\mu ^{\alpha \beta }
+\omega _{\mu \lambda }^\alpha \omega _\nu ^{\beta \lambda }
-\omega _{\nu \lambda }^\alpha \omega _\mu ^{\beta \lambda } \label{g41}
\ed
At this point, the spinor connection $\omega _{\mu \lambda }^\alpha $
is still an independent field.
Covariant derivative of the object with two different kinds of indicies 
is written as:
\bd
\nabla _ \mu  A^\alpha _\mu =
\partial _\mu  A^\alpha _\nu+ \Gamma^\xi_{\mu \nu} A^\alpha_ \xi
+{\omega ^\alpha}_{\mu \beta} A^ \beta _\nu \label{g42}
\ed
The external constraint
\bd
\nabla _ \mu  h^\alpha _\mu =
\partial _\mu  h^\alpha _\nu+ \Gamma^\xi_{\mu \nu} h^\alpha_ \xi
+{\omega ^\alpha}_{\mu \beta} h^ \beta _\nu=0 \label{g43}
\ed
 helps to express the spin connection through veirbeins
\bd
\omega _\mu ^{\alpha \beta }={1 \over 2}h^{\alpha \nu }
(\partial _\mu h_\nu ^\beta -\partial _\nu h_\mu ^\beta ) \nonumber
\ed
\veb
\bd
+{1 \over 4}h^{\alpha \nu }h^{\beta \xi }
(\partial _\xi h_{\nu \gamma }-\partial _\nu h_{\xi \nu })h_\mu ^\nu
 -(\alpha \leftrightarrow \beta ) \label{g44}
\ed
After this preliminary work we can study properties of bosonic and fermionic
fields in an  external gravitational field.

%% file: ikcur1h.tex
\chapter{BOSE FIELDS }
\centerline{\Large \bf  IN CURVED SPACE-TIME}
\vs

As we know already the inclusion of the interaction with gravitational field 
(theory formulated in curved space-time) is accomplished by   replacing of the  
 partial derivative by the  covariant derivative $\partial_\mu \to \nabla_\mu$. 
It is necessary 
to ensure that the Lagrangian is the scalar under the general coordinate 
transformation. Integration is performed over the invariant volume. 
This procedure, based on   general -coordinate covariance, is called the 
minimal interaction for gravity and leads to the action (\ref{g23}). 

However, general coordinate covariance does not forbid   adding to the 
Lagrangian invariant terms which are vanishing in flat space-time.
Such terms describe the non-minimal interaction with gravity.
Therefore the theory under such consideration can be written in the form
\bd
S=\int d^4x\sqrt{-g}\left(L(\phi,\nabla_\mu \phi) 
+ non-min.~~int.\right) \label{h1}
\ed   
From dimensional analysis of  $R$ and $\phi$  conclude that
the  term, describing non-minimal interaction of matter field with 
gravitational field may be written in the  form: 
\bd
(1/2)\xi R\varphi^2 \label{h2}
\ed   
where $\xi$ is a dimensionless parameter (non-minimal coupling constant)
\cite{key17}.

If $m=0$ and $\xi=1/6$ then the action is invariant not only under 
 general-coordinate transformation 
but also under conformal transformations \\
\cite{key16}
\bd
g_{\mu\nu}^{'}(x)=e^{2\chi(x)}g_{\mu\nu},~~
{\varphi^{'}}(x)=e^{\chi(x)}\varphi(x)  \label{h3}
\ed   
where $\chi(x)$ is an arbitrary scalar field 
(parameter of the conformal transformation). 

In the result we will have the action in the form
\bd
S_\varphi=-(1/2)\int d^4x\sqrt{-g}\varphi(x)
\left(-\Box_x+m^2+\xi R\right)\varphi(x) \label{h4}
\ed
where the d'Alembertian operator is $\Box_x=g_{\mu\nu}(x){\nabla^x}_\mu {\nabla^x}_\nu
=\partial_\mu\partial^\mu$.

The generating functional will be 
\bd
Z[R,J]\propto (Det G)^{-1/2}\times \nonumber
\ed
\veb
\bd
\times \exp\left\{ -(i/2)\int d^4 x
g(x)^{1/2} \int d^4 x^{'} {g(x^{'})}^{1/2}J(x)G(x,x^{'})J(x^{'})\right\} \label{h5}
\ed
where $G(x,x^{'})$ is the  Green's function of the scalar field which 
is described by the equation:
\bd
g^{(1/2)}(x)\left(-\Box_x+m^2+\xi R\right)G(x,x^{'})=
\delta(x-x^{'}) \label{h6}
\ed
So, for computation of the generating functional (\ref{h5}) we have to 
define   the  Green's function of the scalar field  
in curved space-time from the equation (\ref{h6}). 

\section{Momentum-space representation of}
\lum
\hspace{20mm}{\Large \bf  the  bosonic  Green's function }
\vsse

In curved space-time we cannot solve the equation (\ref{h6}) and 
compute the  Green's function $G(x,x^{'})$ 
for arbitrary points $x$ and $x^{'}$ of the manifold. 
But we can do so  in  the  particular case of the limit 
of coincidence $(x \simeq x^{'})$. This limit  gives us a possibility to
find the $(Det)$ of the Green's functions to get the effective action.
Thus we will treat the problem of the Green's functions calculations 
in the limit of coincidence.

Let us  select a point of the space-time manifold $x^{'}$ 
 and construct the  tangential space at  this point as we did  
 in the previous section. 
Let  any point $x$ of the manifold have 
the   normal  coordinate $y^\alpha (x)$. This coordinate may 
be treated as a  vector in the  tangent space with the origin at the point $x^{'}$.  
In this tangent space the norm of the vector $y^\alpha (x)$ will be
$(y,y)=y^\alpha (x)y^\beta (x)\eta_{\alpha\beta}$, and
the metric of the manifold may be  written as \cite{petrov1}
  \bd
g_{\mu \nu }(x)=\eta _{\mu \nu }-{1 \over 3}
R_{\mu \alpha \nu \beta }y^\alpha y^\beta
 -{1 \over 6}R_{\mu \alpha \nu \beta ;\gamma }y^\alpha y^\beta y^\gamma
+...\label{h7} 
\ed
and
\bd
g(x)=1-{1 \over 3}R_{\alpha \beta }
y^\alpha y^\beta -{1 \over 6}
R_{\alpha \beta ;\gamma }y^\alpha y^\beta y^\gamma  \nonumber
\ed
\veb  
\bd
+\left( {{1 \over {18}}R_{\alpha \beta }
R_{\gamma \delta }-{1 \over {90}}R^\nu _{\alpha \beta \lambda }
R^\lambda _{\gamma \delta \nu }-{1 \over {20}}
R_{\alpha \beta ;\gamma \delta }} \right)y^\alpha y^\beta y^\gamma y^\delta
 +... \label{h8}
\ed   
where the coefficients are calculated at  the  $(y=0)$   origin of 
the coordinate system.

The second derivative for the  scalar field  is
\bd
\nabla ^{x\mu }\nabla _\mu ^x=\eta ^{\alpha \beta }
\partial _\alpha \partial _\beta 
+{1 \over 3}{{{R_\alpha }^\delta }_\beta }^\gamma y^\alpha y^\beta 
\partial ^x_\delta\partial ^x_\gamma 
 -{2 \over 3}R_\beta ^\alpha y^\beta \partial ^x_\alpha +... \label{h9}
\ed
For our calculations it is convenient to express 
the Green's function $G(x,x^{'})$ as: 
\bd
G(x,x^{'})=g^{-1/4}(x)\Im (x,x^{'})g^{-1/4}(x^{'}) \nonumber
\ed
\veb
\bd
=g^{-1/4}(x)\Im (x,x^{'}) \label{h10}
\ed
Substituting (\ref{h8}),(\ref{h9}) and (\ref{h10}) into (\ref{h6}), we find
\bd
\eta ^{\alpha \beta }\partial _\alpha \partial _\beta \Im
 -\left[ {m^2+\left( {\xi -{1 \over 6}} \right)R} \right]\Im
-{1 \over 3}{R_\alpha }^\beta y^\alpha \partial _\beta \Im \nonumber
\ed
\veb
\bd
 +{1 \over 3}{{{R_\alpha }^\beta }_\gamma }^\delta y^\alpha y^\gamma
 \partial _\beta \partial _\delta \Im 
-\left( \xi -{1 \over 6} \right)R_{;\alpha }y^\alpha \Im \nonumber
\ed
\veb
\bd
+\left( -{1 \over 3}{{R_\alpha}^\beta }_{;\gamma } 
+{1 \over 6}{{R_\alpha} \gamma }^{;\beta } 
\right)y^\alpha y^\gamma \partial _\beta \Im  \nonumber
\ed
\veb
\bd
+{1 \over 6}{{{{R_\lambda} ^\gamma} _\alpha }^\zeta}_{;\beta }
 y^\lambda y^\alpha y^\beta \partial _\gamma \partial _\zeta  \Im
-{1 \over 2}\left( {\xi -{1 \over 6}} \right)
R_{;\alpha \beta }y^\alpha y^\beta \Im + \nonumber
\ed
\veb
\bd
\left( -{1 \over {30}}{R_\alpha }^\beta R_{\beta \gamma }
+{1 \over {60}}{{{R_\alpha }^\beta} _\gamma }^\delta R_{\beta \delta }\right. \nonumber
\ed
\veb
\bd
\left.+{1 \over {60}}{R^{\beta \chi \delta }}_\alpha 
R_{\beta \chi \delta \gamma }
-{1 \over {120}}R_{;\alpha \gamma }
+{1 \over {40}}\Box R_{\alpha \gamma } \right) \Im \nonumber
\ed
\veb
\bd
+\left( -{3 \over {20}}{R^\delta} _{\alpha ;\beta \gamma }
+{1 \over {10}}{{R_{\alpha \beta }}^{;\delta }}_\gamma
 -{1 \over {60}}{{{R^\chi} _\alpha }^\delta} _\beta R_{\chi \gamma }\right. \nonumber
\ed
\veb
\bd
\left.+{1 \over {15}}{R^\chi} _{\alpha \lambda \beta }
{{{R_\chi }^\delta} _\gamma }^\lambda  \right)
y^\alpha y^\beta y^\gamma \partial _\delta  \Im \nonumber
\ed
\veb
\bd
+\left( {1 \over {20}}{{{R^\kappa} _\alpha} ^\chi} _{\beta ;\gamma \delta }
+{1 \over {15}}{R^\kappa }_{\alpha \lambda \beta }
{{{R^\lambda} _\gamma} ^\chi} _\delta  \right)
y^\alpha y^\beta y^\gamma y^\delta \partial _\kappa \partial _\chi \Im
 =-\delta (y), \label{h11}
\ed
where $y^\alpha$ are the coordinates of the point $x$ 
and $\partial_\alpha \Im=(\partial/\partial y^\alpha)$.

We have retained only terms with coefficients involving four 
derivatives of the metric.
These contributions give the ultraviolet divergences that 
arise in the course of renormalization. 

In normal coordinates  with origin at $x^{'}$, $\Im (x,x^{'})$ is a function
of $y$ and $x^{'}$
\bd
\Im (x,x^{'})=\Im (y,x^{'}) \label{h12}
\ed
where $y$ belongs to the small region around $x^{'}$.
In this way the equation (\ref{h6}) may be solved recursively.
Namely, we will introduce the momentum space 
associated with  the point $x^{'}$ $(y=0)$ by making the $n$-dimensional Fourier 
transformation:
\bd
\Im (x,x^{'})=\int \frac{d^n k}{(2\pi)^n}\Im(k)\exp(iky) \label{h13}
\ed   
where $ky=k_\alpha y^\alpha=\eta^{\alpha \beta}k_\alpha y_\beta$ 
and expanding  $\Im (k)$ in a  series:
\bd
\Im (k)=\Im_0 (k)+\Im_1 (k)+\Im _2(k)+...\label{h14}
\ed
or
\bd
\Im_i (x,x^{'})=\int \frac{d^n k}{(2\pi)^n}\Im_i(k) \exp(iky),
~~~i=0,1,2,...\label{h15}
\ed   
where we will assume that the coefficients $\Im (k)$ have   geometrical 
coefficients involving  $i$ derivatives of the mertic. 

On  dimensional grounds, $\Im_i(k)$  are the order $k^{-(2+i)}$,
 so that (\ref{h14}) 
is an asymptotic expansion of $\Im (k)$ in large $k$ (small $y$). 
 
Inserting (\ref{h13}) into (\ref{h11}) we get that the lowest 
order solution ($\sim O(y^2)$) is
\bd
\Im_0 (k)=(k^2+m^2)^{-1} \label{h16}
\ed
and 
\bd
\Im _1(k)=0 \label{h17}
\ed
The function $\Im _2(k)$ ($\sim O(y^4)$) may be found from
\bd
(\eta^{\alpha \beta}\partial_\alpha \partial_\beta-m^2)\Im _2
-\left( {\xi -{1 \over 6}} \right)R \Im_0 \nonumber
\ed
\veb
\bd
-{1 \over 3}{R_\alpha }^\beta y^\alpha \partial _\beta \Im_0
+{1 \over 3}{{{R_\alpha }^\beta }_\gamma }^\delta y^\alpha y^\gamma
 \partial _\beta \partial _\delta \Im_0=0 \label{h18}
\ed
Using (\ref{h13}) one can get that the last two 
terms  of (\ref{h18}) cancel each other.

In another way we can consider that $\Im _0$ is Lorentz invariant of
the form 
\bd
\Im (y)=y_\alpha y^\alpha=\eta^{\alpha \beta}y_\alpha y_\beta \label{h19}
\ed
Then, inserting (\ref{h19}) into (\ref{h18}) we get:
\bd
-{1 \over 3}{R_\alpha }^\beta y^\alpha \partial _\beta \Im_0(y)
+{1 \over 3}{{{R_\alpha }^\beta }_\gamma }^\delta y^\alpha y^\gamma
 \partial _\beta \partial _\delta \Im_0(y)\equiv 0, \label{h20}
\ed
and
\bd
(\eta^{\alpha \beta}\partial_\alpha \partial_\beta-m^2)\Im _2(y)
-\left( {\xi -{1 \over 6}} \right)R \Im_0(y)=0 \label{h21}
\ed
Therefore
\bd
\Im _2(k)=\left( {1 \over 6}-\xi \right)R (k^2+m^2)^{-2}. \label{h22}
\ed
The Lorentz invariance of $\Im_0(y)$ leads to further  simplifications 
of (\ref{h11}).
Namely, the contributions
\bd
\left( -{1 \over 3}{{R_\alpha}^\beta }_{;\gamma } 
+{1 \over 6}{{R_\alpha} \gamma }^{;\beta } 
\right)y^\alpha y^\gamma \partial _\beta \zeta  \Im_0(y)
+{1 \over 6}{{{{R_\lambda} ^\gamma} _\alpha }^\zeta}_{;\beta }
 y^\lambda y^\alpha y^\beta \partial _\gamma \partial _\zeta  \Im_0(y) \equiv 0
\label{addh22}
\ed 
are eliminated.

In the same way
\bd
\left( -{3 \over {20}}{R^\delta} _{\alpha ;\beta \gamma }
+{1 \over {10}}{{R_{\alpha \beta }}^{;\delta }}_\gamma
 -{1 \over {60}}{{{R^\chi} _\alpha }^\delta} _\beta R_{\chi \gamma }\right. \nonumber
\ed
\veb
\bd
\left.+{1 \over {15}}{R^\chi} _{\alpha \lambda \beta }
{{{R_\chi }^\delta} _\gamma }^\lambda  \right)
y^\alpha y^\beta y^\gamma \partial _\delta  \Im_0(y) \nonumber
\ed
\veb
\bd
+\left( {1 \over {20}}{{{R^\kappa} _\alpha} ^\chi} _{\beta ;\gamma \delta }
+{1 \over {15}}{R^\kappa }_{\alpha \lambda \beta }
{{{R^\lambda} _\gamma} ^\chi} _\delta  \right)
y^\alpha y^\beta y^\gamma y^\delta \partial _\kappa \partial _\chi \Im_0(y)
\equiv 0 \label{h23}
\ed
and we have the following equation for $\Im (y)$ 
to the fourth order in derivatives of the metric:
\bd
\eta ^{\alpha \beta }\partial _\alpha \partial _\beta \Im (y)
 -\left[ {m^2+\left( {\xi -{1 \over 6}} \right)R} \right]\Im(y) \nonumber
\ed
\veb
\bd
-\left( \xi -{1 \over 6} \right)R_{;\alpha }y^\alpha \Im(y) \nonumber
\ed
\veb 
\bd
-(1/2)\left( \xi -{1 \over 6} \right)R_{;\alpha\beta }y^\alpha y^\beta\Im(y) \nonumber
\ed
\veb
\bd
\left( -{1 \over {30}}{R_\alpha }^\beta R_{\beta \gamma }
+{1 \over {60}}{{{R_\alpha }^\beta} _\gamma }^\delta R_{\beta \delta }\right. \nonumber
\ed
\veb
\bd
\left.+{1 \over {60}}{R^{\beta \chi \delta }}_\alpha 
R_{\beta \chi \delta \gamma }
-{1 \over {120}}R_{;\alpha \gamma }
+{1 \over {40}}\Box R_{\alpha \gamma } \right) \Im(y)=-\delta(y) \label{h24}
\ed
Substitution of $\Im_2(k)$ (\ref{h22}) instead of $\Im_0(k)$ in 
the identity (\ref{h20}) 
does not change it, thus we can suggest that $\Im_2(k)$ is Lorentz invariant too
and it can be wtitten as $\Im_2(y) \sim (y^\alpha y_\alpha)^2$.  
The equation (\ref{h11}) is simplified to 
\bd
\left[k^2+m^2+\left( \xi -{1 \over 6} \right)R
+i\left( \xi -{1 \over 6} \right)R_{;\alpha}\partial^\alpha \right]\Im (k)+ \nonumber
\ed
\veb
\bd
+\left[-(1/2)\left( \xi -{1 \over 6} \right)R_{;\alpha\beta }
 -{1 \over {30}}{R_\alpha }^\beta R_{\beta \gamma }
+{1 \over {60}}{{{R_\alpha }^\beta} _\gamma }^\delta R_{\beta \delta }\right. \nonumber
\ed
\veb
\bd
\left.+{1 \over {60}}{R^{\beta \chi \delta }}_\alpha 
R_{\beta \chi \delta \gamma }
-{1 \over {120}}R_{;\alpha \gamma }
+{1 \over {40}}\Box R_{\alpha \gamma } \right]
\partial^\alpha \partial^\beta \Im(k)=1 \label{h25}
\ed
where 
\bd
\partial ^\alpha\Im (k)=\partial\Im (k)/ \partial k_\alpha \nonumber
\ed
Making a further recurrent process with this equation we get
\bd
\Im_3(k)=0 \label{h26}
\ed
and
\bd
\Im _4 (k)=i\left(\frac{1}{6}-\xi\right)R_{;\alpha}(k^2+m^2)^{-1}
\partial^\alpha (k^2+m^2)^{-1}+ \nonumber
\ed
\veb
\bd
+\left(\frac{1}{6}-\xi\right)^2 R^2 (k^2+m^2)^{-3}+ \nonumber
\ed
\veb
\bd
+a_{\alpha \beta}(k^2+m^2)^{-1}
\partial^\alpha \partial^\beta (k^2+m^2)^{-1} \label{h27}
\ed
where
\bd
a_{\alpha \beta}=(1/2)\left( \xi -{1 \over 6} \right)R_{;\alpha\beta }
 +{1 \over {30}}{R_\alpha }^\beta R_{\beta \gamma }
-{1 \over {60}}{{{R_\alpha }^\beta} _\gamma }^\delta R_{\beta \delta } \nonumber
\ed
\veb
\bd
-{1 \over {60}}{R^{\beta \chi \delta }}_\alpha 
R_{\beta \chi \delta \gamma }
+{1 \over {120}}R_{;\alpha \gamma }
-{1 \over {40}}\Box R_{\alpha \gamma } \label{h28}
\ed
Now we can write the equation for the  Green 's function in a convenient form. 

Let us introduce  useful equations:
\bd
(k^2+m^2)^{-1}\partial^\alpha (k^2+m^2)^{-1}\equiv
(1/2)\partial^\alpha (k^2+m^2)^{-2} \nonumber
\ed
and
\bd
(k^2+m^2)^{-1}\partial^\alpha \partial^\beta (k^2+m^2)^{-1}\equiv
(1/3)\partial^\alpha \partial^\beta (k^2+m^2)^{-2} \nonumber
\ed
\veb
\bd
-(2/3)\eta_{\alpha \beta}(k^2+m^2)^{-3} \label{h29}
\ed
Using (\ref{h29}) we may write $\Im(k)$ as
\bd
\Im(k) =(k^2+m^2)^{-1}+\left(\frac{1}{6}-\xi\right)R(k^2+m^2)^{-2} \nonumber
\ed
\veb
\bd
(i/2)\left(\frac{1}{6}-\xi\right)R_{;\alpha}(k^2+m^2)^{-2}
\partial^\alpha (k^2+m^2)^{-1} \nonumber
\ed
\veb
\bd
+(1/3)a_{\alpha \beta}\partial^\alpha \partial^\beta (k^2+m^2)^{-2} \nonumber
\ed
\veb
\bd
+\left[\left(\frac{1}{6}-\xi\right)^2R^2
-(2/3)a^\lambda_\lambda \right](k^2+m^2)^{-3} \label{h30}
\ed
Inserting the  last equation into (\ref{h13})  we get:
\bd
\Im(x^{'},y)=\int \frac{d^nk}{(2 \pi)^n} \exp (iky)
\left[1+\gamma_1(x^{'},y)\left(-\frac{\partial}
{\partial m^2}\right)\right. \nonumber
\ed
\veb
\bd
\left.+ \gamma_2(x^{'},y)\left(-\frac{\partial}
{\partial m^2}\right)^2\right](k^2+m^2)^{-1}  \label{h31}
\ed
where, to the fourth order in derivatives of the metric, the coefficients are:
\bd
\gamma_1(x^{'},y)=\left(\frac{1}{6}-\xi\right)R
+(1/2)\left(\frac{1}{6}-\xi\right)R_{;\alpha}y^\alpha
-(1/3)a_{\alpha \beta} y^\alpha y^\beta; \nonumber
\ed
\veb
\bd
\gamma_2(x^{'},y)=(1/2)\left(\frac{1}{6}-\xi\right)^2R^2
-(1/3)a^\lambda_\lambda.  \label{h32}
\ed 
The expression for the  Green's function (\ref{h10}) will be then \cite{bunch1}
\bd
G(x^{'},y)=g^{-1/2}(y) \int \frac{d^n k}{(2 \pi)^n} \exp(iky) \times \nonumber
\ed
\veb
\bd
\times \sum \limits_{j=0}^{2}\gamma_j(x^{'},y)\left(-\frac{\partial}
{\partial m^2}\right)^j(k^2+m^2)^{-1}  \label{h33}
\ed
The equation (\ref{h33}) is very important for further calculations.

\section {The Green's function  and the }
\lum
\hspace{20mm}{\Large \bf  Schwinger-DeWitt method}
\vsse

We have found already that the  geometrical quantities enter directly 
into the structure of  Green's functions for arbitrary fields through
the covariant d'Alembertian operator and non-minimal connection. 
Now we will  treat the equation for the Green's function  with the  
Schwinger-DeWitt method developed  for   flat space-time in 
the previous section.

Let us multiply the equation (\ref{h6}) on the left side by $g^{(1/4)}(x)$ 
and on the right by  $g^{(1/4)}(x^{'})$, and introducing 
\bd
\Im(x,x^{'})=g^{(1/4)}(x)G(x,x^{'})g^{(1/4)}(x^{'})  \label{h34}
\ed
we will rewrite this equation in the form
\bd
\left(-\partial_\mu \partial^\mu +m^2+\xi R\right)\Im(x,x^{'})=
\delta(x,x^{'}),  \label{h35}
\ed
where  $\delta(x,x^{'})=g^{-1/2}(x)\delta(x-x^{'})$ 
is scalar with respect to
general coordinate transformation, 
and product of $\delta$ -functon is
\bd
(1,\delta(x,x^{'}))=\int d^4x \delta(x,x^{'})=1 \label{addh35}
\ed
For simplicity we put $\xi=0$ and get;
\bd
\left(-\Box_x +m^2)\right)\Im(x,x^{'})=\delta (x,x^{'}),  \label{h36}
\ed
or, in operator form
\bd
\hat{F}\Im =1  \label{h37}
\ed
where $F$ is matrix operator.

Let us introduce the representation for $\Im$ in the form
\bd
\Im (x,x^{'};s)=i<x|\int \limits _0 ^\infty ds \exp(is\hat{F})|x^{'}>
=i \int\limits_0^\infty ds f(x,x^{'};s) \exp(-im^2 s)  \label{h38}
\ed
We can get from (\ref{h36}),  that the  function $f(x,x^{'};s)$ is the solution 
of the equation
\bd
\Box_x f(x,x^{'};s)=i\frac{\partial}{\partial s}f(x,x^{'};s)  \label{h39}
\ed
where information about space-time structure 
is included in the  d'Alembertian.

We may turn this equation into an elliptic one by  rewriting 
the equation for amplitude with the replacement $x^{(0)}=ix^{(4)}$
and $s=it$.
We will have
\bd
\frac{\partial}{\partial t}f(x,x^{'};t)=\tilde{\Box_x }f(x,x^{'};t)  \label{h40}
\ed
One can write the solution of the (\ref{h38}) as a simple 
expansion of the  solution  for flat space-time. This solution is
\bd
f(x,x^{'};\bar{s})=(4\pi \bar{s})^{-n/2}
\exp(-|x-x^{'}|^2/4\bar{s}) \label{h41} 
\ed
Returning to the initial variables we get
\bd
f(x,x^{'};s)=(4\pi is)^{-n/2}\exp(-|x-x^{'}|^2/4is) \label{h42} 
\ed
In curved space-time we may expand this solution to a  local asymptotic expansion 
(for $x \simeq x^{'}$ and $s\simeq 0$)
\bd
f(x,x^{'};s) \sim (4 \pi is)^{(-n/2)}\exp(-\sigma(x,x^{'})/2is) 
\sum\limits_{j=0}^\infty \gamma_j(x,x^{'}) (is)^j \label{h43} 
\ed
where $\sigma(x,x^{'})$ is the so-called geodesic interval (half square of geodesic 
distance between points $x$ and $x^{'}$).
In particular 
\bd
f(x,x;s) \sim (4 \pi is)^{(-n/2)} 
\sum\limits_{j=0}^\infty \gamma_j(x) (is)^j \label{h44} 
\ed
The explicit form of $f_j(x,x^{'})$ can be calculated recursively \cite{brown1}\\
 \cite{dt2}, \cite{key17}
 
In the limit of coincidence $x \to x^{'}$ one finds:
\bd
\gamma_0(x^{'})=1;~~\gamma_1(x^{'})=\left(1/6-\xi\right)R; \nonumber
\ed
and
\bd
\gamma_2(x^{'})=\left(1/6-\xi\right)^2 R^2-(1/3)a^\lambda_\lambda. \label{h45}
\ed
From the equations (\ref{h38}), (\ref{h43}) and (\ref{h34}) 
we get the explicit expression for 
Green's function in the Schwinger-DeWitt representation:
\bd
G_{SD}(x,x^{'}) =\frac{i\Delta ^{1/2}(x,x^{'})}{(4 \pi)^{n/2}}
\int\limits_0^\infty ids 
(is)^{(-n/2)}\times \nonumber
\ed
\veb
\bd
\times \exp\left(-ism^2-\sigma(x,x^{'})/2is\right) 
\sum\limits_{j=0}^\infty \gamma_j(x,x^{'}) (is)^j \label{h46} 
\ed
where 
\bd
\Delta (x,x^{'})=-g(x)^{-1/2}det[\partial_\alpha 
\partial_\beta \sigma(x,x^{'})]g(x)^{-1/2} \label{h47}
\ed
is the Van  Vleck determinant 
(in normal coordinates about $x^{'}$ this 
determinant is reduced to $g^{-1/2}(y)$ (\ref{h33})). 

\section {Connection between the two methods}
\vsse

Let us put
\bd
(k^2+m^2)^{-1}=\int \limits_0 ^\infty ids \exp[-is(k^2+m^2)] \label{h48}
\ed
Then integration over the momentum in (\ref{h33}) leads to
\bd
\int \frac{d^n k}{(2 \pi)^n}  \exp[-is(k^2+m^2)+iky]
=i(4 \pi is)^{n/2}\exp(-ism^2-\sigma/2is) \label{h49}
\ed
The resulting  equation for the Green's function will be
\bd
G(y,x^{'})=\frac{i}{(4 \pi)^{n/2}}g(y)^{-1/2}\int \limits_0 ^\infty 
\frac{ds}{(is)^{n/2}}\times  \nonumber
\ed
\veb
\bd
\times \exp\left[ -is(k^2+m^2)+iky \right]
F (x^{'},y,is) \label{h50}
\ed
where
\bd
F (x^{'},y,is)=\sum\limits_{j=0}^2 \gamma_j (x^{'},y)(is)^j \nonumber
\ed
Comparision with the expression of the  Green's function in the form of
(\ref{h46}) gives $\Delta(x,x^{'})=g^{-1/2}(y)$, and
\bd
\gamma_j (x^{'},y)=\gamma_j (x,x^{'}) \label{h51}
\ed 
In this chapter we considered two methods for computing the  Green's function
of a scalar field in curved space-time and we prepared the basis for future finite 
temperature calculations.

%% file: ikfin1i.tex
\chapter{ FINITE TEMPERATURE BOSONS}
\centerline{\Large \bf  IN CURVED SPACE-TIME}
\vs

In the previous section we developed a   mathematical formalism 
which is a   convenient tool for the 
description of thermal  Bose gas in curved space-time.
In this section we  can consider the ensemble of bosons interacting 
with gravity at finite temperature.

Let a total system  ("matter and gravity")  be 
described by the  action 
\bd
S_{tot}=S_g+S_m  \label{i1}
\ed   
The  gravitational action is 
\bd
S_g=\int d^4x\sqrt{g(x)}L_g \label{i2} 
\ed   
with Lagrangian
\bd
L_g={1 \over {16\pi G_0}}(R-2\Lambda _0)
+\alpha _0 R^2+\beta _0R_{\alpha \beta }R^{\alpha \beta }
+\gamma _0 R_{\alpha \beta \chi \delta }R^{\alpha \beta \chi \delta } \label{i3}
\ed   
and the  action for a  matter field is
\bd
S_m=\int  d^4x\sqrt{g(x)}L_{eff,m} \label{i4}  
\ed   
To write the action for the matter field  we use a generating functional $Z[J]$
\bd
Z[0]=\int D\varphi \exp\left( -(i/2)\int d^4x \sqrt{g(x)}
 \varphi(x)(-\Box_x+m^2+\xi R)\varphi(x)\right) \nonumber
\ed
\veb
\bd
\propto DetG(x,x^{'})^{1/2}\exp\left(-(i/2)(J(x),G(x,y)J(y))\right) \label{i5}
\ed   
Then the  functional $W[0]=-i\ln Z[0]$ will be
\bd
W[0]=-(i/2)\ln Det G(x,x^{'}) \nonumber
\ed
\veb
\bd
=-(i/2)\int\limits_0^\infty ids(is)^{-1}\mbox{tr}f(x,x^{'},is)\exp (-im^2 s) \label{i6}
\ed
We may connect the  Green's function and heat kernel with the equation
\bd
 G_{SD}(x,x^{'})= \int\limits _0 ^\infty ids f( x, x^{'},is)
\exp (-im^2 s)\label{i7}
\ed
and write (\ref{i6}) in the following  form
\bd
W[0]=-(i/2)\int d^4x \sqrt{g(x)} \int\limits_{m^2}^\infty dm^2
\mbox{tr} G_{SD}(x,x^{'}) \label{i8}
\ed   
From the  equation (\ref{i8}) we find an important expression  
for the  effective Lagrangian of matter field
\bd
L_{eff,m}=(-i/2)\int\limits d^4x\sqrt{g(x)}
\int  \limits_{m^2}^\infty dm^2\mbox{tr}G(x,x^{'}) \label{i9} 
\ed   
Then the effective action will be:
\bd
S_{eff}=\int d^4x \sqrt{g(x)}L_{eff}(x) \nonumber
\ed
\veb
\bd
=\int d^4x \sqrt{g(x)}\left(L_g (x)+
(-i/2)\int \limits_{m^2}^\infty dm^2 \mbox{tr} G(x,x^{'})\right) \label{i10} 
\ed   
In order to apply the usual formalism of finite temperature quantum 
field theory, we will assume that the  space-time manifold $M_{4)}$ is 
a static manifold with topology $S^{1}\times M_{3)}$ where $S^1$
refers to time coordinate and $M_{3)}$ is the spatial, three 
dimensional section of $M_{4)}$. We will choose $M_{3)}$
without   boundaries \cite{ken1}, then no surface terms 
will appar in the induced action.  

The heat kernel may be expressed as the sum of zero-temperature images \\
  \cite{i17,dow1,dow2}
\bd
f(x,x,is)=\sum\limits_{n=-\infty}^{\infty}
\frac{\exp[-\beta^2 n^2/4is]}{(4\pi is)^{1/2} }
f_{3)}(\tilde {x},\tilde{x},is) \label{i11} 
\ed   
where the sum goes from periodic restrictions and
$f_{3)}(\tilde {x},\tilde{x},is)$ is the solution 
of the three-dimensional equation
\bd
\left(i\frac{\partial}{\partial s}
-{\tilde \Box }_x+\xi R\right)f_{3)}(\tilde {x},\tilde{x},is)=0 \label{i12} 
\ed   
This solution is written in the form of a  series 
\bd
f_{3)}(\tilde {x},\tilde{x},is)=\sum \limits_{n=0}^\infty 
\gamma_j(R)(is)^j \label{i13} 
\ed
Then from the equation (\ref{i7}) in the limit of coincidence $(x \to x^{'})$
\bd
\mathop {\lim}\limits_{x \to x^{'}} G(x,x^{'})=
\mathop {\lim}\limits_{x \to x^{'}} 
\int\limits _0 ^\infty ids f( x, x^{'},is)\exp\{-im^2 s\}. \label{i14} 
\ed
we get the  finite temperature Green's function in the 
Schwinger-DeWitt representation:
\bd
\mathop {\lim}\limits_{x \to x^{'}} {G^ \beta}_{SD}(x,x^{'}) \nonumber
\ed
\veb
\bd
=\frac{i}{(4\pi)^{3/2}}\sum_{n=-\infty}^\infty \sum_{j=0}^\infty
\gamma_j(x^{'})\int_0^\infty ids (is)^{j-3/2}
\exp(-ism^2-n^2\beta^2/4is) \label{i15} 
\ed
Selecting  the  temperature independent part $(n=0)$ we find
\bd
\mathop {\lim}\limits_{x\to x^{'}}{G^ \beta}_{SD} (x,x^{'})=
G_{SD}(x^{'},x^{'})+G_{x^{'}}(\beta) \label{i16} 
\ed   
where the finite temperature contribution $G_{x^{'}}(\beta)$ is
\bd
G_{x^{'}}(\beta)=\frac{i}{(4 \pi)^2}2\sum\limits_{j=0}^\infty 
\sum \limits_{n=1} ^\infty \gamma_j(x^{'})\times   \nonumber
\ed
\veb
\bd
\times \int _0^\infty ids (is)^{j-2}
\exp(-ism^2-n^2\beta^2/4is) \label{i17} 
\ed   
and $G_{SD}(x^{'},x^{'})$ is the limit $(x \to x^{'})$ of 
the Green's function in the  Schwinger-DeWitt  representation (\ref{h46}).

Summation in (\ref{i17}) may be done with (\ref{d25}) and (\ref{d29}).

In the result we find
\bd
G_{x^{'}}(\beta)=\frac{i}{(2\pi)^2}
\sum \limits_{n=1} ^\infty \sum \limits_{j=1} ^\infty 
\gamma_j(x^{'})(\beta n/2m)^{j-3}K_{j-3}(\beta m n) \label{i18} 
\ed   
Total Lagrangian  may be written in the form;
\bd
L_{eff}(\beta)=\left(L_g (x)+
(-i/2)\int \limits_{m^2}^\infty dm^2\mbox{tr}  G_{SD} (x,x^{'})\right)+
(-i/2)\int \limits_{m^2}^\infty dm^2 G_{x^{'}}(\beta) \label{i19}  
\ed   
The first two terms give  the same geometric structure, 
and after renormalizations we will have  the Lagrangian  $\tilde{L}_g$, 
the third one  is temperature contribution 
$f(\beta)$.
\bd
L_{eff}(\beta)=\tilde{L}_g-f(\beta) \label{i20} 
\ed   
where
\bd
\tilde{L}_g=\left(L_g (x)+
(-i/2)\int \limits_{m^2}^\infty dm^2  \mbox{tr} G_{SD}(x,x^{'})\right) \label{i21}
\ed   
and
\bd
f(\beta)=(i/2)\int \limits_{m^2}^\infty dm^2G_{x^{'}}(\beta) \label{i22} 
\ed   
The last expression  can be  written as series
\bd
f(\beta)=\sum\limits_{j=0}^3\gamma_j(R)b_j(\beta m) \label{i23} 
\ed   
where
\bd
b_0 (\beta m)=-\frac{m^2}{2\pi^2 \beta^2}\sum\limits_{n=1}^\infty
(1/n^2)K_2 (\beta m n) \nonumber
\ed
\veb
\bd
b_1 (\beta m)=-\frac{2m}{4\pi^2 \beta}\sum\limits_{n=1}^\infty
(1/n)K_1 (\beta m n) \nonumber
\ed
\veb
\bd
b_2 (\beta m)=-\frac{2}{8\pi^2 }\sum\limits_{n=1}^\infty
K_0 (\beta m n) \label{i24}
\ed   
Using the results of Appendix: (\ref{bb10}), (\ref{bb12}), (\ref{bb14}),
 we may rewrite (\ref{i23}) in the form of  series [Kulikov \& Pronin 1993]
\bd
f(\beta)=b_0(\beta m)+{\gamma }_1(R)b_1(\beta m)
+{\gamma }_2(R^2)b_2(\beta m) \label{i25}
\ed   
where ${\gamma}_j(R)$ are (\ref{h32}), coefficient $b_0(\beta m)$ is
\bd
b_0=(1/\beta)\int \frac{d^3k}{(2\pi)^3} \ln (1-\exp(-\beta \epsilon)); \nonumber
\ed
and
\bd
b_j(\beta m)=\left(-\frac{\partial}{\partial m^2}\right)^j b_0(\beta m) \label{i26}
\ed   
The  expression  $b_0(\beta m)$ in (\ref{i25}) is the density of  Helmholtz free energy 
in flat space
and the following ones  are created by   corrections which are connected with 
the interaction of the heat bosons with gravity, so the equation (\ref{i23}) 
describes the Helmholtz free energy of a  Bose gas in curved space time.

%% file: ikfmioncur.tex
\chapter{FERMI  FIELDS} 
\centerline{\Large \bf IN CURVED SPACE-TIME }
\vs

\section{Momentum-space representation}
\lum
\hspace{25mm}{\Large \bf  for the  Green's function of a  fermion}
\vsse

Now we will get the  Green's function of fermions
in curved space-time to apply the above results for
computation of the effective action of the   system 
"matter field $+$ gravitational field".

As we know already, the fermionic action with Lagrangian (\ref{g39}) 
is written as
\bd
S_\psi=(i/2)\int d^4x \sqrt{g(x)}\bar{\psi}(x)
\left(i\gamma^\mu \nabla_{\mu,x}+m\right)\psi(x) \label{fer1}
\ed   
The generating functional then is
\bd
Z[\bar{\eta},\eta]=\int D \bar{\psi} D \psi \exp \left[ iS_{\psi}
 +i\left(\bar{\eta},\psi\right)+i\left(\bar{\psi},\eta \right)\right] \nonumber
\ed
\veb
\bd
\propto (Det G_F)^{-1/2}
\exp\left[-i\left(\bar{\eta}(x),S_F(x,y)\eta \right)\right] \label{fer2}
\ed   
where the  scalar product includes the square root of the metric.

Connection between the bi-spinor $G_F$ and Fermionic Green's function 
$S_F$ is\\
\cite{key16}:
\bd
S_F(x,y)=(i\gamma^\mu D_\mu+m)G_F (x,y) \label{fer3}
\ed
and the  Green's function satisfies the equation
\bd
(i\gamma^\mu D_\mu +m)S_F (x,y)=-g(x)^{-1/2}\delta (x-y) \hat{1}\label{fer4}
\ed
To solve this equation with momentum space methods we introduce 
Riemann normal coordinates \cite{petrov1} and write the spin 
connection in the form\\
\cite{panan1}
\bd
\Gamma_\mu(y)=(1/16)[\gamma_\alpha,\gamma_\beta]{R^{\alpha \beta}}_{\mu \nu}y^\nu
+O(y^2) \label{fer5}
\ed
and the vierbein field as
\bd
h^\alpha_\mu(y)=\delta^\alpha_\mu
-(1/6)\eta^{\alpha \nu}R_{\mu \nu\beta\zeta}y^\beta y^\zeta +O(y^3) \label{fer6}
\ed
The expression for  $\gamma_\mu (x^{'})$ may be written as 
\bd
\gamma_\mu(x^{'})=h^\alpha_\mu(y)\gamma_\alpha=\delta^\alpha_\mu\gamma_\alpha
-(1/6)\eta^{\alpha \nu}R_{\mu \nu\beta\zeta}y^\beta 
y^\zeta\gamma_\alpha +O(y^3) \label{fer7}
\ed
The spinor derivative appearing in Dirac's equation is written as
\bd
\gamma_\mu D_\mu=\gamma_\mu(x^{'})(\partial_\mu-\Gamma_\mu) \nonumber
\ed
\veb
\bd
=\gamma^\mu\partial_\mu+(1/6)
{{{R^\mu}_\beta}^ \nu}_\zeta y^\beta y^\zeta \partial_\mu- \nonumber
\ed
\veb
\bd
-(1/16)\gamma^\mu[\gamma_\alpha,\gamma_\beta]
{R^{\alpha \beta}}_{\mu\nu}y^\nu \label{fer8}
\ed
The  Fourier transform of $S(x^{'},y)$ is
\bd
S(x^{'},y)=\int\frac{d^nk}{(2 \pi)^n}\exp(iky)S(k) \label{fer9}
\ed
Let
\bd
S(k)=S_0(k)+S_1(k)+S_2(k)+... \label{fer10}
\ed
be an asymptotic representation of $S(k)$ for large $k$.

The values $S_i(k)$ are asymptotic variables of the order $k^{-(1+i)}$.
They may be found with recursion procedure from the equation
\bd
\left[(i\gamma^\mu\partial_\mu+m)+(1/6)
{{{R^\mu}_\beta}^ \nu}_\zeta {y^\beta}{ y^\zeta }\partial_\mu-\right. \nonumber
\ed
\veb
\bd
\left.-(1/16)\gamma^\mu[\gamma_\alpha,\gamma_\beta]
{R^{\alpha \beta}}_{\mu\nu}y^\nu+...\right] S(x^{'},y)=\delta(y) \label{fer11}
\ed
In this case the  momentum space representation of a propagator of a fermion will be
\bd
S(x^{'},y)=\int\frac{d^nk}{(2\pi)^n}\exp(iky) 
\left[\frac{(\gamma \cdot k+m)}{k^2+m^2}
+(1/4)R\frac{(\gamma \cdot k+m)}{(k^2+m^2)^2}-
\right. \nonumber
\ed
\veb
\bd
\left.-(2/3)R_{\alpha\beta}{ k^\alpha }{k^\beta}
\frac{(\gamma \cdot k+m)}{(k^2+m^2)^3}+(i/8){R^{\alpha \beta}}_{\mu\nu}
\frac{\gamma^\mu[\gamma_\alpha,\gamma_\beta]
k^\nu}{(k^2+m^2)^2}+...\right] \label{fer12}
\ed
The momentum space solution of the equation for the bi-spinor    
\bd
\left(\Box_x+1/4 R-m^2\right)G_F(x,x^{'})=
-g(x)^{-1/2}\delta(x-x^{'})\hat{1} \label{fer13}
\ed
where $\Box_x=D^\mu_xD_{\mu,x}$ is the  covariant d'Alembertian of spinor field,
may be obtained with the same momentum space methods as in chapter VI  
\cite{bunch1}.

The result can be written in the form of the following expression
\bd
G_F(x,x^{'})=G_F(x^{'},y) \nonumber
\ed
\veb
\bd
=g^{-1/2}(y)\int\frac{d^nk}{(2\pi)^n}\sum\limits_{j=0}^2 \hat{\alpha}_j(x^{'},y)
\left(-\frac{\partial}{\partial m^2}\right)^j(k^2+m^2)^{-1} \label{fer14}
\ed
where the geometrical coefficients $\hat{\alpha}_j(x^{'},y)$ in the limit of 
coincidence $(x \to x^{'})$ are
\bd
\hat{\alpha}_0(x^{'},y)=\hat{1}; \nonumber
\ed
\bd
\hat{\alpha}_1(x^{'},y)=(1/12)R\cdot \hat{1}; \nonumber
\ed
and
\bd
\hat{\alpha}_2(x^{'},y)=\left(-(1/120){R_\mu}^{;\mu}+(1/288)R^2-\right. \nonumber
\ed
\veb
\bd
\left.-(1/180)R_{\mu \nu}R^{\mu\nu}
+(1/180)R_{\mu\nu\sigma\tau}R^{\mu\nu\sigma\tau}\right)\cdot\hat{1} \nonumber
\ed
\veb
\bd
+(1/48)\Sigma_{[\alpha,\beta]}\Sigma_{[\gamma,\delta]}R^{\alpha\beta\lambda\xi}
{{{R\gamma}_\lambda}^\delta}_\xi \label{fer15}
\ed
where $\Sigma_{[\alpha,\beta]}=(1/4)[\gamma_\alpha,\gamma_\beta]$

\section{The bi-spinor function  in the  }
\lum
\hspace{27mm}{\Large \bf Schwinger-DeWitt representation}
\vsse

In  analogy with the  scalar field we can rewrite equation (\ref{fer14})
in  the  Schwinger-DeWitt representation. 
Since
\bd
(k^2+m^2)^{-1}
=\int \limits _0 ^\infty ids \exp\left[-is(k^2+m^2)\right] \label{fer16}
\ed 
 the equation (\ref{fer14}) will be  
\bd
G_F(x,x^{'})=\frac{i\Delta^{(1/2)}(x,x^{'})}{(4 \pi^{n/2})}
\int \limits _0 ^\infty \frac{ids}{(is)^{n/2}} \exp\left[-is(k^2+m^2)\right]
F(x,x^{'};s)  \label{fer17}
\ed
where
\bd
F(x,x^{'};s)=\sum \limits_{n=0}^\infty \hat{\alpha}_n(x,x^{'})(is)^n  \label{fer18}
\ed
and coefficients $\hat{\alpha}_n(x,x^{'})$ are defined by (\ref{fer15})

As  in the scalar case the  determinant $\Delta^{(1/2)}(x,x^{'})$ is defined  by
the equation 
\bd
\Delta^{(1/2)}(x,x^{'})=g^{-1/2}(y)  \label{fer19}
\ed

%% file: ikfrm1j.tex
\chapter{FINITE TEMPERATURE FERMIONS}
\centerline{\Large \bf IN CURVED SPACE-TIME}
\vs

\section { The Helmholtz free energy of a Fermi gas}
\lum
\hspace{22mm}{\Large \bf in curved space-time}
\vsse

After we constructed the  free energy for a thermal scalar field we can consider a 
thermal  fermi field.
Let the total Lagrangian of the system of  "gravity $+$ fermionic matter" 
be
\bd
S_{tot}=S_g+S_m \label{j1}
\ed
where $S_g$ is (\ref{i2}) and 
\bd
S_m=\int d^4 x \sqrt{g(x)}L_{eff,\psi} \label{j2}
\ed
To write the action for a  spinor field we will use the functional method 
for calculation  of  the generating functional from (\ref{j2}). 
Making the same procedure as in (\ref{i6})-(\ref{i8}), write
\bd
W[0]=i\ln Z[0]=(i/2)\ln DetG_F(x,x^{'}) \nonumber
\ed
\veb
\bd
=\int\limits_0^\infty ids (is)^{-1}\mbox{tr}\hat{f}(x,x^{'},is) \exp(-im^2s) \nonumber
\ed
\veb
\bd
=(i/2)\int d^4x \int_{m^2}^\infty dm^2 \mbox{tr}G_F(x,x^{'}) \label{j3}
\ed
where the  Green's function is expressed by 
\bd
G_F(x,x^{'})=\int_0^\infty ids (is)^{-1}\hat{f}(x,x^{'},is)\exp(-im^2s) \label{j4}
\ed
and  $\hat{f}(x,x^{'},is)$ is the  heat kernel. 

The kernel $\hat{f}(x,x^{'},is)$ is   (in the limit $(x \to x^{'})$) 
  the sum of zero-images \\ 
\cite{i16} antiperiodic in the imaginary time 
\bd
\hat{f}(x,x^{'},is)=\sum\limits_{n=-\infty}^\infty
\frac{\exp[-\beta^2(n-1/2)^2/4is]}{(4\pi is)^{1/2}}
\hat{f}_{3)}(\tilde{x},\tilde{x},is) \label{j5}
\ed
where an equation for $\hat{f}_{3)}(x,x^{'},is)$ is 
\bd
\left( i\frac{\partial}{\partial s}-\tilde{\Box}_{3)}
-(1/4)R\right)\hat{f}_{3)}(\tilde{x},\tilde{x},is)=0 \label{j6}
\ed
and $\tilde{\Box}_{3)}$ is the covariant d'Alembertian on $M_{3)}$.

The solution of (\ref{j6}) is the series
\bd
\hat{f}_{3)}(\tilde{x},\tilde{x},is)
=\sum\limits_{j=0}^\infty\hat{\alpha}_j(x^{'})(is)^j \label{j7}
\ed
where the coefficients $\hat{\alpha}_j(x^{'})$ are determined by (\ref{fer15}).

From the equation (\ref{j4}) we get   the  finite temperature 
Green's function 
\bd
\mathop{\lim}\limits_{x \to x^{'}}G^\beta_{F,SD}(x,x^{'})
=G_{F,SD}(x^{'},x^{'})+{G_F}(\beta) \label{j8}
\ed
where $G_{F,SD}(x^{'},x^{'})$ is (\ref{fer17}) in the limit $(x=x^{'})$ and
\bd
{G_F}(\beta)=\frac{i}{(4\pi)^3/2}\sum\limits_{j=0}^\infty
\sum\limits_{n=-\infty}^\infty
\hat{\alpha}_j(R)\times \nonumber
\ed
\veb
\bd
\times \int \limits_0^\infty ids(is)^{j-3/2}
\exp\left[-ism^2-(\beta^2/4is)(n+1/2)^2\right] \label{j9}
\ed
Summation with respect to (\ref{d25}) with $z=1/2$ gives
\bd
{G_F}(\beta)=\frac{i}{(4\pi)^{3/2}}\sum\limits_{j=0}^\infty
\sum\limits_{n=1}^\infty
\hat{\alpha}_j(R)\times
\ed
\veb
\bd
\times 2\int \limits_0^\infty ids(is)^{j-2}(-1)^n
\exp\left[-ism^2-(n^2\beta^2/4is)\right] \label{j10}
\ed
After integration over the  proper time $(s)$ (\ref{d29}) we get  the 
following  expression for the  finite temperature 
contribution in the  Green's function of a  fermion
\bd
{G_F}(\beta)=\frac{i}{(4\pi)^2}\sum\limits_{j=0}^\infty
\sum\limits_{n=1}^\infty
\hat{\alpha}_j(R)(-1)^j(\beta n/2m)^{j-3} K_{j-3}(\beta m n) \label{j11}
\ed
The  total action of the system will be
\bd
L_{eff}(\beta)=\tilde{ L}_g(R)-f_F(\beta,R) \label{j12}
\ed
where $\tilde{ L}(R)_g $ is the  temperature independent Lagrangian
\bd
\tilde{ L}(R)_g=L_g+(i/2)\mbox{tr} \int\limits_{m^2}^\infty dm^2 
 G_{F,SD}(x,x^{'}), \label{j13}
\ed
and the  finite temperature contribution is expressed in the form of a  series:
\bd
f_F(\beta,R)=(-i/2) \mbox{tr} \int\limits_{m^2}^\infty dm^2{G_F}(\beta)
=\sum\limits_{j=0}^\infty \alpha_j(R)f_j(\beta m) \label{j14}
\ed
with coefficients 
\bd
\alpha_j(R)=(1/2s)\mbox{tr}\hat{\alpha_j(R)}, \label{j15}
\ed
and
\bd
f_0(\beta m)=\frac{m^2\cdot 2s}{2\pi^2 \beta^2}
\sum\limits_{n=1}^\infty \frac{(-1)^n}{n^2} K_2(\beta m n) \nonumber
\ed
\veb
\bd
f_1(\beta m)=\frac{m\cdot 2s}{4\pi^2 \beta}
\sum\limits_{n=1}^\infty \frac{(-1)^n}{n }K_1(\beta m n) \nonumber
\ed
\veb
\bd
f_2(\beta m)=\frac{ 2s}{8\pi^2}
\sum\limits_{n=1}^\infty (-1)^n K_0(\beta m n) \label{j16}
\ed
The  finite, temperature dependent contribution $f(\beta,R)$ represents the density of
Helmholtz free energy in curved space-time.  Using  the  integral representation for the
series of modified Bessel functions (\ref{aa10}),(\ref{aa12}),(\ref{aa14})  
one can write it as [Kulikov \& Pronin 1995]:
\bd
f_F(\beta,R)=f_0(\beta m)+\alpha _1(R)f_1(\beta m)+\alpha _2(R)f_2(\beta
m)+.\;.\;., \label{j17}
\ed 
where the first term is the standard form of the  Helmholtz free energy in Euclidean
space 
\bd
f_0(\beta m)=-{2s \over \beta }\int {{{d^3k} \over {(2\pi )^3}}}\ln
\left( {1+e^{-\beta \varepsilon }} \right) \label{j18}
\ed 
with energy of particle $\varepsilon =\sqrt {\vec{ k}^{2}+m^{2}}$.  

The  factor $2s=4$
reflects  the existence of the four degrees of freedom present in the fermion
field: particles and antiparticles, spin up and spin down. 

The following terms
are geometrical corrections of the Riemann space time structure with respect to
the Euclidean one with temperature coefficients in the form
\bd
f_j(\beta m)=-{2s \over \beta }\int {{{d^3k} \over {(2\pi
)^3}}}\left( {-{\partial 
\over {\partial m^2}}} \right)^j
\ln \left( {1+e^{-\beta \varepsilon }} \right). \label{j19}
\ed 
The method developed above does not allow us  to compute the density of the grand
thermodynamical potential.  Therefore, in the following calculations we will use
the  local momentum space formalism as the most convenient for 
the  construction of local  thermodynamics.

%% file: ikveck.tex
\chapter{THERMODYNAMICS OF GAUGE FIELDS}  
\vs

In this chapter we will apply the   formalism 
of gauge fields in curved space-time developed  in chapter V
to study the properties of thermal photon gas 
in an  external gravitational field. 

\section{The  Green's function of photons}
\vsse

As we know already from chapter IV the total Lagrangian  
for a  vector field in Minkowski space-time
is the sum of three contributions:
 \bd
 L_{tot} = L_m+L_f+L_{gh}, \label{k1}
\ed   
where
 \bd
 L_m = -(1/4)F_{\mu \nu}F^{\mu \nu}, \label{k2}
\ed   
\veb
\bd
 L_f = -(1/2\alpha)(\partial_\mu A^\mu)^2 \label{k3}
\ed   
and  
\bd
 L_{gh} = g^{\mu \nu}(\partial_\mu c)(\partial_\nu c^{*}) \label{k4}
\ed   
It can be  extended to curved space-time with the transformation
\bd
A_\alpha={h_\alpha}^\mu A_\mu,~~ 
\partial_\alpha \to {h_\alpha}^\mu\nabla_\mu \label{k5}
\ed   
where $\nabla_\mu=\partial_\mu+\Gamma_\mu$ and the connection 
$\Gamma_\mu$ is defined by equation
\bd
\Gamma_\mu=(1/2)\Sigma_{\alpha \beta}{h^\alpha}^\nu(x) 
\left[\frac{\partial}{\partial x^{\mu}} {{h^\beta}_nu}(x)\right] \label{k6}
\ed   
where the matrices $\Sigma_{\alpha \beta}$  are (\ref{g7}).  

The strength tensor of electromagnetic field  is
\bd
F_{\mu \nu}=\nabla_\mu A_\nu-\nabla_\nu A_\mu \label{k8}
\ed   
The variation of the action 
\bd
S=\int d^4 x \sqrt{g} \left( L+L_f+L_{gh}\right) \label{k9}
\ed   
gives the equation for the vector field:
\bd
\nabla_\nu F^{\mu \nu}+(1/\alpha)\nabla^\mu (\nabla_\nu A^\nu)=0 \label{k10}
\ed   
In the Feynman gauge $(\alpha=1)$ this equation has the form:
\bd
\nabla_\nu \nabla^\mu A^\nu-\nabla_\nu \nabla^\nu A^\mu+ 
\nabla^\mu \nabla_\nu A^\nu=0 \label{k11}
\ed   
and with the definition of the Riemann tensor (\ref{icor1}) it is
\bd
\nabla_\nu \nabla^\nu A^\mu-{R^\mu }_\nu A^\nu=0 \label{k12}
\ed   
Based on this equation we write the equation for the  Green's function
 \bd
\nabla_\nu \nabla^\nu{ D^\mu}_\tau(x,x^{'})-{R^\mu }_\nu{D^\nu}_\tau(x,x^{'})
=-g^{-1/2}(x)\delta(x-x^{'}){\delta^\mu}_\tau \label{k13}
\ed   
We will be interested in calculations of the Green's function
in the limit $(x \to x^{'})$ to find the effective action. 

We will rewrite the equation (\ref{k13}) in Riemann normal coordinates with 
origin at the point $x^{'}$. For covenience one may define the  Green's function  
${\bar{D}^\mu}_\tau(x,x^{'})$ as
\bd
{ D^\mu}_\tau(x,x^{'})=g^{-1/4}(x){\bar{D}^\mu}_\tau(x,x^{'})g^{-1/4}(x^{'}) \nonumber
\ed
\veb
\bd
=g^{-1/4}(x){\bar{D}^\mu}_\tau(x,x^{'}) \label{k14}
\ed   
where we used the fact $g(x^{'})=1$

For our calculations we will use the Christoffel symbols 
which in the Riemann normal coordinates are
\bd
{\Gamma^\sigma}_{\mu \nu}=-(1/3)({R^\sigma}_{\alpha \beta \gamma}+
{R^\sigma}_{\beta \alpha \gamma})y^\gamma \label{k15}
\ed   
where $y^\gamma$ represents the coordinates of the point $x$ and the Riemann 
tensor is evaluated at $x^{'}$.
The expansion of   equation (\ref{k13}) to the second derivative of the metric
gives
\bd
\eta ^{\alpha \beta }\partial _\alpha \partial _\beta \bar D_\tau ^\mu (y)
+(1/6)R\bar D_\tau ^\mu (y) \nonumber
\ed
\veb
\bd
-(4/3)R_\nu ^\mu \bar D_\tau ^\nu (y)
-(1/3)R_\nu ^\lambda y^\nu \partial _\lambda \bar D_\tau ^\mu (y) \nonumber
\ed
\veb
\bd
+(1/3){{{R^\alpha}_\gamma} ^\beta }_\delta
y^\gamma y^\delta \partial _\alpha
 \partial _\beta \bar D_\tau ^\mu (y)
-(2/3){{{R^\mu}_\gamma} ^\alpha} _\delta  y^\delta 
\partial _\alpha \bar D_\tau ^\gamma (y) \nonumber
\ed
\veb
\bd
+(2/3){R^{\mu \alpha }}_{\lambda \gamma }
y^\gamma \partial _\alpha \bar D_\tau ^\lambda (y)
=-\delta (y)\delta _\tau ^\mu \label{k16} 
\ed   
where $\partial_\alpha=\partial/\partial y^\alpha$.

The momentum space approximation is defined by introducing the quantity 
$D_\tau ^\mu (k)$ defined as
\bd
\bar{D}_\tau ^\mu (x,x^{'})=\bar{D}_\tau ^\mu (x^{'},y)
=\int \frac{d^n k}{(2 \pi) ^n}\bar{D}_\tau ^\mu (k) \exp[iky] \label{k17}
\ed   
This quantity is assumed to have the expansion 
\bd
\bar{D}_\tau ^\mu (k)=\bar{D}_{0,\tau} ^\mu (k)+\bar{D}_{1,\tau} ^\mu (k)
+\bar{D}_{2,\tau} ^\mu (k)+... \label{k18}
\ed   
where $\bar{D}_{i,\tau} ^\mu (k)$  have a geometric coefficients
involving $i$ derivatives of the metric.  On dimensional grounds 
$\bar{D}_{i,\tau} ^\mu (k)$ must be of order $k^{-(2+i)}$ so 
the equation (\ref{k16}) is an asymptotic expansion in large $k$. 

As we can see from (\ref{k16}) the first term in (\ref{k18}) is
\bd
\bar{D}_{0,\tau} ^\mu (k)=\delta^\mu_\tau/k^2, \label{k19}
\ed   
The second one is 
\bd
\bar{D}_{1,\tau} ^\mu (k)=0, \label{k20}
\ed   
The third one is computed from
\bd
\eta ^{\alpha \beta }\partial _\alpha \partial _\beta \bar D_{2,\tau} ^\mu (y)
+(1/6)R\bar D_{0,\tau }^\mu (y) \nonumber
\ed
\veb
\bd
-(4/3)R_\nu ^\mu \bar D_{0,\tau} ^\nu (y)
-(1/3){R_\nu }^\lambda y^\nu \partial _\lambda \bar D_{0,\tau} ^\mu (y) \nonumber
\ed
\veb
\bd
+(1/3){{{R^\alpha}_\gamma} ^\beta }_\delta
y^\gamma y^\delta \partial _\alpha
 \partial _\beta \bar D_{0,\tau }^\mu (y)
-(2/3){{{R^\mu}_\gamma} ^\alpha} _\delta  y^\delta 
\partial _\alpha \bar D_{0,\tau} ^\gamma (y) \nonumber
\ed
\veb
\bd
+(2/3){R^{\mu \alpha }}_{\lambda \gamma }
y^\gamma \partial _\alpha \bar D_{0,\tau }^\lambda (y)+...=0 \label{k21}
\ed   
To simplify calculations  note that 
$y^\alpha \to (-i\partial /\partial k^\alpha)$, and integrate by parts  to find 
that
 \bd
\bar{D}_{2,\tau}^\mu (y)
=\int \frac{d^n k}{(2 \pi) ^n}\bar{D}_{2,\tau}^\mu (k) \exp[iky] 
=\int \frac{d^nk}{(2 \pi)^n} \exp[iky]\times \nonumber
\ed
\veb
\bd
\times \left[ \left\{(1/6)R\delta^\mu_\tau-(2/3)R^\mu_\nu\delta^\nu_\tau\right\}/k^4-
(4/3){R^{\mu\beta}}_{\nu\gamma}
k^\gamma k_\beta\delta^\nu_\tau/k^6 \right] \label{k22}
\ed
The final expression for the photon propagator will be then:
\bd
D_\tau^\mu (y)={g}^{-1/4} (y)\int \frac{d^nk}{(2 \pi)^n} \exp[iky]\times \nonumber
\ed
\veb
\bd
\times \left[ \delta^\mu_\tau/k^2+
\left\{(1/6)R\delta^\mu_\tau-(2/3)R^\mu_\nu\delta^\nu_\tau\right\}/k^4-
(4/3){R^{\mu\beta}}_{\nu\gamma}
k^\gamma k_\beta\delta^\nu_\tau/k^6 \right] \label{k23}
\ed
The  Green's function of the ghost fields  obeys the same equation as a  scalar field,
therefore one can use this equation for our further calculations.  

\section {The thermodynamic  potential of  a  photon gas}
\vsse

As in the case of flat space-time, one must carry out the
 calculations of the free energy of a  photon gas
together with  the ghost contributions.

The generating functional of an  abelian vector field in curved space-time
is written as
\bd
Z[J_\mu ,\eta, \bar{\eta}]=\int D A_\mu D \bar{\eta} D \eta \nonumber
\ed
\veb
\bd
\times \exp\left[ \int  d^4 x \sqrt{g} \left \{ A^{\mu }(D_{\mu \nu})^{-1}A^{\nu}
+\bar{c} D^{-1} c+j_\mu A^\mu +\bar{\eta} c
+\bar{c}\eta\right\} \right] \label{addk24} 
\ed
Integration over the fields with zero sources leads to
the following result for the  logarithm of the  generating functional 
\bd
\ln Z[0]=\ln Det (G)-(1/2) \ln Det(G{\mu \nu}) \label{k24} 
\ed
where the  Green's functions are defined by (\ref{k23}) 
and (\ref{h33})\footnote{To find  the Feynman propagator
of the  ghost field from the boson propagator   we have to 
put the  mass of boson  $m=0$ after calculation. }

From this equation and from the  definition of the  free energy directly follows 
the expression for density of free energy
\bd
f_{ph} (\beta, R) =(-i/2)\mathop{\lim}\limits_{m\to 0}\int\limits_{m^2}^\infty dm^2 
\left\{2 \mbox{tr} G(\beta,x-x^{'})-\mbox{tr} G_{\mu \nu}(\beta,x-x^{'})\right\} \label{k25}
\ed
The final result may be found  after putting a  mass parameter $m^2$ into 
the expressions for the  propagators of the photon and ghost fields and,
after calculation of (\ref{k25}) setting the   mass equal to zero.

After making this procedure we will have
 \bd
f_{ph}(\beta, R)=\sum \limits_j g_j(R)\left(-\frac{\partial}{\partial m^2}\right)^j
\mbox{tr} \ln (\omega^2_n+\epsilon^2) \label{k26}
\ed
 where the  symbol $\mbox{tr}$ means 
\bd
\mbox{tr}=\sum \limits_{n=-\infty}^\infty \int \frac{d^3k}{(2 \pi )^3} \nonumber
\ed
and $g_j(R)$ are geometric coefficients.

After summation in (\ref{k26})  we will have the final result in the form
\bd
f_{ph}(\beta, R)=\int \frac{d^3k}{(2\pi)^3}
\left\{(2/\beta)\ln (1- \exp[\beta \epsilon])\right. \nonumber
\ed
\veb
\bd
\left.-(1/6)(R-2R^\mu_\nu\delta ^\nu_\mu)
[\epsilon( \exp[\beta \epsilon]-1)]^{-1}\right\} \label{k27}
\ed
This expression is the density of the  Helmholtz free energy of a  photon gas 
in an   external gravitational field.

\section {Internal energy and heat capacity of a photon gas}
\vsse

Thermodynamic  properties of the photon gas have been  studied very well 
in flat space-time
\cite{key12,huang1}.   The formalism developed in this section  allows us 
to get the properties of photon gas in curved space-time.

The density of energy of a  photon gas is
\bd
u=\frac{\partial}{\partial \beta}(\beta f)=
\int \frac{d^3k}{(2\pi)^3}
\left\{\frac{2\epsilon}{ \exp[\beta \epsilon]-1}\right. \nonumber
\ed
\veb
\bd
\left.-(1/6)(R-2R^\mu_\nu\delta ^\nu_\mu)
\frac{\partial}{\partial \beta}\int \frac{d^3k}{(2\pi)^3}
\frac{\beta}{\epsilon (\exp[\beta \epsilon]-1)}\right\} \label{k28}
\ed
The first term in (\ref{k28}) corresponds to the results of statistical mechanics,
and the second one is the curved space-time correction to the energy
of photon gas. Integrating over the momentum
\bd
\int \frac{d^3k}{(2\pi)^3}
\frac{1}{\epsilon (\exp[\beta \epsilon]-1)}=\frac{1}{12\beta} \label{k29}
\ed
we get the following  approximation to $O(R^2)$
\bd
u(R)=\sigma T^4+(1/72)(R-2R^\mu_\nu\delta ^\nu_\mu)T^2+... \label{k30}
\ed
where $\sigma=(\pi ^2/15)$.  

The  heat capacity will be
\bd
c(R)=\frac{\partial}{\partial T}u(R)=4\sigma T^3
+(1/36)(R-2R^\mu_\nu\delta ^\nu_\mu)T^2+... \label{k31}
\ed
As  was shown in \cite{alt1}  the effects of curvature for Einstein space 
lead to the results for the  Planck  black body expression which are   equivalent to 
the  results (\ref{k30}).

%% file: ikrenl.tex
\chapter{RENORMALIZATIONS IN }  
\centerline{\Large \bf   LOCAL STATISTICAL MECHANICS }
\vs

\section{Divergences of   finite temperature} 
\lum
\hspace{22mm}{\Large \bf   field models}
\vsse

As one can show by calculations, the contributions $G_{SD}(x,x)$ in the 
expressions (\ref{i16}) and (\ref{j8}) 
are divergent \cite{key16}. To have a clear picture of the model 
under consideration we need to eliminate these divergent contributions. The direct
way to eliminate the divergences is to combine the  gravitational Lagrangian and
the  divergent parts of the matter Lagrangian.               

1) Renormalization of $a$  $bose$ $field$.

As  was shown in (\ref{i21}) the effective Lagrangian for gravitational field
can be written as 
\bd
 \tilde{L}_g = L_g(x)+(-i/2)\int \limits_{m^2}^\infty \mbox{tr} G_{SD}(x,x^{'}) \label{l1}
\ed   
Inserting (\ref{h46}) into the last term of the expression (\ref{l1}), we find
 \bd
(-i/2)\int \limits_{m^2}^\infty \mbox{tr} G_{SD}(x,x^{'}) \nonumber
\ed
\veb
\bd
=\frac{1}{2(4\pi)^2}\sum \limits_{j=0}^\infty \gamma_j(x)
\int \limits_0^\infty ids(is)^{j-3} \exp (-ism^2) \label{l2}
\ed   
We will use the procedure of dimensional regularization  
\cite{hv1}
 to select divergent terms of the expression (\ref{l2}). It gives
\bd
(-i/2)\int \limits_{m^2}^\infty \mbox{tr} G_{SD}(x,x^{'}) \nonumber
\ed
\veb
\bd
=\frac{1}{2}(4\pi)^{-n/2}
\sum \limits_{j=0}^\infty \gamma_j(x)
\int ids(is)^{(j-1-n/2)} \exp (-ism^2) \nonumber
\ed
\veb
\bd
=\frac{1}{2}(4\pi)^{-n/2}\left(\frac{m}{M}\right)^{n-4}
\sum \limits_{j=0}^\infty \gamma_j(x)
m^{4-2j}\Gamma\left(j-n/2\right) \label{l3}
\ed   
where $n$ is a dimension of the space-time and $M$ 
is an arbitrary mass scale. This parameter is introduced 
to preserve the dimension of the Lagrangian to $[L]^{-4}$
for the dimensions $n\neq 4$.  

The functions $\Gamma(z)$ have the poles at $n=4$:
\bd
\Gamma \left( {-{n \over 2}} \right)=
{4 \over {n(n-2)}}\left( {{2 \over {4-n}}-\gamma } \right)+O(n-4) \nonumber
\ed
\veb
\bd
\Gamma \left( {1-{n \over 2}} \right)
={2 \over {2-n}}\left( {{2 \over {4-n}}-\gamma } \right)+O(n-4) \nonumber
\ed
\veb
\bd
\Gamma \left( {2-{n \over 2}} \right)={2 \over {4-n}}-\gamma +O(n-4) \label{l4}
\ed
Therefore the  first three terms of the expression (\ref{l3}) 
are divergent. 

Selecting divergent parts and using the logarithmic expression for
\bd
\left(\frac{m}{M}\right)^{n-4}=1+\frac{1}{2} \ln \frac{m^2}{M^2}+O((n-4)^2) \label{l5}
\ed
  write  the  divergent contributions
in the effective Lagrangian $\tilde{L}_g$ in the form 
\bd
(-i/2)\int \limits_{m^2}^\infty \mbox{tr} G_{SD}(x,x^{'})  \nonumber
\ed
\veb
\bd
=-(4\pi)^{-n/2}\left\{\frac{1}{n-4}
+\frac{1}{2}\left[j+\ln \left(\frac{m^2}{M^2}\right)\right]\right\}\times \nonumber
\ed
\veb
\bd
\times \left[\frac{4m^2}{n(n-4)}-\frac{2}{n-2}m^2\gamma_1(x^{'})
+\gamma_2(x^{'})+...\right] \label{l6}
\ed   
Combining (\ref{l6}) with the  gravitational Lagrangian (\ref{i3}) we can 
redetermine the constants 
of the  gravitational  Lagrangian  (\ref{l1}) as
\bd
{1 \over {8\pi G_R}}\Lambda _R
={1 \over {8\pi G_0}}\Lambda _0-{1 \over {(4\pi )^2}}
{1 \over {(n-4)}}{{m^4} \over 2}\veb
\ed
\veb
\bd
{1 \over {16\pi G_R}}={1 \over {16\pi G_0}}
+{1 \over {(4\pi )^2}}{1 \over {n-4}}m^2
\left( {{1 \over 6}-\xi } \right) \nonumber
\ed
\veb
\bd
\alpha _R=\alpha _0-{1 \over {(4\pi )^2}}
{1 \over {n-4}}\left( {{1 \over 6}-\xi } \right)^2\nonumber
\ed
\veb
\bd
\beta _R=\beta _0+{1 \over {180}}{1 \over {(4\pi )^2}}{1 \over {n-4}}\nonumber
\ed
\veb
\bd
\gamma _R=\gamma _0-{1 \over {180}}{1 \over {(4\pi )^2}}{1 \over {n-4}} \label{l7}
\ed
where $G_0$, $\Lambda_0$, $\alpha_0$, $\beta_0$, $\gamma_0$ are bare constants and 
$G_R$, $\Lambda_R$, $\alpha_R$, $\beta_R$, $\gamma_R$ are 
the physical (finite) constants. 

Since all divergences of the matter field can be included in  $\tilde{L}_g$,
only the  finite temperature contributions observed in (\ref{i20}) remain. 

2) Renormalization of $a$ $fermi$ $field$.

As follows from (\ref{j13})  the effective Lagrangian 
of the gravitational field is
\bd
\tilde{L}_g=L_g-\frac{1}{2}(4 \pi)^{-n/2}\sum\limits_{j=0}^\infty
\mbox{tr} \hat {\alpha}_j(R)\times \nonumber
\ed
\veb
\bd
\times \int\limits _0 ^\infty ids (is)^{j-n/2-1} \exp (-ism^2) \label{l8}
\ed
After  dimensional regularization  procedure (as in scalar field case),
 we get
\bd
{1 \over {8\pi G_R}}\Lambda _R
={1 \over {8\pi G_0}}\Lambda _0+{1 \over {(4\pi )^2}}{{2m^4} \over {n-4}} \nonumber
\ed
\veb
\bd
{1 \over {16\pi G_R}}={1 \over {16\pi G_0}}
-{1 \over {(4\pi )^2}}{1 \over {(n-4)}}{{m^2} \over 6} \nonumber
\ed
\veb
\bd
\alpha _R=\alpha _0+{1 \over {(4\pi )^2}}{1 \over {(n-4)}}{1 \over {144}} \nonumber
\ed
\veb
\bd
\beta _R=\beta _0-{1 \over {(4\pi )^2}}{1 \over {(n-4)}}{1 \over {90}} \nonumber
\ed
\veb
\bd
\gamma _R=\gamma _0-{1 \over {(4\pi )^2}}{1 \over {(n-4)}}{7 \over {720}} \label{l9}
\ed
Thus, the Lagrangian (\ref{j12}) does not include the divergencies in the  finite
temperature loop approximation over the fermi fields.

The  method of dimensional regularization developed here 
is not the only one for  applications 
to field models in background curved space-time. 
The  $\zeta$-function 
method \cite{haw1}, \cite{hurt1} and the method of  covariant geodesic point 
separation
 also lead to a solution of the problem of regularization of 
effective Lagrangians of matter fields and  
calculation of Energy-momentum tensor anomalies \cite{christ1}.

%% file: ikBEin.tex
\chapter{LOCAL QUANTUM STATISTICS AND} 
\centerline{\Large \bf THERMODYNAMICS OF BOSE GAS}
\vs

In this chapter we will continue to develop the methods of local 
statistical thermodynamics 
\cite{kul1} with application to the ideal bose gas in curved
space-time.
In chapter VII working with the  Schwinger-DeWitt proper time formalism
 we found that the density of  Helmholtz free energy in 
curved space-time may be found in  the form of a  series (\ref{i25}) 
including the geometric   structure of   space-time.
However, as in the previous section,
working with this formalism we  could have certain  
difficulties with the introduction of the chemical potential and
the computing of the density of the grand thermodynamical potential.
 
To avoid such problems which will appear on the way of construction of 
local thermodynamics with the Schwinger DeWitt formalism, we turn to   
the local momentum space method \cite{bunch1}, \cite{panan1}
 in quantum field theory in an arbitrary curved space
time and  the imaginary time formalism to introduce the
temperature \cite{i15},  \cite{wein1}. 

 For the  construction of the local quantum statistics and thermodynamics 
of a  bose gas,
we will  use  the  connection between the  partition function of 
ideal quantum systems
and   finite temperature Green's functions, 
which may be found by the local momentum space method. 

\section{Density of grand }
\lum
\hspace{13mm}{\Large \bf   thermodynamical potential}
\vsse

In this section we will analyze the scalar field model with a conserved charge.
The Lagrangian of the model is
\bd
S_m=-(1/2)\int d^4x\sqrt{g(x)}\Phi^{\ast}(x)(-\Box_x+m^2+\xi R)\Phi(x) \label{bos1}
\ed   
where $\Phi=(\phi_1,\phi_2)$ is a dublet of the real fields.

The action written in terms of real fields will be
\bd
S_m=-(1/2)\int d^4x\sqrt{g(x)}\phi^a(x)(-\Box_x+m^2+\xi R)\phi^a(x) \label{bos2}
\ed   
The  total action of the system "matter $+$ gravity" is 
\bd
S_{tot}=S_g+S_m \label{bos3}
\ed   
Now we can write the effective action at finite temperature 
in analogy with (\ref{i20}) as
\bd
L_{eff}(\beta)=\tilde{L}_g-\omega(\beta, \mu ,R) \label{bos4}
\ed   
where $\tilde{L}_g$ is (\ref{i21}) and $\omega(\beta, \mu ,R)$ 
is the density of grand  thermodynamic potential.

The result (\ref{bos4}) may be obtained with the momentum space 
representation for the Green's function of a  boson (\ref{h33}) \cite{lcs1}.

In the momentum space representation, the expression for $L_{eff}$  is
split into two parts
\bd
L_{eff}=-(i/2)\int\limits_{m^2}^\infty dm^2 \mbox{tr} G(x,x^{'})
-\omega(\beta, \mu ,R) \label{bos5}
\ed   
The potential $\omega(\beta, R)$ is
\bd
\omega(\beta,  R)=
(-1/2) \mbox{tr} \int\limits_{m^2}^\infty dm^2\sum \limits_{j=0}^2
\sum \limits_{n=0}^\infty
\gamma_j(x,x^{'})\left(-\frac{\partial}{\partial m^2}\right)^j \times \nonumber
\ed
\veb
\bd
\times \int \frac{d^3k}{(2\pi)^3}({\omega_n}^2+\epsilon^2)^{-1} \nonumber
\ed
\veb
\bd
=(1/2)\sum \limits_{j=0}^2
\gamma_j(x^{'})\left(-\frac{\partial}{\partial m^2}\right)^j
\mbox{tr} \ln ({\omega_n}^2+\epsilon^2) \label{bos6}
\ed   
and coincides with the  Helmholtz free energy (\ref{i22}) and $\omega _n=2\pi n/\beta$.

The symbol $\mbox{tr}$ in  (\ref{bos6}) is determined as
\bd
\mbox{tr} (...)=\sum \limits_{n \neq 0} \int \frac{d^3k}{(2\pi)^3}... \nonumber
\ed
For introducing the chemical potential, we will change the Matsubara 
frequencies $\omega_n \to \omega_n+\mu$ and thermodynamic potential will be
$\omega(\beta, \mu ,R)$.  

Since both positive and negative frequencies are summed, 
we will have
\bd
\mbox{tr} \ln ({\omega_n}^2+\epsilon^2) \to \mbox{tr} \ln [(\omega_n+\mu)^2+\epsilon^2] \nonumber
\ed
\veb
\bd
=\mbox{tr} \left\{\ln \left[{\omega_n}^2+(\epsilon-\mu)^2\right]
+\ln \left[{\omega_n}^2+(\epsilon+\mu)^2\right]\right\} \label{bos7}
\ed   
After summation in (\ref{bos7}) we get
\bd
\omega(\beta, \mu ,R)=\omega_{-}(\beta, \mu ,R)
+\omega_{+}(\beta, \mu ,R) \label{bos8}
\ed   
where
\bd
\omega_{-}(\beta, \mu ,R)=(1/\beta)\sum \limits_{j=0}^2
\gamma_j(x^{'})\left(-\frac{\partial}{\partial m^2}\right)^j
\ln (1-\exp [-\beta(\epsilon-\mu)]) \label{bos9}
\ed   
and
\bd
\omega_{+}(\beta, \mu ,R)=(1/\beta)\sum \limits_{j=0}^2
\gamma_j(x^{'})\left(-\frac{\partial}{\partial m^2}\right)^j
\ln (1-z \exp [-\beta (\epsilon+\mu)]) \label{bos10}
\ed   
So, with  accordance to   (\ref{bos8}) the density of grand thermodynamic potential 
is the series  
\bd
\omega(\beta, \mu ,R)=\sum \limits_{j=0}^2\gamma^{'}_j(x^{'})b_j(\beta m,z) \label{bos11}
\ed   
where
\bd
b_0(\beta m,z)=(1/\beta) \ln (1-z \exp (-\beta \epsilon)); \nonumber
\ed
\veb
\bd
b_j(\beta m,z)=\left(-\frac{\partial}{\partial m^2}\right)^j
b_0(\beta m,z) \label{bos12}
\ed   
and  the  fugacity  is $z=\exp (\beta \mu)$.
The geometrical coefficients $\gamma^{'}_j(R)$ of the equation 
(\ref{bos12}) have the form  (\ref{h45}). 
Renormalizations in the  total Lagrangian (\ref{bos4}) are the same as in chapter XI.

\section{Statistics and thermodynamics of bose gas}
\vsse

We find the  bose distribution function  as the derivative of  the grand thermodynamical potential 
\bd
n_{\vec{k}}=-{{\partial \omega _{\bar k}(\beta ,\mu ,R)}
\over {\partial \mu }} \nonumber
\ed
For occupation numbers with  momentum  $\vec k$ we get 
\bd
n_{\vec{ k}}={1 \over {(z^{-1}e^{\beta \varepsilon _{\vec{ k}}}-1)}}B(\beta
,R),\label{ins1}
\ed 
where the function $B(\beta ,R)$ is described by the formula
\bd
B(\beta ,R)=\ {1+\gamma _1(R){\beta  \over {2\varepsilon
_{\vec{k}}}}\left[{1-(1-ze^{-\beta \varepsilon _{\vec{k}}})^{-1}}\right]+\quad
.\;.\;.} \label{ins2}
\ed 
The function  $B(\beta ,R)$ depends on the curvature, temperature and energy 
of the boson. 

Studying the thermodynamical properties of Bose gases we will 
start with the equation
\bd
\omega(\beta, \mu ,R)=-(1/\beta)\sum \limits_{j=0}^2
\gamma_j(x^{'})\left(-\frac{\partial}{\partial m^2}\right)^j
\ln (1-z \exp [-\beta \epsilon]) \label{bos13}
\ed   
In  the  non-relativistic limit for particle energy 
$\epsilon=\vec{k}^2/2m$ we get from (\ref{bos13}) the equation
 \bd
\omega(\beta,\mu,R)=\sum \limits_{j=0}^2
\gamma_j(x^{'})g_{5/2}(z)\left(-\frac{\partial}{\partial m^2}\right)^j
\lambda^{-3} \label{bos14}
\ed   
where $\lambda=(2\pi/mT)^{1/2}$ is a wavelength of the particle,
and the  function $g_{5/2}(z)$ has the following  form
\bd
g_{5/2}(z)=\sum\limits_{l=1}^\infty \frac{z^l}{l^{5/2}} \nonumber
\ed
\veb
\bd
=-(4/\sqrt{\pi})\int dx x^2\ln \left(1-z \exp (-x^2)\right) \label{bos15}
\ed   
The average number of particles in a certain momentum state $k$ is
obtained as the derivative
\bd
<n_k>=-\frac{\partial}{\partial \mu}\omega(\beta, \mu ,R) \nonumber
\ed
\veb
\bd
=\sum \limits_{j=0}^2
\gamma_j(x^{'})g_{5/2}(z)\left(-\frac{\partial}{\partial m^2}\right)^j
\left(z^{-1} \exp (\beta \epsilon)-1\right) \label{bos16}
\ed   
The  density of particles is
\bd
n=\lambda^{-3}\left[1- \gamma_1(R)(3/4m^2)-\gamma_2(R)(3/16m^4)\right]g_{3/2}(z)
+n_0 \label{bos17}
\ed   
where the  new function $g_{3/2}(z)$ is 
\bd
g_{3/2}(z)=z\frac{\partial}{\partial z}g_{5/2}(z) \nonumber
\ed
and  $n_0=z/(1-z)$ is the average number of particles with zero momentum.
The functions $g_{3/2}(z)$ and  $g_{5/2}(z)$ are   special cases of a more
general class of functions
\bd
g_{n}(z)=\sum \limits_{k=1}^\infty \frac{z^k}{k^n} \label{bos18}
\ed   
In a more simple form the  equation(\ref{bos17}) may be written as
\bd
(n-n_0)\lambda^3=g_{3/2}(z,R) \label{bos19} 
\ed   
where
\bd
g_{3/2}(z,R)=\left[1- \alpha\frac{R}{m^2}+...\right]g_{3/2}(z) \label{bos20}
\ed
is a  function which depends on  curvature, and $\alpha$ is a numerical parameter.

The equation  (\ref{bos19}) connects four values: fugacity, 
temperature, density of the particles and curvature. 

We can solve it graphically and get the dependence 
of the (effective) chemical potential on  curvature, temperature and density.
The function $g_{3/2}(z,R)$  varies with  the   curvature $R$
 as   shown in Fig. I-1.
The graphical solution of the equation (\ref{bos19}) is presented in  Fig. I-2.
In  this Figure  $X$-projection of the point of intersection of the curve 
$g_{3/2}(z,R)+\lambda^3n_0$ and of the line $\lambda^3n$ shows  
the dependence of the fugacity on  curvature. As   follows 
from the graphical picture,  for positive curvature $R<0$ the fugacity
$z(R)>z_0$ (or $\mu(R)>\mu_0$), for $R>0$ we have $z(R)<z_0$ 
(or $\mu(R)>\mu_0$). The fugacity $z_0$ ($\mu_0$-chemical potential) 
corresponds to statistics in  Euclidean space. 
The behaviour  of the  effective chemical potential is shown in  Fig I-3. 
 
\section{Bose-Einstein condensation}
\vsse

At low temperature  there is a significal  number
of particles in the ground state,
 which can be expressed by the equation
\bd
n_0=n-\lambda^{-3}g_{3/2}(z,R) \label{bos21}
\ed   
With the rising of the temperature  the  average number of particles $n_0$ with
zero momentum  is lowered and for 
temperatures  $T>T_{c}$ it  becomes zero.
The temperature $T_{c}$ is the  critical temperature of 
a  Bose  condensation. 
The  critical temperature may be found from the equation with ($z=1$ and $n_0=0$)
\bd
n{\lambda_{c}}^{3}=g_{3/2}(1,R) \label{bos22}
\ed   
The solution of this equation is \cite{i23} 
\bd
{T_{c}(R)}={T_0}\left(1+ \frac{\gamma_1^{'}(R)}{2 m^2}+...\right) \label{bos23}
\ed
The temperature 
\bd
T_0=T_c(R=0)=\frac{2\pi}{m}\left[ \frac{n}{\zeta(3/2)}\right]^{2/3} \label{bos24}
\ed
where  ${\zeta(3/2)}=2.612...$ is the Riemann zeta function,
$n$ is the density of bosons and $m$ is the  mass of boson
is degeneracy temperature (condensation temperature) 
in "flat" space without gravity. 

The  ratio 
\bd
\frac{T_{c}(R)}{T_0}=1+ 
\frac{R}{12 m^2}+... ,\label{bos25}
\ed
is not   one, but depends on the  curvature $R$ of space-time.
 
As we  can see,   $T_{c}(R)>T_0$, 
for  $R>0$  and $T_{c}(R)<T_0$ for   $R<0$.
The  correction  
\bd \frac{\delta T_c(R)}{T_0}=\frac{R}{12 m^2}+O\left((R/m^2)^2\right)
\ed
is  small, therefore effects 
of curvature will be essential for quantum systems 
in strong gravitational fields.

%% file: ikltemfer.tex
\chapter{LOCAL STATISTICS AND} 
\centerline{\Large \bf THERMODYNAMICS OF FERMI GAS }
\vs

The results of the previous chapter XII show  the way to construct thermodynamic 
potentials of quantum systems with a  variable number of  particles.
In this chapter we will develop this formalism for fermi systems.

\section {Grand thermodynamical potential and} 
\lum
\hspace{20mm}{\Large \bf  low temperature properties of Fermi gases}
\vsse

An equivalent to the  Schwinger-DeWitt representation, the momentum space
representation of the  bi-spinor $G_F(x,x^{'})$ 
has the following form (\ref{fer14}):
\bd
G_F(x,x^{'})=G_F(x,y)=g^{-\frac{1}{ 4}}(y)\sum\limits_{j=0}^2 {\hat \alpha
_j(x,y)\left( {-{\partial  \over {\partial m^2}}} \right)^j}G_0(y) \label{tf1}
\ed 
where
\bd
G_0(y)=\int {{{d^4k} \over {(2\pi )^4}}}\,{{e^{iky}} \over {k^2+m^2}}\label{tf2} 
\ed
To introduce temperature, we will extend the time $y^{0}$ coordinate of the
tangential space $\left\{{y^\mu}\right\}$ to the imaginary interval
$[0,-i\beta ]$ and will consider fermionic field to be antiperiodic on
that interval.  Then, in imaginary time formalism, we can write 
\bd
\int {{{d^4k} \over {(2\pi )^4}}}{1 \over {k^2+m^2}}\buildrel
\beta  \over
\longrightarrow {i \over {\beta }}\int {{{d^3k} \over {(2\pi
)^3}}}\sum\limits_{n=-\infty }^\infty  {{1 \over {\omega _n^2+\varepsilon
^2}}} \label{tf3}
\ed 
where $\omega _n=\pi T(2n+1),\quad n=0,\;\pm 1,\;\pm 2,\quad .\;.\;.$
are Matsubara frequencies. 
  To take into consideration
the chemical potential we will  shift   the frequencies by
 $\omega _n \to \omega _n+\mu $   \cite{mor1,kap1}

Making the summation in (\ref{tf3}) we will find 
$G_0(y)$ in the limit of coincidence $x=x'$ in the form
\bd
\mathop{\lim } \limits_{x\to x^{'}} G_0\left( y \right)\mathop
= \limits^\beta {i\over \beta } \int {{{d^3k} \over {\left( {2\pi }
\right)^3}}}\sum\limits_{n=-\infty }^\infty 
{{1 \over {-\left( {\omega _n+\mu }\right)^2 + \varepsilon ^2}}} \nonumber
\ed
\veb
\bd
={{i \over \beta }\int {{{d^3k} \over {\left( {2\pi } \right)^3}}\left\{ {{1
\over {2\varepsilon }}\sum\limits_{n=-\infty }^\infty  {\left( {{{\varepsilon
-\mu } \over {\left( {\varepsilon -\mu } \right)^2-\omega_{n}^2}}+{{\varepsilon
+\mu } \over {\left( {\varepsilon +\mu } \right)^2-\omega_{n}^2}}} \right)}}
\right\}}} \nonumber
\ed
\veb
\bd
={\int {{{d^3k} \over {\left( {2\pi } \right)^3}}}\left\{ {{i \over
{2\varepsilon }}\left( {{1 \over 2}-{1 \over {\exp \beta \left( {\varepsilon -\mu
} \right)+1}}} \right)+{i \over {2\varepsilon }}\left( {{1 \over 2}-{1 \over
{\exp \beta \left( {\varepsilon +\mu } \right)+1}}} \right)} \right\}} \nonumber
\ed
\veb
\bd
=\mathop {\lim }\limits_{x\to x^{'}}\left[ {G_\beta ^+\left( y \right)+G_\beta
^-\left( y \right)} \right] \label{tf4}
\ed
This equation describes the temperature contributions for particles
$(\mu)$ and antiparticles $(-\mu)$  separately.

As a  result one can find the expression for the 
finite temperature contribution to the  Green's function for a fermion
 with non-zero
chemical potential 
\bd
G_\beta (x,\mu )=\int {{{d^3k} \over {(2\pi
)^3}}}\sum\limits_{j=0}^2 {\hat \alpha _j}(R)\left( {{\partial  \over {\partial
m^2}}} \right)^j\left( {1+ze^{-\beta
\varepsilon }} \right)^{-1} \label{tf5}
\ed 
Then the density of grand thermodynamical potential for fermions may be written as
\bd
\omega (\beta ,\mu ,R)=-{i \over
2} \mbox{tr} \int\limits_{m^2}^\infty  {dm^2G_\beta (x,\mu )}=\sum\limits_{j=0}^2 {\alpha
_j}(R)f_j(\beta m;z), \label{tf6}
\ed 
where 
\bd
f_0(\beta m;z)=-{s
\over \beta }\int {{{d^3k} \over {(2\pi )^3}}
\ln (1+ze^{-\beta \varepsilon })} \label{tf7}
\ed
and
\bd
f_j(\beta m;z)=\left( {-{\partial  \over {\partial m^2}}}
\right)^jf_0(\beta m;z),\label{tf8}
\ed
where $z=e^{\beta \mu }$  is the   fugacity and factor  $s=2$ (spin up, down).  

The coincidence of the  finite temperature Schwinger-DeWitt and momentum space methods 
for calculation of the densities of
themodynamical potentials is obvious for $\mu =0$.

Using the equations for thermodynamic potentials one can obtain some
interesting properties of an ideal fermi gas in an external gravitational field :

1) The  Fermi distribution function of the gas in the  gravitational field may be found
from the expression 
\bd
n_{\vec{k}}=-{{\partial \omega _{\bar k}(\beta ,\mu ,R)}
\over {\partial \mu }} \nonumber
\ed
for occupation numbers with  momentum  $\vec k$.

It has the  form 
\bd
n_{\vec{ k}}={1 \over {(z^{-1}e^{\beta \varepsilon _{\vec{ k}}}+1)}}F(\beta
,R),\label{tf9}
\ed 
where the function $F(\beta ,R)$ is described by the formula
\bd
F(\beta ,R)=\ {1+\alpha _1(R){\beta  \over {2\varepsilon
_{\vec{k}}}}\left[{1-(1+ze^{-\beta \varepsilon _{\vec{k}}})^{-1}}\right]+\quad
.\;.\;.} \label{tf10}
\ed 
and depends on curvature, temperature and energy of the fermion. 

2)  We can estimate
the dependence of the chemical potential on the curvature of space time in
non-relativistic approximation.

Let  
\bd
\varepsilon _{\vec{ k}}=\frac{\vec{k}^{2}}{ 2m}
\ed
then from
 \bd
n=-{{\partial\omega (\beta,\mu,R)} \over {\partial\mu }}
\ed
we find the equation
\bd
\frac{n\lambda ^3(T)}{s}=f_{3\over 2}(z,R),\label{tf11}
\ed 
where  $\lambda =\left( {2\pi/ mT} \right)^{1/ 2}$ is the thermal wave
 length of the particle, and
\bd
f_{3/ 2}(z,R)=f_{3/2}(z)\left[\alpha _0-{3 \over4}{{\alpha _1(R)} 
\over {m^2}}-{3 \over {16}}{{\alpha _2(R^2)} \over {m^4}}-...\right]\label{tf12}
\ed
is some function with respect to $z$, and $n$ is the density of Femi gas. 

The function $f_{\frac{3}{2}}(z)$ is
\bd
f_{3/2}(z)=\sum\limits_{n=1}^\infty  {{{z^n} \over {n^{3/
2}}}}\,(-1)^{n+1}={4
\over {\sqrt \pi }}\int\limits_0^\infty  {dx{{x^2}
 \over {z^{-1}\exp (x^2)+1}}}.\label{tf13}
\ed 
 The equation (\ref{tf11}) may be solved with graphical methods. 
As we can see in  Fig. I-4  
the fugacity (chemical potential) depends on the curvature R of the space time.

3)  The explicit expression for the chemical potential 
at low temperatures and high densities ($n\lambda ^3>>1$), where quantum effects
are essential, may be found by the calculations of (\ref{tf11}) with the following
representation of the function $f_{3/2}(z)$ for large $z$ 
\cite{huang1}
\bd
f_{3/2}(z) = {4 \over {3\sqrt \pi }} 
\left[ 
{\left( {\ln z} \right)^{3/2} + {{\pi ^2} \over 8}
				\left( {\ln z} \right)^{-1/2}       
    + \quad \ldots}
\right] + O(z^{-1}).\label{tf14}
\ed 
Inserting (\ref{tf12}) into (\ref{tf11}) and taking into account only the 
first term
in (\ref{tf14}) we find  the Fermi energy of a  gas of fermions  in curved
space-time
\bd
\left( {3n\sqrt \pi \over 4s} \right)^{2/3}\lambda ^2\left[1+{1 \over
{24}}{R \over {m^2}}+\quad .\;.\;.\right]=\beta \varepsilon _F(R),\label{tf15}
\ed 
or
\bd
\varepsilon
_F(R)=\varepsilon _F^{(0)}\left[1+{1 \over {24}}{R \over {m^2}}+\quad
.\;.\;.\right],\label{tf16}
\ed 
where 
\bd
\varepsilon _F^{(0)}=\left( {6\pi ^2n\over s} \right)^{2\over 3}
\left( {1\over 2m} 
\right)
\ed
is the  Fermi energy at the Euclidean space.

Taking into account the second term in (\ref{tf14}) we get the expression 
for chemical
potential
\bd
\mu (T,R)=\varepsilon _F(R)\left\{ {1-{{\pi ^2} \over {12}}\left(
{{T \over {\varepsilon _F(R)}}} \right)^2+\quad .\;.\;.} \right\} \nonumber
\ed
\veb
\bd
=\varepsilon _F^{(0)}\left\{ {1+{1 \over {24}}{R \over {m^2}}-{{\pi ^2}
\over {12}}\left( {{T \over {\varepsilon _F^{(0)}}}} \right)^2+\quad
.\;.\;.}
\right\} \nonumber
\ed
or
\bd
\mu (T,R)=\mu ^{(0)}(T)+{1 \over {24}}{R \over {m^2}}\varepsilon
_F^{(0)}+\quad .\;.\;.\; \label{tf17}
\ed 
which describes the explicit dependence of the chemical
potential of the fermionic gas on temperature T and the curvature R of
space-time.

\mad
\centerline{\Large\bf Conclusion} 
\mad

The thermodynamical potentials of  quantum bose and fermi gases (as  local
 thermodynamical objects  in a curved space-time) were rewritten 
in terms of the finite temperature   Green's functions of bosons and 
fermions by means of the local momentum space formalism.
The   phenomenon of  Bose condensation 
was studied and  the critical temperature of condensation as 
a  functional of curvature was found (\ref{bos23}).  
The  non-thermal character of  Bose and  Fermi distribution functions 
(\ref{ins1}) and (\ref{tf9}) was shown.
The dependence of Fermi energy (\ref{tf16}) and the  chemical potential (\ref{tf17})  
of a fermi gas  on  the curvature of space-time was computed for low temperatures.
The dependence of chemical potential of bose gas 
on the  curvature of space-time was analyzed.
 It was found that  the  temperature  is  a  local thermodynamical  characteristic 
of thermal systems in external gravitational fields.

%% file: ikinterectn.tex
 \begin{center} 
\vspace*{8mm}{\LARGE \bf PART II} 
\end{center}
\vspace{2mm}
\begin{center} 
{\LARGE\bf{INTERACTING FIELDS}} 
\end{center}
\begin{center} 
{\LARGE\bf{AT FINITE TEMPERATURE}} 
\end{center}
\vspace{2mm}
\begin{center} 
{\Large\bf  Introduction } 
\end{center}

\vspace{5mm}
In Part I of this  work we studied ideal quantum systems 
(systems in which particles 
don't interact with each other, but only with external fields) 
at definite temperatures.  We described statistical properties
 and thermodynamical behavior 
of quantum ensembles of these particles in an arbitrary curved space-time.
However  can develop a formalism of  finite temperature field theory for
applications to the systems of interacting particles (bosons or fermions)
and study properties of such systems in external gravitational fields.

Part II is devoted to studying  thermal interacting  quantum systems 
in gravitational fields. 
 In this part we study renormalizability of the finite temperature
self-interacting scalar  $\lambda\varphi^4$ model, and  
consider the phenomenon of non-equality between inertial and
gravitational masses of the  boson in perturbative regime at finite temperature.

In Part II  the following topics are developed:
In chapter XIV  $\lambda\varphi^4$  model at
two-loop perturbative regime is considered. The concepts of renormalizations
and  all necessary counterterms  are described.  
In chapter XV a complete renormalization procedure 
for finite temperature model $\lambda\varphi^4$ in two loop approximation  
of the perturbative scheme is   developed. 
The  Green's function of a boson in a heat bath is computed.

In chapter XVI the  finite temperature 
Hamiltonian of a  boson is constructed.
The  phenomenon  of non-equality between
inertial and gravitational masses of a  boson in non-relativistic aproximation at
high temperature in the heat bath is  described. 

\chapter{TWO-LOOP RENORMALIZATIONS IN}
\centerline{\Large \bf  $\lambda \phi^4$ MODEL}
\vs

We will start with the problem of renormalization procedure for the self-interacting 
scalar model in two-loop approximation of perturbation theory. In our calculations
we will use  the method of counterterms.

Let the Lagrangian of the self-interacting $\lambda\varphi^{4}$ model be
\bd
L={1 \over 2}\left( {\partial \varphi } \right)^2-{1 \over 2}
m_B^2\varphi ^2-{{\lambda _B} \over {4!}}\varphi ^4 \label{new1}
\ed
We may assume that (\ref{new1}) is
\bd
L=L_0+L_I \label{new2} 
\ed
where the Lagrangian 
\bd
L_0={1 \over 2}\left( {\partial \varphi } \right)^2-{1 \over 2}
m_R^2\varphi ^2 \label{new3} 
\ed
describes propagation of free particles,
and the Lagrangian 
\bd
L_I=-{1 \over 2}\delta m^2\varphi ^2-{{\lambda _B} \over {4!}}
\varphi ^4 \label{new4} 
\ed
describes interaction.

Let the coefficient  $\delta m^2$ be the difference of the form 
 \bd
\delta m^2=m_B^2-m_R^2 \nonumber
\ed
and constant $\lambda_B$ is expressed as
\bd
\lambda_B=\mu ^{4-n}(\lambda _R+\delta \lambda ) \label{new5}
\ed
We determine $m_B$ and $m_R$  as bare and renormalizable boson masses and
$\lambda_B$ and $\lambda_R$ as bare and renormalizable constants of interaction.

It is easy to see from (\ref{new3}) and (\ref{new4}) 
that the  free propagator of a  scalar field is

\begin{picture}(5,2.5)
\put(6,1.0){\line(1,0){1}}
\put(4,1.0){\makebox(0,0){$G=i/ (p^2-m_R^2)=$}}
\end{picture}

and its  vertex is 

\begin{picture}(5,2.5)
\put(6,1.0){\line(1,1){0.5}}
\put(6,1.0){\line(1,-1){0.5}}
\put(6,1.0){\circle*{.1}}
\put(6,1.0){\line(-1,1){0.5}}
\put(6,1.0){\line(-1,-1){0.5}}
\put(4,1.0){\makebox(0,0){$-i\lambda _R\mu ^{4-n}=$}}
\end{picture}

Feynman rules for this model have a standard form \cite{its1}.

The Feynman diagrams of the counterterms may be found from the  
Lagrangian of interaction 
(\ref{new4}).  

The  two point counterterm diagram is 

\begin{picture}(8,2.5)
\put(5,1){\makebox(0,0){$-i\delta m^2=$}}
\put(6,1){\line(1,0){1.5}}
\put(6.7,1){\makebox(0,0){$\times$}}
\end{picture}

and the  vertex counterterm is 

\begin{picture}(5,2.5)
\put(6,1.0){\line(1,1){0.5}}
\put(6,1.0){\line(1,-1){0.5}}
\put(6,1.0){\circle{0.2}}
\put(6,1.0){\line(-1,1){0.5}}
\put(6,1.0){\line(-1,-1){0.5}}
\put(4,1.0){\makebox(0,0){$-i\delta \lambda \mu ^{4-n}=$}}
\end{picture}

 Taking into account  these counterterms one   can construct
 contributions of the order $\lambda_B$ to Feynman
 propagator $G$

\begin{picture}(8,3)
\put(1.5,1){\line(1,0){1.5}}
\put(3.4,1){\makebox(0,0){+}}
\put(2.2,0.3){\makebox(0,0){a)}}
\put(4.,1){\line(1,0){1.5}}
\put(4.7,1.5){\circle{1}}
\put(4.7,1){\circle*{.1}}
\put(4.7,0.3){\makebox(0,0){b)}}
\put(6.2,1){\makebox(0,0){+}}
\put(7,1){\line(1,0){1.5}}
\put(7.7,1){\makebox(0,0){$\times$}}
\put(7.7,0.3){\makebox(0,0){c)}}
\end{picture}

Fig. II-1 One loop and counterterm contributions to the self energy of the boson.

Propagator Fig. 1a) has a standard form.  

The self-energy diagram of the first order to $\lambda_B$ for 
Fig. 1-b) may be constructed  from the  above mentioned  graphs  in the form 

\begin{picture}(8,3)
\put(4,1){\makebox(0,0){$Fig.1b~~=~~$}}
\put(5,1.){\line(1,0){1.5}}
\put(5.7,1.5){\circle{1}}
\put(5.7,1.){\circle*{.1}}
\put(7,1){\makebox(0,0){$~~=~~$}}
\end{picture}
\vspace{-5mm}
\begin{eqnarray}
={1 \over 2}{{\left( {-i\lambda _R\mu ^{4-n}} \right)} 
\over {\left( {2\pi } \right)^n}}\int {d^n}k{i \over {k^2-m_R^2}} \nonumber
\ed
\veb
\bd
={1 \over 2}{{\left( {\lambda _R\mu ^{4-n}} \right)}
 \over {\left( {2\pi } \right)^n}}\int {d^n}k{1 \over {k^2-m_R^2}} \label{new6}
\ed
The factor $(1/2)$ is the symmetry factor and $\mu$  is a parameter with dimension of
mass. This parameter is used to absorb the dimension of the coupling constant.

Using the expression \cite{key4} for the  $n$-dimensional integral 
\bd
\int {d^n}k{1 \over {\left( {k^2+2kq-m^2} \right)^\alpha }}
=\left( {-1} \right)^\alpha i\pi ^{{n \over 2}}{{\Gamma 
\left( {\alpha -{n \over 2}} \right)} \over {\Gamma 
\left( \alpha  \right)}}{1 \over {\left( {q^2+m^2} \right)^{\alpha 
-{n \over 2}}}} \label{new7}
\ed
we find

\begin{picture}(8,3)
\put(3,1){\makebox(0,0){$Fig.1b~~=~~$}}
\put(4,1.){\line(1,0){1.5}}
\put(4.7,1.5){\circle{1}}
\put(4.7,1.){\circle*{.1}}
\put(6,1){\makebox(0,0){$~~=~~$}}
\end{picture}
\vspace{-5mm}
\begin{eqnarray}
={{\lambda _R} \over 2}{{\mu ^{4-n}} \over {\left( {2\pi } 
\right)^n}}{{\left( {-i\pi ^{{n \over 2}}} \right)} 
\over {\left( {m^2} \right)^{1-{n \over 2}}}}\Gamma 
\left( {1-{n \over 2}} \right)= \nonumber
\ed
\veb
\bd
={{-i\lambda _R} \over 2}{{m^2} \over {16\pi ^2}}
\left( {{{m^2} \over {4\pi \mu ^2}}} \right)^{{n \over 2}
-1}\Gamma \left( {1-{n \over 2}} \right) \label{new8}
\ed
From the expression for gamma function 
\bd
\Gamma \left( {1-{n \over 2}} \right)={2 \over {n-4}}
+\gamma -1 \label{new9}
\ed
 finally get the divergent part of the self-energy diagram

\begin{picture}(8,3)
\put(3,1){\makebox(0,0){$Fig.1a~~=~~$}}
\put(4,1.){\line(1,0){1.5}}
\put(4.7,1.5){\circle{1}}
\put(4.7,1.){\circle*{.1}}
\put(6,1){\makebox(0,0){$~~=~~$}}
\end{picture}
\vspace{-5mm}
\begin{eqnarray}
=-{{i\lambda _R} \over {16\pi ^2}}{{m^2} \over {n-4}}
+\lambda _R\times finite\;term+O\left( {\lambda _R^2} \right) \label{new10}
\ed
The renormalization of the first order of $\lambda_R$ may be done with the equation
\bd
i\Gamma ^{(2)}=i\left[ {{{p^2-m_R^2} \over i}+(-i\delta m^2)-
{{i\lambda _Rm_R^2} \over {16\pi ^2}}{1 \over {n-4}}+...} \right] \label{new11}
\ed
Let us express term $\delta m^2$  in the form of series \cite{col1}
\bd
\delta m^2=m_R^2\sum\limits_{\nu =1}^\infty  
{\sum\limits_{j=\nu }^\infty  {{{b_{\nu j}\lambda _R^j} \over {(n-4)^\nu }}=}} \nonumber
\ed
\veb
\bd
=m_R^2\left\{ {{{b_{11}\lambda _R} \over {(n-4)}}+{{b_{12}\lambda _R^2} 
\over {(n-4)}}+{{b_{22}\lambda _R^2} \over {(n-4)^2}}+...} \right\} \label{new12}
\ed
where the coefficients $b_{\nu j}$  are the numbers.

In order for vertex $\Gamma ^{(2)}$ in one loop approximation to be finite 
\bd
i\Gamma ^{(2)}=p^2-m_R^2+\left( {\delta m^2+{{\lambda _Rm_R^2} 
\over {16\pi ^2}}{1 \over {n-4}}} \right)=finite\label{new13}
\ed
 assume
\bd
\delta m^2=-{{\lambda _Rm_R^2} \over {16\pi ^2}}{1 \over {n-4}} \label{new14}
\ed
then
\bd
b_{11}=-{1 \over {16\pi ^2}} \label{new15}
\ed
This result  completes the one loop calculations.

Now we will find the vertex corrections for $\Gamma^{(4)}$.

There are four  graphs of the order $\lambda^2_R$:

\vspace{5mm}
\begin{picture}(8,3.2)

\put(0,1.5){\line(1,1){0.5}}
\put(0,1.5){\line(1,-1){0.5}}
\put(0,1.5){\circle{.2}}
\put(0,1.5){\line(-1,1){0.5}}
\put(0,1.5){\line(-1,-1){0.5}}

\put(0,0){\makebox(0,0){a)}}

\put(1.3,1.3){\makebox(0.5,0.5){+}}

\put(3,1.5){\line(-1,1){0.5}}
\put(3,1.5){\line(-1,-1){0.5}}
\put(3,1.5){\circle*{.1}}
\put(3.5,1.5){\circle{1}}
\put(4,1.5){\circle*{.1}}
\put(4,1.5){\line(1,1){0.5}}
\put(4,1.5){\line(1,-1){0.5}}

\put(3.5,0){\makebox(0,0){b)}}
\put(5,1.3){\makebox(0.5,0.5){+}}

\put(6.5,2){\line(-1,1){0.5}}
\put(6.5,1){\line(-1,-1){0.5}}
\put(6.5,2){\circle*{.1}}
\put(6.5,1.5){\circle{1}}
\put(6.5,1){\circle*{.1}}
\put(6.5,2){\line(1,1){0.5}}
\put(6.5,1){\line(1,-1){0.5}}

\put(6.5,0){\makebox(0,0){c)}}
\put(8,1.3){\makebox(0.5,0.5){+}}

\put(9.5,2){\line(-1,1){0.5}}
\put(9.5,1){\line(-1,-1){0.5}}
\put(9.5,2){\circle*{.1}}
\put(9.5,1.5){\circle{1}}
\put(9.5,1){\circle*{.1}}
\put(9.5,2){\line(2,-1){1.5}}
\put(9.5,1){\line(2,1){1.5}}

\put(9.5,0){\makebox(0,0){d)}}

\end{picture}

\vspace*{0.5cm}

\vspace{5mm}
Fig. II-2  Feynman diagrams contributing to the vertex correction $\Gamma^{(4)}$.
 
The vertex counterterm Fig. 2a) was introduced before.

The loop contribution (Fig. 2b)) of the order $\lambda^2_R$   to the vertex function is

\begin{picture}(8,3)
\put(5,1){\line(-1,1){0.5}}
\put(5,1){\line(-1,-1){0.5}}
\put(5,1){\circle*{.1}}
\put(5.5,1){\circle{1}}
\put(6,1){\circle*{.1}}
\put(6,1){\line(1,1){0.5}}
\put(6,1){\line(1,-1){0.5}}
\put(7,1){\makebox(0,0){=}}
\end{picture}
\vspace{-3mm}
\begin{eqnarray}
={1 \over 2}{{\left( {-i\lambda _R\mu ^{4-n}} \right)^2} 
\over {(2\pi )^{2n}}}\int {d^nk{{(i)^2} \over {\left( {k^2-m^2} 
\right)\left[ {\left( {p+k} \right)^2-m^2} \right]}}}= \nonumber
\ed
\veb
\bd
={1 \over 2}{{\lambda _R^2\mu ^{8-n}} \over {(2\pi )^{2n}}}
\int {d^nk\int\limits_0^1 {dx{1 \over {\left[ {\left( {k^2-m^2}
 \right)\left( {1-x} \right)+\left[ {\left( {p+k} \right)^2-m^2} 
\right]x} \right]^2}}}}= \nonumber
\ed
\veb
\bd
={1 \over 2}{{\lambda _R^2\mu ^{8-n}} \over {(2\pi )^{2n}}}
\int\limits_0^1 {dx\int {d^nk{1 \over {\left[ {k^2+2pkx-\left( {m^2-p^2x}
 \right)} \right]^2}}}} \label{new16}
\ed
Taking into account (\ref{new7}) we get

\begin{picture}(8,3)
\put(5,1){\line(-1,1){0.5}}
\put(5,1){\line(-1,-1){0.5}}
\put(5,1){\circle*{.1}}
\put(5.5,1){\circle{1}}
\put(6,1){\circle*{.1}}
\put(6,1){\line(1,1){0.5}}
\put(6,1){\line(1,-1){0.5}}
\put(7,1){\makebox(0,0){=}}
\end{picture}
\vspace{-3mm}
\begin{eqnarray}
={{\lambda _R^2\mu ^{8-n}i\pi ^{{n \over 2}}}
 \over {2(2\pi )^{2n}}}\Gamma \left( {2-{n \over 2}} 
\right)\int\limits_0^1 {dx{1 \over {\left[ {m^2-p^2x(1-x)} 
\right]^{2-{n \over 2}}}}}= \nonumber
\ed
\veb
\bd
=-\frac{i\lambda ^2_R}{16 \pi ^2}\frac{1}{n-4}
+\lambda ^2_R \times finite~~term \label{new17}
\ed
Here we used the representation for $\Gamma \left( {2-{n / 2}} \right)$
 in the form
\bd
\Gamma \left( {2-{n \over 2}} \right)=-{2 \over {n-4}}-\gamma \label{new18} 
\ed
Two other loops have the same divergent contributions,
therefore $\Gamma^{(4)}$ vertex structure can be expressed in the form
\bd
\Gamma ^{\left( 4 \right)}=-i\lambda _R\mu ^{4-n}-{3 \over {16\pi ^2}}
{{\lambda _R^2\mu ^{4-n}} \over {\left( {n-4} \right)}}
-i\delta \lambda \mu ^{4-n}= \nonumber
\ed
\veb
\bd
\Gamma ^{\left( 4 \right)}=-i\lambda _R\mu ^{4-n}-i
\left[ {{3 \over {16\pi ^2}}{{\lambda _R^2\mu ^{4-n}}
 \over {\left( {n-4} \right)}}+\delta \lambda } \right]\mu ^{4-n}
=finite \label{new19}
\ed
From the equation (\ref{new19}) we find, that
\bd
\delta \lambda =-{3 \over {16\pi ^2}}\lambda _R^2 \label{new20}
\ed
Putting
\bd
\lambda _B=\mu ^{4-n}(\lambda _R+\delta \lambda ) \nonumber
\ed
\veb
\bd
=\mu ^{4-n}\left[ {\lambda _R^2+\sum\limits_{\nu =1}^\infty 
 {\sum\limits_{j=\nu }^\infty  {{{a_{\nu j}\lambda _R^j} 
\over {\left( {n-4} \right)^\nu }}}}} \right] \label{new21}
\ed
we find coefficient $a_{12}$:
\bd
a_{12}=-{3 \over {16\pi ^2}} \label{new22}
\ed
Therefore
\bd
\lambda _B=\mu ^{4-n}\left[ {\lambda _R-{3 \over {16}}{{\lambda _R^2} 
\over {(n-4)}}+O\left( {\lambda _R^3} \right)} \right] \label{new23}
\ed
Further we will consider two-loop contributions to two-point vertex $\Gamma^{(2)}$ 

\begin{picture}(8,4)
\put(0,1.){\line(1,0){1.5}}
\put(0.7,1.5){\circle{1}}
\put(0.7,1.){\circle*{.1}}
\put(0.7,2.){\makebox(0,0){$\times$}}
\put(0.7,0){\makebox(0,0){a)}}
\put(1.7,0.9){\makebox(0.5,0.5){+}}
\put(3.,1.){\line(1,0){1.5}}
\put(3.7,1.5){\circle{1}}
\put(3.7,1.){\circle{.2}}
\put(3.7,0){\makebox(0,0){b)}}
\put(5,0.9){\makebox(0.5,0.5){+}}
\put(6,1){\line(1,0){1.5}}
\put(6.9,1.5){\circle{1}}
\put(6.9,1.){\circle*{.1}}
\put(6.9,2.5){\circle{1}}
\put(6.9,2){\circle*{.1}}
\put(6.7,0){\makebox(0,0){c)}}
\put(8,0.9){\makebox(0.3,0.5){+}}
\put(8.5,1.){\line(1,0){1.6}}
\put(9.3,1.){\circle{1}}
\put(8.8,1.){\circle*{.1}}
\put(9.8,1.){\circle*{.1}}
\put(9.3,0){\makebox(0,0){d)}}
\end{picture}

\vspace{5mm}
Fig. II-3  Counterterms and the loop contributions of the order $\lambda^2_R$ 
to the self energy. 

To find counterterms to the order $O(\lambda^3_R)$, 
we will make loop calculation of all these contributions.

Diagram 3(a) gives

\begin{picture}(8,3)
\put(3,1){\makebox(0,0){$Fig.3a~~=~~$}}
\put(4,1.){\line(1,0){1.5}}
\put(4.7,1.5){\circle{1}}
\put(4.7,1.){\circle*{.1}}
\put(6,1){\makebox(0,0){$=$}}
\put(4.7,2){\makebox(0,0){$\times$}}
\end{picture}
\vspace{-5mm}
\begin{eqnarray}
={1 \over 2}{{\left( {-i\lambda _R\mu ^{4-n}} \right)}
 \over {\left( {2\pi } \right)^n}}\left( {-i\delta m^2} \right)\int {d^nk{{(i)^2}
 \over {\left( {k^2-m^2} \right)^2}}=} \nonumber
\ed
\veb
\bd
={1 \over 2}{{\lambda _R\mu ^{4-n}} \over {\left( {2\pi } \right)^n}}
\left( {\delta m^2} \right){{i\pi ^{{n \over 2}}\Gamma 
\left( {2-{n \over 2}} \right)} \over {\left( {m^2} \right)^{2-{n \over 2}}}}= \nonumber
\ed
\veb
\bd
=\frac{\lambda_R}{32 \pi^2}(\delta m^2)\Gamma\left(2-\frac{n}{2}\right)
\left(\frac{4\pi \mu ^2}{m^2}\right) ^{2-\frac{n}{2}} \label{new24}
\ed
Inserting  (\ref{new18}) into (\ref{new24}) we get 

\begin{picture}(8,3)
\put(3,1){\makebox(0,0){$Fig.3a~~=~~$}}
\put(4,1.){\line(1,0){1.5}}
\put(4.7,1.5){\circle{1}}
\put(4.7,1.){\circle*{.1}}
\put(4.7,1.){\circle*{.1}}
\put(4.7,2){\makebox(0,0){$\times$}}
\end{picture}
\vspace{-5mm}
\begin{eqnarray}
={{i\lambda _R} \over {(16\pi ^2)^2}}{{m_R^2} \over {(n-4)^2}}
+{{i\gamma \,\lambda _Rm_R^2} \over {2(16\pi ^2)^2}}{{1}
 \over {(n-4)}}+finite\;terms \label{new25}
\ed
Diagram 3(b) gives

\begin{picture}(8,2.8)
\put(3,1){\makebox(0,0){$Fig.3b~~=~~$}}
\put(4.,1.){\line(1,0){1.5}}
\put(4.7,1.5){\circle{1}}
\put(4.7,1.){\circle{.2}}
\put(6,1){\makebox(0,0){$~~=~~$}}
\end{picture}
\vspace{-5mm}
\begin{eqnarray}
={1 \over 2}{{\left( {-i\delta \lambda \mu ^{4-n}} \right)} 
\over {\left( {2\pi } \right)^n}}\int {d^nk{i \over {k^2-m^2}}=} \nonumber
\ed
\veb
\bd
={1 \over 2}{{\delta \lambda \mu ^{4-n}} \over {\left( {2\pi }
 \right)^n}}\left( -{i\pi ^{{n \over 2}}} \right)
\frac{\Gamma \left( {1-{n \over 2}} \right)} 
{\left( {m^2} \right)^{1-{n \over 2}}}\label{new26}
\ed
Inserting (\ref{new9}) into (\ref{new26}) we find

\begin{picture}(8,3)
\put(3,1){\makebox(0,0){$Fig.3b~~=~~$}}
\put(4.,1.){\line(1,0){1.5}}
\put(4.7,1.5){\circle{1}}
\put(4.7,1.){\circle{.2}}
\put(6,1){\makebox(0,0){$~~=~~$}}
\end{picture}
\vspace{-5mm}
\begin{eqnarray}
={{3i} \over {(16\pi ^2)^2}}{{\lambda _R^2} \over {(n-4)^2}}
+{{3i} \over {2(16\pi ^2)^2}}{{\lambda _R^2} \over {(n-4)}}
\left( {\gamma -1} \right) \label{new27}
\ed
Diagram 3(c) gives:

\begin{picture}(8,4)
\put(3,1){\makebox(0,0){$Fig.3c~~=~~$}}
\put(4,1){\line(1,0){1.5}}
\put(4.9,1.5){\circle{1}}
\put(4.9,1.){\circle*{.1}}
\put(4.9,2.5){\circle{1}}
\put(4.9,2){\circle*{.1}}
\put(6.2,1){\makebox(0,0){$~~=~~$}}
\end{picture}
\vspace{-5mm}
\begin{eqnarray}
={1 \over 4}{{\left( {-i\lambda _R\mu ^{4-n}} \right)^2} 
\over {\left( {2\pi } \right)^{2n}}}
\int {d^nk\int {d^nl{{\left( i \right)^3} \over {\left( {k^2-m^2}
 \right)^2\left( {l^2-m^2} \right)}}}}= \nonumber
\ed
\veb
\bd
={i \over 4}{{\lambda _R^2\mu ^{8-2n}} \over {\left( {2\pi }
 \right)^{2n}}}\int {d^nk{1 \over {\left( {k^2-m^2} \right)^2}}
\int {d^nl{1 \over {\left( {l^2-m^2} \right)}}}}= \nonumber
\ed
\veb
\bd
={i \over 4}{{\lambda _R^2\mu ^{8-2n}} \over {\left( {2\pi } 
\right)^{2n}}}\left\{ {\left( {i\pi ^{{n \over 2}}} \right){{\Gamma 
\left( {2-{n \over 2}} \right)} \over {\left( {m^2} \right)^{2-{n \over 2}}}}}
 \right\}\left\{ {\left( {-i\pi ^{{n \over 2}}} \right){{\Gamma 
\left( {1-{n \over 2}} \right)} \over {\left( {m^2} \right)^{1-{n \over 2}}}}} 
\right\}= \nonumber
\ed
\veb
\bd
={i \over 4}{{\lambda _R^2} \over {(16\pi ^2)^2}}\left( {{{4\pi \mu ^2} 
\over {m^2}}} \right)^{4-n}\Gamma \left( {2-{n \over 2}} \right)\Gamma 
\left( {1-{n \over 2}} \right)= \nonumber
\ed
\veb
\bd
={-{i\lambda _R^2} \over {(16\pi ^2)^2}}{{m_R^2} \over {\left( {n-4} \right)^2}}
-{{i\lambda _R^2} \over {(32\pi ^2)^2}}{{m_R^2} \over {\left( {n-4} \right)}}
\left[ {4\gamma -2} \right]+finite\;terms \label{new28}
\ed
Diagram 3(d) is

\begin{picture}(8,2.8)
\put(3,1){\makebox(0,0){$Fig.3d~~=~~$}}
\put(4.5,1.){\line(1,0){1.6}}
\put(5.3,1.){\circle{1}}
\put(4.8,1.){\circle*{.1}}
\put(5.8,1.){\circle*{.1}}
\put(7,1){\makebox(0,0){$~~=~~$}}
\end{picture}
\vspace{-5mm}
\begin{eqnarray}
={1 \over 6}{{\left( {-i\lambda _R\mu ^{4-n}} \right)^2}
 \over {\left( {2\pi } \right)^{2n}}}\int {d^nk
\int {d^nl{{\left( i \right)^3} \over {\left( {k^2-m^2} \right)
\left( {l^2-m^2} \right)\left[ {\left( {p+k+l} \right)-m^2} \right]}}}}= \nonumber
\ed
\veb
\bd
={{i\lambda _R^2\mu ^{8-2n}} \over {6\left( {2\pi } 
\right)^{2n}}}\int {d^nk\int {d^nl{1 \over {\left( {k^2-m^2} \right)
\left( {l^2-m^2} \right)\left[ {\left( {p+k+l} \right)-m^2} \right]}}}}= \nonumber
\ed
\veb
\bd
={{i\lambda _R^2\mu ^{8-2n}} \over {6\left( {2\pi } \right)^{2n}}}
\pi ^n\Gamma \left( {3-n} \right)\left\{ {-{6 \over {n-4}}(m^2)^{n-3}-{{p^2}
 \over 2}+3m^2} \right\} \label{new29}
\ed
The function $\Gamma(3-n)$  may be written as
\bd
\Gamma \left( {3-n} \right)={1 \over {n-4}}+\gamma -1 \label{new30}
\ed
As the result, the expression for 3(d) will be

\begin{picture}(8,2.8)
\put(3,1){\makebox(0,0){$Fig.3d~~=~~$}}
\put(4.5,1.){\line(1,0){1.6}}
\put(5.3,1.){\circle{1}}
\put(4.8,1.){\circle*{.1}}
\put(5.8,1.){\circle*{.1}}
\put(7,1){\makebox(0,0){$~~=~~$}}
\end{picture}
\vspace{-2mm}
\begin{eqnarray}
=-{{i\lambda _R^2m_R^2} \over {\left( {16\pi ^2} \right)^2}}
\left( {{{m^2} \over {4\pi \mu ^2}}} \right)^{n-4}{1 \over {(n-4)^2}}  \nonumber
\ed
\veb
\bd
-{{i\lambda _R^2} \over {\left( {16\pi ^2} \right)^2}}{1 \over {(n-4)}}
\left\{{ {{p^2} \over {12}}-{{m^2} \over 2}+(\gamma -1)m^2} \right\} \label{new31}
\ed
The complete two loop calculations give us the following
expression for $\Gamma^{(2)}$
\bd
\Gamma^{(2)}={{p^2-m^2} \over i}-i\delta m^2-{{i\lambda _Rm_R^2}
 \over {16\pi ^2(n-4)}}+ \nonumber
\ed
\veb
\bd
+{{3i} \over {(16\pi ^2)^2}}{{\lambda _R^2m_R^2} \over {(n-4)^2}}
+{{3i} \over {2(16\pi ^2)^2}}{{\lambda _R^2m_R^2} \over {(n-4)}}\left( {\gamma -1}
 \right) \nonumber
\ed
\veb
\bd
+{i \over {(16\pi ^2)^2}}{{\lambda _R^2m_R^2} \over {(n-4)^2}}
+{i \over {2(16\pi ^2)^2}}{{\lambda _R^2m_R^2} \over {(n-4)}}\gamma  \nonumber
\ed
\veb
\bd
-{{i\lambda _R^2} \over {(16\pi ^2)^2}}{{m_R^2}
 \over {\left( {n-4} \right)^2}}-{{i\lambda _R^2} 
\over {(32\pi ^2)^2}}{{m_R^2} \over {\left( {n-4} \right)}}
\left[ {4\gamma -2} \right] \nonumber
\ed
\veb
\bd
-{{i\lambda _R^2m_R^2} \over {\left( {16\pi ^2} \right)^2}}
\left( {{{m-R^2} \over {4\pi \mu ^2}}} \right)^{n-4}
{1 \over {(n-4)^2}} \nonumber
\ed
\veb
\bd
-{{i\lambda _R^2} \over {\left( {16\pi ^2} \right)^2}}
{1 \over {(n-4)}}\left\{ {{{p^2} \over {12}}-{{m-R^2} \over 2}
+(\gamma -1)m_R^2} \right\} \label{new32}
\ed
or
\bd
i\Gamma ^{(2)}=p^2-m^2+\delta m^2+{{\lambda _Rm_R^2} \over {16\pi ^2(n-4)}}- \nonumber
\ed
\veb
\bd
-{2 \over {(16\pi ^2)^2}}{{\lambda _R^2m_R^2} \over {(n-4)^2}}
+{1 \over {(16\pi ^2)^2}}{{\lambda _R^2} \over {(n-4)}}
\left[ {{{p^2} \over {12}}-{{m_R^2} \over 2}} \right] \label{new33}
\ed
To make   $\Gamma^{(2)}$ finite in the second order of 
 $\lambda^2_R$, we will put
\bd
\delta m^2=m_R^2\left\{ {{{\lambda _R^{}} \over {(n-4)}}b_{11}
+{{\lambda _R^2} \over {(n-4)}}b_{12}+{{\lambda _R^2} \over {(n-4)^2}}
b_{22}+O(\lambda _R^3)} \right\} \label{new34}
\ed
Then, combining terms in the proper way, we get
\bd
i\Gamma ^{(2)}={{m_R^2} \over {(n-4)}}\left\{ {b_{11}
+{1 \over {16\pi ^2}}} \right\}+ \nonumber
\ed
\veb
\bd
\left[ {1+{1 \over {12(16\pi ^2)^2}}{{\lambda _R^2}
 \over {(n-4)}}} \right]\times \left\{ {p^2-m_R^2
\left[ {1+{1 \over {12(16\pi ^2)^2}}{{\lambda _R^2}
 \over {(n-4)}}} \right]^{-1}\times } \right. \nonumber
\ed
\veb
\bd
\left. {\times \left\{ {1+{{\lambda _R^2} \over {(n-4)^2}}
\left[ {{1 \over {2(16\pi ^2)^2}}-b_{12}} \right]
+{{\lambda _R^2} \over {(n-4)}}\left[ {{2 \over {(16\pi ^2)^2}}-b_{22}} \right]}
 \right\}} \right\} \label{new35}
\ed
It follows from (\ref{new35}) that  coefficient 
 $b_{11}$ is exactly (\ref{new15}).

Two other coefficients 
are found from  the suggestion that $\Gamma^{(2)}$  is analytic at $n=4$.
In the result we get that 
\bd
b_{22}={1 \over {2(16\pi ^2)^2}} \label{new36}
\ed
and $b_{12}$ is the solution of the equation
\bd
{{\lambda _R^2} \over {(n-4)^2}}\left[ {b_{12}
-{5 \over {12(16\pi ^2)^2}}} \right]=0 \label{new37}
\ed
It gives
\bd
b_{12}={5 \over {12(16\pi ^2)^2}} \label{new38}
\ed
After these calculations we will have
\bd
i\Gamma ^{(2)}=\left[ {1+{1 \over {12(16\pi ^2)^2}}{{\lambda _R^2} 
\over {(n-4)}}} \right]\times \left( {p^2-m_R^2} \right) \label{new39}
\ed
or
\bd
\Gamma ^{(2)}=Z\Gamma ^{(2)}_{reg} \label{new40}
\ed
where (wave function) renormalization constant $Z$ is 
\bd
Z=1+{1 \over {12(16\pi ^2)^2}}{{\lambda _R^2} \over {(n-4)}} \label{new41}
\ed
From the calculations  of this section we found that the bare mass $m_B$ and 
coupling constant $\lambda_B$ for the  second order of perturbation theory  are
\bd
m_B^2=m_R^2\left\{ {1+{{\lambda _R} \over {(n-4)}}
\left[ {-{1 \over {16\pi ^2}}+{5 \over {12}}
{{\lambda _R} \over {(16\pi ^2)^2}}} \right]} \right.+ \nonumber
\ed
\veb
\bd
\left. {+{{2\lambda _R^2} \over {(16\pi ^2)^2}}
{1 \over {(n-4)^2}}+O\left( {\lambda _R^3} \right)} \right\}  \label{new42}
\ed
and
\bd
\lambda _B=\mu ^{4-n}\left\{ {\lambda _R-{3 \over {(16\pi ^2)}}
{{\lambda _R^2} \over {(n-4)}}+O\left( {\lambda _R^3} \right)} \right\} \label{new43}
\ed
The expressions (\ref{new42}) and  (\ref{new43}) connect non-renormalizable 
and renormalizable parameters of the model, and the model is renormalized 
in two loop approximation of the perturbative regime.

\chapter{GREEN'S FUNCTION OF  BOSON}
\centerline{\Large \bf IN FINITE TEMPERATURE REGIME }
\vs
	
The aim of this chapter is  to construct
renormalizable Green's function for a boson in a heat bath with a definite 
temperature. For this purpose we will use the  real time representation for the  
finite temperature propagator that will let  us obtain necessary results 
in a natural and elegant way.  

We will repeat  calculations for
the contributions in self energy of the boson in one and 
two loop approximations based on the scheme, developed in the previous chapter.

In contrast to chapter XIV  we will consider that all internal lines of Feynman 
graphs are the  finite temperature propagators of the form
\bd
D(k)=D_0(k)+D_\beta (k)={i \over {k^2-m^2}}
+{{2\pi \delta (k^2-m^2)} \over {e^{\beta |k^0|}-1}}, \label{n1}
\ed
where $\beta^{-1}=T$ is the temperature.

Finite temperature calculations
don't change Feynmann rules for loop calculations
 \cite{mor1}, \cite{key21}.

Fig 1b) gives

\begin{picture}(8,3)
\put(3,1){\makebox(0,0){$Fig.1b~~=~~$}}
\put(4,1.){\line(1,0){1.5}}
\put(4.7,1.5){\circle{1}}
\put(4.7,1.){\circle*{.1}}
\put(4.8,0.5){\makebox(0,0){$(T \neq 0)$}}
\put(6,1){\makebox(0,0){$~~=~~$}}
\end{picture}
\vspace{0.0mm}
\begin{eqnarray}
={1 \over 2}{{\left( {-i\lambda _R\mu ^{4-n}} \right)}
 \over {(2\pi )^n}}\int {d^nk}D(k)= \nonumber
\ed
\veb
\bd
={1 \over 2}{{\left( {-i\lambda _R\mu ^{4-n}} \right)}
 \over {(2\pi )^n}}\int {d^nk}D_0(k)+
{1 \over 2}{{\left( {-i\lambda _R\mu ^{4-n}} \right)} 
\over {(2\pi )^n}}\int {d^nk}D_\beta (k) \nonumber
\ed
or in a more compact form

\begin{picture}(8,3)
\put(3,1){\makebox(0,0){$Fig.1b~~=~~$}}
\put(4,1.){\line(1,0){1.5}}
\put(4.7,1.5){\circle{1}}
\put(4.7,1.){\circle*{.1}}
\put(4.8,0.5){\makebox(0,0){$(T = 0)$}}
\put(6,1){\makebox(0,0){$-$}}
\end{picture}
\vspace{0.0mm}
\begin{eqnarray}
-{1 \over 2}{{\left( {-i\lambda _R\mu ^{4-n}} \right)}
 \over {(2\pi )^n}}\int {d^nk}D_{\beta}(k). \label{n2}
\ed
\vspace{1cm}
The counterterm Fig. 3a) is

\begin{picture}(8,2.3)
\put(3,1){\makebox(0,0){$Fig.3a~~=~~$}}
\put(4,1.){\line(1,0){1.5}}
\put(4.7,1.5){\circle{1}}
\put(4.7,1){\circle*{.1}}
\put(4.7,2){\makebox(0,0){$\times$}}
\put(4.8,0.5){\makebox(0,0){$(T \neq 0)$}}
\put(6,1){\makebox(0,0){$~~=~~$}}
\end{picture}
\vspace{0.0mm}
\begin{eqnarray}
={1 \over 2}{{\left( {-i\lambda _R\mu ^{4-n}} \right)}
 \over {(2\pi )^n}}(-i\delta m^2)\int {d^nk}D_0^2(k)
-\lambda _R\delta m^2\int {{d^4k}\over(2\pi)^4}D_\beta (k)D_0(k), \nonumber
\ed
or

\begin{picture}(8,3)
\put(3,1){\makebox(0,0){$Fig.3a~~=~~$}}
\put(4,1.){\line(1,0){1.5}}
\put(4.7,1.5){\circle{1}}
\put(4.7,1){\circle*{.1}}
\put(4.7,2){\makebox(0,0){$\times$}}
\put(4.8,0.5){\makebox(0,0){$(T = 0)$}}
\put(6,1){\makebox(0,0){$~~-~~$}}
\end{picture}
\vspace{0.0mm}
\begin{eqnarray}
-\lambda _R\delta m^2\int {{d^4k\over(2\pi)^4}}D_\beta (k)D_0(k). \label{n3}
\ed
The contribution of the counterterm Fig. 3b) is

\begin{picture}(8,3)
\put(3,1){\makebox(0,0){$Fig.3b~~=~~$}}
\put(4.,1.){\line(1,0){1.5}}
\put(4.7,1.5){\circle{1}}
\put(4.7,1.){\circle{.2}}
\put(4.8,0.5){\makebox(0,0){$(T \neq 0)$}}
\put(6,1){\makebox(0,0){$~~=~~$}}
\end{picture}
\vspace{0.0mm}
\begin{eqnarray}
={1 \over 2}{{\left( {-i\delta\lambda \mu ^{4-n}} \right)}
 \over {(2\pi )^n}}\int {d^nk}D(k) \nonumber
\ed
\veb
\bd
={1 \over 2}{{\left( {-i\delta \lambda \mu ^{4-n}} \right)} 
\over {(2\pi )^n}}\int {d^nk}(D_0(k)+D_\beta (k)), \nonumber
\ed
and

\begin{picture}(8,3)
\put(3,1){\makebox(0,0){$Fig.3b~~=~~$}}
\put(4.,1.){\line(1,0){1.5}}
\put(4.7,1.5){\circle{1}}
\put(4.7,1.){\circle{.2}}
\put(4.8,0.5){\makebox(0,0){$(T=0)$}}
\put(6,1){\makebox(0,0){$~~-~~$}}
\end{picture}
\vspace{0.0mm}
\begin{eqnarray}
-{\left( {i\delta\lambda } \right)}\int {{d^nk}\over{(2\pi)^n}}D_{\beta}(k).
\ed \label{n4}
\vspace{1cm}
Two loop contribution Fig.  3c) may be written in the form

\begin{picture}(8,3.5)
\put(3,1){\makebox(0,0){$Fig.3c~~=~~$}}
\put(4,1){\line(1,0){1.5}}
\put(4.9,1.5){\circle{1}}
\put(4.9,1.){\circle*{.1}}
\put(4.9,2.5){\circle{1}}
\put(4.9,2){\circle*{.1}}
\put(5,0.5){\makebox(0,0){$(T \neq 0)$}}
\put(6.2,1){\makebox(0,0){$~~=~~$}}
\end{picture}
\vspace{0.0mm}
\begin{eqnarray}
={1 \over 4}{{(-i\lambda _R\mu ^{4-n})^2}
 \over {(2\pi )^{2n}}}\int {d^nk}\int {d^nq}(D_0(k)+D_\beta (k))^2
(D_0(q)+D_\beta (q)) \nonumber
\ed
\veb
\bd
={1 \over 4}{{(-i\lambda _R\mu ^{4-n})^2} 
\over {(2\pi )^{2n}}}\int {d^nk}\int {d^nq\left[ {D_0^2(k)D_0(q)}
 \right.+D_0^2(k)D_\beta (q)+} \nonumber
\ed
\veb
\bd
\left. {+2D_0(k)D_\beta (k)D_0(q)+2D_0(k)D_\beta (k)D_\beta (q)
+D_\beta ^2(k)D_0(q)+D_\beta ^2(k)D_\beta (q)} \right]. \nonumber
\ed
Then

\begin{picture}(8,3.8)
\put(3,1){\makebox(0,0){$Fig.3c~~=~~$}}
\put(4,1){\line(1,0){1.5}}
\put(4.9,1.5){\circle{1}}
\put(4.9,1.){\circle*{.1}}
\put(4.9,2.5){\circle{1}}
\put(4.9,2){\circle*{.1}}
\put(5,0.5){\makebox(0,0){$(T = 0)$}}
\put(6.2,1){\makebox(0,0){$~~-~~$}}
\end{picture}
\vspace{0.0mm}
\begin{eqnarray}
-{{\lambda _R^2} \over 4}\int {{{d^4q} \over {(2\pi )^4}}D_\beta (q)}
\left\{ {{{\mu ^{4-n}} \over {(2\pi )^n}}\int {d^nk}D_0^2(k)} \right\} \nonumber
\ed
\veb
\bd
-{{\lambda _R^2} \over 2}\left\{ {\int {{{d^4q} \over {(2\pi )^4}}
D_0(k)D_\beta (k)}} \right\}\left\{ {{{\mu ^{4-n}} \over {(2\pi )^n}}
\int {d^nq}D_0(q)} \right\} \nonumber
\ed
\veb
\bd
-{{\lambda _R^2} \over 2}\left\{ {\int {{{d^4q} \over {(2\pi )^4}}
D_\beta (q)}} \right\}\left\{ {\int {{{d^4k} \over {(2\pi )^4}}}.
D_0(k)D_\beta (k)} \right\} \label{n5}
\ed
Finally the contribution Fig.  3d) will be

\begin{picture}(8,2.8)
\put(3,1){\makebox(0,0){$Fig.3d~~=~~$}}
\put(4.5,1.){\line(1,0){1.6}}
\put(5.3,1.){\circle{1}}
\put(4.8,1.){\circle*{.1}}
\put(5.8,1.){\circle*{.1}}
\put(7,1){\makebox(0,0){$~~=~~$}}
\put(5.3,0){\makebox(0,0){$(T \neq 0)$}}
\end{picture}
\bd
{1 \over 6}{{(-i\lambda _R\mu ^{4-n})^2}
 \over {(2\pi )^{2n}}}\int {d^nk}D_\beta (k)\times \nonumber
\ed
\veb
\bd
\times\left[ {\int {d^nq}D_\beta (q)D_0(q-p-k)} \right.
+\int {d^nq}D_\beta (q)D_0(k-p-q) \nonumber
\ed
\veb
\bd
+{\int {d^nq}D_\beta (q)D_0(q+p+k)}+\int {d^nq}D_\beta (q)D_0(q-k+p) \nonumber
\ed
\veb
\bd
+\left. {\int {d^nqD_0(q)D_0(q-p-k)+\int {d^nqD_0(q)D_0(q+p+k)}}} \right] \nonumber
\ed
or

\begin{picture}(8,2.6)
\put(3,1){\makebox(0,0){$Fig.3d~~=~~$}}
\put(4.5,1.){\line(1,0){1.6}}
\put(5.3,1.){\circle{1}}
\put(4.8,1.){\circle*{.1}}
\put(5.8,1.){\circle*{.1}}
\put(7,1){\makebox(0,0){$~~+~~$}}
\put(5.3,0){\makebox(0,0){$(T = 0)$}}
\end{picture}
\bd
+{1 \over 2}{{(-i\lambda _R\mu ^{4-n})^2}
 \over {(2\pi )^{2n}}}\int {d^nk}D_\beta (k)\times \nonumber
\ed
\veb
\bd\times\left[ {\int {d^nq}D_\beta (q)D_0(q+p+k)} 
+\int {d^nq}D_0 (q)D_0(k+p+q)\right] \nonumber
\ed
So, we get

\begin{picture}(8,2.8)
\put(3,1){\makebox(0,0){$Fig.3d~~=~~$}}
\put(4.5,1.){\line(1,0){1.6}}
\put(5.3,1.){\circle{1}}
\put(4.8,1.){\circle*{.1}}
\put(5.8,1.){\circle*{.1}}
\put(7,1){\makebox(0,0){$~~-~~$}}
\put(5.3,0){\makebox(0,0){$(T = 0)$}}
\end{picture}
\bd
-{{\lambda _R^2} \over 2}\int {{{d^4k} \over {(2\pi )^4}}}
D_\beta (k)\left\{ {\int {{{d^4q} \over {(2\pi )^4}}}.
D_\beta (q)D_0(k+q+p)} \right\} \nonumber
\ed
\veb
\bd
-{{\lambda _R^2} \over 2}\int {{{d^4k} 
\over {(2\pi )^4}}}D_\beta (k)\left\{ {{{\mu ^{4-n}} 
\over {(2\pi )^n}}\int {d^nq}D_0(q)D_0(k+q+p)} \right\} \label{n6}
\ed
Now we can find counterterms. 

Assuming the sum of the finite temperature 
contributions of the equations (\ref{n3}) and (\ref{n5})  zero,
 \bd
\left\{ {\lambda _R\delta m^2+{{\lambda _R^2} \over 2}
{{\mu ^{4-n}} \over {(2\pi )^n}}\int {d^nq}D_0(q)} \right\}
\int {{{d^4k} \over {(2\pi )^4}}}D_0(k)D_\beta (k)=0. \nonumber
\ed
we find the expression for $\delta m^2$ in the form
\bd
{\delta m^2=-{{\lambda _R} \over 2}{{\mu ^{4-n}}
 \over {(2\pi )^n}}\int {d^nq}D_0(q)}. \label{n7}
\ed
The divergent part of this counterterm will be
\bd
{\delta m_{div}^2=-{{\lambda _R} \over {16\pi ^2}}{{m_R^2}
 \over {(n-4)}}}. \label{n8}
\ed
The following counterterm $\delta\lambda$  may be found by the summation of the 
finite temperature contributions of the equations (\ref{n4}), (\ref{n5}) 
and (\ref{n6}) 
\bd
\int {{d^4k} \over {(2\pi )^4}}D_\beta (k)
\left\{ i\delta \lambda 
+{{\lambda _R^2} \over 2}
\left[ {{\mu ^{n-4}} \over {(2\pi )^n}}
\int {d^nqD_0^2(q)} \right]\right. \nonumber
\ed
\veb
\bd
\left.+\lambda _R^2
\left[ {{\mu ^{n-4}} \over {(2\pi )^n}}
\int {d^nqD_0(q)D_0(k+q+p)} \right] \right\}=0. \nonumber
\ed
For zero external momentum $p$ the divergent part of the  $\delta\lambda$
will be determined by the  divergent part of the integral
\bd
{\delta \lambda ={3 \over 2}(i\lambda _R^2)
\left[ {{{\mu ^{n-4}} \over {(2\pi )^n}}\int {d^nqD_0^2(q)}} \right]} \label{n9}
\ed
and the divergent contribution will be
\bd
{\delta \lambda_{div} =-{{3\lambda _R^2} \over {16\pi ^2}}
{1\over {(n-4)}}}.\label{n10}
\ed
The  temperature counterterms (\ref{n8}) and (\ref{n10}) have the same structure as the 
counterterms which annihilate the divergent parts of zero temperature  loop 
contributions.

It is easy to see that at $T=0$ the sum of the loop contribution 
and the counterterm gives zero:

\begin{picture}(8,3)

\put(3,1.){\line(1,0){1.5}}
\put(3.7,1.5){\circle{1}}
\put(3.7,1.){\circle*{.1}}

\put(5.3,1){\makebox(0,0){+}}
\put(6.,1){\line(1,0){1.5}}
\put(6.7,1){\makebox(0,0){$\times$}}

\put(8,1){\makebox(0,0){$=$}}
\end{picture}
\vspace{-5mm}
\begin{eqnarray}
{={{(-i\lambda _R\mu ^{4-n})} \over {2(2\pi )^n}}
\int {d^nk}D_0(k)-\delta m^2=0}. \nonumber
\ed
It leads us to the equation (\ref{n7}).

Loop contributions and counterterm at $T=0$ in $\Gamma^{(4)}$ (Fig.II-2)
gives the equation
\bd
{{3 \over 2}{{\lambda _R^2\mu ^{4-n}} \over {(2\pi )^n}}
\int {d^nk}D_0(k)D_0(k+p)-i\delta \lambda =0}, \nonumber
\ed
which coincides with (\ref{n9}).

From the  mentioned above analysis we can conclude that the following loops 
have the same divergent structure:

\begin{picture}(8,4)
\put(0,1.){\line(1,0){1.5}}
\put(0.7,1.5){\circle{1}}
\put(0.7,1.){\circle*{.1}}
\put(2.3,1){\makebox(0,0){+}}
\put(1.7,0.0){\makebox(0,0){$T=0$}}
\put(3.,1){\line(1,0){1.5}}
\put(3.7,1){\makebox(0,0){$\times$}}
\put(5.4,1){\makebox(0,0){$\leftrightarrow$}}
\put(6,1){\line(1,0){1.5}}
\put(6.9,1.5){\circle{1}}
\put(6.9,1.){\circle*{.1}}
\put(6.9,2.5){\circle{1}}
\put(6.9,2){\circle*{.1}}
\put(8,1){\makebox(0,0){+}}
\put(8,0){\makebox(0,0){$T \neq 0$}}
\put(8.5,1.){\line(1,0){1.6}}
\put(9.3,1.){\circle{1}}
\put(8.8,1.){\circle*{.1}}
\put(9.8,1.){\circle*{.1}}
\end{picture}
\vspace*{8mm}

and

\begin{picture}(8,4)
\put(0,1){\line(1,1){0.5}}
\put(0,1){\line(1,-1){0.5}}
\put(0,1){\circle{.2}}
\put(0,1){\line(-1,1){0.5}}
\put(0,1){\line(-1,-1){0.5}}
\put(1,1){\makebox(0,0){+}}
\put(1,0){\makebox(0,0){$T=0$}}
\put(1.4,1){\makebox(0,0){$3 \times$}}
\put(2,1){\line(-1,1){0.5}}
\put(2,1){\line(-1,-1){0.5}}
\put(2,1){\circle*{.1}}
\put(2.5,1){\circle{1}}
\put(3,1){\circle*{.1}}
\put(3,1){\line(1,1){0.5}}
\put(3,1){\line(1,-1){0.5}}
\put(3.7,1){\makebox(0,0){$\leftrightarrow$}}
\put(4.1,1){\line(1,0){1.2}}
\put(4.7,1.5){\circle{1}}
\put(4.7,1.){\circle{.2}}
\put(6,1){\makebox(0,0){+}}
\put(6.3,1){\line(1,0){1.2}}
\put(6.9,1.5){\circle{1}}
\put(6.9,1.){\circle*{.1}}
\put(6.9,2.5){\circle{1}}
\put(6.9,2){\circle*{.1}}
\put(8,1){\makebox(0,0){+}}
\put(7,0){\makebox(0,0){$T \neq 0$}}
\put(8.5,1.){\line(1,0){1.6}}
\put(9.3,1.){\circle{1}}
\put(8.8,1.){\circle*{.1}}
\put(9.8,1.){\circle*{.1}}
\end{picture}

\vspace{10mm}
The Green's function $D^{'}(p)$ of the boson is the object which takes into 
account virtual processes of  creation and  anihilation of the additional 
particles when  this boson  moves through the vacuum. 

The graph composing $D^{'}(p)$ may be divided into two distinct and unique 
classes of 
proper and improper graphs (Fig.II-4)\footnote{The proper graphs cannot be
 divided into two disjoint  parts by the removal of 
a single line, whereas the improper ones can be disjoint \cite{key6}}.

\begin{picture}(8,3)
\put(0,1){\makebox(0,0){$D^{'}(p)=$}}
\put(1,1){\circle{.2}}
\put(1,1){\line(1,0){0.5}}
\put(1.5,1){\circle{.2}}
\put(1.2,0.2){\makebox(0,0){$D(p)$}}
\put(2,1){\makebox(0,0){+}}
\put(2.5,1){\line(1,0){0.5}}
\put(2.7,0.2){\makebox(0,0){$D(p)$}}
\put(2.5,1){\circle{.2}}
\put(3.0,1){\circle{.2}}
\put(3.5,1){\oval(1,0.7)}
\put(4,1){\circle{.2}}
\put(4,1){\line(1,0){0.5}}
\put(4.3,0.2){\makebox(0,0){$D(p)$}}
\put(4.5,1){\circle{.2}}
\put(5,1){\makebox(0,0){+}}
\put(5.5,1){\circle{.2}}
\put(5.5,1){\line(1,0){0.5}}
\put(5.7,0.2){\makebox(0,0){$D(p)$}}
\put(6,1){\circle{.2}}
\put(6.5,1){\oval(1,0.7)}
\put(7,1){\circle{.2}}
\put(7.5,1){\circle{.2}}
\put(7,1){\line(1,0){0.5}}
\put(7.1,0.2){\makebox(0,0){$D(p)$}}
\put(7.8,0.2){\makebox(0,0){$D(p)$}}
\put(7.5,1){\line(1,0){0.5}}
\put(8,1){\circle{.2}}
\put(8.5,1){\oval(1,0.7)}
\put(9,1){\circle{.2}}
\put(9,1){\line(1,0){0.5}}
\put(9.2,0.2){\makebox(0,0){$D(p)$}}
\put(9.5,1){\circle{.2}}
\put(10,1){\makebox(0,0){+}}
\put(10.8,1){\makebox(0,0){$.~~.~~.$}}
\end{picture}

\vspace{1cm}
Fig.II-4  Green's function $D^{'}(p)$ of the boson as  sum of proper self-energy 
insertions.

In  accordance with Fig.II-4 the Green's function $D^{'}(p)$ is obtained by the 
summing  of the series:
\bd
D^{'}(p)=D(p)+D(p)\left( {{{\Sigma (p)} \over i}} \right)D(p) \nonumber
\ed
\veb
\bd
+D(p)\left( {{{\Sigma (p)} \over i}} \right)D(p)D(p)
\left( {{{\Sigma (p)} \over i}} \right)D(p)+... \nonumber
\ed
\veb
\bd
=D(p){1 \over {1+i\Sigma (p)D(p)}}={i \over {p^2-m^2-\Sigma (p)}} \label{n11}
\ed
In this equation  $\Sigma(p)$  is the sum of all two point improper
graphs. All divergent improper graphs and their counterterm graphs (Fig.II-1,II-3)
may be divided into two parts of the first and the second orders with respect 
to $\lambda_{R}$. We will define them as $\Sigma_1$ and $\Sigma_2$ self-energy 
graphs. 

One  can write finite contributions in $\Sigma_2$ in the form

\begin{picture}(8,4)
\put(2,1){\makebox(0,0){$(-i)\Sigma_2=$}}
\put(3.3,1){\line(1,0){1.2}}
\put(3.9,1.5){\circle{1}}
\put(3.9,1.){\circle*{.1}}
\put(3.9,2.5){\circle{1}}
\put(3.9,2){\circle*{.1}}
\put(5,1){\makebox(0,0){+}}
\put(5.5,1.){\line(1,0){1.6}}
\put(6.3,1.){\circle{1}}
\put(5.8,1.){\circle*{.1}}
\put(6.8,1.){\circle*{.1}}
\put(7.8,1){\makebox(0,0){$=$}}
\end{picture}
\vspace{0.0mm}
\begin{eqnarray}
=-{{\lambda _R^2} \over 2}\int {{{d^4q} \over {(2\pi )^4}}}D_\beta (q)
\left\{ {\int {d^4k}D_\beta (k)D_0(k+q+p)} \right\} \nonumber
\ed
\veb
\bd
-{{\lambda _R^2} \over 2}\int {{{d^4q} \over {(2\pi )^4}}}D_\beta (q)
\left\{ {{{\mu ^{4-n}} \over {(2\pi )^n}}
\int {d^4k}D_0(k)D_0(k+q+p)} \right. \nonumber
\ed
\veb
\bd
\left. {-{{\mu ^{4-n}} \over {(2\pi )^n}}
\int {d^4k}D_0(k)D_0(k+p)} \right\}_{p=0} \label{n12}
\ed
Let us introduce functions
\bd
F_\beta ={1 \over 2}\int {{{d^4q} \over {(2\pi )^4}}}D_\beta (q) \label{n13}
\ed
and
\bd
iI_\beta (p)=\int {{{d^4q} \over {(2\pi )^4}}}D_\beta (q)D_0(q+p). \label{n14}
\ed
Then we get 
\bd
\Sigma _2=\lambda _R^2F_\beta I_0(0)+{{\lambda _R^2} \over 2}G_\beta \label{n15} 
\ed
where
\bd
G_\beta =(F_\beta ,(I_\beta +I_0))_{finite} \label{n16}
\ed
The first equation in (\ref{n16}) is a scalar product of the form
\bd
(F_\beta ,I_{\beta})=\int {{{d^4k} 
\over {(2\pi )^4}}}D_\beta (k)\int {{{d^4q} 
\over {(2\pi )^4}}}D_0(q)D_\beta (k+q)  \label{n117}
\ed
and the following one is
\bd
(F_\beta ,I_0)_{finite}=\int {{{d^4k} 
\over {(2\pi )^4}}}D_\beta (k)\left\{\int {{{d^4q} 
\over {(2\pi )^4}}}{D_0(q)D_0 (k+q)}\right\}_{finite}  \label{n18}
\ed
The contribution of the first order in $\Sigma_1$ may 
be found from (\ref{n2}).

This contribution is
\bd
\Sigma _1={{\lambda _R} \over 2}\int {{{d^4q} \over {(2\pi )^4}}}
D_\beta (q)  
={\lambda _R}F_{\beta}.\label{n19}
\ed
The temperature contribution in the boson's mass\footnote{For finite temperature
quantum electrodynamics  the   non-equality of fermionic  masses 
$\delta m_\beta /m \sim\alpha(T/m)^2$  is described by a similar
equation  \cite{per1}} 
will have the following form
\bd
m^2(T)=m_R^2+\Sigma _1+\Sigma _2 \nonumber
\ed
\veb
\bd
=m_R^2+\lambda _{R}F_\beta+\lambda _R^2F_\beta I_\beta (0)
+{{\lambda _R^2} \over 2}\lambda _R^2G_\beta  \label{n20} 
\ed
As the result the Green's function will be
\bd
D^{'}(p)=\frac{i}{p^2-m^2(T)}  \label{n21} 
\ed
Thus we have  computed the  finite temperature Green's function of a  boson 
in two-loop approximation in the form of Feynman propagator with 
finite temperature dependent mass parameter (\ref{n20}). 

\chapter{THERMAL  PROPERTIES OF BOSON}
\vs
	
In chapter XV we showed that the model is renormalizable in each order 
of the  perturbative regime, and found the   finite temperature propagator of a 
 boson in 
a heat bath in two loop approximation.   We also got the expression for the  finite 
temperature mass of  a  boson. These results may help us to get an  effective Hamiltonian 
of the particle and  to study its finite temperature 
behavior in gravitational fields.

\section {Effective Hamiltonian of the boson}
\lum
\hspace{22mm}{\Large \bf   in non-relativistic approximation}
\vsse

After   renormalization   the  pole of boson propagator (\ref{n21})  
 may be written as 
\bd
E=\left[{\vec{p}}^{~2}
+m_R^2+\lambda _{R}F_\beta+\lambda _R^2F_\beta I_\beta (0)
+{{\lambda _R^2} \over 2}\lambda _R^2G_\beta\right]^{1\over2}  \label{n22} 
\ed
We can rewrite the  equation (\ref{n22}) in non-relativistic approximation in
the following form
\bd
E=m_R\left\{ {1+\lambda _Rf(\beta m_R)+o(\lambda _R^2)} \right\}^{{1 \over 2}}
\left[ {1+{{\vec{p}^{~2}} \over {m_R^2\left\{ {1+\lambda _Rf(\beta m_R)
+o(\lambda _R^2)} \right\}}}} \right]^{{1 \over 2}} \nonumber
\ed
\veb
\bd
=m_R+{1 \over 2}\lambda _Rm_Rf(\beta m_R)
+{\vec{p}^{~2}\over{2m_R}}
{\left( 1+{1 \over 2}\lambda _Rf(\beta m_R) \right)}^{-1}
+o(\lambda _R^2)  \label{n23}
\ed
Here the function $f(y)$ is connected with the function $F_\beta (y)$  (\ref{n13}) 
in the following way
\bd
F_\beta(\beta m_R) ={1 \over 2}\int {{{d^3k} \over {(2\pi )^3}}
{1 \over {\varepsilon \left( {e^{\beta \varepsilon }-1} \right)}}}
=m_R^2f(\beta m_R) \nonumber
\ed
The function $f(y)$ has an asymptotic form (for $(y\ll1)$) \cite{i15}:
\bd
f(y)={1 \over {(2\pi)^2}}\int\limits_1^\infty  {dx}
{{\sqrt {x^2-1}} \over {e^{xy}-1}} \nonumber
\ed
\veb
\bd
=\frac{1}{24y^2}
-{1 \over {8\pi y}}+O\left( {y^2\ln y^2} \right),  \label{n24}
\ed
which is very useful for the analysis of the  high temperature
behavior of the model.

\section{Inertial and gravitational masses of a boson}
\vsse

For our following calculations we will consider that the quantum system 
interacts with the gravitational field which is described by the metric
\bd
g_{\mu \nu }=\eta _{\mu \nu }+{\Phi\over2}{\delta_{0\mu}}\delta_{0\nu},  \label{n25}
\ed
where $\Phi$ is a gravitational potential.

The second term in (\ref{n25})
describes a small correction to the Minkowski metric which is connected with the  
presence of  the gravitational field \\
\cite{land2}, \cite{miz1}.

In order to write the  Hamiltonian of the boson in the presence of gravitational 
field one can consider that temperature $T$ changes according to 
Tolmen's law \cite{tol1}.
\bd
T={{T_0} \over {1+\Phi }},  \label{n26}
\ed
where $T_0$ is the temperature with $\Phi=0$.

The finite temperature Hamiltonian with precision to the first leading 
term of the series (\ref{n24}) will be
\bd
H={{\vec{p}^{~2}} \over 2}\left( {m_R+{\lambda_{R} \over {48}}
{{T_0^2} \over {(1+\Phi )^2m_R}}}
 \right)^{-1}+m_R+{\lambda_{R} \over {48}}{{T_0^2}
 \over {m_R(1+\Phi )^2}}+m_R\Phi +...  \label{n27}
\ed  
The last term of the equation (\ref{n27}) describes energy of boson's interaction  
with gravitational field.

Let us rewrite the Hamiltonian (\ref{n27}) as
\bd
H={{\vec{p}^{~2}} \over 2}\left( {m_R+{{\lambda _R} \over {48}}{{T_0^2} 
\over {(1+\Phi )^2m_R}}} \right)^{-1}
+\left( {m_R-{{\lambda _R} \over {24}}{{T_0^2} \over {m_R}}} \right)\Phi
 +...  \label{n28}
\ed
The acceleration of the boson in a gravitational field  may be found from
the quantum mechanical  relation
\bd
\vec{a}=-\left[ {H,\left[ H,{\vec{r}} \right]} \right]=
-\left( {1-{{\lambda _R} \over {12}}{{T_0^2} 
\over {m_R^2}}} \right)\nabla \Phi,  \label{n29} 
\ed
and the mass ratio  will be
\bd
{{m_g} \over {m_i}}=1-{{\lambda _R} \over {12}}{{T_0^2} \over {m_R^2}}  \label{n30}
\ed
so the  inertial and gravitational masses
of the boson in the heat bath are seen to be unequal. 

One can estimate the value of $\delta m_g/m_i$ for some gravitational source. 
Let the source of gravitational field be the Sun ( $1.989\times10^{30} kg$)
then the relation (\ref{n30}) for the combined boson (Cooper pair with 
mass $m_b=1Mev$) 
in the heat bath with temperature $300 K$  gives the following corrections for 
non-equality between masses
\bd
\frac{\delta m_g}{m_i}=\frac{\lambda _R}{12}\frac{T_0^2}{m_R^2}
\sim \lambda  \times 10^{-17} \label{n31}
\ed
or
\bd
10^{-21}< \frac{\delta m_g}{m_i} <10^{-17} \label{n32}
\ed
for the range of the coupling constant $10^{-3}<\lambda <10^{-2}$.

From the analysis we made in this chapter
one may conclude that thermal  interaction of the bosons
in a gravitational field causes  non-equality between inertial 
and gravitational masses. Non-thermal systems do not demonstrate such properties.

The calculations for non-equality between inertial and gravitational mass of electron
were made for thermal quantum electrodynamics by Donoghue\\
 \cite{don1}.
His  result for massive fermions has the same functional structure as 
the equation (\ref{n30}).

%% file: iktopol.tex
 \begin{center} 
\vspace*{8mm}{\LARGE \bf PART III} 
\end{center}
\vspace{2mm}
\begin{center} 
{\LARGE\bf{NON-LINEAR MODELS IN}} 
\end{center}
\begin{center} 
{\LARGE\bf{TOPOLOGY NON-TRIVIAL }} 
\end{center}
\begin{center} 
{\LARGE\bf{ SPACE-TIME}} 
\end{center}
\vspace{2mm}
\begin{center} 
{\Large\bf  Introduction } 
\end{center}

\vspace{5mm}

The usual method of generating spontaneous symmetry breaking in quantum field 
theory is to introduce multiplets of vector  or scalar  fields which develop 
a nonvanishing vacuum expectation values \cite {top1}. 
This mechanism is not necessary.
The general features of spontaneous symmetry breaking 
are independent of whether the Goldstone  or Higgs particle is associated with an 
elementary field or with a composite field. 
However it is   possible to develop a dynamical theory of 
elementary particles in which the origin of the  spontaneous symmetry 
breaking is  dynamical \cite {top3}.
The finite mass appears in analogy with the  phenomenon of superconductivity. 
This part III deals with the research of  \cite {top4}, \cite{top5}, \cite{km1}
in the field of dynamical symmetry breaking and phase transitions.
Models with dynamical mass generation are interesting for the following reasons:

First: Initially massless fermi fields expose $\gamma_5$ invariance, which is 
violated by the dynamical mass generation. 
Thus the initial symmetry will be broken, 
and this  broken symmetry may appear in experimental measurements \cite {top6}.

Second: Dynamical mass of particles in such models is a function of
 a constant of interaction of primary fields. Perhaps this will indicate how 
to solve the puzzle of the origin of particle mass. 

Third: Higgs mechanism introduces into the  theory additional parameters which are
connected 
with Higgs-Goldstone fields. These are  additional difficulties of the model.
\cite{kaku3}.

Distinctive and important characteristics of the  models with dynamic symmetry 
violation  are the introduction of   bound states of particles, and 
calculations in non-perturbative regime. Estimations of the  condensate 
may be obtained from the sum rule \cite{top9}, or from Monte-Carlo method \cite{top7},
 or for the semiclassical 
instanton solutions of the Yang-Mills equations \cite{top8}. 

To develop further the theory one should take into account also 
the topology of space-time \cite{toms1}, \cite{isham1}, \cite{ford1}. 
One should consider 3-D and 4-D spinor models with non-linearity 
of the type $(\bar{\psi}\psi)^2$.
 The non-linear Gross-Neveu model in 3-D space-time in non-trivial topology is studied 
in chapter XVII. 3-D Heisenberg-Ivanenko  model in non-Euclidean space-time
and 4-D  Heisenberg-Ivanenko  model in Riemann space-time are studied in
chapter XVIII. The influence of topology  and curvature on the generation of 
dynamical mass   
of fermions and the effects of violation of symmetry 
in these models are also  studied in  chapter XVIII.   

\chapter{NON-PERTURBATIVE EFFECTS}
\centerline{\Large \bf IN GROSS-NEVEU MODEL }
\vs
	
	Quantum Field Theory may be essentially symplified in the limit of high 
interanal symmetries such as O(N), SU(N) and so on.  
In some appropriate cases field models can be solved strictly 
in the limit of large flavor numbers N \cite{top10}.  Asymptotically free 
Gross-Neveu model without dimensional parameters in the Lagrangian 
is one of such models. This model is renormalizable in 3-D dimensions.  
The solution of this model shows that the phenomenon of dimensional 
transmutation \cite{top11} has place and there is a gap in mass spectrum 
of the model. 
Studying this model one will find the connection between topological
characteristics of space time and the behavior of the solution of the
mass gap equation and also the behavior of dinamical mass of the model.
N-flavor Gross-Neveu model is described by  the Lagrangian
\bd
L=\bar \psi _ii\hat \partial \psi _i+{{g^2} \over 2}
\left( {\bar \psi _i\psi _i} \right)^2 \label{f1}
\ed
This Lagrangian is invariant under the discrete transformations:
\bd
\psi \to \gamma _5\psi,~~\bar{\psi} \to -\bar{\psi}\gamma _5 \label{f2} 
\ed
The generating functional of the model
\bd
Z[\eta,\bar{\eta}]=\int D\psi D\bar{\psi}
\exp \left[i\int d^nx\left(i\bar{\psi}\partial \psi
+(1/2) g^2(\bar{\psi}\psi)^2
+\bar{\eta}\psi+\bar{\psi}\eta\right)\right]    \nonumber
\ed
may be rewritten in the form:
\bd
Z[\eta,\bar{\eta},\sigma]=\int D\psi D\bar{\psi}D\sigma   \nonumber
\ed
\veb
\bd
\times \exp \left[i\int d^nx\left(i\bar{\psi}\partial \psi
- g(\bar{\psi}\psi)\sigma-(1/2)\sigma^2
+\bar{\eta}\psi+\bar{\psi}\eta\right)\right] \label{f3} 
\ed
Here we introduced the new field $\sigma$ and used a useful relation
\bd
\int D\sigma
\exp \left[-i(1/2)\left( \sigma, \sigma \right)
-i\left(g\bar{\psi}\psi,\sigma\right)\right]
\propto \exp \left[(i/2)g^2 \left (\bar{\psi}\psi\right) ^2 \right] \label{f4} 
\ed
Since 
\bd
m\bar{\psi}\psi \to m\left(-\bar{\psi}{\gamma_5}^2\psi \right)
=-m\bar{\psi}\psi \label{f5} 
\ed
we must  put $m=0$ for symmetry of the model.

Therefore a new Lagrangian may be written in the form
\bd
L_\sigma =\bar \psi _ii\hat \partial \psi _i-\sigma 
\left( {\bar \psi _i\psi _i} \right)-{1 \over {2g^2}}\sigma ^2 \label{f6}
\ed
and the symmetry of (\ref{f6}) is
\bd
\psi \to \gamma _5\psi,~~\bar{\psi} \to-\gamma _5\bar{\psi},
~~\sigma \to -\sigma    \label{f7}  
\ed
The generating functional of the model after integration over the matter 
fields will be 
\bd
Z[0]=\int {D\bar \psi D\psi D\sigma }\exp \,i\int {L_\sigma \left( x \right)}dx   \nonumber
\ed
\veb
\bd
=\int {D\sigma \cdot }Det(i\hat \partial -\sigma )
\exp [{{-i} \over {2g^2}}(\sigma ,\sigma )] \label{f8}
\ed
Then effective action is written as
\bd
\Gamma[\sigma ]=-i\ln \,Det(i\hat \partial -\sigma )
-{1 \over {2g^2}}(\sigma ,\sigma ) \label{f9}
\ed
and the effective potential is
\bd
V_{eff}={{iN} \over 2}(\mbox{Tr} \hat 1)\int {{{d^2k}
 \over {\left( {2\pi } \right)^2}}}\ln \left( {k^2-\sigma ^2} \right)
+{1 \over {2g^2}}\sigma ^2 \label{f10}
\ed
The conditions for the energy to be minimal are
\bd
\left({{\delta V_{eff}} \over {\delta \sigma }}\right)_{|\sigma 
=\sigma _c}=0,\quad \left({{\delta ^2V_{eff}} 
\over {\delta \sigma ^2}}\right)_{|\sigma =\sigma _c}>0 \label{f11}
\ed
From these  conditions one can get the gap equation for definition of $\sigma _c$
  ($\sigma _c$ defines the minimum of the effective potential) in the form
\bd
{1 \over \lambda }=\mbox{Tr} \hat 1\int {{{d^2\bar k}
 \over {\left( {2\pi } \right)^2}}}{1 \over {\bar k^2+\sigma _c^2}} \label{f12}
\ed
where constant  $\lambda =g^2N$.

\section  {Trivial case. Euclidean space time.}
\vsse

Let us find the solution of the gap equation (\ref{f12}).

Ultraviolet cut-off of the integral gives:
\bd
\int_{-\Lambda}^{\Lambda} {{{d^2\bar k}
\over {\left( {2\pi } \right)^2}}}{1 \over {\bar k^2+\sigma _c^2}}   \nonumber
\ed
\veb
\bd
=(1/2\pi)\int_{0}^{\Lambda} \frac{2\pi dk^2}{\bar k^2+\sigma _c^2} =(1/4\pi)
\ln \frac{\Lambda^2}{\sigma^2} \label{f13}
\ed
Then, after regularization of (\ref{f13}) we get the equation 
\bd
{1 \over {\lambda \left( \Lambda  \right)}}
={1 \over {2\pi }}\ln {{\Lambda ^2} \over {\sigma ^2}} \label{f14}
\ed
Let the subtraction point be $\mu =\sigma $,
then the renormalized coupling constant  may be written as
\bd
{1 \over {\lambda \left( \mu  \right)}}
={1 \over {2\pi }}\ln {{\mu ^2} \over {\sigma ^2}}\label{f15}
\ed
To eliminate the parameter $\mu$ one can use the methods 
of the renormalization group.
The $\beta$-function in one loop approximation is
\bd
\beta \left( {\lambda _R(\mu )} \right)=-{1 \over \pi }
\left( {\lambda _R(\mu )} \right)^2 \label{f16}
\ed
then Gell-Mann Low equation 
\bd
{{d\lambda _R} \over {\lambda _R^2}}=-{1 \over \pi }{{d\xi } \over \xi } \label{f17}
\ed
with initial condition $\lambda \left( {\xi _0} \right)=\lambda _0$
determines the behavior of the coupling constant with respect to 
scaling of the momentum:
\bd
\lambda _R( t )=\frac{\lambda _0}{1+(\lambda _0 /\pi )t} \label{f18}
\ed
where $t=\ln ( \xi /\xi _0)$

The dynamical mass can be found from (\ref{f11}) in the form
\bd
\sigma _c(triv.)=\mu \exp (-{\pi  \over {\lambda _R\left( \mu  \right)}})
=\mu \exp \left( {-{\pi  \over {\lambda _0}}} \right)=const.\label{f19}
\ed

\section{ Non-trivial topology of space time.}
\vsse

In this case the constant of interaction can be written as
\bd
{1 \over {\lambda \left( {\mu ,L} \right)}}
={1 \over {2\pi }}\left[\ln {{\mu ^2} \over {\sigma _c^2}}+f(L,\sigma _c)
\right] \label{f20}
\ed
where $f(\sigma _c,L)$ is some function which depends on topological parameter L.

The  $\beta$ function is
\bd
\beta \left( {\lambda _R(\mu ,L)} \right)
=-\frac{1}{\pi }\left( {\lambda _R(\mu ,L)} \right)^2 \label{f21}
\ed
and the solution of Gell-Mann-Low equation with 
$\lambda _R\left( {\xi _0,L} \right)=\lambda _0(L)$
is expressed by the equation:
\bd
\lambda _R\left( {t,L} \right)=\frac{\lambda _0(L)}
{1+(\lambda _0(L) /\pi )t} \label{f22}
\ed
The dynamical mass is governed by the equation
\bd
\sigma _c=\mu \exp \left( {-{1 \over {\lambda _0(\mu ,L)}}
-{{f(\sigma _c,L)} \over 2}} \right) \label{f23}
\ed
or
\bd
\sigma _c=\sigma _c(triv)\exp \left( {-{{f(\sigma _c,L)} \over 2}}
 \right) \label{f24}
\ed
One can see that the dynamical mass depends on function $f(\sigma _c,L)$.
 
The explicit expression of the function $f(\sigma _c,L)$ is 
\bd
f_{(\pm )}(L,\sigma _c)=\pm {1 \over {\pi ^2}}\int\limits_0^\infty  
{dx{1 \over {\sqrt {x^2+(L\sigma _c)^2}}}}
\left( {\exp \sqrt {x^2+(L\sigma _c)^2}-1} \right)^{-1} \label{f25}
\ed
for  topologies of the cylinder $(+)$ and the  Mobius strip $(-)$.
Therefore non-Euclidean structure of space time leads 
to redefinition of the gap equation (\ref{f12}) for the Gross-Neveu model. 
That gives us the possibility to define the dependence of the 
fermionic mass on topology of the space time.
Dynamical violation of the $\gamma_5$ symmetry occurs when $\sigma$
is not equal zero.
The method developed above is very useful when the  number of "flavors" 
of the fundamental fields is big ($N \to \infty$). 
This method does not work for a small "flavor" number $N$. 
In Quantum  Field Theory  
there is another method, based on an analogy with superconductivity. This is  
a  Mean Field Method \cite{km1},
\cite{key14}, which works very well for any number 
of "flavors" of the particles. The idea of the method is based on 
effective potential\footnote{Effective potential is the generating functional 
for  (1PI) Green's functions \cite{jak1,rs1}}
 calculations,  that allows us to take  into consideration the 
effects of topology \cite{toms1} and curvature for self-interacting 
and gauge models \cite{ish1}.

 In this work the Mean Field Method is used in dynamical modeling 
of the behavior
of  elementary  particles  and is  based on the idea that the masses of  
compound particles (e.g. nucleons) are  generated by the  self-interaction of 
some fundamental  fermion fields  through the same mechanism 
as  superconductivity. Here the combined particles are treated 
as the quasi-particles excitations. 
The  Mean Field Method also leads to the  mass gap equation, and the solution gives 
the dynamical mass of the particle. In the following section we will treat 
the problem of non-Euclidean space-time structure in 
 models with a dynamical mass.      

\chapter{ $(\bar{\psi}\psi)^2$ NON-LINEAR SPINOR MODELS}
\vs

\section{Dynamical mass and}
\lum 
\hspace{33mm}{\Large \bf  symmetry breaking}
\vsse

In the construction of the unified  models of the elementary
particles one
can admit the possibility that the mass of combined particles 
appears as the result of self-interaction of certain fundamental fields, 
for instance, quarks and leptons from preons, or  compound 
fermions in technicolor models \cite{ro1,ta1}, \cite{ito1},
\cite{sd1}. 
 Following this idea we can get the gap equation. 
Its solution   can predict  the dynamical 
mass of the compound particles.  As in the previous case of 
Gross-Neveu model, we will study here the effects of 
non-trivial topology and, also, geometry of background space-time.

In this section we will consider  the phenomenon of the dynamical 
generation of mass of particles  in the application 
to the models in non-trivial  space-time.

Let us consider Heisenberg-Ivanenko \cite{top12} non-linear spinor model
with Lagrangian
\bd
L=\bar \psi i\hat \partial \psi +{{g_0^2}
 \over {2\mu _0^2}}\left( {\bar \psi \psi } \right)^2 \label{f26}
\ed
where $g_0^2$ is a massless parameter and  parameter $\mu_0^2$  has 
 dimension which is connected with the dimension of space time.
The symmetry of the Lagrangian (\ref{f26}) is
\bd
\psi \to \gamma _5\psi, ~~\bar \psi \to -\bar \psi \gamma _5 \label{f27}
\ed
We can rewrite this Lagrangian in the new form
\bd
L_\sigma =\bar \psi i\hat \partial \psi -g_0\sigma 
\left( {\bar \psi \psi } \right)-{{\mu _0^2} \over 2}\sigma ^2 \label{f28}
\ed
The symmetry of the last one is (\ref{f7}).

The equation of motion for the   $\sigma$ field is
\bd
\sigma =-{{g_0} \over {\mu _0^2}}\left( {\bar \psi \psi } \right)  \label{f29}
\ed
thus we may assume that the  field $\sigma$ is a collective field.  

Let us  consider that   $\sigma$  is $\sigma \to \sigma +\tilde \sigma $,
where 
\bd
\sigma =(g_0 /{\mu _0}^2)<\psi \bar \psi >\ne 0   \nonumber
\ed
is the background field and $\tilde \sigma $  is quantum 
fluctuations of the $\sigma $-field.

Then 
\bd
L_\sigma =\bar \psi (i\hat \partial -g_0\sigma )\psi 
-g_0\left( {\bar\psi  \tilde{\sigma} \psi } \right)
-{{\mu _0^2} \over 2}\sigma ^2
-{{\mu _0^2} \over 2}{\tilde{\sigma }}^2-{\mu_0}^2\sigma\tilde \sigma \label{f30}
\ed
The last term of (\ref{f30}) may be eliminated because of the 
redefinition of the sources
of the quantum field $\tilde \sigma$, and the  Feynman graphs will be

\begin{picture}(8,3.5)
\put(2,1){\line(-1,1){1}}
\put(2,1){\line(-1,-1){1}}
\put(2,0.99){\line(1,0){1}}
\put(2,1.01){\line(1,0){1}}
\put(1,1.5){\makebox(0,0){$\psi$}}
\put(1,0.5){\makebox(0,0){$\bar \psi$}}
\put(3.5,1){\makebox(0,0){$\tilde \sigma$}}
\put(1.5,1){\makebox(0,0){$g_0$}}
\put(2,1){\circle*{.1}}
\put(2,0){\makebox(0,0){$a)$}}
\put(5,1.01){\line(1,0){1.5}}
\put(5.8,1.3){\makebox(0,0){$i\hat k-g_0\sigma_0$}}
\put(5.8,0){\makebox(0,0){$b)$}}
\put(8,0.99){\line(1,0){1}}
\put(8,1.01){\line(1,0){1}}
\put(8.5,1.3){\makebox(0,0){$(i/\mu_0^2)$}}
\put(8.5,0){\makebox(0,0){$c)$}}
\end{picture}

\vspace{10mm}

Fig. III-1  Feynman graphs including a  collective field.

Graph $1 a)$ describes the interaction of fermi field with collective field,
$1 b)$ is the  propagator of fermi field, and $1 c)$ is the  
propagator of collective field.

In tree approximation with respect to  $\sigma$-field we can write 
an effective action:
\bd
\Gamma_{eff} [\sigma]=(-i/2)\ln Det(i\hat k-g_0\sigma)
-\frac{\mu_0^2}{2}(\sigma,\sigma) \label{f31}
\ed
The effective potential of this model will be
\bd
V_{eff}={i \over 2}(\mbox{Tr} \hat 1)\int {{{d^nk} 
\over {\left( {2\pi } \right)^n}}}\ln 
\left( {k^2-(g_0\sigma )^2} \right)+{{\mu _0^2} \over 2}\sigma ^2 \label{f32}
\ed
Minimum  $V_{eff}$ gives the gap equation 
\bd
\sigma _c=(g_0 / \mu _0)^2\sigma _cI\,(g_0\sigma _c) \label{f33}
\ed
or, in another form,
\bd
m=s\lambda _0mI\left( m \right) \label{f34}
\ed
where $m$ is the dynamical mass of fermionic field $m=g_0\sigma _c$,
$s$ is the dimension of $\gamma$- matrices 
and $\lambda _0=(g_0^2/ \mu _0)^2$

Now one can solve the equation (\ref{f34}) for different space time topologies.

\section{ Model  with topologies} 
 \lum 
\hspace{33mm} {\Large \bf $R_1 \times R_1 \times S_1$ 
and $R_1 \times  Mobius~strip$}
\vsse

The solution of this gap equation is connected with 
the calculation of the function  I(m) of  the equation (\ref{f34}).

Let us consider two types of topologies:

1)  $R_1\times R_1\times S_1$ with $\psi(x,y,0)=\psi(x,y,L)$\\ 
and 

2) $R_1\times Mobius~strip$ with $\psi(x,y,0)=-\psi(x,y,L)$ (Fig.III- 2)

We can find for 3-D space time that 
\bd
I(m)=\Lambda -F_{(\pm )}(L,m) \label{f35}
\ed
where
\bd
F_{(\pm )}(L,m)={1 \over {\pi L}}\ln (1\;_+^-\;e^{-Lm}) \label{f36}
\ed
	with (+) for $R_1\times R_1\times S_1$
 topology and (-) for $R_1\times Mobius\;strip$  topology.

The gap equation will be
\bd
m=m\lambda _0\Lambda \left( {1-{1 \over {\pi L\Lambda }}
\ln (1\;_+^-\;e^{-Lm})} \right) \label{f37}
\ed
The analysis of this expression can be made by the theory 
of bifurcations\\
 \cite{hk1}. 
For this purpose let us write the gap equation (\ref{f37}) in the form 
\bd
m=f_{(\pm )}(m,\bar \lambda ) \label{f38}
\ed
where the functions $f_{(\pm )}(m,\bar \lambda )$ 
for topologies $(+)$and $(-)$  are 
\bd
f_{(\pm )}(m,\bar \lambda )=m\bar \lambda 
\left( {1-{1 \over {\pi L\Lambda }}\ln (1\;_+^-\;e^{-Lm})} \right) \label{f39}
\ed
The  equation (\ref{f38}) has stable m=0 solutions, if $f_m(0,\bar \lambda )<1$.

For $f_m(0,\bar \lambda )>1$ the equation (\ref{f38}) has no stable 
trivial solutions.  

One can see that there is a stable m=0 solution for 
(-) topology with the critical parameter 
\bd
L_c={{\ln 2} \over \pi }{{\lambda _0} \over {\lambda _0\Lambda -1}}
\approx {{\ln 2} \over {\pi \Lambda }} \label{f40}
\ed
for $\lambda _0\Lambda =\bar \lambda >1$

The gap equation for topology (+) has no stable trivial solutions.  

The dynamical mass in this case is a smooth function with respect to parameter L:
\bd
m_{(+)}=-{1 \over L}\ln \left( {1-f(\Lambda )
\exp \left( {-{{\pi L} \over {\lambda _0}}} \right)} \right) \label{f41}
\ed
for  $\lambda _0\Lambda =\bar \lambda <1$

The solution for topology $R_1\times Mobius\;strip$  is
\bd
m_{(-)}=-{1 \over L}\ln \left( {\exp \left( {{L \over {L_c}}\ln 2} \right)-1}
 \right)\label{f42}
\ed
The restoration of the symmetry takes place when $L=L_c$.
In this case the condensate function of the bound state equals zero
\bd
\sigma _c={{g_0} \over {\mu _0^2}}<\psi \bar \psi >=0 \label{f43}
\ed
Now we can consider  the models with  the  more complicated 
space-time structures of Klein bottle and Torus topologies.

\newpage

\section{Torus topology $M_{3)}=R_1 \times S_1 \times S_1$} 
\lum 
\hspace{35mm}{\Large \bf and topology $M_{3)}=R_1\times Klein\;bottle$}
\vsse

We can introduce  topologies $R_1\times S_1\times S_1$
and $R_1\times Mobius~strip$ as an identification of space points
for wave functions Fig. III-3:
\bd
\psi(x,0,0)=\psi(x,L,L)~and~\psi(x,0,0)=-\psi(x,L,L)
\ed 
For simplicity  consider that these topologies have only one parameter
 $L=L^{'}$.
The gap equations for the topologies will be
\bd
m=m\bar {\lambda} s\left\{1+\sqrt {m /(\Lambda ^2L)}
( \pi)^{-3 /2}
[\sum\limits_{n=1}^\infty  {1 \over {\sqrt {2n}}}K_{-1/2}(2Lmn)\right.   \nonumber
\ed
\veb
\bd
\left.+\sum\limits_{n,k=1}^\infty  K_{-1/2}
(2Lm\sqrt {n^2+k^2})\left( {n^2+k^2} \right)^{-1/4}]\right\} \label{f44}
\ed
and
\bd
m=m\bar {\lambda} s\left\{1+\sqrt {m /(\Lambda ^2L)}
( \pi)^{-3 /2}
[\sum\limits_{n=1}^\infty  {1 \over {\sqrt {2n}}}K_{-1/2}(2Lmn)\right.   \nonumber
\ed
\veb
\bd
\left.+\sum\limits_{n,k=1}^\infty (-1)^k K_{-1/2}
(2Lm\sqrt {n^2+k^2})\left( {n^2+k^2} \right)^{-1/4}]\right\} \label{f45}
\ed
The solutions of (\ref{f44}) and (\ref{f45}) give dynamical masses for 
these topologies\\
 \cite{top13}.

The results of this paragraph show that the application of the  Mean Field Method
to non-linear models  gives non-renormalized solutions, 
though we can obtain some information about  the influence 
of topology on dynamical mass behavior.
The evaporation of condensate and restoration of chiral symmentry 
proceed in different ways and are ruled by topology. There are topologies 
in which these phenomena do not take place. We believe that these results
are  important in the bag models, because the energy of bag is dependent 
on non-perturbative effects and boundary conditions\\
 \cite{guid1}. 

\section{Non-linear spinor  $(\bar \psi \psi)^2$ model in}
\lum  
\hspace{23mm}{\Large \bf  Riemann space-time at finite temperature.}
\vsse

In this section we will  treat the problem of dynamical mass generation 
of the  non-linear spinor model in 4-D Riemann space-time at finite temperature.
We will get the  finite temperature effective potential  
and find out information about the   influence of curvature of 
the background gravitational field and temperature 
on the value of  dynamical fermionic mass.

Let the total Lagrangian of the model be
\bd
L=L_g+L_m \label{f46}
\ed
where the  gravitational Lagrangian is\footnote{For clear understanding 
of the problem, we study the fermionic system in a weak gravitational field.} 
\bd
L_g={1 \over {16\pi G_0}}(R-2\tilde \Lambda _0) \label{f47}
\ed
and the  Lagrangian of matter field $L_m$ is
\bd
L_m=\bar \psi i\bar \gamma ^\mu \nabla _\mu \psi +{{g_0^2} 
\over {2\mu _0^2}}\left( {\bar \psi \psi } \right)^2 \label{f48}
\ed
$\nabla _\mu $ is a covariant derivative.

The first  term of (\ref{f48}) describes 
kinetics of the fermi field and its interaction with the  gravitational field,
 and the second one describes the interaction of the fields.

The effective potential of the model is written from the action
\bd
\Gamma_{eff}[\sigma]=-{i \over 2}\ln Det[\nabla ^2
+{1 \over 4}R-(g_0\sigma )^2]-{{\mu _0^2} \over 2}(\sigma ,\sigma ) \label{f49}
\ed
After making ultraviolet regularization and renormalizations of the model we get
\bd
L_{tot}=L_g+L_m=\left( {L_g+L_m(\infty )} \right)
+L_m(\beta )   \nonumber
\ed
\veb
\bd
=L_g^{ren}+L_m(\beta ) \label{f50}
\ed
The renormalized gravitational constant  $G_R$ 
and the   $\tilde{\Lambda} _R$-term are:
\bd
{1 \over {8\pi G_R}}\tilde \Lambda _R
={1 \over {8\pi G_0}}\tilde \Lambda _0
+{1 \over {16\pi ^2}}\Lambda ^2-{1 \over {16\pi ^2}}m^2\ln \Lambda ^2   \nonumber
\ed
\veb
\bd
{1 \over {16\pi G_R}}={1 \over {16\pi G_0}}
+{1 \over {16\pi ^2}}\ln \Lambda ^2 \label{f51}
\ed
where $\Lambda$ is a cut-off parameter, and $m$ is a dynamical mass.

In the  calculations (\ref{f50}) and (\ref{f51}) there was used the cut-off 
method  of regularization of divergent integrals  \cite{key7}.

The effective potential can be written from (\ref{f49}) in the form
\bd
V_{eff}[\sigma ]=\sum\limits_{j=1}^2 {\hat 
\alpha _j(R)f^j(\beta g_0\sigma )+{{\mu _0^2} \over 2}}\sigma ^2 \label{f52}
\ed
where
\bd
f^0(\beta g_0\sigma )={{2m^2} \over {\pi ^2\beta ^2}}
\sum\limits_{n=1}^\infty  {{{\left( {-1} \right)^n}
 \over {n^2}}}K_{-2}(\beta g_0\sigma )   \nonumber
\ed
\veb
\bd
=-{4 \over \beta }
\int {{{d^3k} \over {\left( {2\pi } \right)^3}}
\ln \left( {1+e^{-\beta \varepsilon }} \right)} \label{f53}
\ed
and
\bd
f^j(\beta g_0\sigma_0 )={1 \over {4g_0}}
\left( {-{\partial  \over {\partial \sigma ^2}}}
 \right)^jf^0(\beta g_0\sigma ) \label{f54}
\ed
The  coefficients $\hat \alpha _j(R)$ are 
\bd
\hat \alpha _0=1,\quad \hat \alpha _0={1 \over {12}}R,\quad ....\label{f55}
\ed
The solution of the gap equation
\bd
\frac{\partial}{\partial \sigma } 
 V_{eff}[\sigma ]_{|\sigma=\sigma_c}
=0 \label{f56}
\ed
gives, in high temperature approximation (\ref{bb15}), the expression 
for dynamical mass without redefinition of 
the coupling constant and temperature
\bd
m^2(R,T)=(g_0 \sigma_c)^2 = A/\lambda + b\cdot T^2 + C\cdot R+...\label{f57}
\ed
where the  constants 
\bd
A=32\pi^2/3.84,~~
B=A/24,~~
C=0.7/12,~~
\lambda=(g_0/\mu_0)^2   \nonumber
\ed
are numerical positive coefficients.
As we can see from the equation (\ref{f57}), the effective dynamical mass
of fernion is a positive function for any temperature and curvature.

%% file: ikintrcs.tex
 \begin{center} 
\vspace*{8mm}{\LARGE \bf PART IV} 
\end{center}
\vspace{2mm}
\begin{center} 
{\LARGE\bf{TOPOLOGICALLY MASSIVE}} 
\end{center}
\begin{center} 
{\LARGE\bf{GAUGE THEORIES}} 
\end{center}

\vspace{2mm}
\begin{center} 
{\Large\bf  Introduction } 
\end{center}

\vspace{5mm}

3-D non-Abelian gauge models describing   systems of point particles carrying 
  non-Abelian charge   have been under investigation for over 
two decades. These models (so called topologically massive models)
 have a number of interesting features:

1) For vector fields, these models possess  single, parity-violating, massive, 
spin 1 excitations, in contrast to  single, massless, spin 0 exitations 
in the Maxwell  theory, and to a pair of spin 1 degrees
 of freedom in gauge non-invariant  models with a mass \\
\cite{int2} 

2) For second-rank tensor fields, describing gravity, the topological 
model leads to a single, parity-violating, spin 2 particle, whereas a 
conventional  (gauge non-invariant) mass term gives rise to a spin 2 doublet.
Furthermore, the topological term is of third-derivative order, yet the 
single propagating mode is governed by the Klein-Gordon equation.
 Einstein gravity, which is trivial and without propagation in three dimensions,
becomes a dynamical theory with  propagating particles.

3) Particles interacting via the Abelian Cherm-Simons term (CS-term)
acquire anomalous spin and fractional statistics.  They are
 called anyons \cite{int4}.
 Anyons  play a role in the fractional quantum Hall effect \cite{arov1},
[Lee \& Fisher 1989] and perhaps also in high temperature superconductivity
[Lauglin 1988].

This  is not a complete list of interesting properties of gauge theories 
in an odd number of dimensions.
An  important part of these models is the   Chern-Simons action (CS-term).  
Some interesting aspects of quantum field models 
arising from the topology of odd-dimensional manifolds are  discussed in
chapter XIX. The origin of the  CS-term of vector type that is induced  
by   gauge interaction  of 3-D fermions is studied in chapter XX. 
In this chapter  
the influence of topology and temperature effects are also considered.
The origin of the  induced gravitational CS-term at finite temperature 
is considered in chapter XXI.
   
\chapter{INTRODUCTION}
\centerline{\Large \bf TO  TOPOLOGICAL FIELD MODELS }
\vs

As an  introduction to odd dimensional topological field models 
let us consider  their origin and topological significance.

\mad
{\large \it Topological aspect of the model}
\mad

From gauge-invariant fields in even dimensions we may construct  
 gauge invariant 
Pontryagin densities:
\bd
 P_{(2)}=-(1/2 \pi )\epsilon ^{\mu \nu}{^*F}^{\mu \nu} \nonumber
\ed
\veb
\bd 
 P_{(4)}=-(1/16 \pi ^2) \mbox{tr} ^*F^{\mu \nu}F_{\mu \nu}\label{int1}
\ed
whose integrals over the even dimensional space are    invariants
that measure the topological content of the model. 

These gauge invariant objects can also be written  as total derivatives of gauge
invariant quantities
\bd 
P_n=\partial_\mu X^\mu _n \label{int2}
\ed
The two and four-dimensional expressions are 
\bd
X^\mu _2=(1/2\pi)\epsilon^{\mu \nu}A_\nu \nonumber
\ed
\veb
\bd 
X^\mu _4=(1/2\pi)\epsilon^{\mu \alpha \nu \beta}
\mbox{tr} (A_\alpha F_{\beta \gamma}-(2/3)A_\alpha A_\beta A_\gamma)  \label{int3}
\ed
The Chern-Simons (CS) secondary characteristic class is obtained by integrating one 
component of $X^ \mu _n$ over the $(n-1)$ dimensional space which does 
not include that component.

Therefore the  3-D action $S_{CS}$ is proportional to
\bd 
S_{CS} \sim \int dx^0dx^1dx^2 X^ 3 _4 \label{int4}
\ed 
A  topological massive  term (we will name it CS term) can be added  to the fundamental action for 
a gauge fields, but unlike the ways in which
gauge fields are usually given a mass, no gauge symmetry is broken
by its introduction.

\mad
{\large \it Quantum aspects of 3-D field theory}
\mad

Let us consider  non-Abelian quantum model with topological
mass term. The  Lagrangian of this model is
\bd 
L=L_0+L_{CS} +L_{gauge}  \label{int5}
\ed
where $L_0$ is the usual action for non-Abelian gauge field
\bd 
L_0=-(1/2)\mbox{tr}(F_{\mu \nu}F^{\mu \nu})  \label{int6} 
\ed
with
\bd 
F_{\mu \nu}=\partial_\mu A_\nu-\partial_\nu A_\mu +g[A_\mu, A_\nu] \label{int7}
\ed
$L_{CS}$ is CS term
\bd 
L_{CS}=-im\epsilon ^{\mu \nu \rho}
\mbox{tr} (A_\mu \partial_\nu A_\rho-(2/3)A_\mu A_\nu A_\rho)  \label{int8}
\ed
and $L_{gauge}$ includes the gauge-fixing term 
\bd 
L_{gauge}=
(\partial_\mu\bar{\eta}^a)(\partial^\mu \eta^a)
+gf_{abc}(\partial_\mu\bar{\eta}^a)D^\mu \eta  \label{int9}
\ed
We introduce  $SU(N)$ gauge group here with matrix notation:
$A_\mu=A^a_\mu\tau^a$,  where $\tau^a$ are anti-Hermitian 
matrices in the fundamental representation:
\bd 
[\tau^a, \tau^b]=f^{abc}\tau^c,~~~ \mbox{tr} (\tau^a \tau^b)
=-(1/2)\delta^{ab} \label{int10}
\ed
and  $f^{abc}$ are the structure constants of $SU(N)$.

The  theory is defined in three space-time dimensions with Euclidean signature 
$(+~+~+)$.
The coupling of the CS term is imaginary in Euclidean space-time and real 
in Minkowski space-time.

\mad
{\large \it Properties of the model}:
\mad

For an odd number of dimensions, the operation of parity, P, can be defined
 as reflection in all axes:
\bd 
x^\mu  \mathop{\to}\limits_{P}-x^\mu ,
~~~A_\mu  \mathop{\to}\limits_{P} -A_\mu \label{int11}
\ed
The usual gauge Lagrangian is even under parity,
\bd 
L_0 \mathop{\to}\limits_{P}+L_0 \label{int12}
\ed
but the  CS term is odd
\bd 
L_{CS} \mathop{\to}\limits_{P}-L_{CS} \label{int13}
\ed
Under gauge transformation 
\bd 
 A_\mu \to \Omega^{-1}\left\{(1/g)\partial_\mu +A_\mu\right\}\Omega \label{int14}
\ed
The  Lagrangian $L_0$ is invariant but $L_{CS}$ is not:
\bd
\int d^3x L_{CS} \to \int d^3x L_{CS} \nonumber
\ed
\veb
\bd 
+(im/g)\int d^3x \epsilon^{\mu \nu \rho}\partial_\mu 
\mbox{tr} [(\partial_\nu\Omega)\Omega^{-1}A_\rho] +8\pi^2 (m/g^2)i\omega \label{int15}
\ed
where:
\bd 
\omega=(1/24)\int d^3x \epsilon^{\mu \nu \rho}
 \mbox{tr}[\Omega^{-1}(\partial_\mu\Omega)\Omega^{-1}
(\partial_\nu\Omega)\Omega^{-1}(\partial_\rho\Omega)] \label{int16}
\ed
The set of gauge transformations is divided into global gauge rotations,
$\partial_\mu \Omega=0$, and all others, for which we assume that 
$\Omega (x) \to 1$ as $x^\mu \to \infty$. Integrating  over global 
gauge rotations requires the system to have a total color 
charge equal to zero. In this case, $A_\mu (x)$ falls off faster than $1/|x|$  
as $x^\mu \to \infty$, and the second term on the right-hand  side
of (\ref{int15}), which is a  surface integral, vanishes. 

The last term in (\ref{int15}) does not vanish in general.
The $\omega$ of (\ref{int16}) is a winding number, which labels 
the homotopy class of $\Omega (x)$ \cite{nak1} 

Changing variables $A \to A^U$, where $A^U$ is a gauge transformation  of $A$, 
implies that the vacuum average of the value $\hat{Q}$ is 
\bd 
<Q>=\exp |i8\pi^2 (m/g^2)\omega(U)|<Q> \label{int17}
\ed
This invariance gives us a quantization condition for the dimensionless ratio
\bd 
4\pi^2 (m/g^2)=n,~~~n=0,\pm 1,\pm 2,   \label{int18}
\ed
Therefore, for the theory to be invariant under certain large gauge 
transformations    (for a non-Abelian gauge group), 
which are not continiously deformable to the identity, 
the ratio of the CS mass $m$ 
and the gauge coupling $g^2$ must be quantized \cite{int6}.

Further we can study the problem of massive excitations.

For spinor electrodynamics in three dimensions we have
\bd 
L=L_g+L_f +L_{int}  \label{int19}
\ed
where
\bd 
L_g=-(1/4)F^{\mu \nu}F_{\mu \nu}+
(\mu/4)\epsilon ^{\mu \nu }F_{\mu \nu}A_\alpha, ~~
F_{\mu \nu}=\partial_\mu A_\nu-\partial_\nu A_\mu \label{int20}
\ed
\veb
\bd 
L_f=i\bar{\psi}\hat{\partial}\psi-m\bar{\psi}\psi \label{int21}
\ed
\veb
\bd 
L_{int}=-J^\mu A_\mu,~~ J^\mu=-e\bar{\psi}\gamma^\mu \psi \label{int22}
\ed
The coupling constant $e$ has dimension $(mass)^{-1/2}$.

The equations of motion will be
\bd 
\partial_\mu F^{\mu \nu}+
(\mu /4)\epsilon ^{\nu \mu \alpha}F_{\mu \alpha }=J^\mu \label{int23}
\ed
\veb
\bd 
(i\bar{\partial}+e\bar{A}-m) \psi =0\label{int24}
\ed
One can introduce the dual field strength tensor in 3-D space-time 
\bd 
^*F^\mu=(1/2)\epsilon ^{\mu \alpha \beta}F_{\alpha \beta}~~
F^{\alpha \beta}={\epsilon ^{ \alpha \beta \mu}}{^*}F_\mu \label{int25}
\ed
The  Bianchi identity follows from (\ref{int23}): 
\bd 
{\partial_\mu}{^*F^\mu}=0 \label{int26}
\ed
and the equation (\ref{int23}) may be written in a dual form
\bd 
{\partial_\alpha}{^*F_\beta}-{\partial_\beta}{^*F_\alpha}-
\mu F_{\alpha \beta}=-\epsilon _{\alpha \beta \mu}J^\mu \label{int27}
\ed
or 
\bd 
(\Box+ \mu^2){^*F^\mu}=\mu\left(\eta^{\mu \nu}
-\epsilon^{\mu \nu \alpha} \frac{\partial_\alpha}{\mu}\right)J_\nu \label{int28}
\ed
This equation demonstrates that the gauge excitations are massive.

\mad
{\large \it 3-D gravity. Connection with topology}
\mad

The results of the topological part of the introduction gives us a hint how to 
construct the  topological term for three dimensional gravity 
from a four-dimensional  $^*RR$ Pontryagin density.
\bd 
^*RR=(1/2)\epsilon^{\mu \nu \alpha \beta}R_{\mu \nu \rho \sigma}
{R_{\alpha \beta}}^{\rho \sigma}=\partial_\mu X^\mu \label{int29}
\ed 
Let us find $X^\mu$ from (\ref{int29}). To do this we will rewrite $^*RR$
in the following way
\bd
^*RR=(1/2)\epsilon^{\mu \nu \alpha \beta}R_{\mu \nu \rho \sigma}
{R_{\alpha \beta}}^{\rho \sigma}
=(1/2)\epsilon^{\mu \nu \alpha \beta}R_{\mu \nu ab}
{R_{\alpha \beta}}^{ab} \nonumber
\ed
\veb
\bd
=(1/2)\epsilon^{\mu \nu \alpha \beta}R_{\mu \nu ab}
\left\{
\partial_\alpha {\omega_\beta}^{ab}-\partial_\beta {\omega_\alpha}^{ab}
+{\omega_\alpha}^{ac}{\omega_{\beta c}}^{b}
-{\omega _\beta }^{ac}{\omega _{\alpha c}}^{b}
\right\} \nonumber
\ed
\veb
\bd
=\epsilon^{\mu \nu \alpha \beta}R_{\mu \nu ab}
\left\{
\partial_\alpha {\omega_\beta}^{ab}
+{\omega_\alpha}^{ac}{\omega_{\beta c}}^{b}
\right\} \label{int30}
\ed
In our calculations we used the following expression for the Ricci connection
\bd
R_{\mu \nu \rho \sigma}
=\partial_\alpha {\omega_{\beta a b}}-\partial_\beta {\omega_{\alpha a b}}
+{\omega_{\alpha a}}^{c}{\omega_{\beta c}}_{b}
-{\omega _{\beta a}}^{c}{\omega _{\alpha c}}_{b} \label{int31}
\ed
where $\omega_{\mu ab}$ is  3-D spin connection. 

Then the expression for Pontryagin density will be
\bd 
^*RR=\epsilon^{\mu \nu \alpha \beta}R_{\mu \nu ab} 
~\partial_\alpha {\omega_\beta}^{ab}
+\epsilon^{\mu \nu \alpha \beta}R_{\mu \nu ab}
~ {\omega_\alpha}^{ac}{\omega_{\beta c}}^{b} \label{int32}
\ed 
Let us find these two contributions separately.

The first contribution  to (\ref{int32}) gives
\bd
\epsilon^{\mu \nu \alpha \beta}R_{\mu \nu ab} 
~\partial_\alpha {\omega_\beta}^{ab}=
\epsilon^{\mu \nu \alpha \beta}\partial_\alpha (R_{\mu \nu ab} 
{\omega_\beta}^{ab})-
\epsilon^{\mu \nu \alpha \beta}
{\omega_\beta}^{ab}\partial_\alpha R_{\mu \nu ab} \nonumber
\ed
\veb
\bd
=\epsilon^{\mu \nu \alpha \beta}\partial_\alpha (R_{\mu \nu ab} 
{\omega_\beta}^{ab}) \nonumber
\ed
\veb
\bd 
-\epsilon^{\mu \nu \alpha \beta}
{\omega_\beta}^{ab}\partial_\alpha 
\left\{
\partial_\mu \omega_{\nu ab}-\partial_\nu \omega_{\mu ab}
+\omega_{\mu a}^c \omega_{\nu cb}\omega_{\nu a}^c \omega_{\mu cb}
\right\} \label{int33}
\ed 
\veb
\bd 
=\epsilon^{\mu \nu \alpha \beta}\partial_\alpha (R_{\mu \nu ab} 
{\omega_\beta}^{ab})-2\epsilon^{\mu \nu \alpha \beta}
{\omega_\beta}^{ab}\partial_\alpha 
\left\{
\omega_{\mu a}^c \omega_{\nu cb}
\right\} \label{int34}
\ed 
The second contribution to (\ref{int32}) will be
\bd
\epsilon^{\mu \nu \alpha \beta}R_{\mu \nu ab}
~ {\omega_\alpha}^{ac}{\omega_{\beta c}}^{b}= \nonumber
\ed
\veb
\bd
-\epsilon^{\mu \nu \alpha \beta}
 \left\{
\partial_\mu \omega_{\nu ab}-\partial_\nu \omega_{\mu ab}
+\omega_{\mu a}^c \omega_{\nu cb}-\omega_{\nu a}^c \omega_{\mu cb}
\right\}
{\omega_\alpha}^{ac}{\omega_{\beta c}}^{b} \nonumber
\ed
\veb
\bd 
=2\epsilon^{\mu \nu \alpha \beta}
\partial_\mu \omega_{\nu ab}
{\omega_\alpha}^{ac}{\omega_{\beta c}}^{b}
+2\epsilon^{\mu \nu \alpha \beta}\omega_{\mu a}^c \omega_{\nu cb}
{\omega_\alpha}^{ac}{\omega_{\beta c}}^{b} \label{int35}
\ed 
On the other hand the second one  to (\ref{int35}) is zero, then we have
\bd
^*RR=\epsilon^{\mu \nu \alpha \beta}\partial_\alpha (R_{\mu \nu ab} 
{\omega_\beta}^{ab}) \nonumber
\ed
\veb
\bd 
-2\epsilon^{\mu \nu \alpha \beta}
{\omega_\beta}^{ab}\partial_\alpha 
\left\{
\omega_{\mu a}^c \omega_{\nu cb}
\right\}+2\epsilon^{\mu \nu \alpha \beta}
\partial_\mu \omega_{\nu ab}
{\omega_\alpha}^{ac}{\omega_{\beta c}}^{b} \label{int36}
\ed 
One can find that 
\bd
-2\epsilon^{\mu \nu \alpha \beta}
{\omega_\beta}^{ab}\partial_\alpha 
\left\{
\omega_{\mu a}^c \omega_{\nu cb}
\right\}
+2\epsilon^{\mu \nu \alpha \beta}
\partial_\mu \omega_{\nu ab}
{\omega_\alpha}^{ac}{\omega_{\beta c}}^{b}\ \nonumber
\ed
\veb
\bd 
=2\epsilon^{\mu \nu \alpha \beta}
\partial_\mu \omega_{\nu a}^b
{\omega_\alpha b}^{c}{\omega_{\beta c}}^{a} \label{int37}
\ed 
and
\bd 
2\epsilon^{\mu \nu \alpha \beta}
\partial_\mu \omega_{\nu a}^b
{\omega_\alpha b}^{c}{\omega_{\beta c}}^{a}
=(2/3)\epsilon^{\mu \nu \alpha \beta}
\partial_\mu 
\left\{
\omega_{\nu a}^b
{\omega_\alpha b}^{c}{\omega_{\beta c}}^{a}
\right\}  \label{int38}
\ed 
In the result we will have
\bd 
^*RR=\partial_\mu \epsilon^{\mu \nu \alpha \beta}
\left\{
{\omega_\nu}^{ab}R_{\alpha \beta ab}+
(2/3)\omega_{\nu a}^b
{\omega_{\alpha b}}^{c}{\omega_{\beta c}}^{a}
\right\}  \label{int39}
\ed 
Therefore
\bd 
X^\mu=\epsilon^{\mu \nu \alpha \beta}
\left\{
{\omega_\nu}^{ab}R_{\alpha \beta ab}+
(2/3){\omega_{\nu a}}^b
{\omega_{\alpha b}}^{c}{\omega_{\beta c}}^{a}
\right\}  \label{int40}
\ed 
Let parameter $\mu$ be equal to zero, 
then determining  $\epsilon ^{0 \nu \alpha \beta}=\epsilon ^{\nu \alpha \beta}$
with $\nu, \alpha, \beta=1,2,3$ we get the  CS action in the form 
\bd 
S_{CS}\sim \int d^3x X^0=\int d^3x\epsilon^{\nu \alpha \beta}
\left\{
{\omega_\nu}^{ab}R_{\alpha \beta ab}+
(2/3){\omega_{\nu a}}^b
{\omega_{\alpha b}}^{c}{\omega_{\beta c}}^{a}
\right\}  \label{int41}
\ed 
As we can see from (\ref{int41}) the  CS term is of the third derivative order  
in contrast to the  first one as in the vector case (\ref{int3}).

\mad
{\large \it Nonlinear theory of gravity}
\mad

We can construct total gravitational action in the form 
\bd 
S_{tot}=(1/k^2) \int d^3x \sqrt{g}R+(1/k^2\mu)S_{CS}\label{int42}
\ed 
The sign of the Einstein part  is opposite to the conventional
one in four dimensions. The Einstein part of the action has 
coefficient $k^{-2}$ with the dimension of mass, while the topological part has 
a dimensionless coefficient ($\mu$ has a dimension of mass). 

Now we can find some interesting properties from this action.

Varying (\ref{int42}) with respect to the metric, we get field equations
\bd 
\Theta^{\mu \nu}\equiv G^{\mu \nu} +(1/\mu)C^{\mu \nu}=0 \label{int43}
\ed 
where the second rank  Weyl tensor $C^{\mu \nu}$ is  
\bd 
C^{\mu \nu}=(1/\sqrt{g})
\epsilon ^{\mu \alpha \beta} D_\alpha \tilde{R}^\nu _\beta \label{int44}
\ed 
and 
\bd 
\tilde{R}_{\alpha \beta} =R_{\alpha \beta}-(1/4)g_{\alpha \beta}R,
~~ R=R^\alpha_\alpha\label{int45}
\ed 
The components of the Einstein tensor $G^{\alpha \beta}$ are 
\bd 
G^{\alpha \beta} =R^{\alpha \beta}-(1/2)g^{\alpha \beta}R,
\label{int46}
\ed 
and the components of the Riemann tensor ${R^\alpha}_{\beta \gamma \delta}$ are
\bd 
{R^\alpha}_{\beta \gamma \delta}=
\partial _\delta{\Gamma^\alpha}_{\beta \gamma }
-\partial _\gamma {\Gamma^\alpha}_{\beta \delta}
+{\Gamma^\alpha}_{\mu \gamma }{\Gamma^\mu}_{\beta \delta}
-{\Gamma^\alpha}_{\mu \delta}{\Gamma^\mu}_{\beta \gamma} \label{int47}
\ed 
From (\ref{int43}) we get first order form for field equations
\bd 
{K_{\mu \nu}}^{\lambda \sigma} (\mu) R_{\lambda \sigma}=0 \label{int48}
\ed 
where  ${K_{\mu \nu}}^{\lambda \sigma}$ is operator of the form
\bd 
{K_{\mu \nu}}^{\lambda \sigma} =(\delta^\lambda _\mu \delta^\lambda _\nu 
-(1/2)g_{\mu \nu}g^{\lambda \sigma})+
\frac{1}{\mu \sqrt{g}}{\epsilon _\mu}^{\alpha \beta}
(\delta^\lambda _\beta  \delta^\sigma _\nu 
-(1/2)g^{\lambda \sigma}g^{\nu \beta})  \label{int49}
\ed 
Operator ${K_{\mu \nu}}^{\lambda \sigma}(\mu)$ may be multiplied by
 ${K_{\mu \nu}}^{\lambda \sigma}(- \mu)$
to yield a second-order equation for Ricci tensor.

From (\ref{int48}) we find
\bd 
{K_{\alpha \beta}}^{\mu \nu}(-\mu){K_{\mu \nu}}^{\lambda \sigma}(\mu)
R_{\lambda \sigma}=0  \label{int50}
\ed 
that is 
\bd 
(D_\alpha D^\alpha +\mu^2)R_{\mu \nu}=-g_{\mu \nu}R^{\alpha \beta}
R_{\alpha \beta}+
3R^\alpha _\beta R_\alpha ^\beta \label{int51}
\ed 
This exhibits the massive character of the excitations.

After this short review of topological field models we will consider 
3-D  fermionic models interacting with vector and with tensor fields.
These models may give effective induced topological action of the  CS type.

%% file: ikveccs.tex
\chapter{INDUCED CHERN-SIMONS MASS TERM}
\centerline{\Large \bf IN TOPOLOGICALLY NON-TRIVIAL  SPACE-TIME }
\vs
	
In gauge theories with fermions, the  a topological mass term is induced 
by fermionic interactions  \cite {vcs1}.
 If the topology of space-time is not trivial, topological parameters will appear 
in the  effective action and in the  CS Lagrangian. In this chapter we show 
 that the  CS term depends on topological parameters in such a way that the 
topological gauge invariance of the total action can be maintained.

Let us consider the case of a massive fermion field interacting with external
gauge field in 3-D space-time.

\section{ Euclidean space-time. Trivial topology}
\vsse
 
The  Euclidean action of this quantum system is given by
\bd
S=\int d^3 x \bar{\psi}(x)(i\hat{\partial}_x+m+g\hat{A}(x))\psi(x) \label{vcs1}
\ed
 The corresponding  generating functional is
\bd
Z=\int D\bar{\psi}(x) D \psi(x)exp\left[-\int d^3x
\bar{\psi}(x)(i\hat{\partial}_x+m+g\hat{A}(x))\psi(x)\right] \label{vcs2}
\ed
Here $A_\mu(x)=A_\mu^a (x)T^a$, and $T^a$ are generators of the 
gauge transformations. We will choose to work with two-component Dirac spinors.

Euclidean Dirac matrices have the   algebra:
\bd
\{\gamma_\mu, \gamma_\nu \}=-\delta_{\mu \nu} \label{vcs3}
\ed
This equation is satisfied by the matrices
\bd
\gamma_1=i\sigma_1,~~\gamma_2=i\sigma_2,~~\gamma_3=i\sigma_3 \label{vcs4}
\ed
where $\sigma_i$ are the Pauli matrices.

Furthermore, they satisfy
\bd
\gamma_\mu \gamma_\nu=-\delta_{\mu \nu}-\epsilon_{\mu \nu \rho}\gamma_\rho, \nonumber
\ed
\veb
\bd
\mbox{Tr} (\gamma_\mu \gamma_\nu \gamma_\rho)=2 \epsilon_{\mu \nu \rho} \nonumber
\ed
\veb
\bd
\mbox{Tr} (\gamma_\mu \gamma_\nu \gamma_\rho \gamma_\lambda \gamma_\sigma)
=-2(\delta _{\mu \nu} \epsilon_{\rho \lambda \sigma}
+\delta _{\rho \lambda} \epsilon_{\mu \nu \sigma}
-\delta _{ \nu \sigma} \epsilon_{\rho \lambda \mu}
+\delta _{\mu \sigma} \epsilon_{\rho \lambda \nu}) \label{vcs5}
\ed
The integration in (\ref{vcs2}) over the fermionic fields gives
\bd
Z[A_\mu]=Det(i\hat{\partial}+m+g\hat{A}))=exp(-S_{eff}) \label{vcs6}
\ed
Therefore we can write the expression for the effective action
\bd
S_{eff}=-\ln Det(i\hat{\partial}+m+g\hat{A}))
=-\mbox{Tr} \ln (i\hat{\partial}+m+g\hat{A}) \label{vcs7}
\ed
Now one  can split (\ref{vcs7})  in two parts
\bd
S_{eff}=-\mbox{Tr} \ln (i\hat{\partial}_x+m+g\hat{A}(x)) \nonumber
\ed
\veb
\bd
=\mbox{Tr} \ln \hat{S}_f-\mbox{Tr} 
\ln \left[1+g\hat{S}_f\hat{A}\right]  \label{vcs8}
\ed
where
\bd
\hat{S}_f=\frac{1}{i\hat{\partial}+m}
=\frac{i\hat{\partial}+m}{\partial^2 +m^2}  \label{vcs9}
\ed
is  a  fermionic propagator. 

The first term of the equation (\ref{vcs8}) is a divergent gauge-independent 
contribution, and the second one may be expanded in a power series \cite{vcs3}:
\bd
S_{eff}=-\mbox{Tr} \ln \left[1+g\hat{S}_f\hat{A}\right] =\mbox{Tr} \sum_{n=1}^{\infty}
\frac{1}{n}[g\hat{S}_f\hat{A}] \label{vcs10}
\ed
Let us consider the term which is 
quadratic in $A_\mu (x)$. The corresponding action is given by
\bd
S^{(2)} _{eff}=\frac{1}{2}g^2 \mbox{Tr} \left[ 
\frac{i\hat{\partial}+m}{\partial^2 +m^2}\hat{A}
\frac{i\hat{\partial}+m}{\partial^2 +m^2}\hat{A}\right]  \label{vcs11}
\ed
The terms with  two and four $\gamma$ matrices contribute to the wave function 
renormalization of the gauge boson and will not be taken into consideration
in further calculations.

The other two terms are linear with respect to mass $m$ and give
\bd
S^{(2)} _{eff}=\frac{1}{2}g^2 m \mbox{Tr} \left[ 
\frac{i\hat{\partial}}{\partial^2 +m^2}\hat{A}
\frac{1}{\partial^2 +m^2}\hat{A}
+\frac{1}{\partial^2 +m^2}
\hat{A}\frac{i\hat{\partial}}{\partial^2 +m^2}\hat{A}\right] \nonumber
\ed
\veb
\bd
=\frac{img^2}{2}\mbox{tr} _D \gamma_\mu \gamma_\nu \gamma_\rho
\mbox{tr} \left[\frac{\partial_\mu}{\partial^2 +m^2}A_\nu
\frac{1}{\partial^2 +m^2}A_\rho
+\frac{1}{\partial^2 +m^2}A_\mu\frac{\partial_\nu}{\partial^2 +m^2}A_\rho 
\right] \label{vcs12}
\ed
where $\mbox{Tr}=\mbox{tr} _D \mbox{tr}$  is the trace of  both the $\gamma$ matrices and the  
vector field, and 
also includes the integrations. 

To write $\mbox{tr}$ in the form
$\mbox{tr}\left[f(\partial)g(x)\right]$
one can use the relation
\bd
\phi(x)\frac{1}{\partial^2+m^2}=\frac{1}{\partial^2+m^2}\phi(x)
+\frac{[\partial^2,\phi(x)]}{(\partial^2+m^2)^2}
+\frac{[\partial^2,[\partial^2,\phi(x)]]}{(\partial^2+m^2)^3}
+... \label{vcs13}
\ed 
Then
\bd
S^{(2)}_{eff}=img^2~\epsilon_{\mu \nu \rho}
\mbox{Tr}\left[\frac{1}{(k^2 +m^2)^2}
(\partial_\nu A_\mu )A_\rho)\right] \nonumber
\ed
\bd  
=img^2~\epsilon_{\mu \nu \rho}
\int d^3 x <x|\frac{1}{(\partial^2+m^2)^2}|x>
tr\{(\partial_\nu A_\mu )A_\rho\}  \label{vcs14}
\ed
In another form
\bd  
S^{(2)}_{eff}=-img^2~\epsilon_{\mu \nu \rho}
\int d^3 x \left[\int \frac{d^3k}{(2\pi)^3}\frac{1}{(k^2+m^2)^2}\right]
\{(\partial_\mu A_\nu )A_\rho\}  \label{vcs15}
\ed
The additional contribution to the CS term comes from the terms which are 
cubic in $A_\mu$'s.
We can rewrite them as
\bd
S^{(3)} _{eff}=\frac{1}{3}g^3 m \mbox{Tr} \left[ 
\frac{i\hat{\partial}+m}{\partial^2 +m^2}\hat{A}
\frac{\hat{i\partial}+m}{k^2 +m^2}\hat{A} 
\frac{\hat{i\partial}+m}{\partial^2 +m^2}\hat{A}\right]  \label{vcs17}
\ed 
Noticing that the terms having even number 
of $\gamma$ matrices do not contribute to
the CS term, we have:
\bd
S^{(3)} _{eff}=-\frac{1}{3}g^3 m Tr\left[ 
\frac{\hat{\partial}}{\partial^2 +m^2}\hat{A}
\frac{\hat{\partial}}{\partial^2 +m^2}\hat{A}
\frac{1}{\partial^2 +m^2}\hat{A}\right. \nonumber
\ed
\veb
\bd
+\frac{\hat{\partial}}{\partial^2 +m^2}\hat{A}
\frac{1}{\partial^2 +m^2}\hat{A}
\frac{\hat{\partial}}{\partial^2 +m^2}\hat{A} \nonumber
\ed
\veb
\bd
+\frac{1}{\partial^2 +m^2}\hat{A}
\frac{\hat{\partial}}{\partial^2 +m^2}\hat{A}
\frac{\hat{\partial}}{\partial^2 +m^2}\hat{A} \nonumber
\ed
\veb
\bd
\left.-
m^2\frac{1}{\partial^2 +m^2}\hat{A}
\frac{1}{\partial^2 +m^2}\hat{A}
\frac{1}{\partial^2 +m^2}\hat{A}
\right] \label{vcs18}
\ed 
Using the identity  (\ref{vcs13}) find
\bd
S^{(3)}_{eff}=-\frac{mg^3}{3}
\left[\frac{}{}\mbox{tr} _D \gamma_\mu \gamma_\nu \gamma_\rho 
\gamma_\lambda \gamma_\sigma\right.
\times \nonumber
\ed
\veb
\bd
\times \mbox{tr} \left\{ \frac{\partial_\mu \partial_\rho }{(\partial^2+m^2)^3}A_\nu
A_\lambda A_\sigma + 
\frac{\partial_\mu \partial_\lambda }{(\partial^2+m^2)^3}A_\nu A_\rho A_\sigma + 
\frac{\partial_\nu \partial_\lambda }{(\partial^2+m^2)^3}A_\mu A_\rho A_\sigma
\right\}
\nonumber
\ed
\veb
\bd
\left. -m^2 \mbox{tr} _D \gamma_\mu \gamma_\nu \gamma_\rho
 \mbox{tr} \frac{1}
{(\partial^2+m^2)^3}A_\mu A_\nu A_\rho \right] \label{vcs19} 
\ed 
Tracing with respect to $\gamma$ matrices, find
\bd
S^{(3)}_{eff}=-\frac{2mg^3}{3}
\left[\int \frac{d^3k}{(2\pi) ^3}\frac{1}{(k^2+m^2)^2}\right]
\mbox{tr} \int d^3x \epsilon ^{\mu  \nu  \rho} 
 A_\mu A_\nu A_\rho  \label{vcs20}  
\ed
Combining (\ref{vcs15}) with (\ref{vcs20}) we get \cite{vcs3}
the induced Chern-Simons term in the form
\bd
S^{CS}_{eff}=-img^2\left[\int \frac{d^3k}{(2\pi) ^3}\frac{1}{(k^2+m^2)^2}\right] 
\mbox{tr} \int d^3x 
\epsilon _{\mu  \nu  \rho} 
\left[\partial_\mu A_\nu A_\rho
-\frac{2}{3}ig A_\mu A_\nu A_\rho \right] \label{addcs20} 
\ed 
or, after integration over the momenta
\bd
S^{CS}_{eff}=-\frac{ig^2}{8 \pi}\frac{m}{|m|} \mbox{tr} \int d^3x 
\epsilon _{\mu  \nu  \rho} 
\left[\partial_\mu A_\nu A_\rho
-\frac{2}{3}ig A_\mu A_\nu A_\rho \right] \label{vcs21} 
\ed 
Now we can apply the  method we have developed  for a  gauge model 
in 3-D non-trivial space-time. 

\section{ Non-trivial topology}
\vsse

a)  Model with space-time topology  $\Sigma=R^{(2)}\times mobius~strip$.

To introduce topology $\Sigma =R^{(2)} \times S^1$ with parameter $\zeta$
we rewrite the expression (\ref{vcs15}) in the following way.
Momentum integration in (\ref{vcs15}) will be integration with respect 
to two dimensions ($(t,x) \to R^{(2)}$),and the third integration will be 
transformed in the sum, 
because the  fermionic propagator is antisymmetric with respect to  the
 selected axis ($y \to S^1$).

We can consider that $k^2=\vec{k}^2+\omega^2_n$ 
with $\omega_n=(2\pi /\zeta)(n+1/2)$ and $n=0,\pm1,\pm2,...$. 
Integrating in 2-D momentum space  we  find that
 \bd
\int\limits^\sim \frac{d^3k}{(2\pi)^3}\frac{1}{(k^2 +m^2)^2}
=\int \frac{d^2 k}{(2\pi)^2}\frac{1}{\zeta}
\sum\limits_{n=-\infty}^\infty \frac{1}{((\vec{k})^2+\omega^2_n +m^2)^2} \nonumber
\ed
\veb
 \bd
=\frac{1}{8\pi |m|}tanh\left\{\frac{|m| \zeta}{2}\right\}  \label{vcs22}
\ed 
where 
\bd
 \int\limits^\sim \frac{d^3k}{(2\pi)^3}
=\frac{1}{\zeta}\sum \limits_{n=-\infty}^\infty 
\int \frac{d^2k}{(2\pi)^2} \nonumber
\ed
In the process of calculations we used the useful equation 
\bd
\sum\limits_{n=-\infty}^\infty \frac{y}{(y^2+(n+1/2)^2)}
=\pi tanh (\pi y)  \label{vcs23}
\ed 
Then the expression  (\ref{addcs20}) in this topology will be
\bd
S^{(CS)}_{eff}=-img^2\int \frac{d^2 k}{(2\pi)^2}\frac{1}{\zeta}
\sum\limits_{n=-\infty}^\infty 
\frac{1}{((\vec{k})^2+\omega^2_n +m^2)^2} \nonumber
\ed
\veb
\bd 
\mbox{tr} \int \epsilon_{\mu \nu \rho}\left\{\partial_\nu A_\mu A_\rho 
-\frac {2i}{3}g A_\nu A_\mu A_\rho \right\}\label{vcs24}
\ed
and the  induced CS term at non-trivial 3-D space-time is written as
\bd
S^{CS}_{eff}(\zeta)=-\frac{ig^2}{8\pi}\frac{m}{|m|}
tanh \left\{\frac{|m|\zeta}{2}\right\}
\mbox{tr} \int d^3 x\epsilon_{\mu \nu \rho}(\partial_\mu A_\nu A_\rho-\frac{2}{3}ig  
 A_\mu A_\nu A_\rho ) \label{vcs28}
\ed
The relation between CS terms is the function of the form 
\bd
\frac{S^{CS}_{eff}(\zeta)}{S^{CS}_{eff}}=
tanh \left\{\frac{|m|\zeta}{2}\right\} \label{vcs29}
\ed
b) Model with space-time topology  $\Sigma=R^{(2)}\times S^1$.

For this space-time topology the propagator of the fermionic field 
will be periodic
at the boundary points of interval $[0,\zeta]$, that leads 
to the modification of equation (\ref{vcs21}) with respect to new frequencies
$\omega_n=2\pi n/\zeta$ with $n=0,\pm1,\pm2,...$.

Taking into account the  summation formula
  \bd
\sum \limits_{n=-\infty}^\infty \frac{y}{y^2+n^2}
=\pi coth (\pi y)  \label{sum2}
\ed 
we find, that
\bd
\int \limits_\Sigma \frac{d^3k}{(2\pi)^3}\frac{1}{(k^2 +m^2)^2}
=\int \frac{d^2 k}{(2\pi)^2}\frac{1}{\zeta}
\sum\limits_{n=-\infty}^\infty \frac{1}{((\vec{k})^2+\omega^2_n +m^2)^2} \nonumber
\ed
 \veb
\bd
=\frac{1}{8\pi |m|}coth\left\{\frac{|m| \zeta}{2}\right\}  \label{vcs30}
\ed 
Using this equation  and the equation (\ref{vcs20}), get the  induced CS term in  the new  
topology
\bd
S^{CS}_{eff}(\zeta)=-\frac{ig^2}{8\pi}\frac{m}{|m|}
coth\left\{\frac{|m|\zeta}{2}\right\}
\mbox{tr} \int d^3 x\epsilon_{\mu \nu \rho}(\partial_\mu A_ \nu A_\rho-\frac{2}{3}ig  
 A_\mu A_\nu A_\rho ) \label{vcs31}
\ed
The rotio of the  CS terms will be the function of the topological parameter $\zeta$:
\bd
\frac{S^{CS}_{eff}(\zeta)}{S^{CS}_{eff}}=
coth\left\{\frac{|m|\zeta}{2}\right\} \label{vcs32}
\ed
These results show that the relations (\ref{vcs29}) and (\ref{vcs32}) are 
smooth functions of the  topological parameter $\zeta$.

%% file: ikGCS.tex
\chapter{ GRAVITATIONAL CHERN-SIMONS }
\centerline{\Large \bf MASS TERM }
\vspace{3mm}
\centerline{\Large \bf AT FINITE TEMPERATURE }
\vs
	
In this chapter we will consider fermions interacting with
an  external gravitational field
at finite temperature. 
In  the  previous chapter XX we learned that the interaction of
fermions with external gauge bosons leads to the induced action of the Chern-Simons type.
In the same way the interaction of fermions with external gravitational fields
may leads to the effective gravitational Chern-Simons term
\cite{vcs1},\\
\cite{gcs1}.   
Now we will develop a formalism of  calculations  of the  effective gravitational 
 CS action  from  the action for massive
fermions interacting  with an external gravitational field.   

\section{ Induced gravitational Chern-Simons}
\lum
\hspace{18mm}{\Large \bf  mass term} 
\vsse

Let us introduce the action for massive fermions connected with an external 
gravitational field as:
\bd  
S=\int d^3x \sqrt{g}\bar{\psi}(x)(iD+m)\psi(x) \label{gcs1}
\ed
where $D=\gamma^\mu D_\mu=\gamma^\mu (\partial_\mu +\omega_{ab\mu}\sigma^{ab})$
 with $\gamma^\mu(x)=h^\mu _a(x)\gamma^a $.

The  function $\omega_{ab\mu}$ is the local Lorentz connection 
and $\sigma^{ab}=(1/8)[\gamma^a,\gamma^b]$ is the commutator of $\gamma$ matrices.

$\gamma$-matrices with Latin indices are constructed from Pauli matrices
\bd  
\gamma^a=i\sigma^a ~~,\sigma^a (a=1,2,3) \label{gcs2}
\ed
and obey the  relations:
\bd
 \{\gamma^a,\gamma^b \}=-2\delta^{ab},
~~~\gamma_a\gamma_b=-\delta_{ab}-\epsilon _{abc}\gamma^c  \label{gcs3}
\ed
\veb
\bd  
\mbox{tr} (\gamma^a\gamma^b)=-2i\delta^{ab},
 ~~~ \mbox{tr} (\gamma^a\gamma^b\gamma^c)=2\epsilon^{abc}  \label{gcs4}
\ed
We can find the effective action by integrating (\ref{gcs1})
with  respect  to  the   fermionic field from the equation
\bd  
exp[-S_{eff}]=\int D\bar{\psi}(x)\psi(x) \exp [-S]=Det(-iD-m)\label{gcs5}
\ed
To find  the  one-loop effective action  we 
will rewrite  (\ref{gcs5}) 
in the following form \cite{ojima1}:
\bd
S_{eff}=-ln~ Det\frac{(-iD-m)}{(-iD+i\omega-m)} \nonumber
\ed
\veb
\bd
=-\ln~Det\frac{(-iD-m)(-iD-i\omega-m)}{(-iD+i\omega-m)(-iD-i\omega-m)} \nonumber
\ed
\veb
\bd  
=-\ln~Det(DD-D\omega+im\omega+m^2) +logDet[(D-\omega)^2]+m^2 \label{gcs6}
\ed
where we put $\omega=\gamma^\mu \omega_\mu=\gamma^\mu \sigma^{ab}\omega _{ab \mu}$.

Now we can use the momentum space formalism  we developed in part I of this work 
 for    calculation of the  first determinant of the expression  (\ref{gcs6}).
We can simplify the problem  by
 noticing that the second term of (\ref{gcs6})
does not give contributions to  the induced CS term.

For  the calculation of the first term let us introduce normal coordinates
and write the metric at the origin ($x^{'}$) of these coordinates. We will use variable
$y=x-x^{'}$ for the definition of the point of manifold in tangent space.
The  tangent space $Y$ will be "flat" with metric
$g_{ab}=-\eta_{ab}=diag(-1,-1,-1)$. 
In the origin of these coordinates the tetrad functions  and the metric
 will be the  series:
\bd  
h^\mu _a(x)=h^\mu _a(x^{'})-(1/6){R^\mu}_{\nu a \sigma}y^\mu y^\sigma
+...\label{gcs7}
\ed
\veb
\bd  
g_{\mu \nu}(x)=g_{\mu \nu}(x^{'})-(1/3)R_{\mu \alpha \nu \beta}y^\mu y^\nu
-(1/6){R _{\mu \alpha \nu \beta}}_{;\lambda}
y^\mu y^\nu y^\lambda +... \label{gcs8}
\ed
The covariant derivative in a tangent space is written as
\bd
D=\gamma_\mu (x)D_\mu (x) \nonumber
\ed
\veb
\bd  
=\left[\frac{}{}\gamma^a h^\mu _a(x^{'})-(1/6)\gamma^a
{R^\mu}_{\nu a \sigma}
y^\mu y^\sigma+...\right]\left( \frac{\partial}
{\partial y^\mu}+\omega_{ab\mu }(x^{'})\sigma^{ab}\right) \label{gcs9}
\ed
The effective action can be represented in the normal coordinates as
\bd
S_{eff}=\int d^3 x\sqrt{g} S_{eff}(x) \nonumber
\ed
\veb
\bd  
=\mathop{\lim}\limits_{x \to x^{'}}
\int d^3 x^{'}\sqrt{g( x^{'})} S_{eff}( x^{'})=
\mathop{\lim}\limits_{y \to 0}\int d^3 x\sqrt{g} S_{eff}(x,y) \label{gcs10}
\ed
In proper time formalism $ln Det$ is
\bd
S_{eff}[\omega]=\ln~Det(DD-D\omega+im\omega+m^2)= \nonumber
\ed
\veb  
\bd
\mathop{\lim}\limits_{x \to x^{'}}
\int d^3x <x|\int \limits_0^\infty
\frac{ds}{s}\mbox{tr} \exp \left\{-i\left( D^\mu D_\mu+\frac{R}{4}
-D\omega +im\omega +m^2\right)\right\}s|x^{'}> \label{gcs11} 
\ed
In accordance with (\ref{gcs10}) and (\ref{gcs11}) the density $S_{eff} (x)$
may be written in the momentum-space form 
\bd
 S_{eff}(x)=\mathop{\lim}\limits_{y \to 0}
\int \limits_0^\infty \frac{ds}{s}\int \frac{d^3 k}{(2\pi)^3} \exp [~iky~]
\exp [-i(m^2-\Delta_\mu \Delta^\mu )s]\times \nonumber
\ed
\veb
\bd  
\mbox{tr} \exp \left\{-i\left[\left( D^\mu D_\mu+\frac{R}{4}
-D\omega \right)  +(i \Delta_\mu D^\mu
+iD_\mu \Delta^ \mu-i\Delta \omega +im\omega ) \right]s \right\} \label{gcs12} 
\ed
where  $\mathop{\lim}\limits_{y \to 0} \Delta_\mu (y)=k_\mu$.

Notice that if we  like to obtain the contribution of finite terms as
$(m \to 0) $, it is sufficient to extract the terms which are proportional 
to $s^2$ with a factor $m$ or proportional to $s^3$ with a factor $m^3$
(or $k^2$).  Thus we get the term in $S_{eff}[\omega]$, 
 which contains the induced CS term.

The second and the third orders of the expansion (\ref{gcs12})
give
\bd
 S_{eff}(x)=\mathop{\lim}\limits_{y \to 0}
\int \frac{d^3 k}{(2\pi)^3} \exp [~iky~]\int \limits_0^\infty \frac{ds}{s}
\exp [-i(m^2-\Delta_\mu \Delta^\mu)s](im)\times \nonumber
\ed
\veb
\bd  
\times \mbox{tr} \left[\frac{1}{2!}\left( \omega D\omega +
D\omega \omega  \right)s^2- \right. \nonumber
\ed
\veb
\bd
\left. - \frac{1}{3!}(\Delta \omega\Delta \omega \omega +
\Delta \omega \omega \Delta \omega+\omega\Delta \omega\Delta \omega
+m^2\omega\omega\omega)s^3 \right]  \label{gcs13}
\ed
where $\Delta(y)=\gamma^\mu \Delta_\mu(y)$.

In the limit $(y \to 0)$ the equation  (\ref{gcs13}) is
\bd
 S_{eff}(x)=(im)\int \frac{d^3 k}{(2\pi)^3}\int \limits_0^\infty \frac{ds}{s}
\exp [-i(m^2+k^2)s]\times \nonumber
\ed
\veb
\bd  
\times \mbox{tr} \left[\left( \omega \hat{\partial}\omega 
 +\omega \omega\omega\right)~s^2- \right.
\nonumber
\ed
\veb
\bd
\left.- \frac{1}{3!}(\hat{k} \omega\hat{k} \omega \omega +
\hat{k}\omega \omega \hat{k}\omega+\omega \hat{k} \omega \hat{k}\omega
+m^2\omega\omega\omega)s^3 \right]  \label{gcs14}
\ed
Noticing, that
\bd
\mbox{tr}(\hat{k} \omega\hat{k} \omega \omega +
\hat{k}\omega \omega \hat{k}\omega+\omega \hat{k} \omega \hat{k}\omega
+m^2\omega\omega\omega)=(k^2+m^2)\mbox{tr}\omega\omega\omega  \label{gcs15}
\ed
rewrite (\ref{gcs14}) as
\bd
 S_{eff}(x)=(im)\left[\mbox{tr}\omega
\hat{\partial}\omega+\frac{2}{3}\mbox{tr}\omega\omega\omega \right]
\int \frac{d^3 k}{(2\pi)^3}\int \limits_0^\infty
ds(s)
\exp [-is(m^2+k^2)]  \label{gcs16}
\ed
Applying the equation for the  $\Gamma(\nu )$ function 
\bd  
\int \limits_0^\infty ds s^{\nu-1}\exp [-sm^2]
=\frac{\Gamma(\nu)}{|m|^{2\nu}} \label{gcs17}
\ed
and computing \mbox{tr} of $\gamma$ matrices,  find
\bd  
S_{eff}^{CS}[\omega]=-\frac{i}{64 \pi}\frac{m}{|m|}\int d^3 x \epsilon^{\mu \nu\rho}
\left(\partial_\mu {\omega^b}_{a \nu}{\omega^a}_{b \rho}+
\frac{2}{3}{\omega^a}_{b \nu}{\omega^b}_{c \mu}{\omega^c}_{a \rho}\right)
\label{gcs18}
\ed
This is the expression for gravitational CS term \cite{ojima1}
which we get using  momentum-space method.

\section{Induced gravitational Chern-Simons}
\lum
\hspace{13mm}{ \Large \bf mass term  at finite temperature}
\vsse

The calculations of finite temperature gravitational  action of CS type
are  based on the computation of   $\ln Det$ (\ref{gcs5}). 
In Part I we developed the formalism of such 
 calculations with  the help of the momentum space methods.
For these calculations we will replace the  
integration procedure over three momenta in the expression (\ref{gcs16}) by 
the summation over Matsubara frequencies and integration over the  
two remaining momenta.

Let us introduce temperature in the  tangent 3-D space to the  curved manifold 
as (part I,(\ref{b47})):
\bd  
\int \frac{d^3k}{(2 \pi)^3}F(k^2_1,k^2_2,k^2_3,s)
\mathop{\longrightarrow}\limits^{T\neq 0}
\frac{1}{\beta}\sum \limits_{n=-\infty}^{\infty}
\int \frac{d^2k}{(2 \pi)^2}F(\omega^2_n,k^2_2,k^2_3,s) \label{gcs19}
\ed
where $\beta^{-1}=T$ is temperature, and $\omega^2_n=(2\pi /\beta)(n+1/2)$.

To introduce temperature in the model, rewrite integral over $(s)$ in (\ref{gcs16})
in the form 
\bd  
\int \frac{d^3p}{(2\pi)^3}\int ds~s \exp [-is(k^2+m^2)]= 
-\int \frac{d^3p}{(2\pi)^3}\frac{1}{(k^2+m^2)^2}   \label{gcs20}
\ed
Using (\ref{gcs19}) and  the summation formula (\ref{vcs23}), find  
the finite temperature  CS term 
\bd  
S_{eff}[\omega]=-\frac{i}{64 \pi}\frac{m}{|m|}
tanh\frac{ |m|}{2T}
\int d^3 x \epsilon^{\mu \nu\rho}
\left(\partial_\mu {\omega^b}_{a \nu}{\omega^a}_{b \rho}+
\frac{2}{3}{\omega^a}_{b \nu}{\omega^b}_{c \mu}{\omega^c}_{a \rho}\right)
\label{gcs22}
\ed
Comparing (\ref{gcs18}) and (\ref{gcs22}) we get the following result  
\bd  
\frac{S_{eff}[T \neq 0]}{S_{eff}[T=0]}=
tanh\frac{\beta |m|}{2} \label{gcs23}
\ed
The expression (\ref{gcs22}) shows that the structure of the  induced CS term at finite 
temperature is exactly the same as at zero temperature, and the functional relations 
 between the  CS terms in tensor (\ref{gcs23}) and vector types of interactions 
(\ref{vcs29}) are the same.

%% file: ikapp1m.tex
\renewcommand{\theequation}{XXII.\arabic{equation}}
 \begin{center} 
\vspace*{8mm}{\LARGE \bf APPENDIX} 
\end{center}
\vspace{2mm}
\begin{center} 
{\LARGE\bf{SUMS  OF BESSEL FUNCTIONS}} 
\end{center}

\vspace{3mm}

\begin{center} 
{\Large  \bf I.  Type I sums of the modified Bessel functions} 
\end{center}
\vspace{5mm}

\setcounter{equation}{0}

The integral representations for series of modified Bessel
functions  may be calculated  from the following
summation formula:
\bd
\sum\limits_{n=-\infty }^\infty  
{\left[ {y^2 + 
				\left( {n + 1\over 2} \right)^2}
    \right]^{-1}}
={{\pi}\over{y}} {\tanh (\pi y)}.\label{aa1}
\ed 
In proper time representation the left side of  (\ref{aa1})  
can be written in the form
\bd
\sum\limits_{n=-\infty }^\infty 
				{\left[ {y^2
				+ \left( {n+1\over 2}\right)^2}
				\right]^{-1}} \nonumber
\ed
\veb
\bd
= \int\limits_0^\infty  {d\alpha \exp
\left( {-\alpha y^2} \right)}
\sum\limits_{n=-\infty }^\infty  
{\exp \left( {-\alpha\left( {n+{1 \over 2}} \right)^2} \right)}. \label{aa2}
\ed 
Taking into account the  equation
\bd
\sum\limits_{n=-\infty }^\infty  {\exp \left\{ {-\alpha \left( {n-z}
\right)^2}
\right\}}=\sum\limits_{n=-\infty }^\infty  {\left( {{\pi  \over \alpha }}
\right)}^{1\over 2}
\exp \left( {-{{\pi ^2} \over \alpha }n^2-2\pi izn} \right),\label{aa3}
\ed 
we can write (\ref{aa2}) as
\bd
 \sum\limits_{n=-\infty }^\infty 
    {\left[ {y^2 + \left( {n+1\over 2}\right)^2}
\right]^{-1}}
=\sum\limits_{n=-\infty }^\infty 
{(-1)^n}\int\limits_0^\infty 
{d\alpha \left( {{\pi  \over \alpha }} \right)^{1\over 2}\exp \left( {-\alpha
y^2-{{\pi ^2} \over \alpha }n^2} \right)} \nonumber
\ed
\veb
\bd
={\pi  \over y}+2\sum\limits_{n=1}^\infty  {(-1)^n\int\limits_0^\infty  {d\alpha
\left( {{\pi  \over \alpha }} \right)^{1\over 2}\exp \left( {-\alpha y^2-{{\pi ^2}
\over \alpha }n^2} \right)}}.\label{aa4}
\ed 
The right side of (\ref{aa1}) is
\bd
{\pi  \over y}\tanh (\pi y)={\pi  \over y}-{{2\pi } \over {y\left(
{e^{2\pi y}+1}
\right)}}. \label{aa5}
\ed 
Therefore, we get from (\ref{aa4}) and (\ref{aa5})  the following useful equation 
\bd
{1 \over {z\left( {e^z+1} \right)}}=-{1 \over {2\pi
^2}}\sum\limits_{n=1}^\infty  {(-1)^n\int\limits_0^\infty  {d\alpha \left( {{\pi 
\over \alpha }} 
\right)^{1\over 2}\exp \left( {-\alpha {{z^2} \over {4\pi ^2}}-{{\pi
^2} \over \alpha }n^2}
\right)}}. \label{aa6}
\ed 
Moreover, we may consider that $z^2=g_a(x^2)$ is the function of
variable $x\in R^3$ with a parameter $a$, namely $z^2=x^2+a^2$.

Integrating  with respect to $x$ one can get
\bd
\int {{{d^3x} \over {\left( {2\pi } \right)^3}}}\left[ {\sqrt
{x^2+a^2}\left( {\exp \left( {\sqrt {x^2+a^2}} \right)+1} \right)}
\right]^{-1} \nonumber
\ed
\veb
\bd
=-{1 \over 2}\sum\limits_{n=1}^\infty 
{(-1)^n\int\limits_0^\infty  {d\alpha \alpha ^{-2}\exp \left( {-\alpha
{{a^2} \over {4\pi ^2}}-{{\pi ^2} \over
\alpha }n^2} \right)}}.\label{aa7}
\ed 
The modified Bessel function may be written as
\bd
\int\limits_0^\infty  {d\alpha \cdot \alpha ^{\nu -1}\cdot \exp \left(
{-\gamma
\alpha -{\delta \over \alpha} } \right)}=2\left( {\delta \over \gamma }
\right)^{\nu \over 2}K_\nu \left(2\sqrt {\delta \,\gamma }\right).\label{aa8}
\ed 
Then  (\ref{aa7}) will be
\bd
\int {{{d^3x} \over {\left( {2\pi } \right)^3}}}\left[ {\sqrt
{x^2+a^2}\left( {\exp \left( {\sqrt {x^2+a^2}} \right)+1} \right)} \nonumber
\right]^{-1}
\ed
\veb
\bd
= -{1 \over 2}\left(
{{a \over {\pi ^2}}} \right)\sum\limits_{n=1}^\infty  {{{(-1)^n} \over
n}K_1}(an). \label{aa9}
\ed 
Scaling $x$ and $a$ with a parameter $\beta $ as $(x,a)=\beta
(k,m)$  write (\ref{aa9}) in the form
\bd
\int {{{d^3k} \over {\left( {2\pi } \right)^3}}\,{2 \over {\varepsilon
\left( {\exp
\left( {\beta \varepsilon } \right)+1} \right)}}}=-{m \over {\beta \pi
^2}}\sum\limits_{n=1}^\infty  {{{(-1)^n} \over n}K_1}(\beta mn), \label{aa10}
\ed 
where $\varepsilon =\sqrt {\vec{k}^2+m^2}$. 

Differentiating the equation  (\ref{aa7}) with   respect to parameter $a$ we get 
\bd
\int {{{d^3x} \over {\left( {2\pi } \right)^3}}\left(
{-{\partial 
\over {\partial a^2}}} \right)}\left[ {\sqrt {x^2+a^2}\left( {\exp \left( {\sqrt
{x^2+a^2}}
\right)+1} \right)} \right]^{-1}\hfil \nonumber
\ed
\veb
\bd
\hfil = -{1 \over {4\pi ^2}}\sum\limits_{n=1}^\infty 
{(-1)^nK_0}(an), \label{aa11} 
\ed 
or, in new variables,
\bd
\int {{{d^3k} \over {\left( {2\pi } \right)^3}}\left( {{\partial 
\over {\partial m^2}}} \right){1 \over {\varepsilon \left( {\exp \left( {\beta
\varepsilon }
\right)+1} \right)}}}={1 \over {4\pi ^2}}\sum\limits_{n=1}^\infty 
{(-1)^nK_0}(\beta mn). \label{aa12}
\ed 
Integrating in (\ref{aa7}) with respect to parameter $(a)$ and
using the equation 
\bd
\int\limits_{a^2}^\infty  {da^2\left[ {\sqrt {x^2+a^2}\left( {\exp
\left( {\sqrt {x^2+a^2}} \right)+1} \right)} \right]^{-1}} \nonumber
\ed
\veb
\bd
=2\ln \left(
{1+\exp\left(-
\sqrt {x^2+a^2}\right)}
\right) \label{aa13}
\ed  
we find with  new variables the following equation 
\bd
-{1 \over \beta }\int {{{d^3k} \over {\left( {2\pi } \right)^3}}}\ln
\left( {1+\exp (-\beta \varepsilon )} \right)={{m^2} \over {2(\beta \pi
)^2}}\sum\limits_{n=1}^\infty  {{{(-1)^n} \over {n^2}}K_2}(\beta mn). \label{aa14}
\ed 
High temperature asymptotes $(\beta{m}\ll{1})$ of the equations 
(\ref{aa9}), (\ref{aa11}) and (\ref{aa14}) are 
\bd
{{2m^2} \over {(\beta \pi )^2}}\sum\limits_{n=1}^\infty  {{{(-1)^n}\over
n^2}K_2}(\beta mn)=-{{7\pi ^2} \over {180\beta ^4}}
+{{m^2} \over {12\beta ^2}}+{1
\over 8}m^4\left( {\ln {{\beta m} \over {4\pi }}+\gamma -{3 \over 4}}
\right) \label{aa15}
\ed
\veb
\bd
-{m \over {\beta \pi ^2}}\sum\limits_{n=1}^\infty  {{{(-1)^n} \over
n}K_1}(\beta mn)={1 \over {12\beta ^2}}+{1 \over 4}m^2\left( {\ln \left(
{{{\beta m} \over {4\pi }}} \right)+\gamma -{1 \over 2}}
\right) \label{aa16}
\ed
\veb
\bd
{1 \over {2\pi ^2}}\sum\limits_{n=1}^\infty 
{(-1)^nK_0}(\beta mn)={1
\over 4}\left( {\ln \left( {{{\beta m} \over {4\pi }}} \right)+\gamma }
\right). \label{aa17}
\ed

\newpage

\begin{center} 
{\Large  \bf II.  Type II sums of the modified Bessel functions} 
\end{center}

Let us start with  the sum:
\bd
\sum\limits_{n=-\infty }^\infty  
 \left(y^2 + 	n^2 \right)^{-1}
={{\pi}\over{y}} {\coth (\pi y)}.\label{bb1}
\ed 
The   proper time representation the left side of (\ref{bb1}) will be
\bd
\sum\limits_{n=-\infty }^\infty 
				\left( y^2
				+ n^2\right)^{-1}
= \int\limits_0^\infty  {d\alpha \exp
\left( {-\alpha y^2} \right)}
\sum\limits_{n=-\infty }^\infty  
{\exp \left( {-\alpha n^2} \right)}.\label{bb2}
\ed 
Taking into account the  equation (\ref{aa3}) and putting $(z=0)$ we find
\bd
\sum\limits_{n=-\infty }^\infty  {\exp \left\{ {-\alpha n^2}
\right\}}=\sum\limits_{n=-\infty }^\infty  {\left( {{\pi  \over \alpha }}
\right)}^{1\over 2}\exp \left( -{{\pi ^2} \over \alpha }n^2 \right),\label{bb3}
\ed 
Then (\ref{bb2}) may be written in the form
\bd
 \sum\limits_{n=-\infty }^\infty 
    \left( y^2 + n^2\right)^{-1}
=\sum\limits_{n=-\infty }^\infty 
\int\limits_0^\infty 
d\alpha \left( \frac{\pi}{\alpha } \right)^{1\over 2}\exp \left( -\alpha
y^2-\frac{\pi ^2}{\alpha} n^2 \right) \nonumber
\ed
\veb
\bd
={\pi  \over y}+2\sum\limits_{n=1}^\infty  
\int\limits_0^\infty  d\alpha
\left( \frac{\pi}{\alpha } \right)^{1\over 2}\exp \left( -\alpha y^2-\frac{\pi ^2}
{\alpha }n^2 \right).\label{bb4}
\ed 
The right side of  (\ref{bb1}) is
\bd
{\pi  \over y}\coth (\pi y)={\pi  \over y}+{{2\pi } \over {y\left(
{e^{2\pi y}-1}
\right)}}.\label{bb5}
\ed 
Therefore, we get from (\ref{bb4}) and (\ref{bb5}) the following useful equation 
\bd
{1 \over {z\left( {e^z-1} \right)}}={1 \over {2\pi
^2}}\sum\limits_{n=1}^\infty  {\int\limits_0^\infty  {d\alpha \left( {{\pi 
\over \alpha }} 
\right)^{1\over 2}\exp \left( {-\alpha {{z^2} \over {4\pi ^2}}-{{\pi
^2} \over \alpha }n^2}
\right)}}. \label{bb6}
\ed 
Let $z^2=g_a(x^2)$ be a function of
variable $x\in R^3$ with a parameter $a$ of the form $z^2=x^2+a^2$.

After  integration with  respect to  $x$ we get
\bd
\int {{{d^3x} \over {\left( {2\pi } \right)^3}}}\left[ {\sqrt
{x^2+a^2}\left( {\exp \left( {\sqrt {x^2+a^2}} \right)-1} \right)}
\right]^{-1}\\[.15in] \nonumber
\ed
\veb
\bd
={1 \over 2}\sum\limits_{n=1}^\infty 
{\int\limits_0^\infty  {d\alpha \alpha ^{-2}\exp \left( {-\alpha
{{a^2} \over {4\pi ^2}}-{{\pi ^2} \over
\alpha }n^2} \right)}}.\label{bb7}
\ed 
The modified Bessel function is written as
\bd
\int\limits_0^\infty  {d\alpha \cdot \alpha ^{\nu -1}\cdot \exp \left(
{-\gamma
\alpha -{\delta \over \alpha} } \right)}=2\left( {\delta \over \gamma }
\right)^{\nu \over 2}K_\nu \left(2\sqrt {\delta \,\gamma }\right).\label{bb8}
\ed 
and (\ref{bb7}) will be
\bd
\int {{{d^3x} \over {\left( {2\pi } \right)^3}}}\left[ {\sqrt
{x^2+a^2}\left( {\exp \left( {\sqrt {x^2+a^2}} \right)-1} \right)}
\right]^{-1} \nonumber
\ed
\veb
\bd
= {1 \over 2}\left(
\frac{a}{\pi ^2} \right)\sum \limits_{n=1}^\infty  \frac{1}{n}K_1(an).\label{bb9}
\ed 
Scaling $x$ and $(a)$ with a parameter $\beta $ as $(x,a)=\beta
(k,m)$    write (\ref{bb9}) in the form
\bd
\int {{{d^3k} \over {\left( {2\pi } \right)^3}}\,{2 \over {\varepsilon
\left( {\exp
\left( {\beta \varepsilon } \right)-1} \right)}}}={m \over {\beta \pi
^2}}\sum\limits_{n=1}^\infty  \frac{1}{n}K_1(\beta mn),\label{bb10}
\ed 
where $\varepsilon =\sqrt {\vec{k}^2+m^2}$. 

Differentiating the equation  (\ref{bb7}) with respect to parameter $a$ we find 
\bd
\int {{{d^3x} \over {\left( {2\pi } \right)^3}}\left(
{-{\partial 
\over {\partial a^2}}} \right)}\left[ {\sqrt {x^2+a^2}\left( {\exp \left( {\sqrt
{x^2+a^2}}
\right)-1} \right)} \right]^{-1}\hfil \nonumber
\ed
\veb
\bd
\hfil = {1 \over {4\pi ^2}}\sum\limits_{n=1}^\infty 
{K_0}(an),\label{bb11} 
\ed 
or, in new variables,
\bd
\int {{{d^3k} \over {\left( {2\pi } \right)^3}}\left( {{\partial 
\over {\partial m^2}}} \right){1 \over {\varepsilon \left( {\exp \left( {\beta
\varepsilon }
\right)-1} \right)}}}=-{1 \over {4\pi ^2}}\sum\limits_{n=1}^\infty 
{K_0}(\beta mn).\label{bb12}
\ed 
Integrating  (\ref{bb7}) with respect to parameter $(a)$ and
using the equation 
\bd
\int\limits_{a^2}^\infty  {da^2\left[ {\sqrt {x^2+a^2}\left( {\exp
\left( {\sqrt {x^2+a^2}} \right)-1} \right)} \right]^{-1}} \nonumber
\ed
\veb
\bd
=2\ln \left(
{1-\exp\left(-
\sqrt {x^2+a^2}\right)}
\right) \label{bb13}
\ed  
we find with  new variables the following equation 
\bd
-{1 \over \beta }\int {{{d^3k} \over {\left( {2\pi } \right)^3}}}\ln
\left( {1-\exp (-\beta \varepsilon )} \right)={{m^2} \over {2(\beta \pi
)^2}}\sum\limits_{n=1}^\infty  \frac{1}{n^2}K_2(\beta mn).\label{bb14}
\ed 
High temperature asymptotes $(\beta{m}\ll{1})$ of the equations (\ref{bb9}),
 (\ref{bb11}) and (\ref{bb14}) are
\bd
{{2m^2} \over {(\beta \pi )^2}}\sum\limits_{n=1}^\infty  
\frac{1}{n^2}K_2(\beta mn) \nonumber
\ed
\veb
\bd
=\frac{\pi ^2}{90 \beta ^4}-\frac{m^3}{24 \beta^2}
+\frac{m^3}{12 \pi \beta}+\frac{m^4}{64}\left[ \ln \frac{m^2 \beta^2}{16 \pi^2}
-\frac{3}{2}+2\gamma\right] \nonumber
\ed
\veb
\bd
{m \over {4\beta \pi ^2}}\sum\limits_{n=1}^\infty  
\frac{1}{n}K_1 (\beta mn), \label{bb15}
\ed
\veb
\bd
=\frac{1}{24\beta^2}-\frac{m}{8 \pi \beta}-\frac{m^2}{32 \pi^2}
\left[ \ln \frac{m^2 \beta^2}{16 \pi ^2}-1+2\gamma \right] \label{bb16}
\ed
and
\bd
{1 \over {4\pi ^2}}\sum\limits_{n=1}^\infty 
K_0(\beta mn) \nonumber
\ed
\veb
\bd
=\frac{1}{16 \pi m \beta}+\frac{1}{32 \pi^2}
 \left[ \ln \frac{m^2 \beta^2}{16 \pi ^2}-\frac{1}{2}+2\gamma \right] \label{bb17}
\ed

%% file: ikpic1.tex
\vspace*{8mm}
\begin{center}
{\Large \bf GRAPHICS}
\end{center}

\vspace*{10mm}
\begin{picture}(7,6)

\put(3,3){\line(1,0){3}}
\put(6,1){\line(0,1){3}}
\put(3,1){\vector(0,1){4}}
\put(3,1){\vector(1,0){4}}
\put(3.02,1){\vector(0,1){4}}
\put(3,1.02){\vector(1,0){4}}

\put(6,0.5){\makebox(0,0){$1$}}
\put(3,0.5){\makebox(0,0){$0$}}
\put(2.3,3){\makebox(0,0){$g_{3/2}(1)$}}
\put(3,5.5){\makebox(0,0){$g_{3/2}(z,R)$}}
\put(7,3.5){\makebox(0,0){$R<0$}}
\put(7,2.5){\makebox(0,0){$R>0$}}
\put(7,3){\makebox(0,0){$R=0$}}
\put(7.5,1){\makebox(0,0){$z(R)$}}

\put(2.3,3.5){\makebox(0,0){$g_{3/2}(1,R)$}}
\put(3,3.5){\line(1,0){3}}
\put(2.3,2.5){\makebox(0,0){$g_{3/2}(1,R)$}}
\put(3,2.5){\line(1,0){3}}

\end{picture}
\vspace*{5mm}
\begin{center}
Fig. I-1  Graphical expression of the function $g_{3/2}(z,R)$
\end{center}

\newpage
\vspace*{20mm}
\begin{picture}(9,7)

\put(3,2.3){\line(1,0){2.8}}
\put(6.5,1){\line(0,1){5}}
\put(3,1){\vector(0,1){5}}
\put(3,1){\vector(1,0){5}}
\put(3.02,1){\vector(0,1){5}}
\put(3,1.02){\vector(1,0){5}}

\put(4.8,1){\line(0,1){1.3}}
\put(5.3,1){\line(0,1){1.3}}
\put(5.8,1){\line(0,1){1.3}}
\put(3,5){\line(1,0){3.5}}
\put(2.9,6.5){\makebox(0,0){$g_{3/2}(z,R)+\lambda^3 n_0$}}
\put(2.3,5){\makebox(0,0){$2.612$}}

\put(1.7,5.5){\makebox(0,0){$g_{3/2}(1,R)+\lambda^3 n_0$}}
\put(3,5.5){\line(1,0){3.5}}
\put(1.7,4.5){\makebox(0,0){$g_{3/2}(1,R)+\lambda^3 n_0$}}
\put(3,4.5){\line(1,0){3.5}}

\put(6.6,0.5){\makebox(0,0){$1$}}
\put(3,0.5){\makebox(0,0){$0$}}
\put(8.5,1){\makebox(0,0){$z(R)$}}
\put(4.8,0.5){\makebox(0,0){$z_1$}}
\put(5.3,0.5){\makebox(0,0){$z_0$}}
\put(5.8,0.5){\makebox(0,0){$z_2$}}

\put(7,5.5){\makebox(0,0){$R<0$}}
\put(7,4.5){\makebox(0,0){$R>0$}}
\put(7,5){\makebox(0,0){$R=0$}}

\put(6.5,5.5){\line(-1,0){0.1}}
\put(6.5,4.5){\line(-1,0){0.1}}

\put(2.3,2.3){\makebox(0,0){$\lambda^3 n$}}

\end{picture}
\vspace*{5mm}
\begin{center}
Fig.I-2  Graphical solution for bosons.\\
Solution of the equation (\ref{bos19}) for different curvatures 
and fixed temperature 
and density $z_1$ for $R<0$,
$z_0$ for $R=0$ and $z_2$ for $R>0$
\end{center}

\newpage
\vspace*{30mm}
\begin{picture}(8,6)

\put(2.5,1){\vector(0,1){4}}
\put(2.5,3){\vector(1,0){6}}
\put(2.52,1){\vector(0,1){4}}
\put(2.5,3.02){\vector(1,0){6}}

\put(5,3){\line(0,-1){0.15}}
\put(6,3){\line(0,-1){0.15}}
\put(7,3){\line(0,-1){0.15}}

\put(2.2,3){\makebox(0,0){$0$}}

\put(9.5,3){\makebox(0,0){$(\lambda^3 n)^{-1}$}}

\put(2.5,5.5){\makebox(0,0){$\mu_{eff}(R)$}}
\put(5,3.5){\makebox(0,0){$T_c^{'}$}}
\put(6,3.5){\makebox(0,0){$T_c$}}
\put(7,3.5){\makebox(0,0){$T_c^{''}$}}

\put(5.8,1.3){\makebox(0,0){$R<0$}}
\put(7,1.3){\makebox(0,0){$R=0$}}
\put(8,1.3){\makebox(0,0){$R>0$}}

\end{picture}

\vspace*{5mm}

\begin{center}
Fig.I- 3 Chemical potential $\mu _{eff}(R)$ as a 
functional of a curvature of space-time.
\end{center}

\newpage
\vspace*{20mm}
\begin{picture}(7,6)

\put(3,1){\vector(0,1){4}}
\put(3,1){\vector(1,0){5}}
\put(3.02,1){\vector(0,1){4}}
\put(3,1.02){\vector(1,0){5}}

\put(4,1){\line(0,1){2}}
\put(5,1){\line(0,1){2}}
\put(6,1){\line(0,1){2}}
\put(3,3){\line(1,0){3}}

\put(3,0.5){\makebox(0,0){$0$}}
\put(8.5,1){\makebox(0,0){$z(R)$}}
\put(4,0.5){\makebox(0,0){$z_1$}}
\put(5,0.5){\makebox(0,0){$z_0$}}
\put(6,0.5){\makebox(0,0){$z_2$}}

\put(3,5.5){\makebox(0,0){$f_{3/2}(z,R)$}}
\put(4.5,1.6){\makebox(0,0){$R<0$}}
\put(5.5,1.6){\makebox(0,0){$R=0$}}
\put(6.5,1.6){\makebox(0,0){$R>0$}}
\put(2.2,3){\makebox(0,0){$(n/s)\lambda^3(T)$}}

\end{picture}

\vspace*{5mm}
\begin{center}
Fig.I-4  Graphical solution for fermions.\\
Solution of the equation (\ref{tf11})
for different curvatures: $z_1$ for $R>0$,
$z_0$ for $R=0$ and $z_2$ for $R<0$
\end{center}

\newpage
\vspace*{20mm}
\hspace{2.8cm}$R_1\times R_1\times S_1$\hspace{2cm}$R_1\times Mobius~strip$

\begin{picture}(6,5.5)
\put(2,1){\line(1,0){2}}
\put(2,1){\line(0,1){4}}
\put(2,5){\line(1,0){2}}
\put(4,1){\line(0,1){4}}
\put(2,2){\vector(1,0){2}}
\put(4,2){\vector(-1,0){2}}
\put(2,4){\vector(1,0){2}}
\put(4,4){\vector(-1,0){2}}

\put(3,0.5){\makebox(0,0){$L$}}

\put(6,1){\line(1,0){2}}
\put(6,1){\line(0,1){4}}
\put(6,5){\line(1,0){2}}
\put(8,1){\line(0,1){4}}
\put(6,1){\vector(1,2){2}}
\put(6,5){\vector(1,-2){2}}
\put(6,3){\vector(1,0){2}}
\put(7,0.5){\makebox(0,0){$L$}}
\end{picture}

\vspace*{5mm}
Fig. III-1 Topologies of cylinder and Mobius strip in $(y,z)$
spaces.\\
 Identification of the points
$\psi(x,y,0)=\psi(x,y,L)$ and $\psi(x,y,0)=-\psi(x,y,L)$
\vspace{1cm}

\newpage
\vspace*{20mm}
\hspace{2.5cm}$R_1\times R_1\times S_1$\hspace{2.5cm}$R_1\times Mobius~strip$

\begin{picture}(6,5.5)
\put(2,1){\line(1,0){2}}
\put(2,1){\line(0,1){4}}
\put(2,5){\line(1,0){2}}
\put(4,1){\line(0,1){4}}
\put(2,2){\vector(1,0){2}}
\put(4,2){\vector(-1,0){2}}
\put(2,4){\vector(1,0){2}}
\put(4,4){\vector(-1,0){2}}
\put(3,1){\vector(0,1){4}}
\put(3,5){\vector(0,-1){4}}
\put(3,0.5){\makebox(0,0){$L^{'}$}}
\put(1.5,3){\makebox(0,0){$L$}}
\put(6,1){\line(1,0){2}}
\put(6,1){\line(0,1){4}}
\put(6,5){\line(1,0){2}}
\put(8,1){\line(0,1){4}}
\put(6,1){\vector(1,2){2}}
\put(6,5){\vector(1,-2){2}}
\put(6,3){\vector(1,0){2}}
\put(7,1){\vector(0,1){4}}
\put(7,5){\vector(0,-1){4}}
\put(5.5,3){\makebox(0,0){$L$}}
\put(7,0.5){\makebox(0,0){$L^{'}$}}
\end{picture}

\vspace*{5mm}

Fig. III-2 Topologies of torus and Klein bottle in $(y,z)$
spaces.\\
 Identification of the points
$\psi(x,,0)=\psi(x,L,L')$ and $\psi(x,0,0)=-\psi(x,L,L')$